\documentclass{aa}  
\usepackage{graphicx}
\usepackage{xcolor}
\usepackage{rotating}
\usepackage{tikz}
\usepackage{txfonts}
\usepackage{supertabular}
\usepackage{longtable}
\usepackage{tabularx}
\usepackage{lscape}
\usepackage{soul}
\usepackage{appendix}
\usepackage{float}
\usepackage{natbib}
\usepackage{hyperref}
\bibpunct{(}{)}{;}{a}{}{,} % to follow the A&A style
\makeatletter
\newcommand\footnoteref[1]{\protected@xdef\@thefnmark{\ref{#1}}\@footnotemark}
\makeatother
\usepackage[para]{footmisc}
\begin{document}

   \title{Bird's eye view of molecular clouds in the Milky Way}
   \subtitle{I. Column density and star formation from sub-parsec to kiloparsec scales}
   \author{Andri Spilker
          \inst{1}
          \and
          Jouni Kainulainen
          \inst{1}
          \and
          Jan Orkisz
          \inst{1}
          }

   \institute{Chalmers University of Technology, Department of Space, Earth and Environment, SE-412 93 Gothenburg, Sweden\\
              \email{andri.spilker@chalmers.se}
             }
   \date{September 2, 2021} %Received ----; accepted -----}

% \abstract{}{}{}{}{} 
% 5 {} token are mandatory
  \abstract
  {Describing how the properties of the interstellar medium are combined across various size scales is crucial for understanding star formation scaling laws and connecting Galactic and extragalactic data of molecular clouds.}
  % aims heading (mandatory)
  {We describe how the statistical structure of the clouds and its connection to star formation changes from sub-parsec to kiloparsec scales in a complete region within the Milky Way disk.}
  % methods heading (mandatory)
  {We built a census of molecular clouds within 2 kpc from the Sun using data from the literature. We examined the dust-based column density probability distributions (N-PDFs) of the clouds and their relation to star formation as traced by young stellar objects (YSOs). We then examined our survey region from the outside, within apertures of varying sizes, and describe how the N-PDFs and their relation to star formation changes with the size scale.}
  % results heading (mandatory)
  {We present a census of the molecular clouds within 2 kpc distance, including 72 clouds and YSO counts for 44 of them. The N-PDFs of the clouds are not well described by any single simple model; use of any single model may bias the interpretation of the N-PDFs. The top-heaviness of the N-PDFs correlates with star formation activity, and the correlation changes with Galactic environment (spiral- and inter-arm regions). We find that the density contrast of clouds may be more intimately linked to star formation than the dense gas mass fraction. The aperture-averaged N-PDFs vary with the size scale and are more top-heavy for larger apertures. The top-heaviness of the aperture N-PDFs correlates with star formation activity up to roughly 0.5 kpc, depending on the environment. Our results suggest that the relations between cloud structure and star formation are environment specific and best captured by relative quantities (e.g. the density contrast). Finally, we show that the density structures of individual clouds give rise to a kiloparsec-scale Kennicutt-Schmidt relation as a combination of sampling effects and blending of different galactic environments. }
  {}

    \maketitle
%
%________________________________________________________________

\section{Introduction}

Star formation rates, efficiencies, and timescales are linked to the complex distribution and energetics of gas in galaxies. This complex distribution manifests itself over a wide range of size scales, from galaxy scales down to sub-parsec (sub-pc) scales. Studies of external galaxies can probe the distribution of gas at galactic scales (kiloparsec; kpc), and today, even down to the individual molecular clouds \citep[tens of pc; e.g. the PHANGS survey:][see also \citealt{faesi2018}]{hughes2013probability,leroy2016portrait,sun2018cloud,sun2020molecular,leroy2021phangs}. In the Large Magellanic Cloud, the internal structure has even been studied down to sub-pc scales \citep{sawada2018internal}. Except for these few exceptions, the sub-pc internal structure of molecular clouds can only be studied for the clouds in our own galaxy, through dust emission, dust extinction and molecular line data (e.g. \citealt{goldsmith2008taurus,kainulainen2009probing,lada2010star,heiderman2010star,schneider2013pdfs,kainulainen2014unfolding,stutz2015evolution,lane2016orionA,pety2017orionb}). It is therefore difficult to reconcile how the properties of gas at galactic (kpc) scales connect with those within clouds (sub-pc). At the heart of this problem is the behaviour of the density structure and gas energetics as a function of spatial scale, which gives rise to canonical relations such as the Kennicutt-Schmidt (KS) relation \citep{kennicutt1998global,schmidt1959rate}. Decoding the information carried by the scale-dependence holds one key to understanding the regulation of star formation in galaxies \citep[e.g.][]{elmegreen2002star,kravtsov2003origin, kruijssen2014uncertainty,leroy2016portrait}.

In order to study the scale dependence of the gas distribution, we need a tool to describe the distribution. One simple tool for the purpose is provided by the column density probability distribution functions (N-PDFs). The N-PDFs can be tied to analytical theories of star formation because their shape depends on the physical processes in the clouds and on their evolution \citep[reviewed in][ see also e.g. \citealt{burkhart2019self}]{hennebelle2012turbulent,padoan2014ppvi}. Simulations of molecular clouds that are dominated by supersonic turbulence and are not significantly affected by gravity predict log-normal N-PDFs, while self-gravitating gas is expected to have N-PDFs with power-law shapes at high densities \citep[\citealt{federrath2013sfe}; see also e.g.][]{klessen2000gravitational,federrath2008density,ballesteros2009gravitational,kritsuk2011comparing}. These are crucial aspects in the theory of star formation and highlight the utility of N-PDFs as a tool for quantifying the internal structure of clouds and relate it to star formation.  

\begin{figure}
   \centering
        \includegraphics[width=0.49\textwidth]{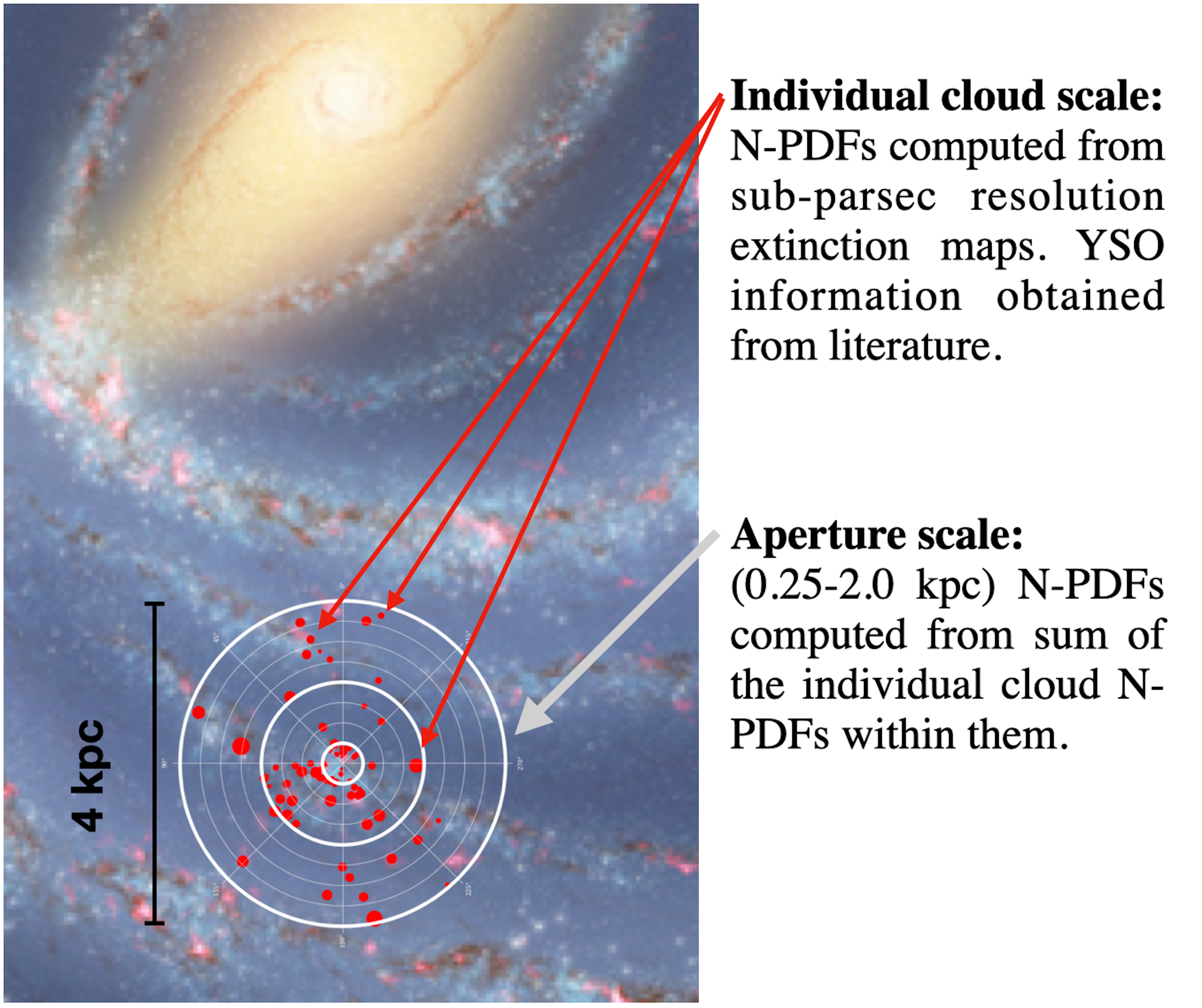}
    \caption{Illustration of the survey area from a bird's eye perspective, and of the two ways in which we analysed our cloud sample. The picture shows the molecular cloud sample (filled red circles; the size of the circles scales with the physical size of the clouds; see also Fig. \ref{fig:Damexy}). The empty white circles represent apertures of various sizes. Background image credit: NASA/JPL-Caltech/R. Hurt (SSC/Caltech).}
    \label{fig:bird}
\end{figure}

N-PDFs have so far mostly been studied using small samples or individual clouds in the solar neighbourhood \citep{kainulainen2009probing,kainulainen2014unfolding,schneider2013pdfs,Alvesdeoliveira2014chamaeleon,lombardi2015molecular}, through various approaches of probing gas farther in the galactic disk \citep{kainulainen2013high, abreuvicente2015, csengeri2016}, or by covering large portions of external galaxy disks \citep{hughes2013probability}. The most common shape of N-PDFs within the Milky Way is debated. Previous works have established a habit of describing them with log-normal functions, power laws, or a combination of the two \citep[e.g.][]{kainulainen2009probing,schneider2013pdfs,kainulainen2013connection}. Some studies have argued that all N-PDFs are power laws \citep[i.e.][]{lombardi2015molecular}, while others argued that when field selection is tightly constrained to the cold molecular zone, N-PDFs are best described by log-normals \citep[][]{brunt2015power}. Outside the Milky Way, \citet{hughes2013probability} reported that the CO N-PDFs of kpc regions of M51, M33 and the LMC are best described by log-normals. Extragalactic studies cover entire galaxies (several kpc), while Galactic studies so far only cover an area of some hundred pc; only the solar neighbourhood closer than $\sim$250 pc has been studied in a complete manner so far \citep{kainulainen2009probing,kainulainen2014unfolding}. As a result, it is not yet possible to understand the connection between the statistics describing the internal cloud structure and those describing the galactic-scale gas distribution. To tie them together, an overlap in the scales between the Galactic and extragalactic works is needed. 

In this work, we aim to bridge part of the gap between Galactic and extragalactic observations of molecular cloud structure and star formation. We use sub-pc resolution observations available for nearby molecular clouds to describe the internal structure of the dense interstellar medium (ISM) from the sub-pc to kpc scales. To do this, we assembled a census that is as complete as possible of the molecular clouds in a portion of our own Galaxy. This portion is a 2 kpc radius circle of the Milky Way disk, centred on the Sun. This is the first time that the molecular cloud density structure and star formation in such a large region of the Milky Way has been studied in a complete manner. The range of spatial scales probed by our survey significantly overlaps with ISM studies of the nearby galaxies. In addition, it probes the sub-pc scales that are not accessible outside the Milky Way. This enables us to describe how the statistical structure of dense ISM and its connection to star formation changes with size scale. This represents a step forward in understanding what substructure might be present within the beam or resolution element of extragalactic observations and how this substructure connects with the larger-scale statistics of the galaxy gas reservoirs.

\section{Data and methods}
\label{sec:method}

Our goal is to analyse the column density statistics and star formation rates of molecular clouds in a complete volume of our 2 kpc survey area. For this, we need 1) to compile a census of molecular clouds that is as complete as possible, and 2) obtain column density and star formation data for these clouds from literature. After it was collected, the sample of clouds was examined in two different ways. First we studied the sample as individual clouds, and then from a bird’s-eye perspective through apertures of various sizes (illustrated in Fig. \ref{fig:bird}).

To achieve this, we gather a sample of clouds (Sect. \ref{sec:msample}) and search for star formation metrics for them (Sect. \ref{sec:data_yso}). Then we obtain column density maps covering the clouds (Sect. \ref{sec:mext}) and use them to derive N-PDFs (Sect. \ref{sec:mpdf}). These N-PDFs are fitted (Sect. \ref{sec:mfit}) and analysed individually. After this, we use our census to simulate a face-on view of the Galactic disk (Sect. \ref{sec:mbird}).

%\onecolumn
\begin{figure*}%[h!]sidewaysfigure}%}%
    \centering
    \includegraphics[width=\textwidth]{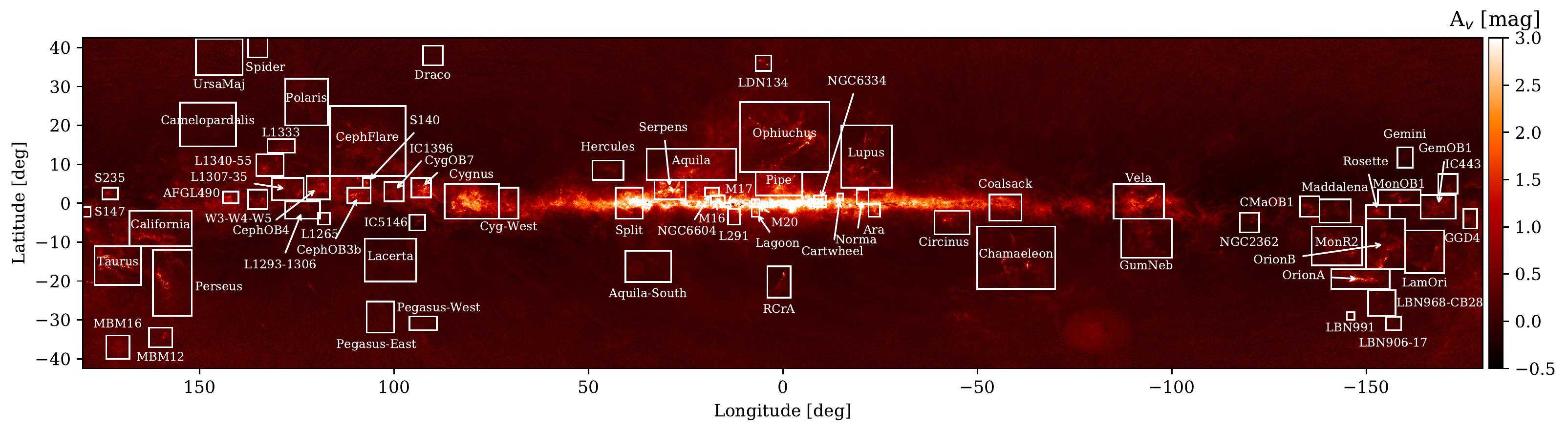}
    \caption{Census of the 72 molecular clouds within 2 kpc from us shown in Galactic coordinates, superimposed on the NICEST extinction map of the Milky Way from \citet{juvela2016allsky}.}
    \label{fig:Dameclouds}
\end{figure*}%sidewaysfigure}%

\begin{figure}%[h!]
   \centering
        \includegraphics[width=0.49\textwidth]{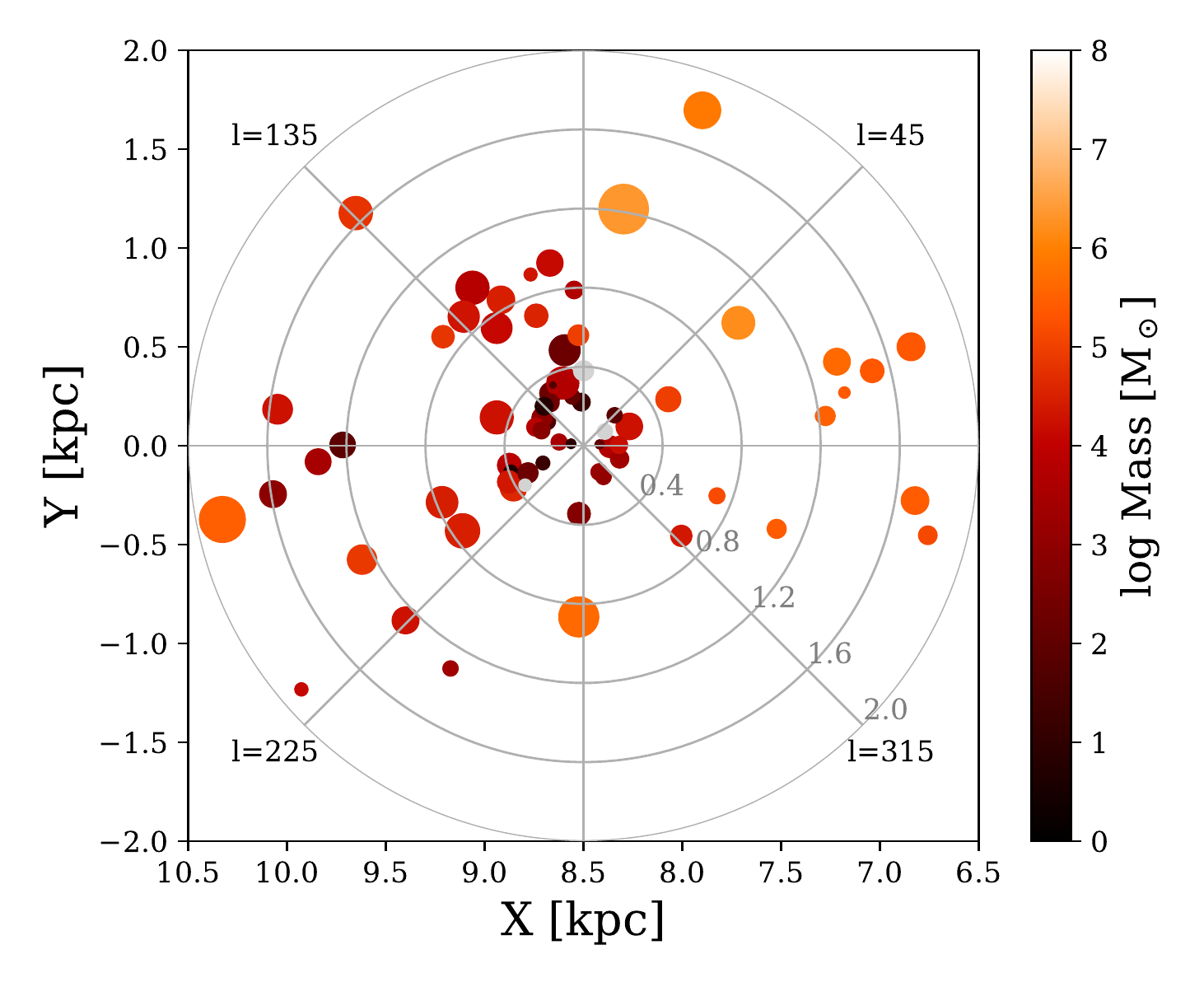} 
    \caption{Molecular cloud sample of the survey area seen from a bird's eye perspective. The figure is centred on the Sun, and the Galactic centre is at (0,0). The colour corresponds to the molecular mass of the cloud, and the size corresponds to the physical size of the cloud. Clouds with a maximum extinction below 3 mag are marked in grey. The thin grey circles are spaced by 0.4 kpc and are examples of what solar centred apertures would contain. The angles marked with $l$ mark the Galactic longitude.}
    \label{fig:Damexy}
\end{figure}

\subsection{Molecular cloud sample}
\label{sec:msample}
The sample of molecular clouds was compiled by searching the literature surveys of extinction and CO in the Galaxy. We aimed to include all major molecular structures ($M \gtrsim 10^4$ M$_\odot$) within 2 kpc from the Sun. We also included the lower-mass clouds found in literature. Our search procedure consisted of three steps. First, we included all clouds that we were able to identify in the \citet{dame2001milky} CO survey of the Milky Way. This survey is a composite of 37 separate regions that are described and numbered in Table 1 of their paper, and many of these regions appear as clouds in our sample. We chose to include as clouds those regions that were coherent in position and velocity at a distance $<$ 2 kpc. 

Second, we added sightlines from \citet{zucker2020compendium,zucker2019large}, who inferred distances and extinction for a large sample of nearby molecular clouds using stellar photometric data and Gaia DR2 parallax measurements. Many of these sightlines point towards small cores or \textsc{Hii} regions. In these cases, we included the dense structure surrounding the sightline into our cloud and adopted the naming of \citet{zucker2020compendium} for the entire structure. When sightlines were close together with similar distances, we combined them into one cloud. In these cases, we chose the name of the dominating structure in mass or combined the names. 

As a third step, we compared our cloud sample to recent 3D dust maps by \citet{kh2018detection}, \citet{lallement2018three,lallement2019gaia}, and \citet{green20193d} to search for further missing clouds. For the most part, our sample covers the same structures as seen in the 3D dust maps. However, for two structures prominent in \citet{green20193d}, we were unable to find a corresponding cloud in our sample, even though the structures were coherent in extinction and CO emission. We therefore added them to our sample. These were the Cygnus-West cloud and the Split. The Split was discovered and named by \citet{lallement2019gaia}. 
%\end{enumerate}

With this procedure, we assembled a sample of 72 clouds. For this sample, we obtained dust-based column density maps and distances. We adopted the newly derived distances from \citet{zucker2020compendium} for all the clouds for which they are available (80\% of the sample), and for the remaining clouds, we used the most recent distance estimate in the literature. The location of the molecular clouds in Galactic coordinates is shown in Fig. \ref{fig:Dameclouds}, and Table \ref{tab:mastertable} lists the cloud names, their properties, and references. The location of the clouds is shown in Fig. \ref{fig:Damexy} in a face-on view.

%------------------------------------------
\subsection{Star formation measures}
\label{sec:data_yso}
%------------------------------------------

We searched the literature for studies of young stellar objects (YSOs) in the clouds and found YSO counts for 44 of our 72 clouds. Most of the YSO counts come from dedicated observations in the near- and mid-infrared (NIR and MIR, see references in Table \ref{tab:mastertable}). We note that the YSO data set is not homogeneous; the mass completeness is higher for nearby clouds than for the more distant ones. We discuss this issue and its possible effects in Sect. \ref{sec:discuss_inhomo}. With the YSO counts in hand, we computed the star formation rate (SFR) as 
\begin{equation}
    \textrm{SFR}=\frac{M_{\mathrm{YSOs}}}{t_{\mathrm{YSO}}}=\frac{N_{\mathrm{YSOs}} \cdot 0.5\mathrm{M}_\odot}{2\cdot10^6 \ \textrm{yr}},
\end{equation}
where $N_{\mathrm{YSOs}}$ is the number of YSOs in the cloud, 0.5M$_\odot$ is the mean mass of YSOs \citep{chabrier2003review}, and $t_{\mathrm{YSO}}=2\cdot10^6$ yr is the mean lifetime of YSOs. We note that the mean mass of YSOs may vary between star-forming regions, and the lifetime is uncertain by a factor of 2 \citep{padoan2014ppvi}. The star formation efficiency (SFE) is computed as 
\begin{equation}
    \textrm{SFE}= \frac{M_{\mathrm{YSOs}}}{M_{\mathrm{YSOs}}+M_{A_\mathrm{V}>1\textrm{mag}}},
\end{equation}
where $M_{A_\mathrm{V}>1\textrm{mag}}$ is the mass of the cloud above 1 mag. Gas is expected to be molecular above 1-3 mag \citep{heyer2015molecular}, and we here used the lower limit to include as much gas as possible.

\begin{figure*}
    \centering
    \begin{minipage}[b]{0.9\textwidth}
        \includegraphics[width=\textwidth]{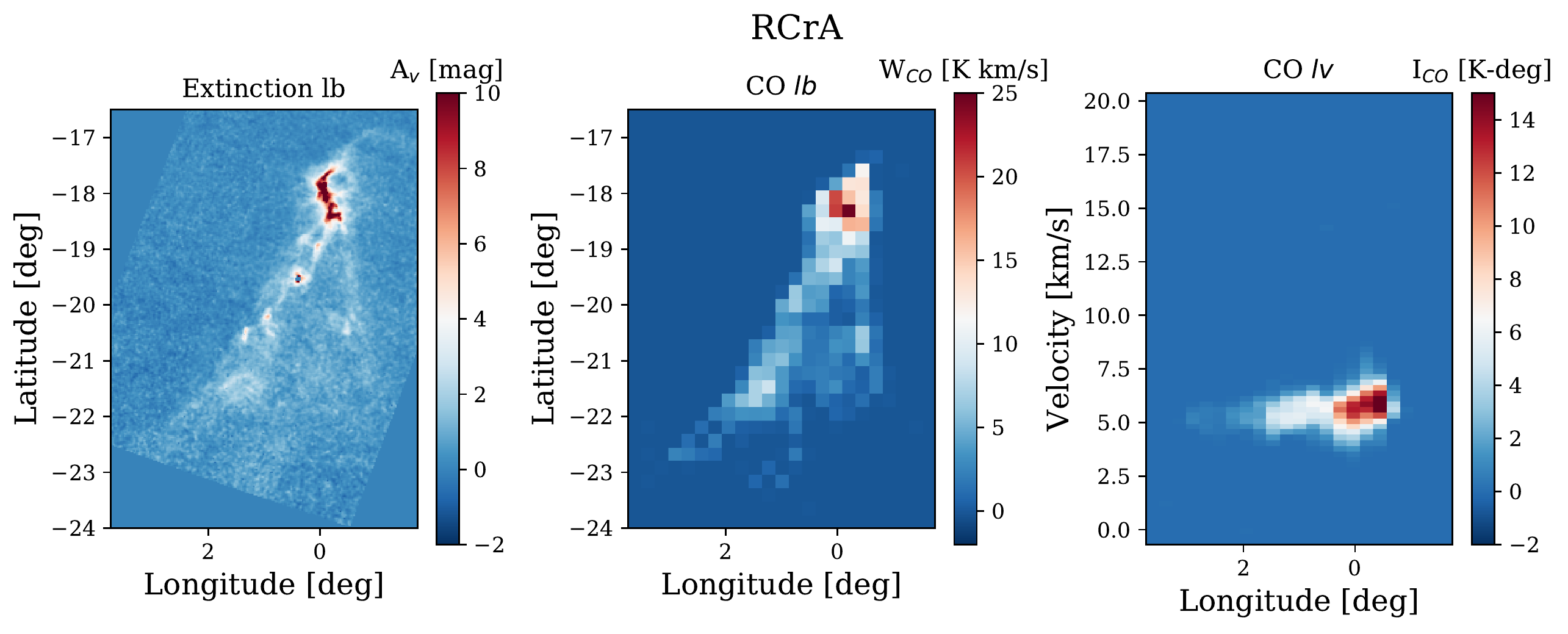} %png}
    \end{minipage}
    \begin{minipage}[b]{0.9\textwidth}
        \includegraphics[width=\textwidth]{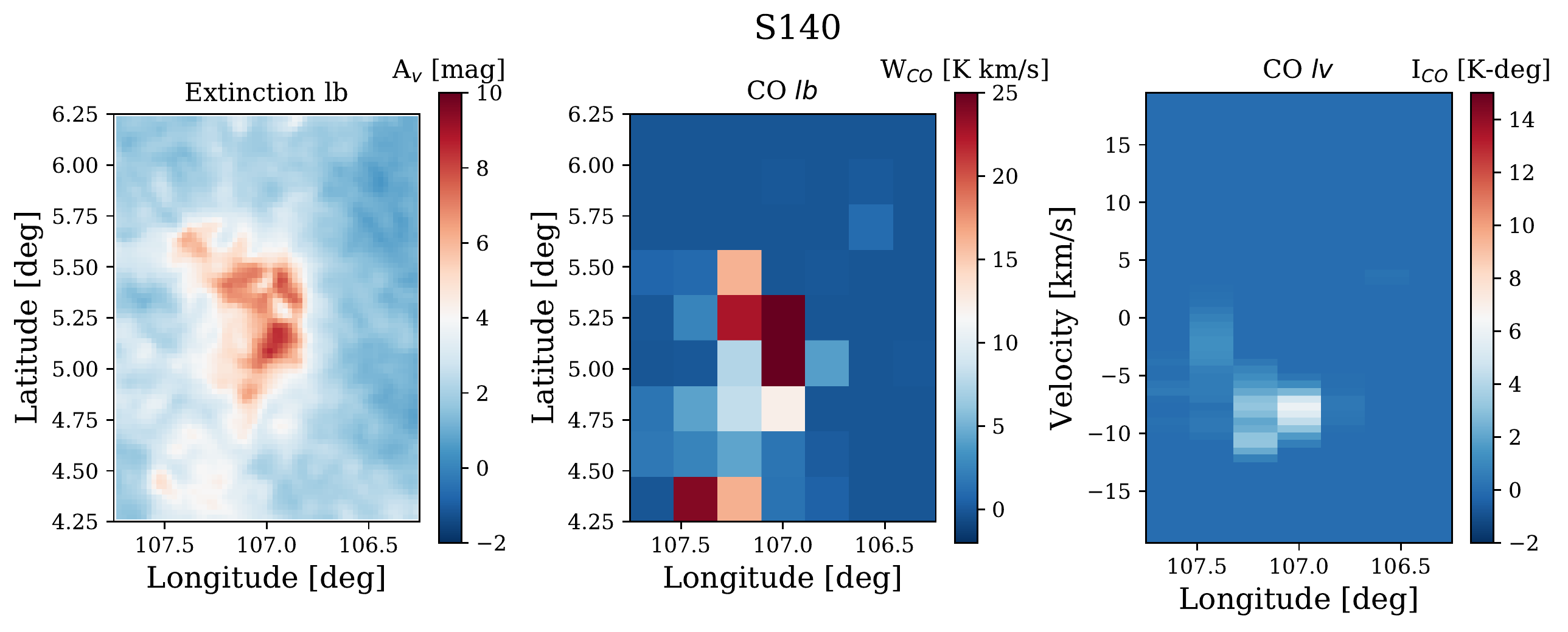} %png}
    \end{minipage}
    \begin{minipage}[b]{0.9\textwidth}
        \includegraphics[width=\textwidth]{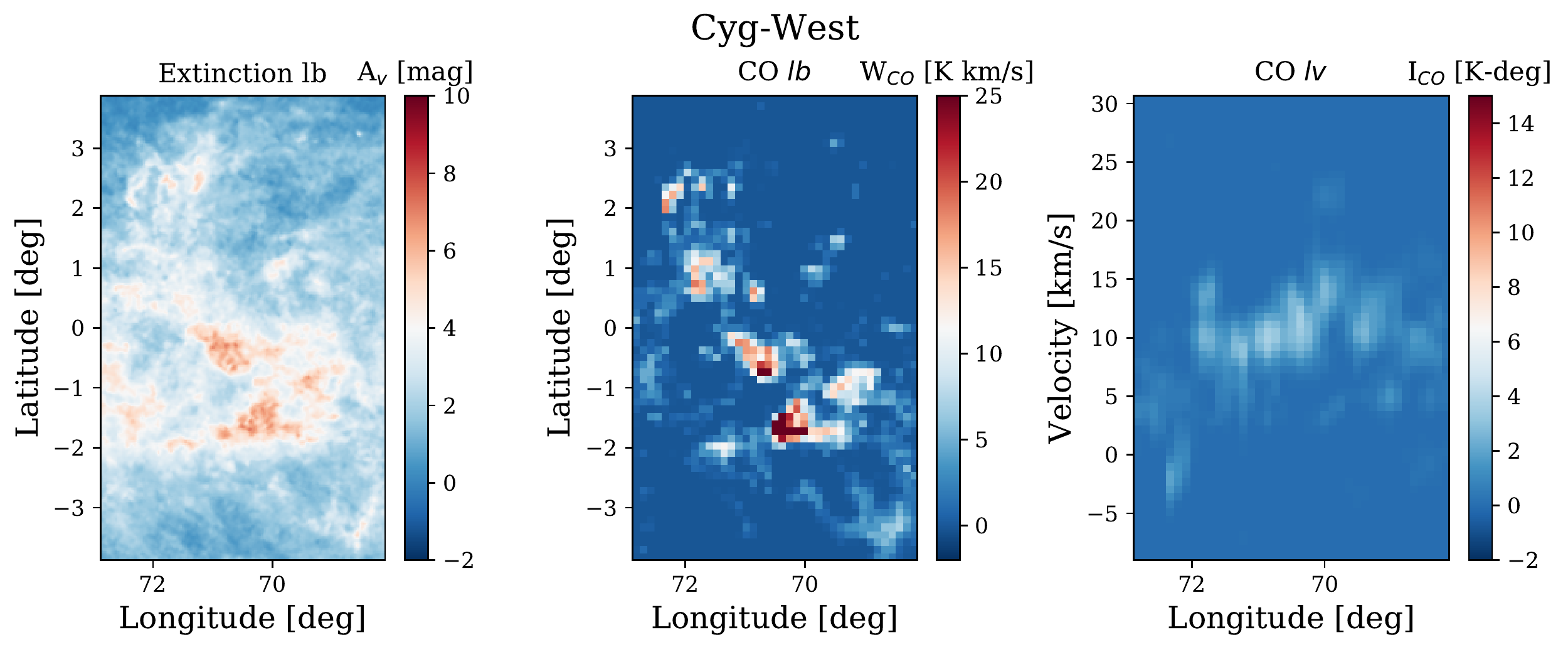} %png}%Cartwheel_4pan.png}
    \end{minipage}
    \caption{Three of the molecular clouds in our sample, seen in extinction and in CO position and velocity. CO is taken from \citet{dame2001milky} survey, and the extinction is adopted from \citet{kainulainen2009probing,kainulainen2014unfolding} for RCrA and from \citet{juvela2016allsky} for S140 and Cyg-West. Extinction maps for all the clouds are shown in Fig. \ref{fig:cloudext}-\ref{fig:cloudext4}.
    }
    \label{fig:exampleclouds}
\end{figure*}

\subsection{Column density maps}
\label{sec:mext}

We obtained column density maps for the clouds in our sample mostly from the literature. For the majority of the clouds, we exploited column density maps based on dust extinction, and therefore used $A_\mathrm{V}$ as the basic unit in our analyses. When necessary, we used the conversion between gas column density and extinction from \citet{guver2009relation},
\begin{equation}
    N_\mathrm{H} \textrm{ [cm}^{-2}] = (2.21 \pm 0.09) \times 10^{21} A_\mathrm{V} \textrm{ [mag]}.
    \label{eq:conversion}
\end{equation}

The masses were computed from the extinction maps by multiplying each pixel value reaching an extinction of 3 mag with the conversion factor and the physical size of the pixel ($M_{A_\mathrm{V}>3 \mathrm{mag}} = \Sigma_i A_\mathrm{V}(p_i[A_\mathrm{V}>3 \mathrm{mag}]) \cdot \tan(\Delta l)\cdot \tan(\Delta b) \cdot $ distance$^2 \cdot $ conversion). We chose the limit of 3 mag to be above the limit of the last closed contour for most clouds. In practice, three different types of column density maps were used to cover the clouds in our sample. This was done to use the best maps available for each individual cloud. 

%\begin{enumerate}
    %\item 
For most clouds in our sample, we used extinction maps derived using the NICEST near-infrared dust extinction mapping technique \citep{lombardi2009nicest} and data from 2MASS \citep{skrutskie2006two}. For most nearby clouds within $\sim$0.7 kpc distance, these maps are available from our earlier works \citep{kainulainen2009probing,kainulainen2014unfolding, kainulainen2017relationship,bieging2018}. New maps were adopted for clouds for which those studies provide no data, either by applying NICEST or by cropping a map from \citet{juvela2016allsky} (wich also uses NICEST). \citet{juvela2016allsky} used a background colour determination that is well suited for automated full-sky work and no dedicated foreground star removal process; in case of some individual clouds, we considered that adjustments in these result in more accurate maps, and we therefore used our own implementation of NICEST for these clouds. In summary, the dominant set of column density maps (66 out of 72) consists of NICEST extinction maps for all clouds for which we found it reasonable to use these data.

For four clouds, no adequate dust extinction maps were available (NGC6334, M16, M17, and M20), and we instead used column density maps based on dust emission derived from \textit{Herschel} data by \citet{marsh2015temperature,marsh2017multitemperature} using the PPMAP technique. We did not use extinction maps for these clouds because the clouds are in the Galactic plane, and the standard extinction mapping techniques perform very poorly because of foreground stars and the diffuse dust component along the line of sight. The PPMAP data describe the total column density derived from the \textit{Herschel} data. We performed a subtraction of a diffuse component from the data. The estimate for the diffuse component was derived by choosing relatively cloud-free columns from the column density data over the entire PPMAP data coverage, creating smoothed vertical profiles for these columns, and then performing a spline-interpolation in the horizontal direction between these columns. The resulting diffuse dust component varies roughly between a few to $\sim$15 mag between the Galactic latitudes of $b=[-1,+1]$. 
    
For Cartwheel and Cygnus, we used the extinction maps derived in Appendix \ref{app:cygnus_and_cartwheel}. These clouds are close to the Galactic plane, but are unfortunately not covered by the \emph{Herschel} data. Our new extinction maps for these clouds implement a method to account for foreground stars and diffuse dust component, but we acknowledge that their quality is poorer than the extinction maps of the nearby clouds for which NICEST maps are available.
%\end{enumerate}

It is important to recognise that different column density mapping techniques differ in how they treat the extinction or emission contribution from the diffuse dust component in the Galactic plane. The reliability of the techniques can therefore vary significantly in different regions of the sky. For example, NICEST is highly reliable at high Galactic latitudes, but fails for distant clouds in the plane because of foreground sources and overlap of dust structures. In our approach, we tried to alleviate these problems by using the maps that were most suitable for each region, even if this meant that our data set was composed of maps derived with different techniques. Using different techniques should be a smaller problem than using clearly erroneous maps for some regions. We also note that currently, no single technique or data is available that could be used reliably in all conditions of our survey area. We argue that the use of several types of column density maps should not affect our conclusions, which are based on trends present in the full sample of clouds.  

The column density maps of three clouds are shown as examples together with CO maps from \citet{dame2001milky} in Fig. \ref{fig:exampleclouds}, while a thumbnail collection of all column density maps is shown in Figs. \ref{fig:cloudext}--\ref{fig:cloudext4}. The coverage of the fields was chosen by eye based on previous studies of the clouds. The choice of the field may affect the shape of the low column density part of the N-PDFs and hence their physical interpretation \citep{kainulainen2013high,alves2017shapes,lombardi2015molecular,kortgen2019shape,chen2018anatomy}. In this work, we focus on the high column density part of the N-PDF that is more intimately linked to star formation. How much diffuse material is included in the clouds, and thereby the shape of the N-PDF at very low column densities, is therefore not of high importance here. We also note that our extinction maps inherently have different physical resolutions for different clouds. The resolution varies from about $2.5\arcmin$ for the nearby clouds to about $30\arcsec$-$2.5\arcmin$ for the distant clouds in the Galactic plane. This causes a difference in physical resolution between a factor of a few to ten. However, resolution studies in the context of N-PDFs have found that resolution does not strongly affect the high-column density characteristics of the N-PDFs \citep{schneider2015}, provided that the N-PDF is still sampled reasonably. In Appendix \ref{app:resolution} we show that degrading the resolution is unlikely to alter the results of this work significantly.

We caution that it is difficult to assign distances to all individual structures in the column density maps. It is possible that some fields contain gas components at different distances. However, based on examination of CO data from \citet{dame2001milky} (see Fig. \ref{fig:exampleclouds}), the fields we chose are relatively coherent structures in velocity as well. We therefore conclude that while some individual fields may be affected by multiple gas components, our analysis, which is based on a large sample of clouds, should not be significantly affected by these effects.

\begin{figure*}
    \centering
    \begin{minipage}[b]{0.44\textwidth}
        \includegraphics[width=\textwidth]{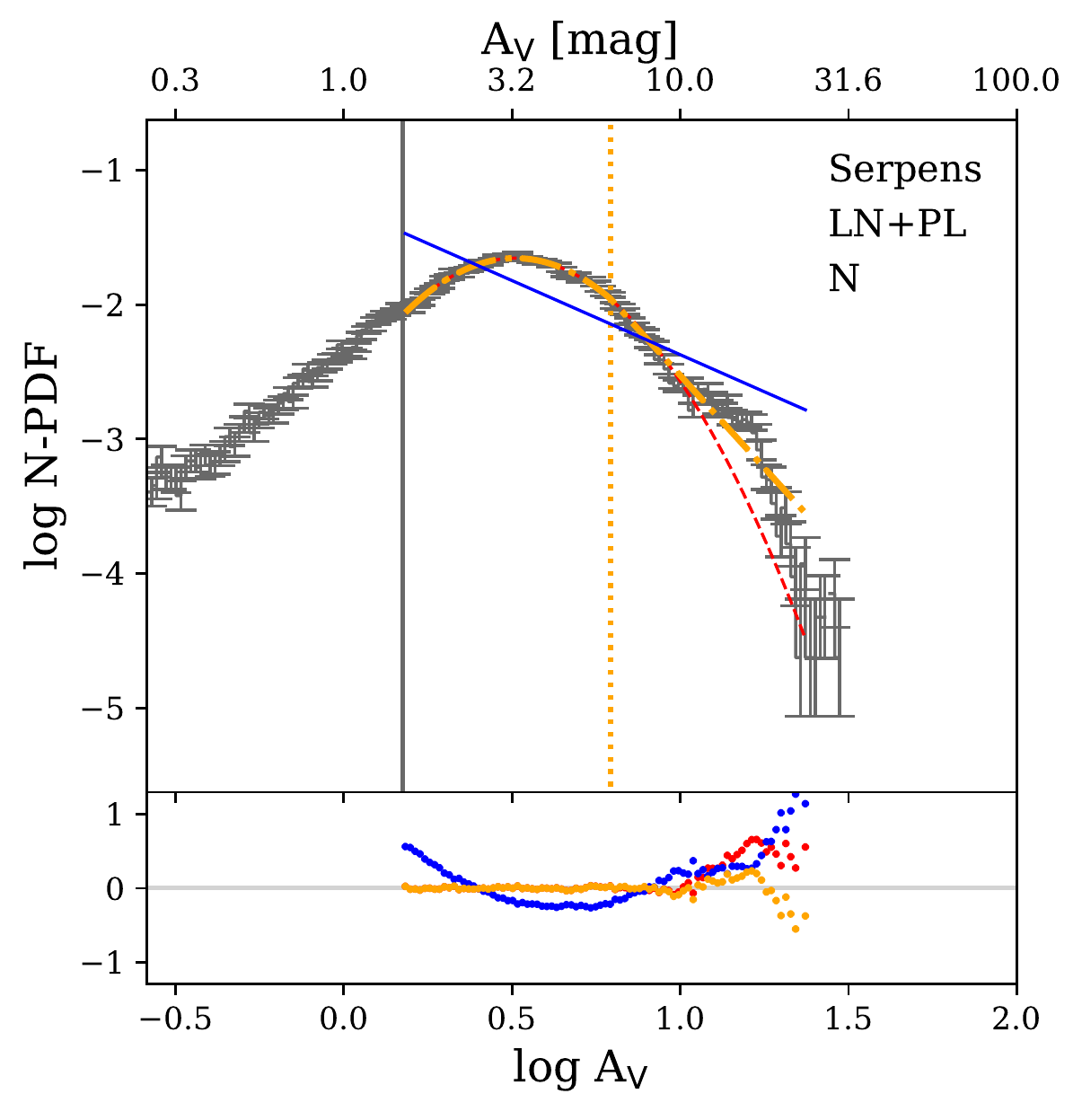}
    \end{minipage}
    \begin{minipage}[b]{0.44\textwidth}
        \includegraphics[width=\textwidth]{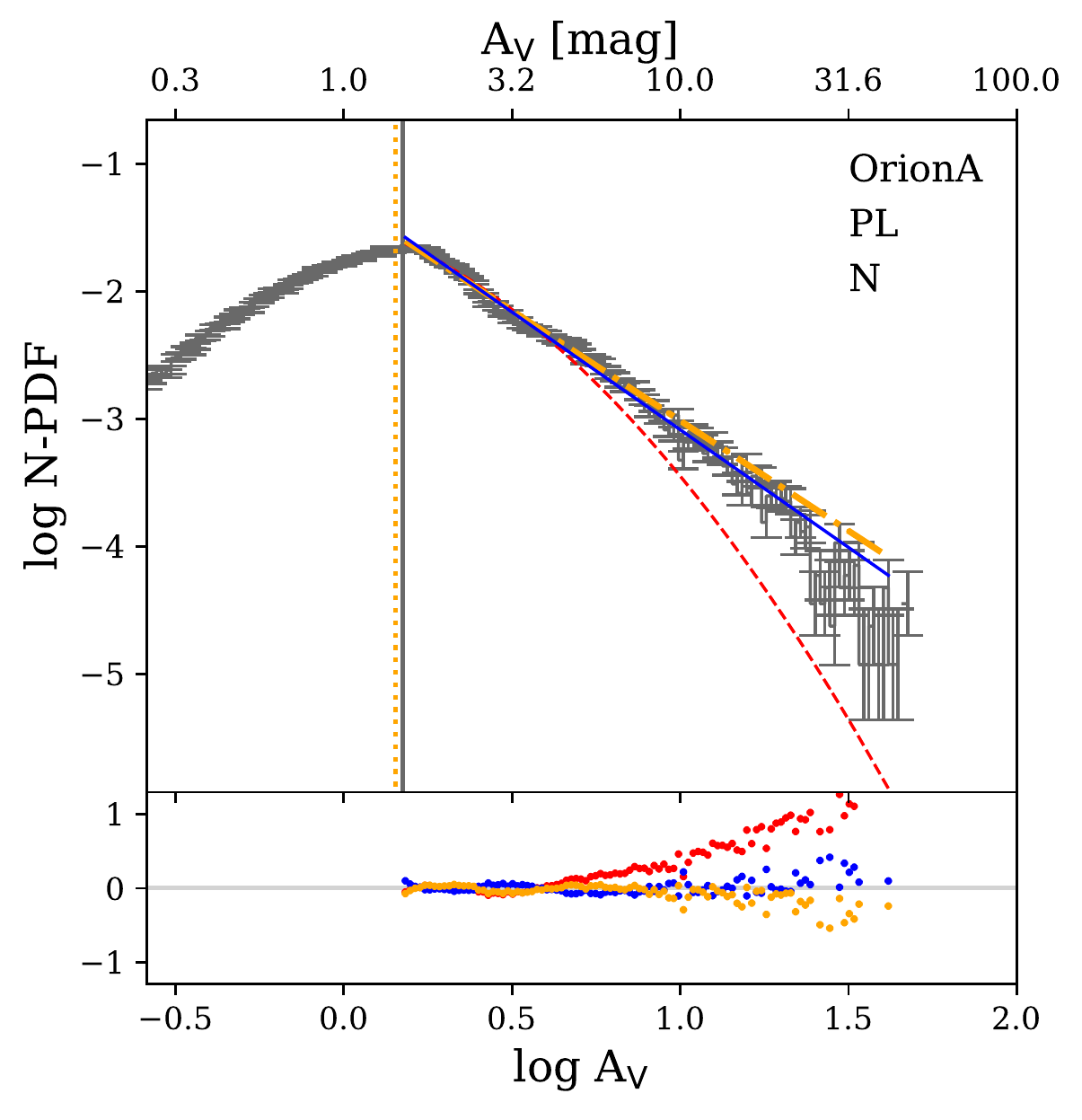}
    \end{minipage}
    \begin{minipage}[b]{0.44\textwidth}
        \includegraphics[width=\textwidth]{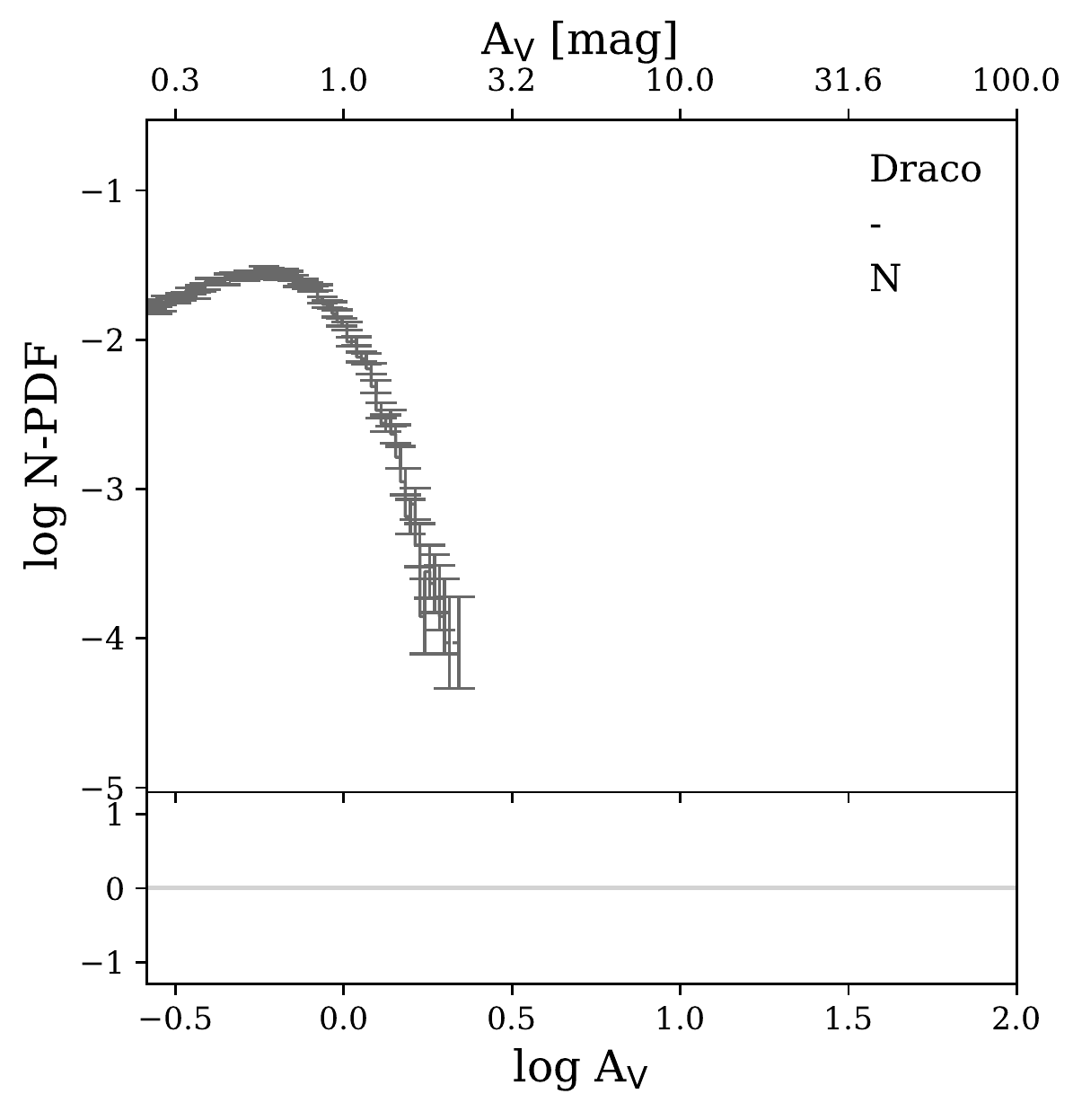}
    \end{minipage}
    \begin{minipage}[b]{0.44\textwidth}
        \includegraphics[width=\textwidth]{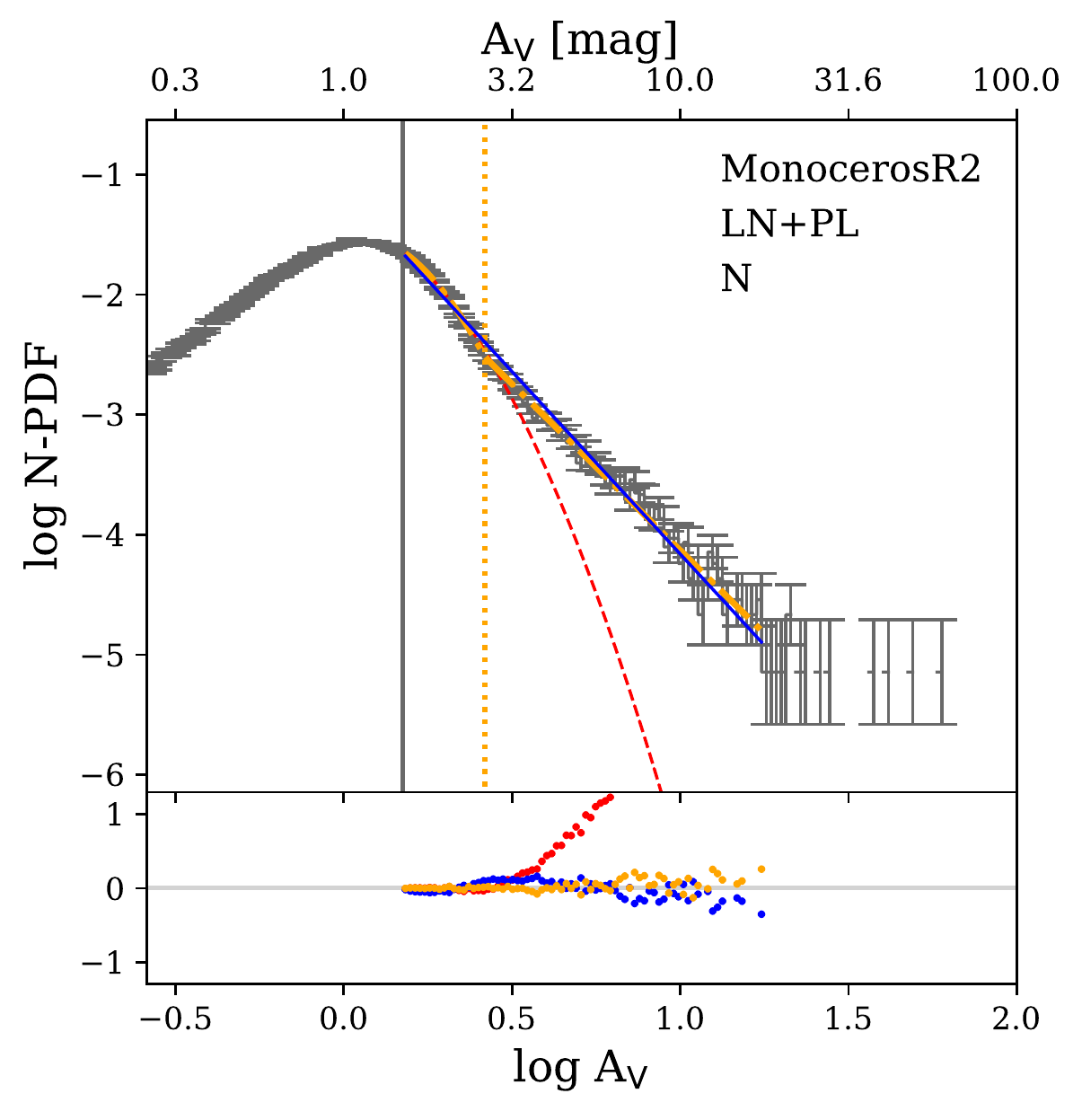}
    \end{minipage}
    \caption{Examples of the N-PDFs of individual clouds. Those used in the analysis are shown in dark grey, and the column density map used is noted in the legend: N for NICEST (all of these). The blue, red, and yellow lines show the fits of PL, LN, and LN+PL models, respectively. The transition point of the LN+PL model is shown with a vertical dotted yellow line. All fits were performed above the extinction at the last closed contour, which is marked with the vertical grey line. The residuals of the fitted models are shown below the N-PDFs, and the best-fit model is noted in the legend. 
    }
    \label{fig:examplePDFs}
\end{figure*}

\subsection{Column density PDFs}
\label{sec:mpdf}

We used the column density maps to compute N-PDFs of the clouds by counting the pixels in intervals of $\log A_\mathrm{V}$. The N-PDFs were computed as 
\begin{equation}
    P(a \leq \log{A_\mathrm{V}} \leq b) = \frac{\Sigma_i p_i[a \leq \log{A_\mathrm{V}} \leq b]}{\Sigma_i p_i}\frac{1}{\Delta (\log{A_\mathrm{V}})}.
\end{equation}
Here, $P(a \leq \log{A_\mathrm{V}} \leq b)$ is the probability density of a pixel having an extinction $A_\mathrm{V}$ between $a$ and $b$, $p_i$ is the number of pixels, and $p_i[a \leq A_\mathrm{V} \leq b]$ is the number of the pixels with extinction in the bin interval. We used bins between $\log(A_\mathrm{V})= 0.6-2.3$, with a bin size $\Delta (\log{A_\mathrm{V}})=$ 0.0085. The \textit{Herschel}-based column density maps reach values higher than 100 mag, but due to the high uncertainties at these densities and because the extinction-based maps do not reach these densities, we assigned these pixels a magnitude of 100 mag for the mass calculation and ignored these pixels in the N-PDFs. Examples of the derived N-PDFs are shown in Fig. \ref{fig:examplePDFs}, and all the N-PDFs are shown in Figs. \ref{fig:Allfits1}-\ref{fig:Allfits4}. The errors on the N-PDFs were calculated as $\sigma_y =\frac{d f(N_{pix})}{dN_{pix}}\sigma_{N_{pix}} = \frac{\sqrt{N_{pix}}}{\Sigma(N_{pix})\cdot \log(N_{pix})}$.

\subsection{Quantification of the N-PDF shapes}
\label{sec:mfit}
The observed N-PDFs show a variety of shapes (as shown in Fig. \ref{fig:examplePDFs} and Figs. \ref{fig:Allfits1}-\ref{fig:Allfits4}). It is not possible to use one model to describe all of them. Therefore we adopted three models to quantify the N-PDF shapes; we describe them below. 
\begin{enumerate}
    \item The log-normal function (hereafter, LN model) 
    \begin{equation}
        P(A_\mathrm{V}) = \frac{a}{\sigma \sqrt{2\pi}A_\mathrm{V}}\exp \left(\frac{-(\ln(A_\mathrm{V})-\mu)^2}{2\sigma ^2}\right).
    \end{equation}
    In this model, $\sigma$ is the standard deviation of the LN function, $a$ is the amplitude, and $\mu$ is the mean of the logarithmic extinction (the location of the peak of the N-PDF of logarithmic $A_\mathrm{V}$, i.e. $P(\log A_\mathrm{V})$). The fitting was made with the Python package LMFIT \citep{newville_matthew_2014_11813}. The free parameters in this fit were $a$, $\sigma,$ and $\mu$. 
    
    \item The power law (hereafter, PL model),
    \begin{equation}
        P(\log(A_\mathrm{V})) = \alpha \log(A_\mathrm{V}) + \beta. 
    \end{equation}
    The slope of the power law is denoted by $\alpha$, and $\beta$ is the intercept. The power-law fitting was made with the Python Scipy package \texttt{linregress} \citep{2020SciPy-NMeth}.  This was also used to determine the fitting parameters and the significance of parameter correlations below. 
    
    \item A piecewise combination of LN and PL functions (hereafter, LN+PL model), 
    
    \begin{align}
        P(A_\mathrm{V}) = \left\{ \begin{array}{cc} 
            \frac{a}{\sigma \sqrt{2\pi}A_\mathrm{V}}\exp \left(\frac{-(\ln(A_\mathrm{V})-\mu)^2}{2\sigma ^2}\right) \notag & \hspace{5mm} \text{if } A_\mathrm{V} \leq x_0 \\
            \left(\frac{A_\mathrm{V}}{x_0}\right)^\alpha \cdot \frac{a}{\sigma \sqrt{2\pi}x_0}\exp \left(\frac{-(\ln(x_0)-\mu)^2}{2\sigma ^2}\right) \notag & \hspace{5mm} \text{if } A_\mathrm{V}>x_0. \\
            \end{array} \right.
    \end{align}
    This is a combination of the two previous models, with a transition point $x_0$ from log-normal at low column density to power law at high column density. The intercept of the PL part is determined from the LN part. The fit was performed using \texttt{scipy optimize brute} \citep{2020SciPy-NMeth}, and then the output was used as an initial guess for LMFIT \citep{newville_matthew_2014_11813}. Thirty grid points were used in each dimension of the brute-force fit. The brute-force optimisation was used as a first step because LMFIT was unsuccessful with this high number of fitting parameters ($a$, $\sigma$, $\mu$, $x_0$, and $\alpha$). 
\end{enumerate}
    
All fits were made from the extinction value at the last closed contour. The last closed contour was determined by eye. In all models, the weight of the data was given as one over the square of the standard error (statistical weighting). Which shape fit the N-PDF of each cloud best was decided by individual examination of the $\chi^2$ values and the residual distribution. The fit with the lowest $\chi^2$ value and the highest normality, linearity, and homoscedasticity of the residuals was chosen as the best fit for each cloud. Fig. \ref{fig:examplePDFs} shows all the models fit to Serpens, OrionA, and MonR2. The best-fit model for each is noted in the legend, and the distribution of residuals is shown below the N-PDFs. MonR2 is an example where the PL and LN+PL both fit the data quite well, but the LN+PL model was chosen due to the more normally distributed residuals and a slightly lower $\chi^2$ value. For some N-PDFs, the residuals were clearly not normally, linearly, and homogeneously distributed for any of the models, and none of the models were chosen. The N-PDF was then marked with a questionmark.

\subsection{Bird's eye view}
\label{sec:mbird}

With our sample of clouds and their N-PDFs in hand, we had a census of the molecular clouds and their density structure in a well-defined region of the Milky Way. This enabled us to perform an experiment in which we viewed this region from a bird's eye perspective. To our knowledge, this is the first time that this has been done. The goal is to understand how the column density statistics would appear when viewed from outside the Milky Way at lower resolution. This experiment is the first step in trying to understand how the true N-PDFs of clouds, observed at sub-pc resolution, combine to yield the column density statistics at much coarser resolution. The setup is illustrated in Fig. \ref{fig:bird}.

The fundamental assumption we made in the experiment is that the N-PDFs of the clouds are similar when viewed from ``above'', that is, from the face-on perspective of the Milky Way, to what we observe from our location within the Milky Way. While this may be unlikely for individual clouds \citep[e.g.][]{ballesteros2011gravity,chira2016viewingangles}, we hypothesise that in a large sample such as ours, projection effects cancel out and the trends found by viewing the N-PDFs from within the disk remain when they are viewed from the face-on perspective. Under this hypothesis, we used the N-PDFs derived for the clouds as proxies of their N-PDFs when viewed from the face-on perspective. Some simulation studies indicate that molecular clouds might be flattened in the plane of the disk in which they reside \citep[e.g. ][]{benincasa2013giant,kruijssen2019dynamical}, but as we are not aware of any conclusive or quantitative studies on this in the solar neighbourhood, our assumption seems a reasonable premise to conduct an experiment. 

The aperture N-PDFs were derived in a similar way as for the individual clouds for apertures of various sizes. In this derivation, the cloud coordinates ($l$, $b$, $distance$)$\rightarrow$($x$,$y$) alone determine whether the entire
N-PDF of a cloud is included in the aperture, that is, the clouds were considered point sources. The N-PDFs of individual clouds within the apertures were weighted by the areas of the clouds and summed,
\begin{equation}
    P(\log A_\mathrm{V})_{aperture} = \frac{\Sigma_{clouds} (A_{cloud} \cdot P(\log A_\mathrm{V})_{cloud})}{\Sigma_{clouds} A_{cloud}}.
\end{equation}
Here $A_{cloud}=A_{pix}\cdot N_{pix}$ is the area of each cloud, and $P(\log A_\mathrm{V})_{cloud}$ is the N-PDF of that cloud.

\begin{figure}
    \centering
    \includegraphics[width=0.49\textwidth]{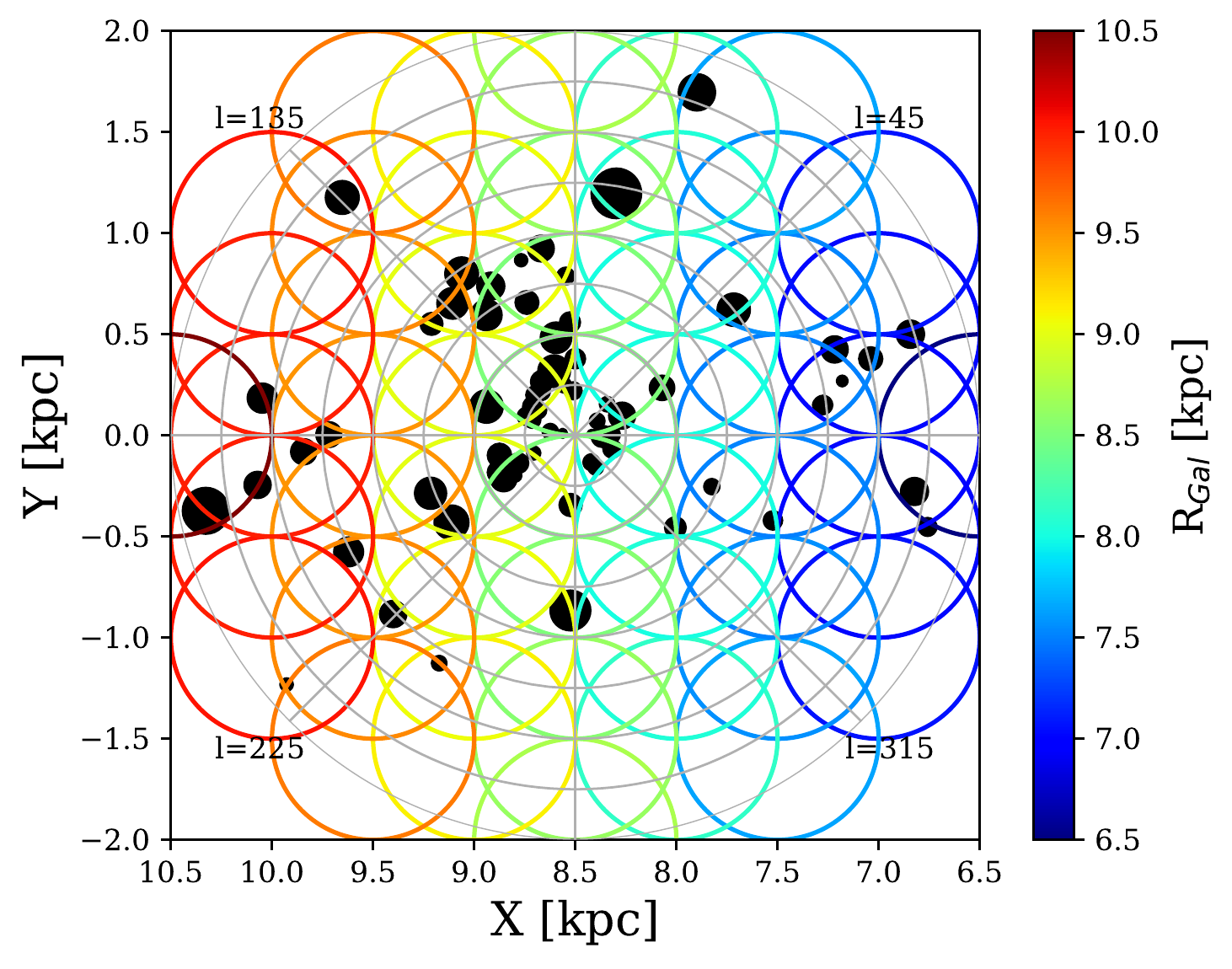}
    \caption{Coverage for apertures of $R=0.5$ kpc, moved with Nyquist sampling across our survey area. The clouds are shown in black, and the colours of the apertures correspond to Galactic radius (see colour bar).}
    \label{fig:move_cov}
\end{figure}

We constructed maps of aperture properties for which we sampled the entire 2 kpc radius survey area. This was performed for apertures of radius 1, 0.5, and 0.25 kpc to study the dependence the aperture-averaged N-PDFs on the location and scale. The apertures were moved with Nyquist sampling intervals, as illustrated for $R=0.5$ kpc apertures in Fig. \ref{fig:move_cov}. For each aperture position, we computed aperture N-PDFs from the N-PDFs of the clouds within. To illustrate the link to the scales probed in extragalactic works, we note that the aperture sizes we used correspond to angular resolutions of $25-6\arcsec$ at the distance of M51 (8.4 Mpc), comparable to \citet{leroy2017cloud}, for example. We also examined our full 2 kpc aperture, which is centred on Sun. We note that our experiment is different from extragalactic observations in two important ways. Firstly, the apertures in this study only contain the cloud-like high-density component of the molecular ISM; extragalactic studies detect \emph{\textup{all}} emission from within the observed areas, regardless of whether it originates from cloud-like structures or diffuse gas. Secondly, the aperture N-PDFs in this work contain small-scale structure that is inaccessible in extragalactic studies; thus, the dynamic ranges of the N-PDFs between this work and extragalactic works are not the same. The aperture N-PDFs are therefore not directly comparable to extragalactic N-PDFs. We emphasise that our goal in this paper is not to perform this comparison; we used our experiment to study the variety of the aperture N-PDFs and their dependence on the scale and galactic environment in the Milky Way.

%-----------------------------
%-----------------------------
\section{Results}
\label{sec:Results}
%-----------------------------
%-----------------------------

The main results of our work are the currently most complete sample of molecular clouds within a distance of 2 kpc (Sect. \ref{sec:results_sample}), the description of the N-PDFs of these clouds and their relation to star formation (Sect. \ref{sec:results_individual}), and the column density characteristics of the aperture N-PDFs and their relation to star formation, as seen from a bird’s eye perspective (Sect. \ref{sec:results_bird}).

%---------------------------------------------------------------------------------
\subsection{Highly complete sample of molecular clouds in the solar neighbourhood}
\label{sec:results_sample}
%---------------------------------------------------------------------------------

The first result of this work is a collection of what we consider to be the most complete sample to date of the main molecular clouds or complexes within 2 kpc from the Sun, amounting to 72 clouds or regions. The names, coordinates, masses, YSO counts, and references to the relevant literature information are given in Table \ref{tab:mastertable}. The column density maps of the clouds are shown in Figs. \ref{fig:cloudext}-\ref{fig:cloudext4}. 

As for completeness, the sample is unlikely to miss clouds larger than $\sim10^4$ M$_\odot$ that contain a significant fraction of molecular gas, or clouds that have more than $\sim$ 100 YSOs. These clouds would be easily detectable in our survey area. At distances closer than some 1 kpc, the mass completeness limit is lower, probably closer to $10^3$ M$_\odot$. Insight into the completeness of the sample can be gained by considering its total mass against the mass recovered by CO surveys. The average CO profile of the Milky Way indicates a CO-derived molecular gas surface density value of $\Sigma_\mathrm{mol} \approx 1.5$ M$_\odot$ pc$^{-2}$ at the radius of the Sun \citep{kennicutt2012star,nakanishi2006three}. When this value is used, the total molecular mass within a 2 kpc radius circle from the Sun is $1.9 \cdot 10^7$ M$_\odot$. The total mass of our cloud sample above 3 mag is $8.8 \cdot 10^6$ M$_\odot$, about 47\% of this. The limit of 3 mag corresponds to the region in which we expect all the gas to be molecular \citep{heyer2015molecular}. Molecular gas is also present at lower extinctions, however. If we instead choose a threshold of 1 mag \citep[i.e.][]{pety2017orionb}, the total mass is $1.5 \cdot 10^7$ M$_\odot$, 81\% of the estimate from the average CO profile.

It is interesting to place our cloud sample in the context of extragalactic works, even knowing that the data are not directly comparable (cf. Sect. \ref{sec:mbird}). This helps understanding what our sample represents. Compared to molecular clouds identified in M51 by \cite{colombo2014pdbi}, our density of clouds is far lower. \cite{colombo2014pdbi} quoted a completeness limit of $3.6\cdot10^5$ M$_\odot$ and a density of 19 clouds/kpc$^2$ in inter-arm regions, while above the same mass limit, we only have about 0.5 clouds/kpc$^{2}$ (our total cloud density is 6 clouds/kpc$^2$). It is difficult to ascertain the origin of this difference, which might be either observational or physical. Significant observational differences exist in the sensitivity to large structures of relatively low-column density between our data and \cite{colombo2014pdbi}. CO-emitting gas might be detected as large complexes in M51 but might be invisible to us if they do not contain significant amounts of high column density gas. The mean inter-arm surface density of the clouds in \cite{colombo2014pdbi} is 34 M$_\odot$/pc$^2$, while within our survey area, we did not detect any clouds with low surface densities like this (the mean surface density of our clouds is 88 M$_\odot$/pc$^2$). If these CO clouds exist in our survey region but do not have high extinction signatures, we would probably not include them. Finally, the difference can also be physical: Colombo et al. studied the inner ($R_{gal}\sim 0-4$ kpc) region of M51, while our survey area is 6-10 kpc from the centre of the Milky Way. In this region, M51 is considerably more molecule rich than the solar neighbourhood \citep{schinnerer2013pdbi}, and the molecular gas fraction is higher.

Another approach to completeness can be taken by considering CO clouds that were previously identified in the survey area of the Milky Way. \citet{miville2017physical} identified CO clouds from the entire Milky Way; the total mass of clouds in their catalogue within 2 kpc from the Sun is $8.4 \cdot 10^6$ M$_\odot$. The total mass of our sample above 3 mag is 105\% of this, and above 1 mag, it is 184\%. Given the associated uncertainties, the total mass we derive is similar to that derived from CO clouds. More qualitatively, the clouds in our sample correspond relatively well to the recent 3D dust maps of the local Galactic neighbourhood by \citet{lallement2018three,lallement2019gaia} and \citet{green20193d}, for instance. 

That the total mass of the sample is 47\%--81\% of the estimate from the average CO profile can likely be understood, in addition to the large uncertainties, by considering that a large fraction of CO emission originates not from massive cloud-like structures, but from the more diffuse, extended gas component \citep[cf.][]{romanduval2016diffusegas} or from smaller clouds \citep[cf.][]{miville2017physical}. This might also be part of the reason for the discrepancy between clouds identified in M51 and underlines the fact that our dust based data best trace the high column density component of the molecular gas, which is organised into relatively massive clouds.

%------------------------------------------
\subsection{Individual clouds}
\label{sec:results_individual}
%------------------------------------------

We first used our sample to study the properties of individual molecular clouds. Specifically, we analysed the shape of their N-PDFs and their relation to the Galactic location of the clouds and the star formation rates as measured by the YSO content.

%----------------------------------------------
% FIGURE:The Pie diagrams of the model frequencies
\begin{figure*}
    \centering
        \includegraphics[width=0.55\textwidth,trim=10 39 0 0,clip]{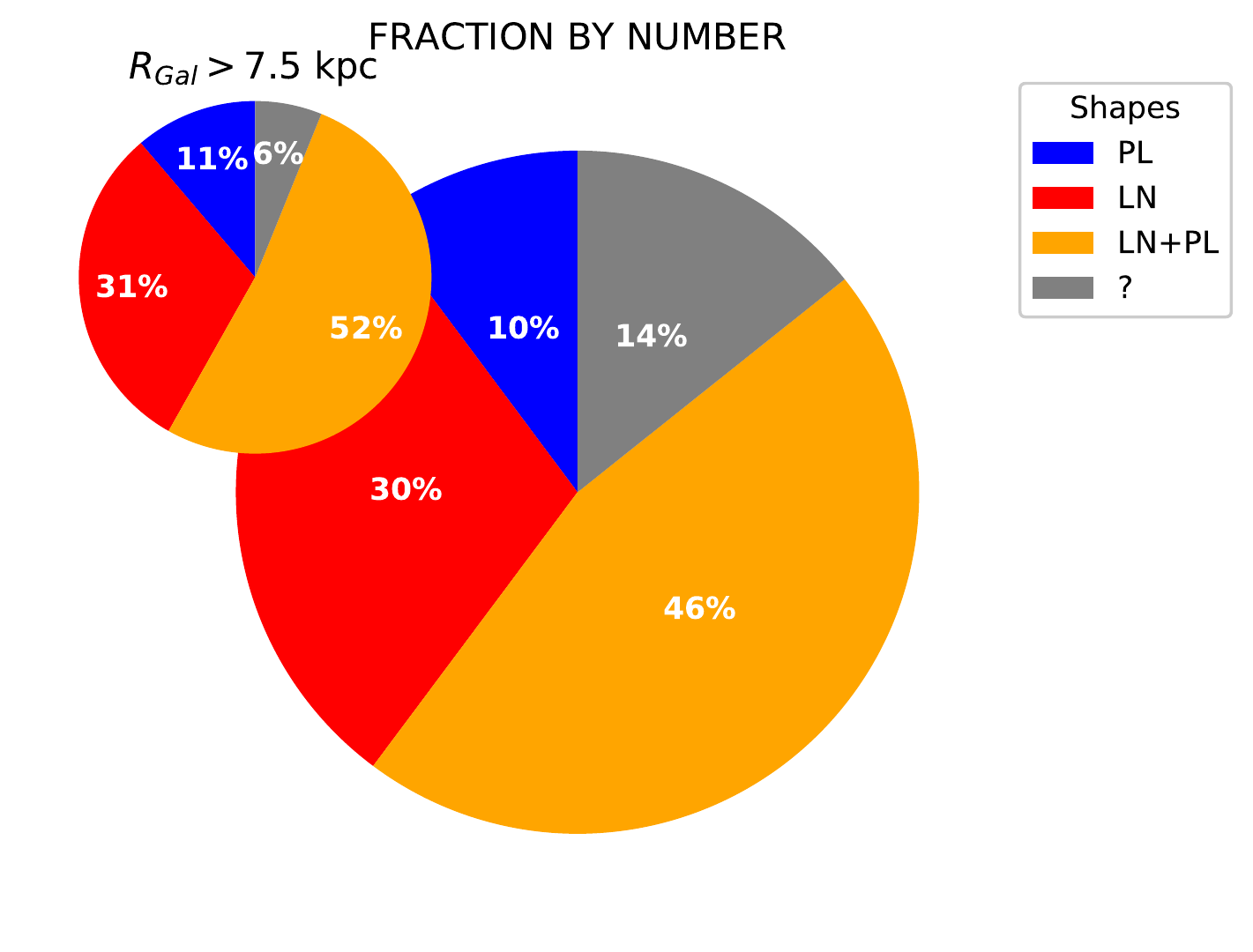}
        %12_n-eps-converted-to.pdf}
        \includegraphics[width=0.44\textwidth,trim=10 39 10 0,clip]{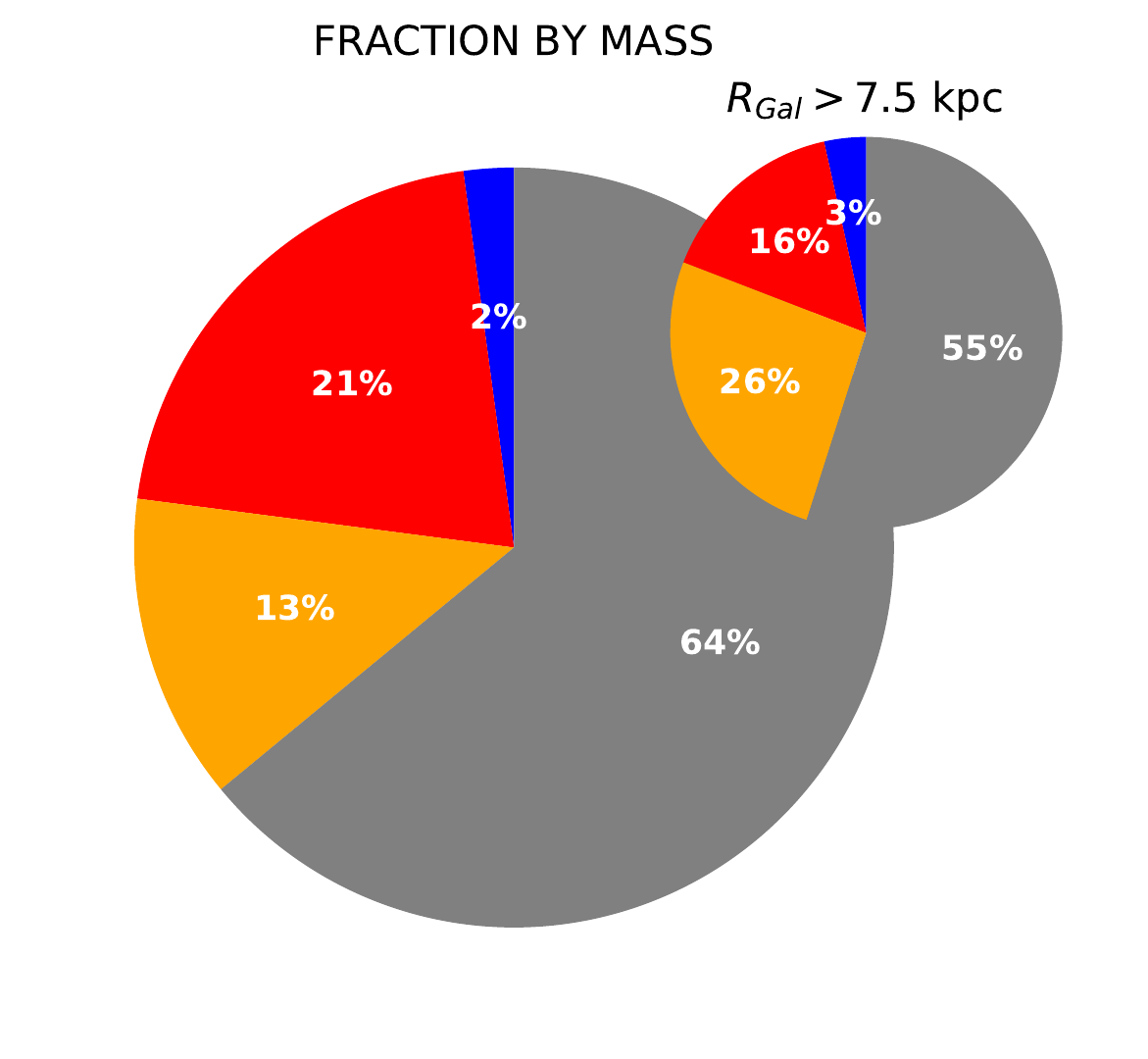}
    \caption{Fractions of cloud N-PDFs that are best fit by our three models, the PL, LN, and combination (LN+PL). The grey sector (marked with a questionmark) shows the clouds that are poorly fit by any of the three models. \emph{Left:} Fraction by number, and \emph{right:} fraction by mass.  The insets show the same pie diagrams only for clouds with $R_{Gal}>7.5$ kpc. 
    }
    \label{fig:pie}
\end{figure*}

%----------------------------------------------
%----------------------------------------------
% FIGURE:Histograms of fit parameters
\begin{figure*}%[h!]
    \centering
    \includegraphics[width=0.33\textwidth]{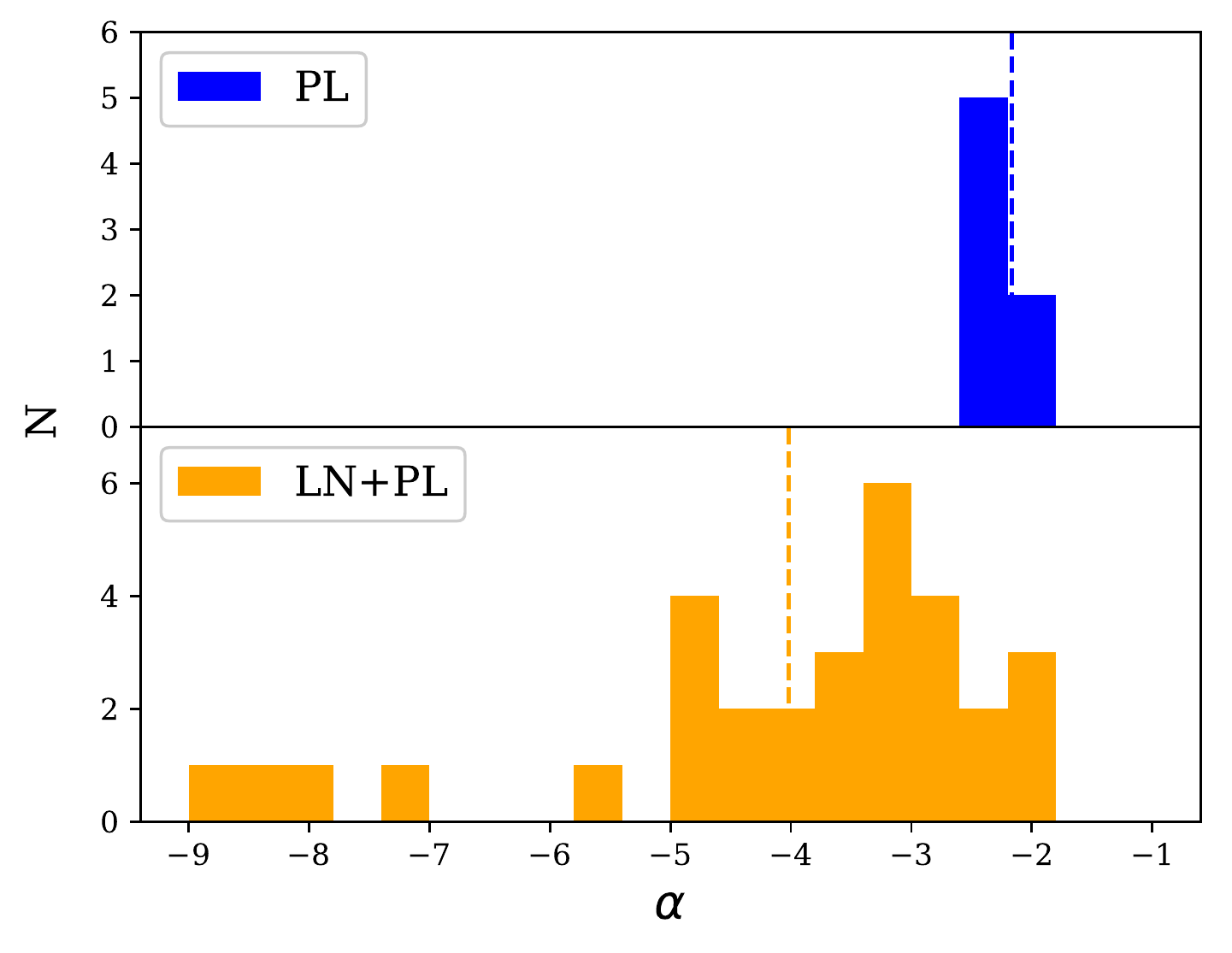}
    \includegraphics[width=0.33\textwidth]{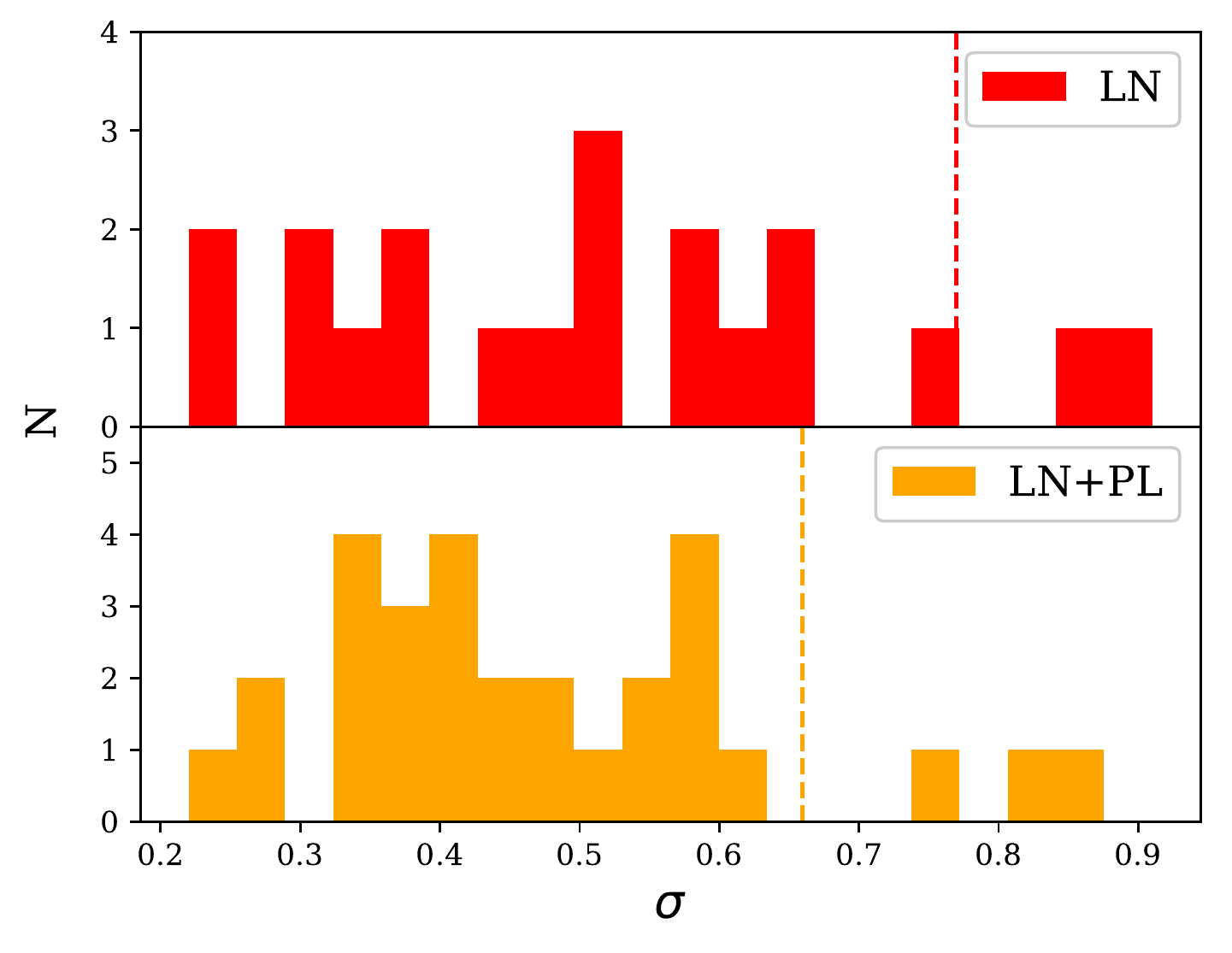}
    \includegraphics[width=0.33\textwidth]{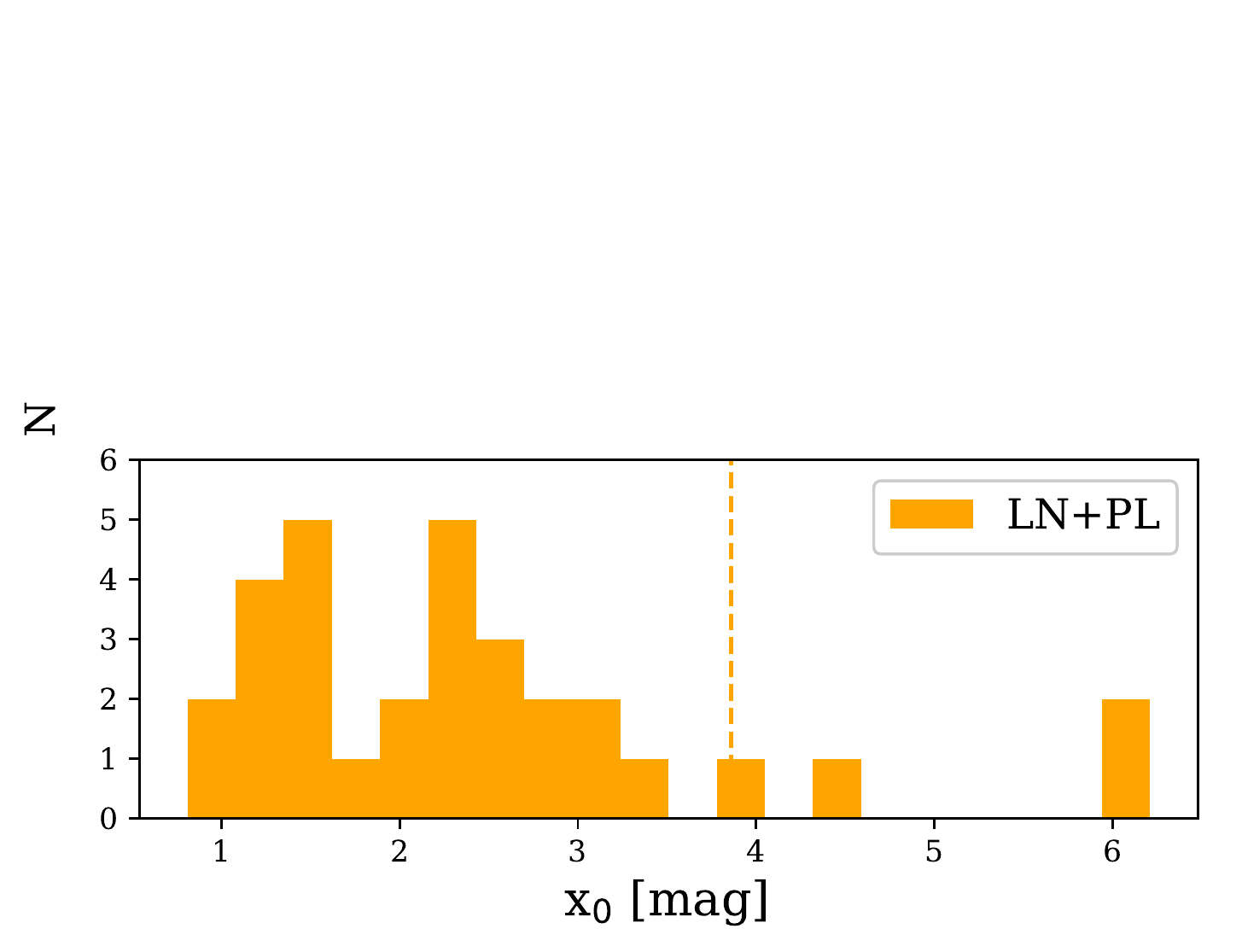}
    \caption{Distributions of the best-fit model parameters of the N-PDFs of individual clouds. Dashed lines show the mass-weighted average of the sample. M16 is the widest LN cloud, and Pipe and Serpens are LN+PL clouds with high values for the transition point. The fit parameters for all clouds are shown in Table \ref{tab:mastertable_shapes}.} 
    \label{fig:shapehist2}
\end{figure*}

%----------------------------------------------

%--------------------------------------------------
\subsubsection{Variety of molecular cloud N-PDFs}
\label{sec:results_npdfs}

Our census enables us to describe the variety of a large set of N-PDFs in our survey region. As individual examples, we show the N-PDFs of four clouds in Fig. \ref{fig:examplePDFs}, while the full collection of N-PDFs can be found in Figs. \ref{fig:Allfits1}-\ref{fig:Allfits4}. The N-PDFs show a variety of shapes, some resembling  a functional form such as a power law well (e.g. Orion A), and some not (e.g. Serpens). They show a variety of the relative amount of dense gas, that is, top-heavy and bottom-heavy N-PDFs (e.g. Draco vs. MonocerosR2). 

Most of the N-PDFs can be fitted reasonably well with one or several, of the models we adopted (LN, PL, and LN+PL). To describe the degree to which these forms are present in the sample, we assigned the best-fitting model for each cloud. However, we immediately note that some clouds can be fitted almost equally well with two models. Our goal in choosing one best fit for each N-PDF is to obtain a first-look understanding of the occurrence of the different shapes. The left panel of Fig. \ref{fig:pie} shows the fraction of clouds that are best fit with each model. The LN+PL model was found best for 46\% of the clouds, while 30\% of the clouds are best fit by the LN model. Ten percent of the clouds were best fit with the PL model. The frequency distributions of all fit parameters are shown in Fig. \ref{fig:shapehist2}, and the fit parameters for all clouds are listed in Table \ref{tab:mastertable_shapes}. A significant minority, 14\%, was not adequately fit by any of the three models. Four clouds did not reach an extinction of 3 mag; we did not attempt to fit these clouds because only very little molecular gas is present. They appeared log-normal or Gaussian, likely dominated by the uncertainty of the extinction data (the PDF of the error kernel). 

Even though most clouds are reasonably well fitted with one of the adopted simple models, two important points hamper their usefulness. First, a significant fraction of the total gas mass may be in N-PDFs that are poorly fitted by any of the models. To illustrate this, the right panel of Fig. \ref{fig:pie} shows the mass fraction fitted with each model, indicating that $\sim$64 \% of all gas is in N-PDFs poorly fit by any of the three models. This result is driven by the massive clouds in the Galactic plane, dominated by Cygnus. The insets in Fig. \ref{fig:pie} show the same pie diagrams only for clouds with $R_{\mathrm{Gal}}>7.5$ kpc. By setting this limit, we removed the clouds that are located in the vicinity of the Sagittarius spiral arm, where the properties of N-PDFs possibly change (Sect. \ref{sec:result_ind_gal_dist}). For clouds outside the spiral arm environment, the fraction of unfitted clouds is smaller, but still over half of the mass falls in this category. Second, the occurrence of the three models in the sample is not uniform. For example, the PL model provides the best fit only to relatively top-heavy N-PDFs, whereas more bottom-heavy N-PDFs tend to be better described by the LN+PL model. As noted above, the most massive clouds tend to have complex N-PDFs. This implies that analysing any sample of N-PDFs using only one model framework may lead to biases in the ability of that model to describe the sample.

\subsubsection{Empirical description of the N-PDF shapes}
\label{sec:empirical}

\begin{figure}%[h!]
    \centering
    \includegraphics[width=\columnwidth]{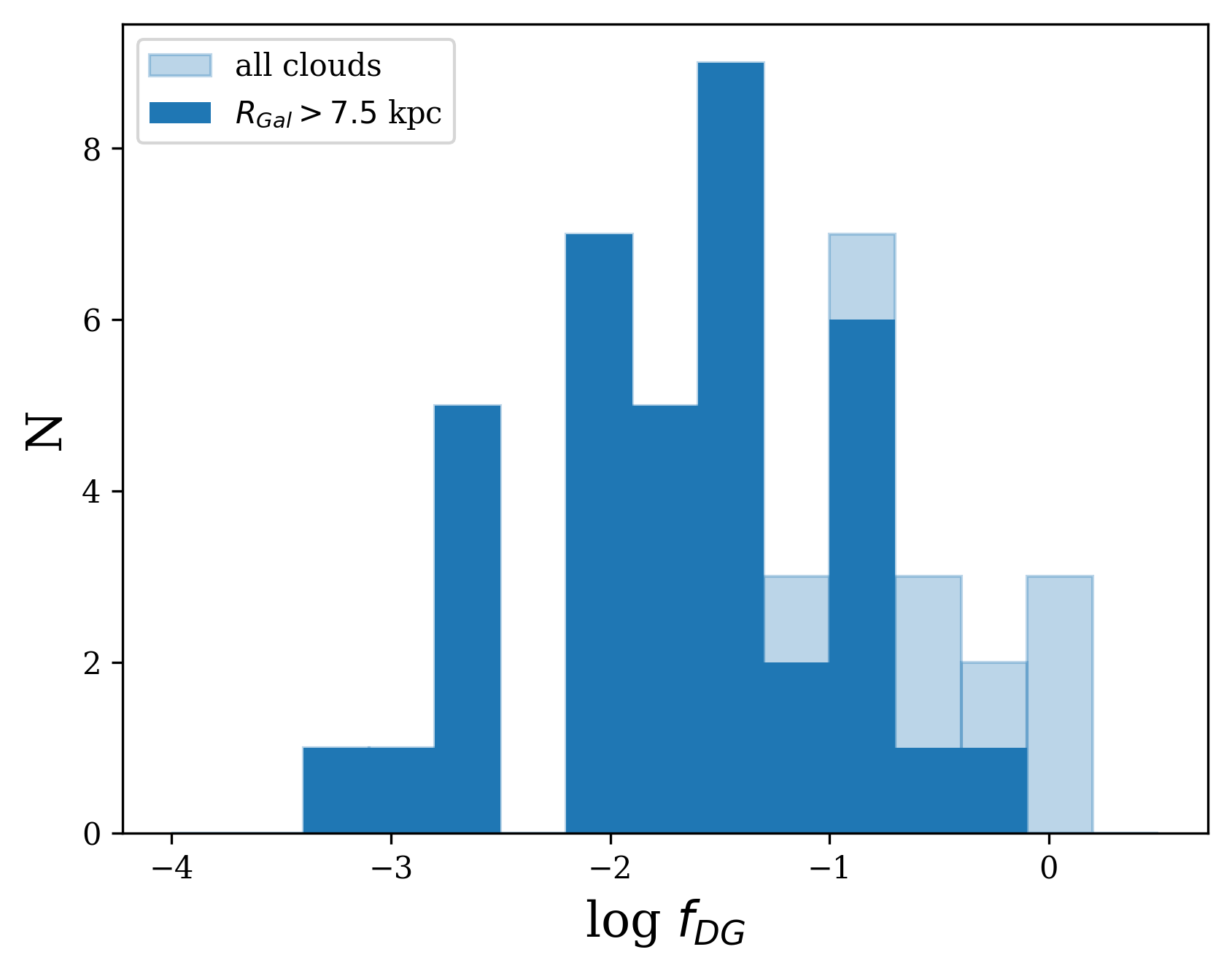}
    \includegraphics[width=\columnwidth]{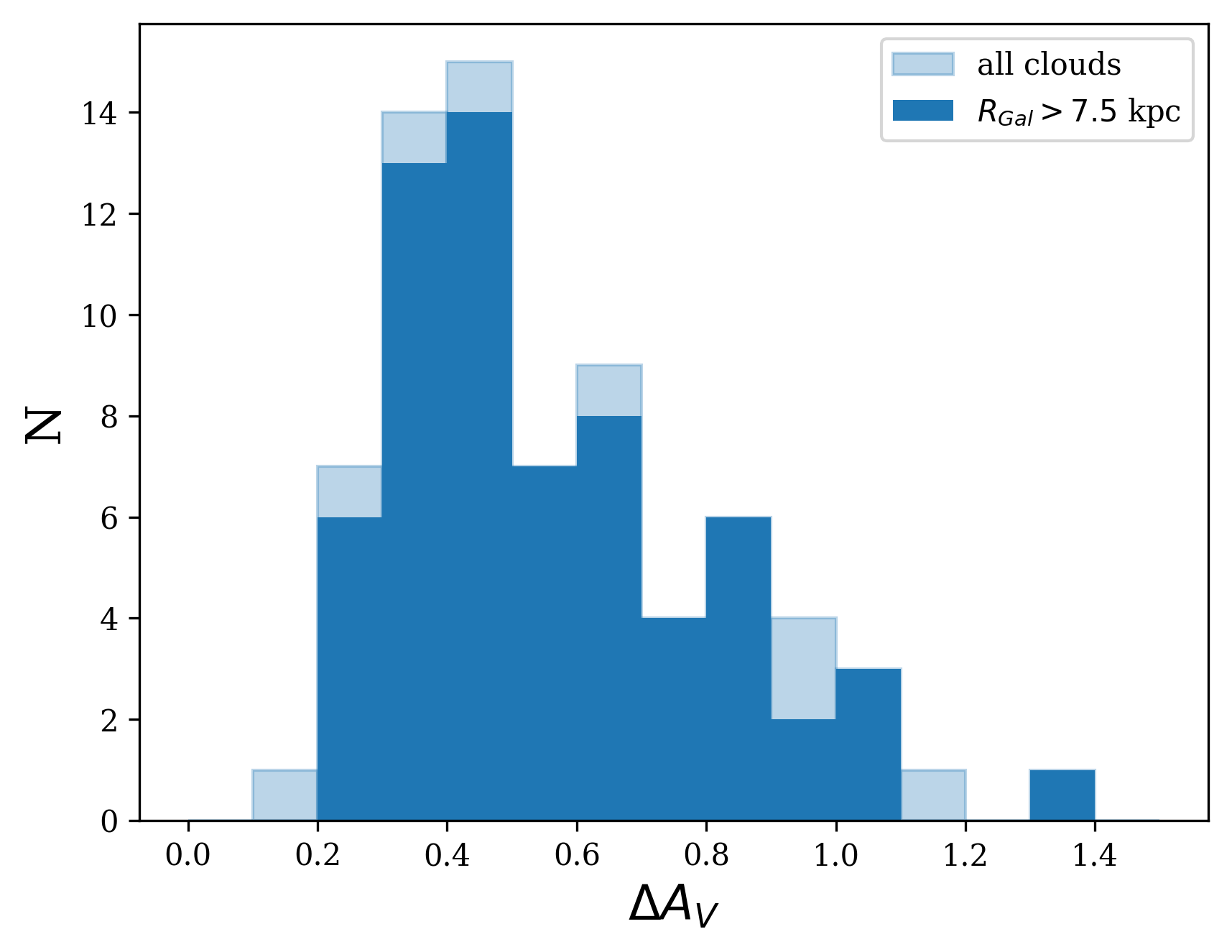}
    \caption{Frequency distribution of dense gas measures for the individual clouds. All clouds are shown in light blue, and clouds with $R_{\mathrm{Gal}}>7.5$ kpc are overplotted in darker blue. \emph{Top:} Dense gas mass fraction, $f_\mathrm{DG}$. \emph{Bottom:} Density contrast, $\Delta A_\mathrm{V}$ (see Sect. \ref{sec:empirical} in text).} 
    \label{fig:densegasfrac_histograms}
\end{figure}

These problems related to the use of simple models mean that it is beneficial to also describe the N-PDFs empirically, without any assumptions for the underlying model. For this purpose, we present two measures to quantify the N-PDF shapes, aimed at describing the relative amount of dense gas in the clouds, or their top-heaviness. The first measure is the dense gas mass fraction, $f_\mathrm{DG}$, the ratio of the mass of gas found at column densities higher than $A_\mathrm{V}^\mathrm{dense}=8$ mag and the mass found above $A_\mathrm{V}^\mathrm{all}=1$ mag, 
\begin{equation}
    f_\mathrm{DG} = \frac{\int_{A_\mathrm{V}=8}^\infty A_\mathrm{V} P(A_\mathrm{V}) \ \mathrm{d}A_\mathrm{V}}{\int_{A_\mathrm{V}=1}^\infty A_\mathrm{V} P(A_\mathrm{V}) \ \mathrm{d}A_\mathrm{V}}.
\end{equation}
$A_\mathrm{V}^\mathrm{dense}$ refers to high column density (dense) gas and $A_\mathrm{V}^\mathrm{all}$ to all gas (acknowledging that using any threshold always means that some of the gas is not accounted for). This measure is commonly used in the literature to describe the dense gas mass fraction \citep[e.g.][]{lada2012star,kainulainen2013connection,evans2014}. The frequency distribution of the resulting dense gas mass fractions is shown in Fig. \ref{fig:densegasfrac_histograms}. It peaks at values of about a few percent. For a biased sample of nearby clouds, a value of $\langle f_\mathrm{DG}\rangle = 0.1 \pm 0.06$ has been reported \citep{lada2010star, lada2012star}; for the same sample, we derive $\langle f_\mathrm{DG}\rangle = 0.07 \pm 0.05$ using our data, indicating general agreement. These arithmetic means are naturally dominated by the high $f_\mathrm{DG}$ values. Most clouds have lower dense gas fractions than the arithmetic mean. Furthermore, most of the \emph{\textup{mass}} in the clouds in the solar environment is in clouds with lower $f_\mathrm{DG}$; we find a mass-weighted mean of $\langle f_\mathrm{DG} \rangle = 0.03$ for our clouds.

The dense gas mass fraction $f_\mathrm{DG}$ uses absolute column density values to define what dense gas is. Depending on the purpose, it may be advantageous to consider a \emph{\textup{relative}} measure of the dense gas fraction instead. To do this, we defined as the second measure of the N-PDF shapes the density contrast, $\Delta A_\mathrm{V}$, as the column density contrast between the peak of the N-PDF and the column density above which the densest 5\% of the cloud mass resides. Formally, the $\Delta A_\mathrm{V}$ is defined with the help of the column density $A_\mathrm{V,}'$
\begin{equation}
    \frac{\int_{A_\mathrm{V}'}^\infty A_\mathrm{V} P(A_\mathrm{V}) \ \mathrm{d}A_\mathrm{V}}{\int_\mathrm{peak}^\infty A_\mathrm{V} P(A_\mathrm{V}) \ \mathrm{d}A_\mathrm{V}} = 0.05,
\end{equation}
where `peak' refers to the most common $A_V$ value (maximum mode) of the N-PDF above the last closed contour. Then, the density contrast is computed as
\begin{equation}
    \Delta A_\mathrm{V} = \log \frac{A_\mathrm{V}'}{A_\mathrm{V}^\mathrm{peak}}.
\end{equation}
The frequency distribution of the resulting density contrasts is shown in Fig. \ref{fig:densegasfrac_histograms}. It shows that most clouds have low density contrasts; the distribution declines roughly linearly with increasing density contrast (the density contrast is a logarithmic quantity).
We employ both the dense gas mass fraction and the density contrast further in Sect. \ref{sec:results_ind_npdfs_and_sf} to examine the relation of the cloud N-PDF shapes and star formation activity. 

%------------------------------% FIGURE: Dense gas fraction histograms

%-----------------------------------------------------
\subsubsection{Galactic distribution of the N-PDF shapes}
\label{sec:result_ind_gal_dist}
%-----------------------------------------------------

We now examine the N-PDFs of individual clouds in the context of the galactic environment. Our 2 kpc radius survey region extends over potentially differing galactic environments. While the solar neighbourhood is generally considered to be located in a less prominent part of the main spiral structure, the major Sagittarius spiral arm enters the 2 kpc aperture at the side of the Galactic centre (see Fig. \ref{fig:bird}). It is interesting to consider whether the differences we see in the N-PDFs correlate with these environments. 

The N-PDFs of a few individual clouds stand out from our sample. They have wide distributions, peak at high extinction, and reach high column densities. Those that stand out are Cygnus, M16, M17, M20, Cartwheel, and NGC6334. These clouds are all located in the direction of the Galactic centre and the Sagittarius spiral arm. Cartwheel and Cygnus are the only clouds that have extinction maps as derived in Appendix \ref{app:cygnus_and_cartwheel}, and M16, M17, M20, and NGC6334 are the only ones that use the PPMAP dust emission map from \citet{marsh2015temperature}. The different column density maps may be part of the explanation for the differences in the N-PDFs of these clouds. To determine this, we derived N-PDFs for these clouds using the extinction map of \citet{juvela2016allsky} for comparison (Figs. \ref{fig:Allfits1}-\ref{fig:Allfits4}), although this map has serious issues in the Galactic plane due to foreground contamination and is not entirely suitable for the clouds there. This comparison shows that many of the N-PDFs have similar shapes as the \emph{Herschel}-based ones, but an offset in $A_\mathrm{V}$. The offset is likely due to the different treatment of the background in the methods. For many clouds, the shape differs as well, however; this is especially true for the clouds in the Galactic plane and towards the Galactic centre. Even though the N-PDFs of the six spiral arm clouds derived from the \citet{juvela2016allsky} data look different, they also peak at high extinction. This indicates that while the exact N-PDF shape may depend on the data that are used, this does not hold in the same way for the average values such as the mean extinction. Overall, the comparisons suggest that the N-PDFs of spiral arm clouds show differences to non-spiral arm clouds also when only dust-extinction-based data are used (acknowledging that the reliability of this is hard to quantify because of the known problems in the extinction map).

\begin{figure}
    \centering
    \includegraphics[width=0.9\columnwidth]{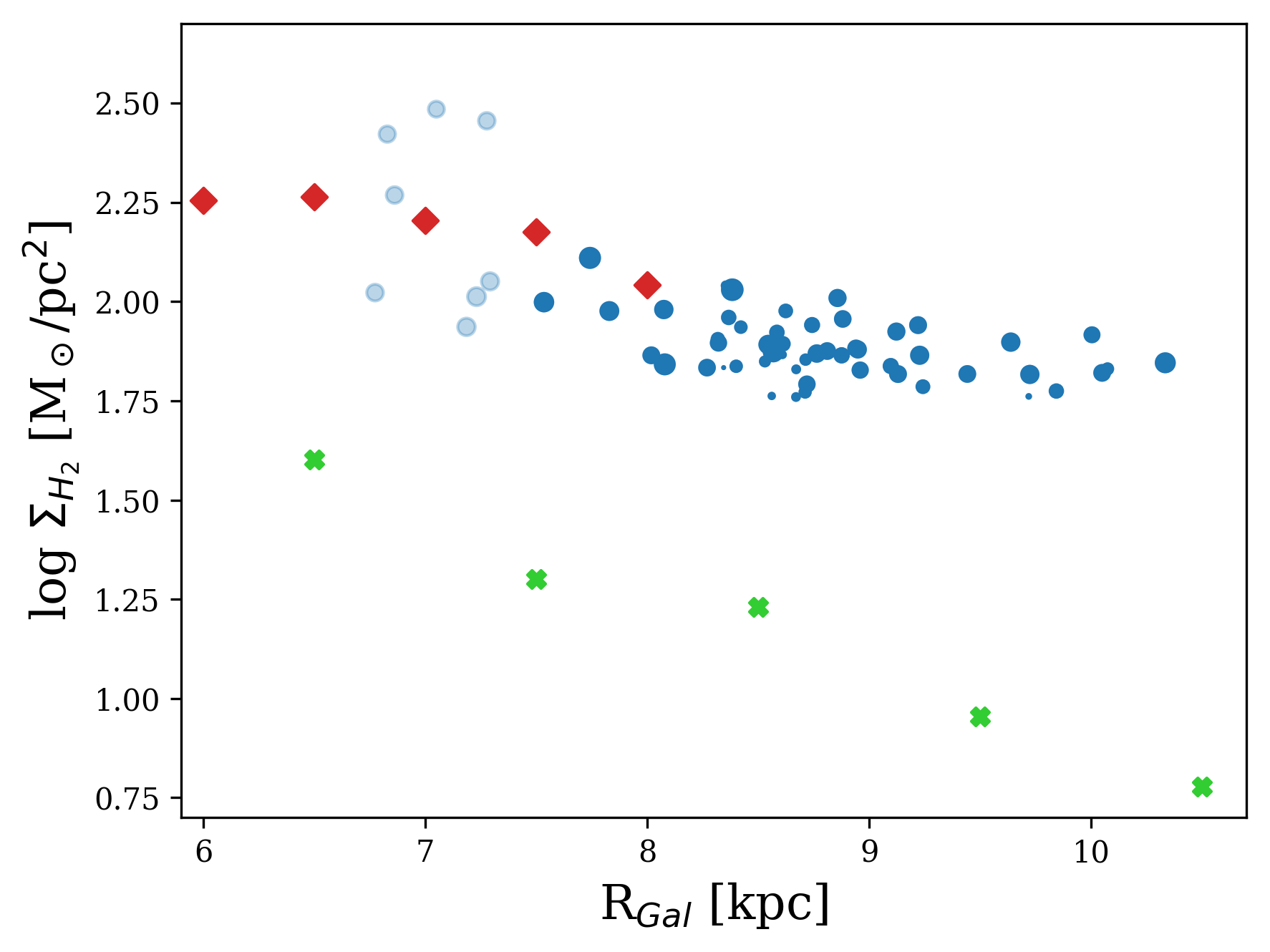}
    %width=0.87\columnwidth]{newfigs/jouni/Fig_Sigma_vs_Rgal-eps-converted-to.pdf} others 0.8
    \includegraphics[width=0.9\columnwidth]{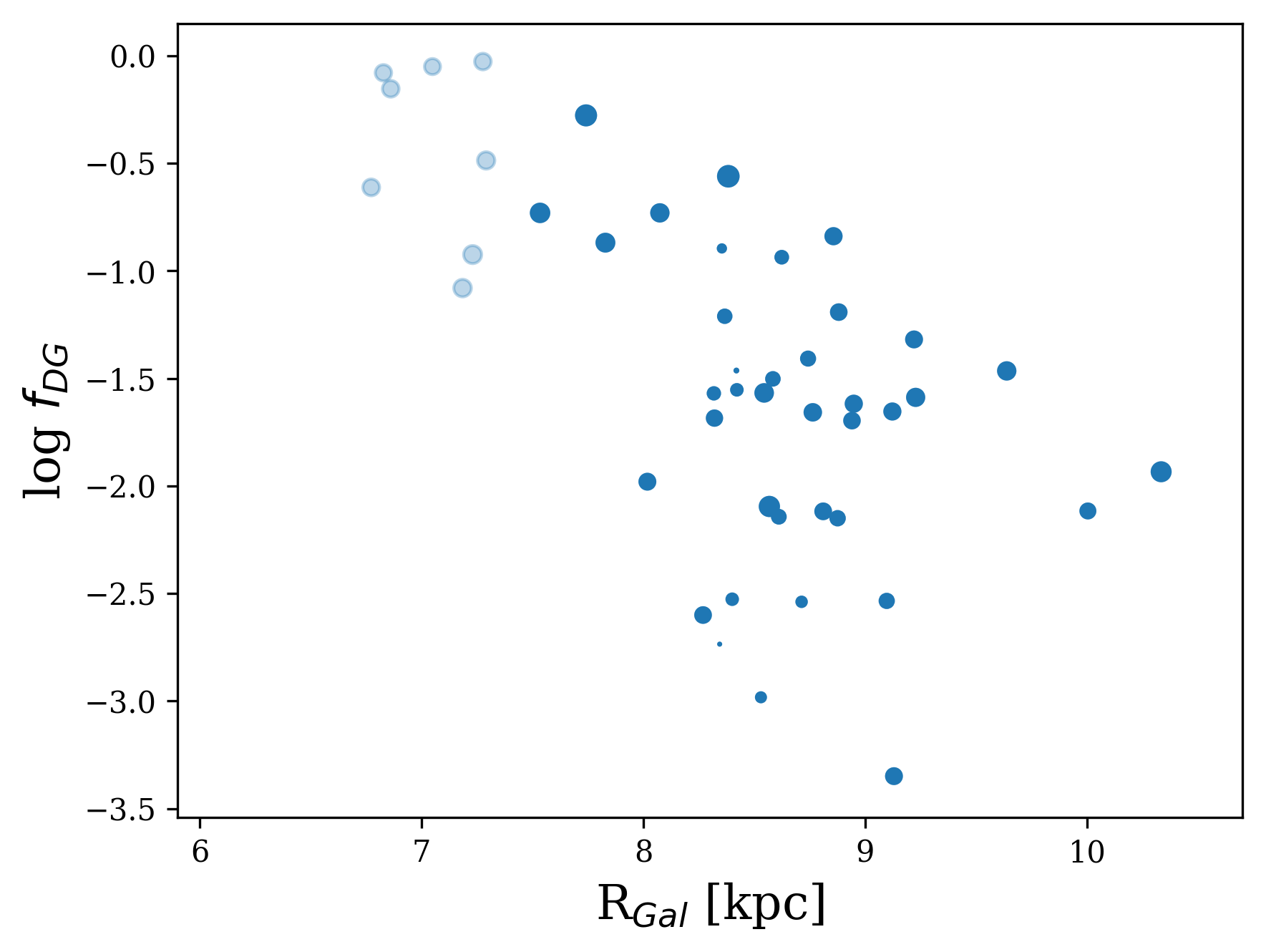}
    %Cloud_fdg_Rgal_pc-eps-converted-to.pdf}
    \includegraphics[width=0.9\columnwidth]{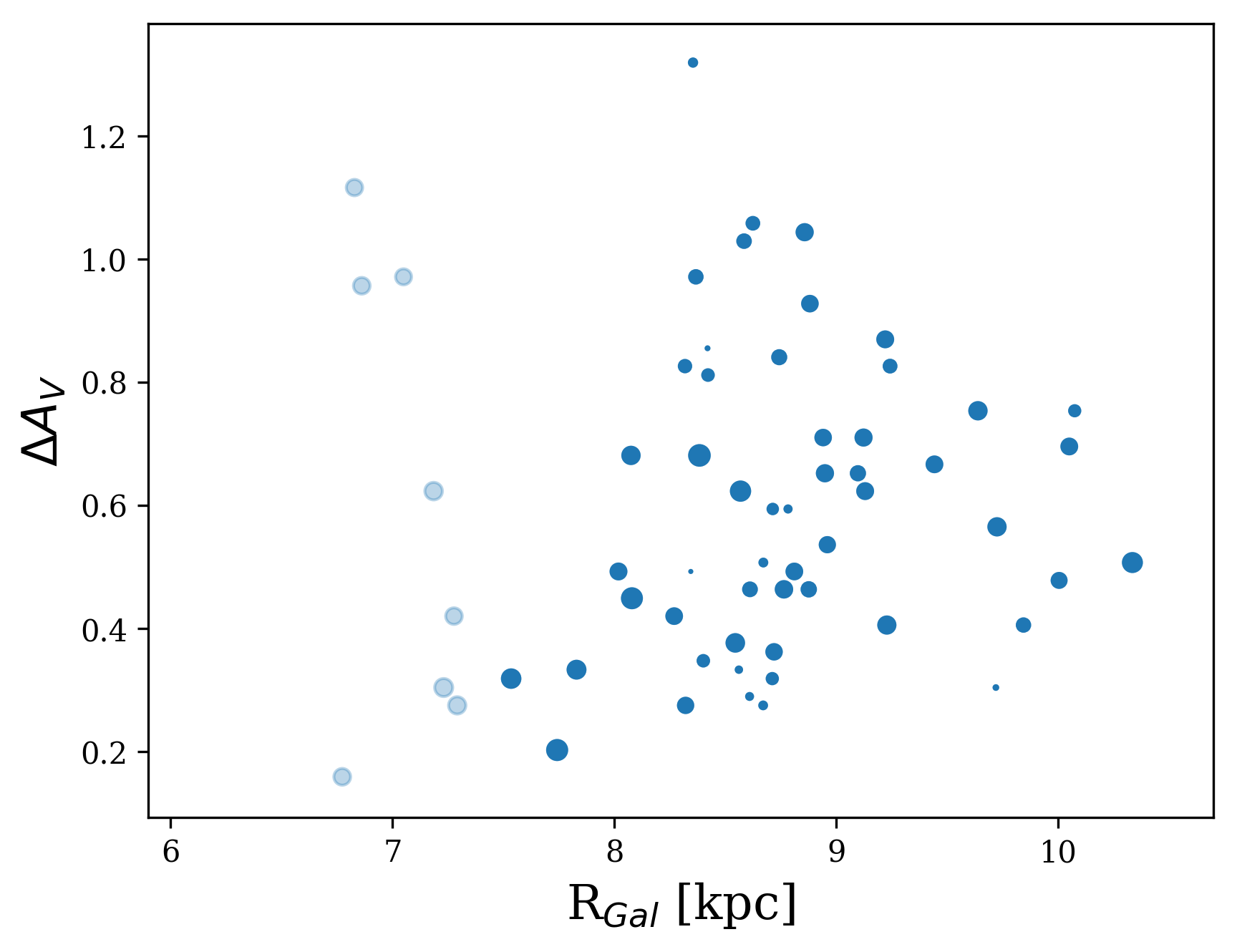}
    %Cloud_delAv_Rgal_pc-eps-converted-to.pdf}
    \caption{Dense gas measures of clouds as a function of galactocentric radius (blue circles). Clouds with $R_{\mathrm{Gal}}<7.5$ kpc are shown in lighter blue, and size corresponds to the cloud area. \emph{Top: }Mean surface density of the individual clouds as a function of the galactocentric radius. The green crosses show the data points from \citet{miville2017physical}, and the red diamonds shows the data from \citet{roman2010physical}. \emph{Middle: } Dense gas mass fraction, $f_\mathrm{DG}$, as a function of the galactocentric radius. \emph{Bottom: }Density contrast, $\Delta A_\mathrm{V}$, as a function of the galactocentric radius.} 
    \label{fig:galactic_densgas_profile}
\end{figure}

It is possible that the high mean column densities of these six clouds capture at least in part a real underlying trend. We studied the mean surface densities and the dense gas measures of these clouds as a function of the galactocentric radius (Fig. \ref{fig:galactic_densgas_profile}). We find that there is a possible correlation; the clouds closer to the Galactic centre have higher mean column or surface densities (p-value: $10^{-9}$, R-value: $-$0.66) and higher dense gas fractions (p-value: $10^{-6}$, R-value: $-$0.62), but no correlation is seen for $\Delta A_{\mathrm{V}}$ (p-value: 0.8). The R and p values here and in further correlation tests were computed using the Python Scipy package \texttt{linregress}, which uses the Wald test \citep{2020SciPy-NMeth}.

%-------------------------------------------------
% FIGURE: Color-coded (NYSO) N-PDFs

\begin{figure*}
    \centering
        \includegraphics[width=0.34\textwidth]{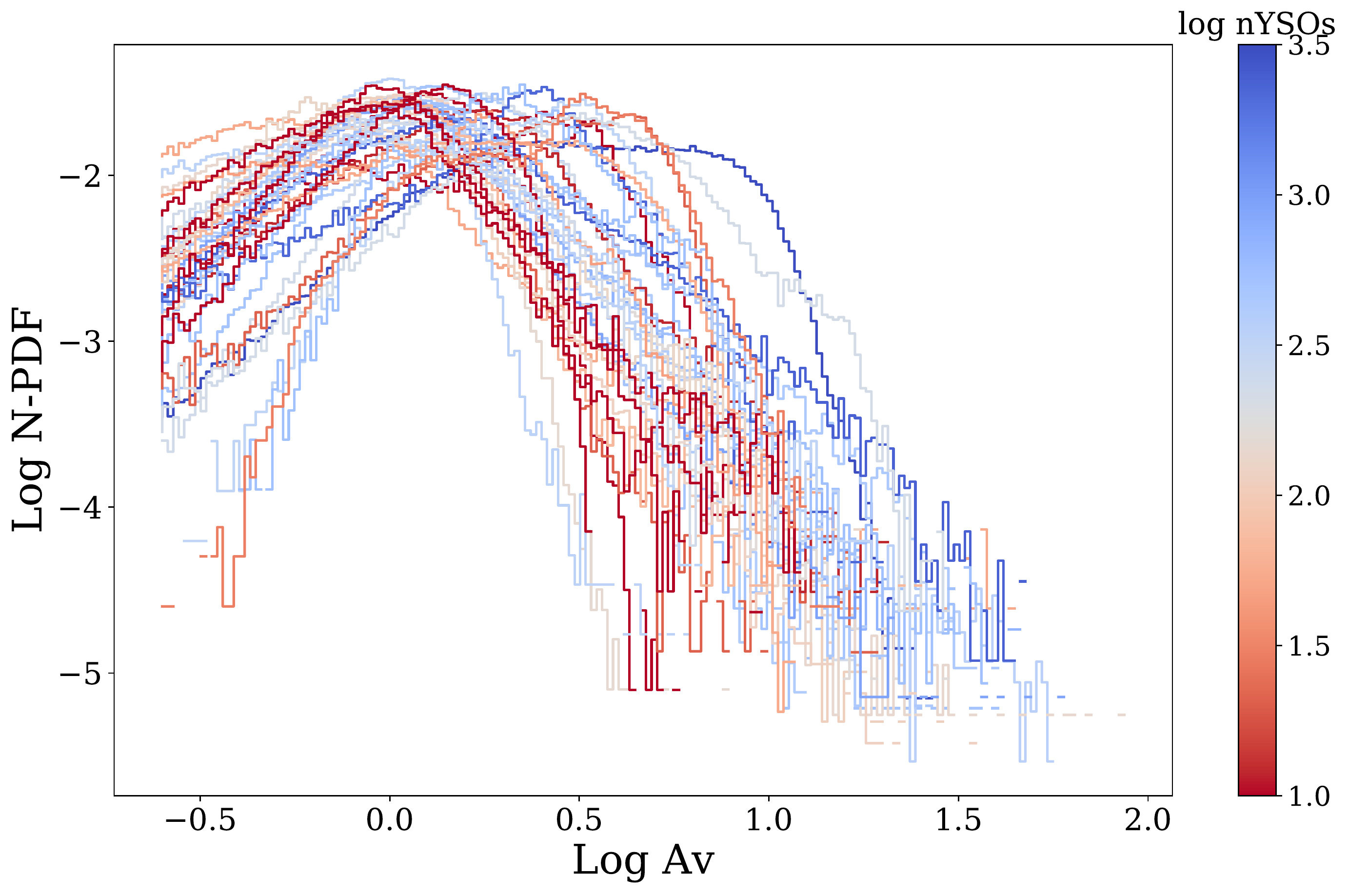}
        \includegraphics[width=0.32\textwidth, trim=55 55 40 30, clip]{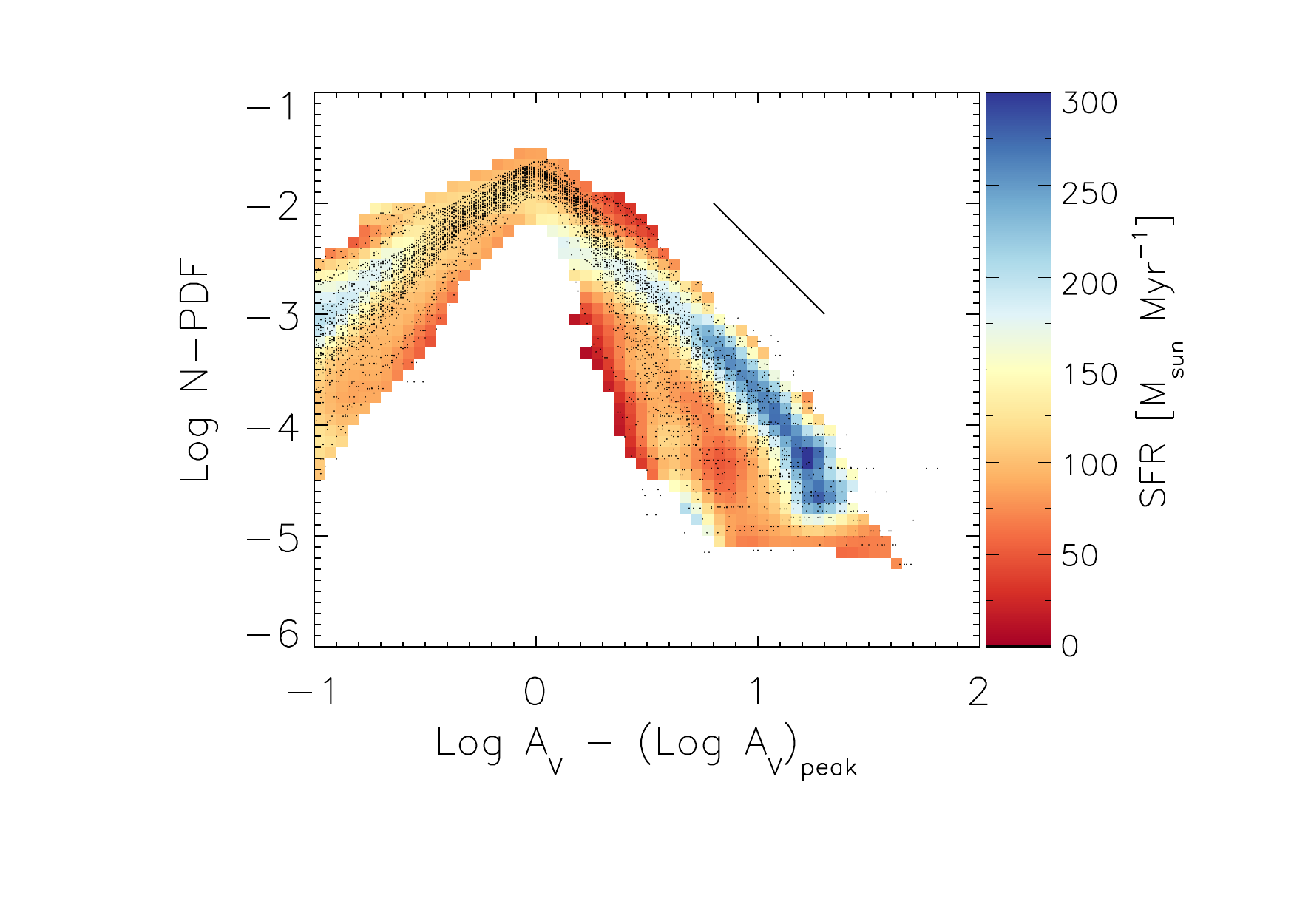}
        \includegraphics[width=0.32\textwidth, trim=55 55 40 30, clip]{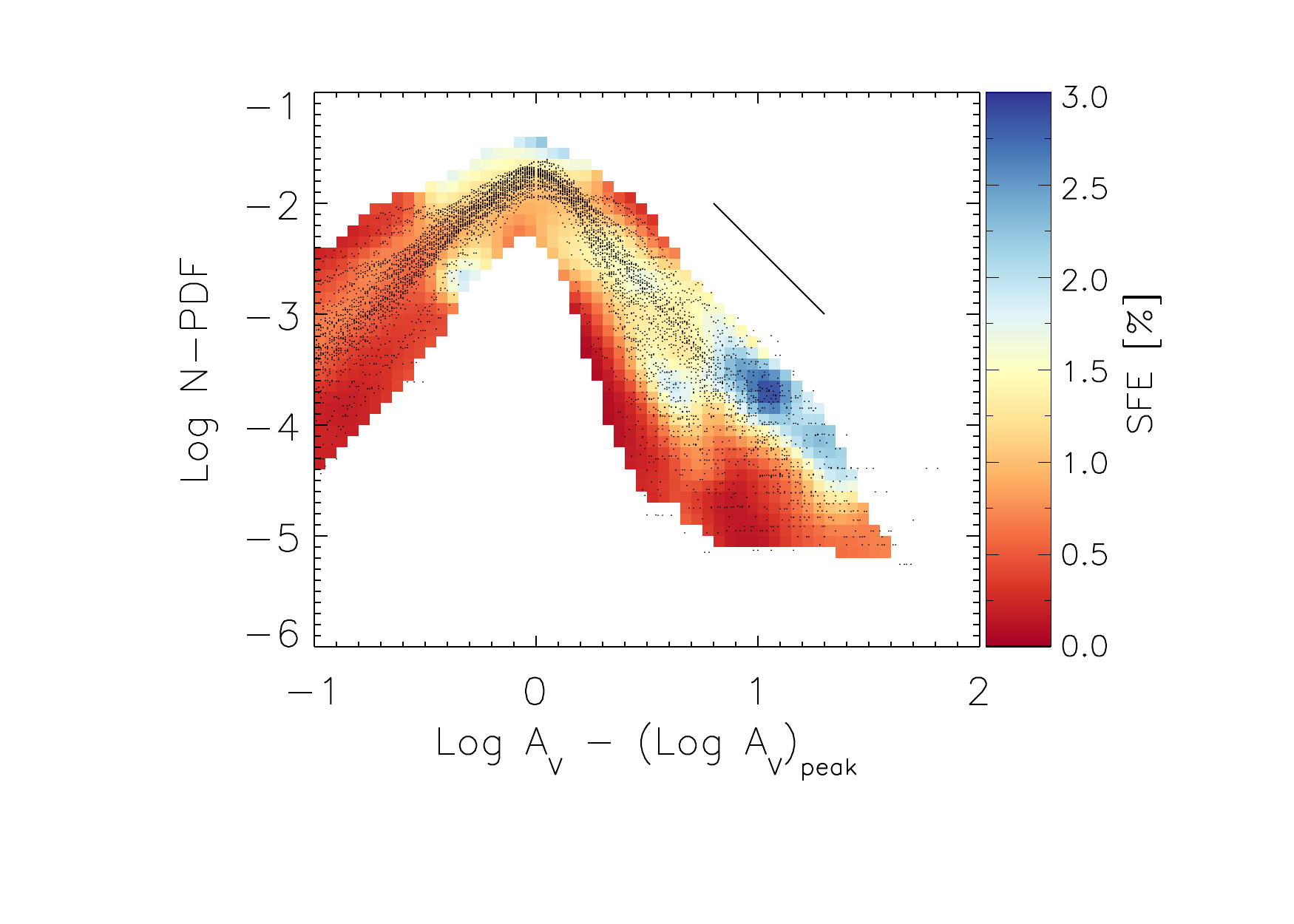}
    \caption{N-PDFs of the individual clouds with $R_{\mathrm{Gal}}>7.5$ kpc for which the SFR is known. \emph{Left:} N-PDFs colour-coded with the SFR. \emph{Centre:} Peak-matched N-PDFs with background colour-coding that shows the mean SFR of the N-PDFs at a given location. \emph{Right:} The same for the SFE. The black lines denote a power law with slope -2.}
    \label{fig:PDFs_YSOcolor}
\end{figure*}

%--------------------------------------------------------------------
%------------------------------------------------------
% FIGURE: Dense gas mass fraction and SFR/SFE

\begin{figure*}
    \centering
    \includegraphics[width=0.9\columnwidth]{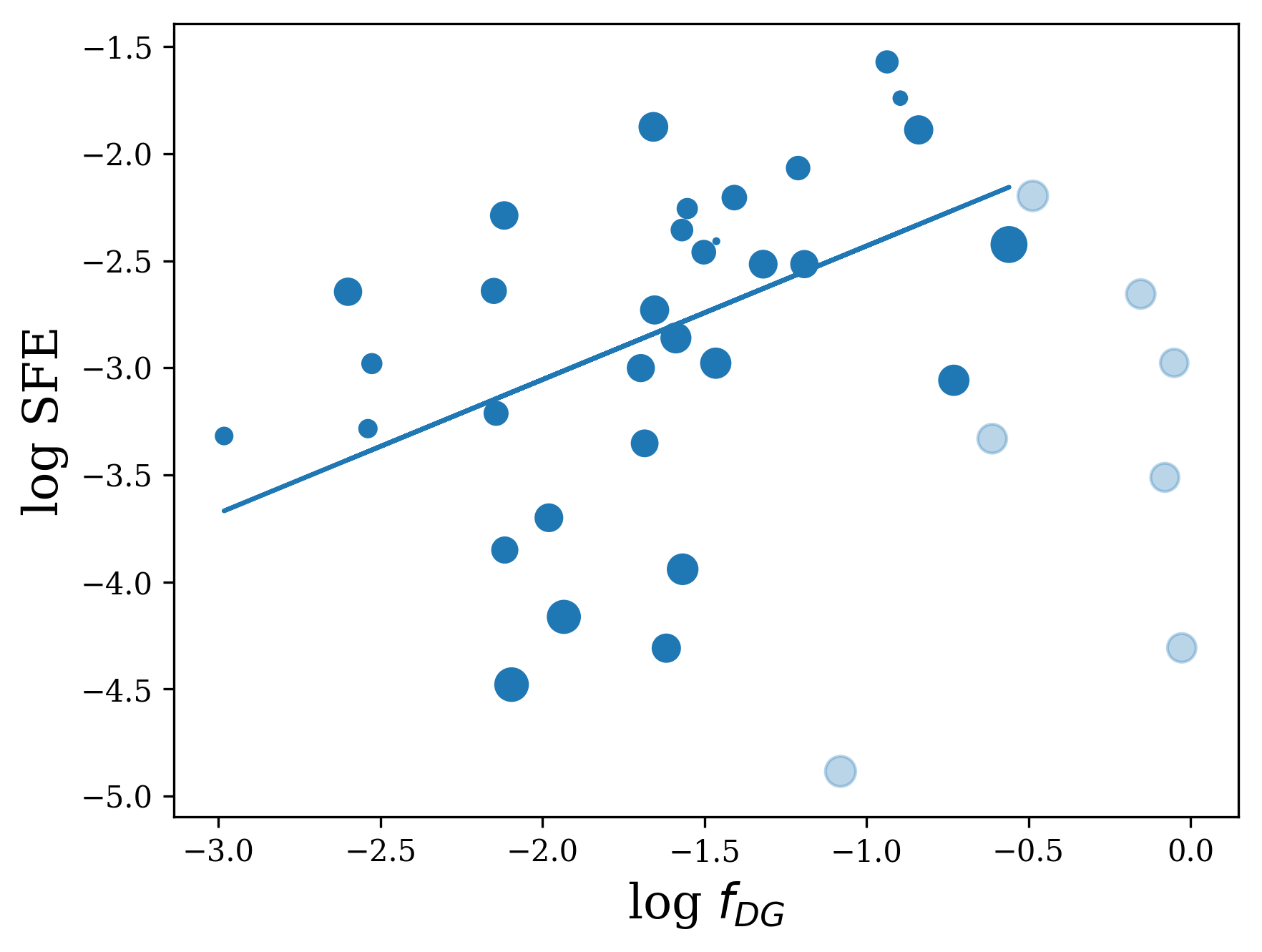}
    \includegraphics[width=0.9\columnwidth]{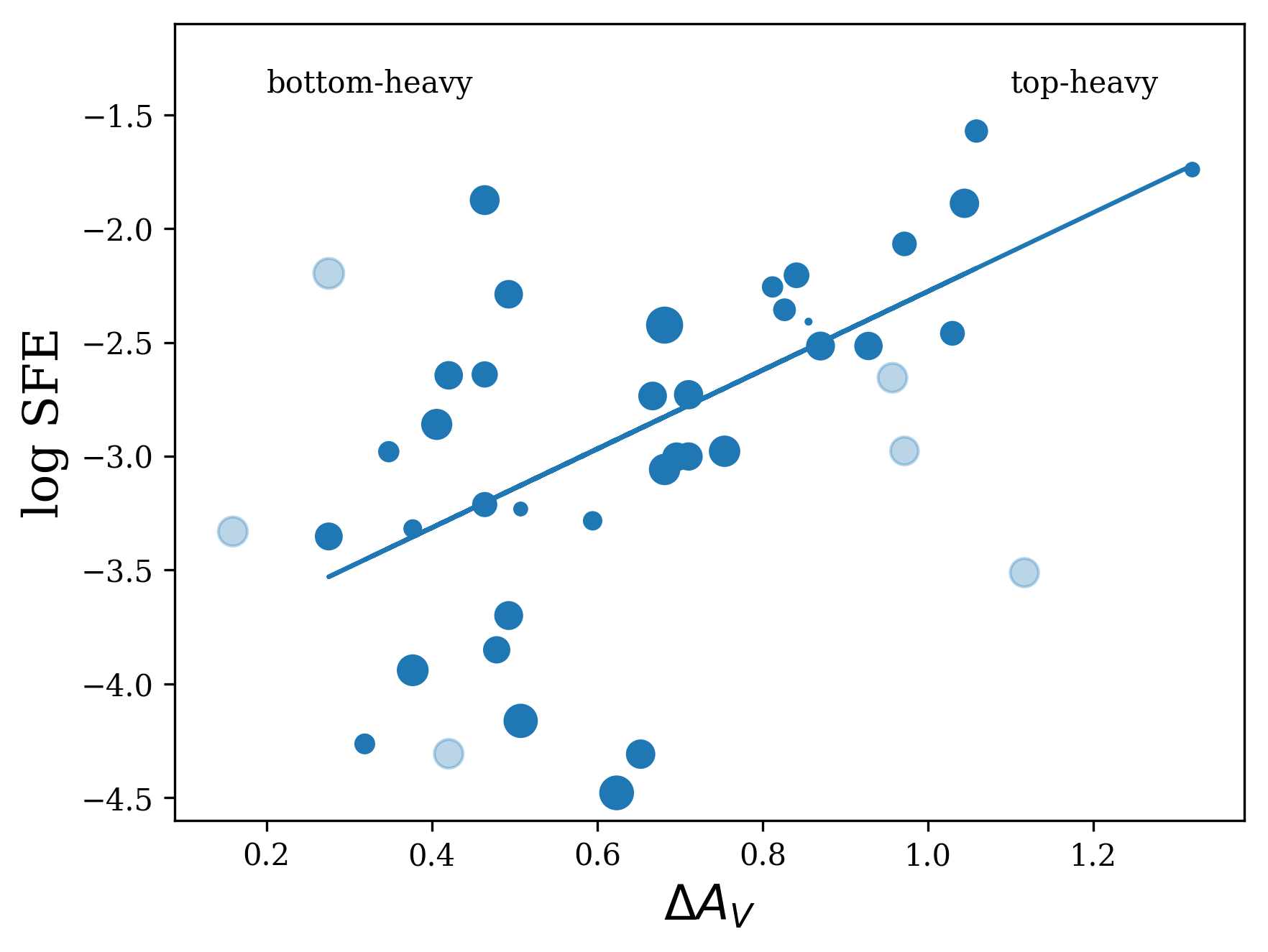}
    \caption{Relation of SFE and the dense gas measures of individual clouds. Light blue shows clouds with $R_{\mathrm{Gal}}<7.5$ kpc kpc. The symbol size corresponds to the cloud area $>3$mag. \emph{Left:} Using the dense gas mass fraction, $f_\mathrm{DG}$, measured using two column density thresholds, $A_\mathrm{V}^\mathrm{dense} = 8$ mag and $A_\mathrm{V}^\mathrm{all} = 1$ mag. The line shows a fit with $a = 0.6\pm0.2$, similar to the relation found previously for nearby molecular clouds \citep[e.g.][]{kainulainen2014unfolding}. \emph{Right:} Using the relative density contrast, $\Delta A_\mathrm{V}$. The line shows the fit of SFE $\propto 10^{a\Delta A_\mathrm{V}}$ with $a = 1.7\pm0.4$. }
    \label{fig:SF_vs_densegasfrac}
\end{figure*}

The trend of higher surface density and $f_{\mathrm{DG}}$ of clouds towards the galactic centre could be directly linked to the galactocentric profile of molecular cloud surface density in the Milky Way. This was studied by  \citet{miville2017physical}, for example, who found a gradient in annulus-averaged surface densities of CO clouds at $R_\mathrm{Gal}\gtrsim$5 kpc (see their Fig. 9, also shown here in the top panel of Fig. \ref{fig:galactic_densgas_profile}). A similar gradient is seen in the surface densities of clouds based on CO and $^{13}$CO data of the Galactic ring survey (GRS), but with a clearly different calibration of masses \citep[][also shown in Fig. \ref{fig:galactic_densgas_profile}]{roman2010physical}. Using higher transition $^{13}$CO (3-2) data on a coverage similar to the GRS, \citet{rigby2019chimps} found a slight increase in the mean density with decreasing galactocentric radius. The gradient we see in the surface densities of our clouds is roughly consistent with the gradients of these works. It therefore seems possible that our dust-based data reflect the same correlation, as opposed to being dominantly hampered by mapping or confusion issues.

%-------------------------------------------------------------------------
\subsubsection{Cloud N-PDFs and star formation}
\label{sec:results_ind_npdfs_and_sf}
%-------------------------------------------------------------------------

% First topic: N-PDFs and SF (the "color-coded" N-PDFs)

Figure \ref{fig:PDFs_YSOcolor} shows N-PDFs of all the clouds outside the spiral arm environment for which we have YSO information, colour-coded with the number of YSOs in the cloud (proxy for the SFR). The figure indicates that the clouds that have more YSOs occupy a different region of the N-PDFs compared to those with fewer YSOs. This originates from the correlation between the amount of dense gas and the SFR of the clouds: clouds with higher star formation rates tend to have more top-heavy N-PDFs. This correlation has commonly been reported in previous studies of smaller samples \citep[e.g.][]{kainulainen2009probing,schneider2013pdfs,lombardi2015molecular}. We emphasise here that this correlation seems to be driven by the changes in the \emph{\textup{relative}} amount of dense gas. We demonstrate this by shifting the N-PDFs by their logarithmic peak value, hence examining their relative shapes. These shifted N-PDFs are shown in the second panel of Fig. \ref{fig:PDFs_YSOcolor}, where the colour scale is the mean SFRs of the clouds whose N-PDFs go through the given pixel. The figure indicates a clear correlation between the relative N-PDF shape and SFR: the clouds with more top-heavy N-PDFs have higher SFRs. The same is seen in SFE: clouds with top-heavy N-PDFs have higher SFEs (third panel Fig. \ref{fig:PDFs_YSOcolor}). These results strengthen the picture that both SFR and SFE are sensitive to how exactly the gas is distributed in the clouds.

% Second topic: SFR/SFE and the simple models.

Ideally, it would be desirable to link the star formation activities of the clouds to the parameters of the models used to quantify the N-PDF shapes (LN, PL, and LN+PL). However, this is difficult because only relatively few clouds fitted by each model have YSO information. The LN+PL model is the best-fitting model for most clouds, but no significant correlations are detected between the model parameters and the star formation rate. Quantifying the N-PDF shapes with the simple models is further complicated by the fact that the best-fit model depends on the SFR of the cloud. The PL model is the best fit only for clouds that have relatively high SFRs, while the LN+PL model can be the best model for clouds over a wide range of SFRs. As a result of these issues, it is not straightforward to use the model parameters as a simple, generally applicable proxy of the cloud SFR. 

% Third topic: The relationships between the empirical measures of dense gas and SFR/SFE

We also quantified the link between the N-PDF shapes and star formation using the empirical model-independent dense gas measures (dense gas mass fraction, $f_\mathrm{DG}$, and density contrast, $\Delta A_\mathrm{V}$; see Sect. \ref{sec:results_npdfs}). Figure \ref{fig:SF_vs_densegasfrac} shows the relations between SFE and these measures. The relation between SFE and $f_\mathrm{DG}$ is characterised by the concentration of clouds roughly along a relation that has a power-law slope of $\sim$1.5; this relation has been claimed by previous studies and hypothesised to be fundamentally linked to the ability of clouds to form stars \citep[e.g.][]{lada2012star}. This relation breaks down when clouds with $R_{\mathrm{Gal}}<7.5$ kpc are included (in the spiral-arm environment; see Fig. \ref{fig:SF_vs_densegasfrac}). This suggests that the relation between the dense gas and star formation is specific for a given galactic environment.

The relation between SFE and $\Delta A_\mathrm{V}$ establishes a correlation between star formation and the relative amount of dense gas, or top-heaviness of the N-PDF, in our sample (Fig. \ref{fig:SF_vs_densegasfrac}, right panel). While the scatter in the relation is high, the correlation is moderate ($R=0.42$) and significant ($p$-value = 0.019; 31 clouds). A fit of a line, SFE $\propto 10^{a\Delta A_\mathrm{V}}$, to the clouds with $R_{\mathrm{Gal}}<7.5$ kpc yields the slope of $1.7\pm 0.4$. When all clouds are included, the correlation remains, but with a lower intercept.

%%%%%%%%%%%%%%%%%%%%%%%%%%%%%%%%%%%%%%%%%%%%%%%%%%%%%
\subsection{Bird's eye view}%Density PDF}
\label{sec:results_bird}
%%%%%%%%%%%%%%%%%%%%%%%%%%%%%%%%%%%%%%%%%%%%%%%%%%%%%

We now place ourselves in the position of an observer outside the Milky Way, viewing our survey area through apertures of various sizes (see Sect. \ref{sec:mbird}). In this setup, we study the shapes of the N-PDFs within the apertures, the dependence of the shapes on the scales they are measured and on the Galactic environment, and the relation between the aperture N-PDFs and star formation activity. We use apertures with radius 0.25-2 kpc.

%----------------------------------------------------------------
\subsubsection{Variety of aperture N-PDFs}
\label{sec:results_aps}
%----------------------------------------------------------------

\begin{figure}%[h!]
    \centering
        \includegraphics[width=0.49\textwidth]{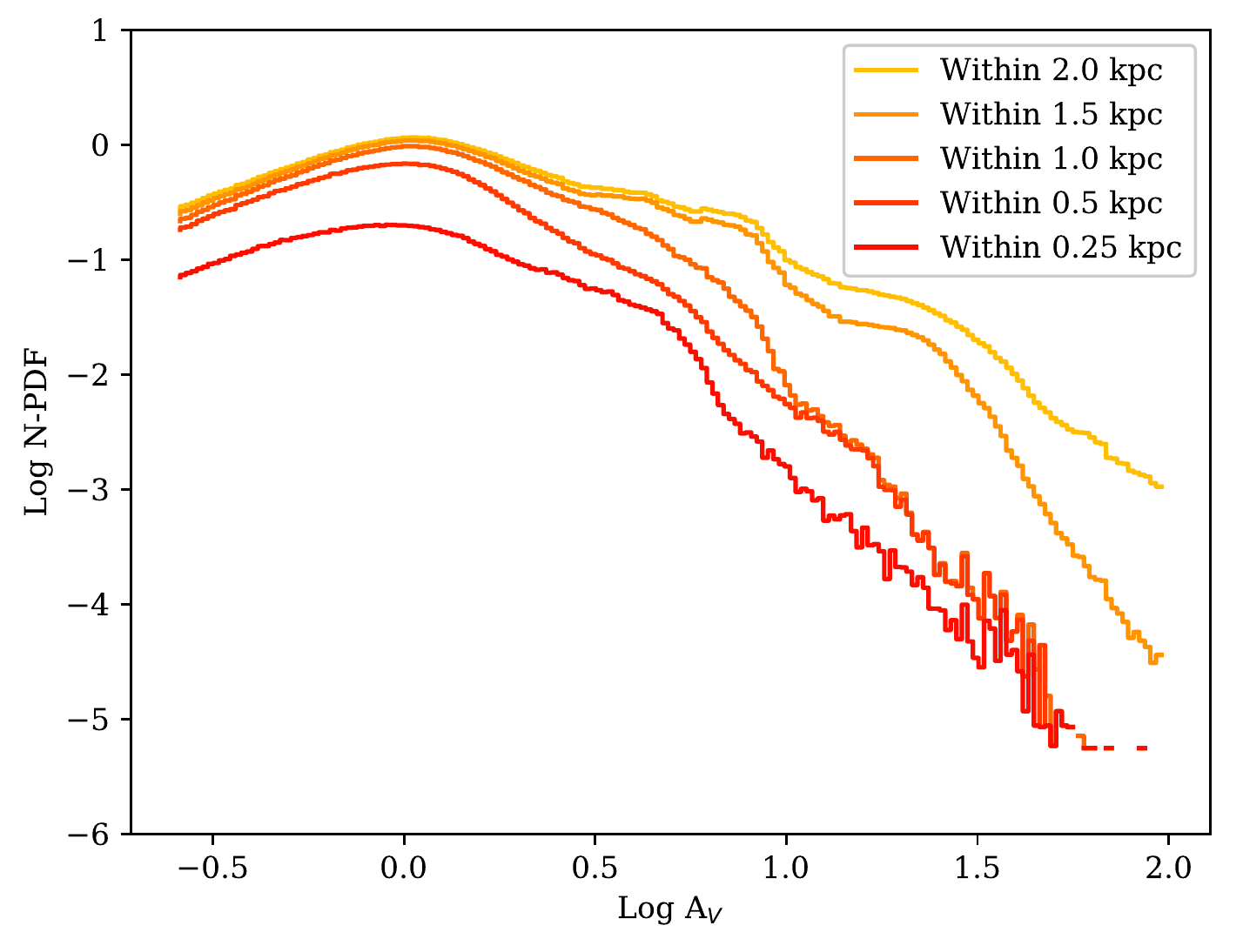}
    \caption{Aperture N-PDFs of apertures centred on the Sun, with radii between 0.25-2 kpc.}
    \label{fig:PDFs_ap}
\end{figure}

Aperture N-PDFs were derived for the entire survey area, with aperture radii of 0.25-2 kpc (as described in Sect. \ref{sec:mbird}). Table \ref{tab:ap_shapes_new} lists the derived properties within the apertures separately for the entire survey area and the area at galactocentric radius larger than 7.5 kpc (i.e. the region unrelated to the spiral arm environment). To illustrate the aperture N-PDFs, Fig.  \ref{fig:PDFs_ap} shows as a special case the N-PDFs centred on the Sun. The shapes of the aperture N-PDFs decay roughly in a power-law manner above a few magnitudes of $A_\mathrm{V}$. Even though single power-law models do not result in statistically acceptable fits of the N-PDFs, we used them to obtain a rough quantification of the shapes. The resulting slopes span the range $\alpha \approx [-2.7, -3.4]$ (steeper slopes for smaller apertures, see Table \ref{tab:ap_shapes_new}). Generally, the aperture N-PDFs are relatively top-heavy compared to the spectrum of individual cloud N-PDFs because the active star-forming clouds dominate the gas content within the apertures by mass; the mass of the gas with low star formation activity is low.

% FIGURE: Aperture N-PDFs

\setcounter{table}{2}

%\input{W_aperture_table}
% Table of the aperture N-PDF properties
\begin{table*}
\caption{Properties of the aperture N-PDFs.} % title of Table
\label{tab:ap_shapes_new}      % is used to refer this table in the text
\centering                          % used for centering table
\begin{tabular}{c c c c c c c c c c}        % centered columns (4 columns)
\hline\hline                 % inserts double horizontal lines
R  & N\tablefootmark{a} & M$_{A_\mathrm{V} > 3 \ \mathrm{mag}}$   &  $\alpha$\tablefootmark{b} & $\overline{\Sigma}$\tablefootmark{c} & SFR & $\frac{\mathrm{SFR}}{A}$\tablefootmark{d} & SFE &  $f_\mathrm{DG}$ & $\Delta A_\mathrm{V}$\\   
\hline
kpc & & $10^6$ M$_\odot$ & & M$_\odot$ pc$^{-2}$ & $10^3$M$_\odot$ Myr$^{-1}$ & $10^3$M$_\odot$ Myr$^{-1}$kpc$^{-2}$ & \% & \% &  \\
\hline      
2.00\tablefootmark{e} & 1 & 8.8 & -1.9 & 0.7  & 9.7 & 0.8  & 0.13 &  23  & 1.45 \\ 
1.00    &       13      &       6       $\pm$   2       &        -2.7   $\pm$     0.2    &         1.2   $\pm$     0.3   &         7     $\pm$     2     &         2.1    $\pm$     0.6   &       0.10    $\pm$   0.02    &       17      $\pm$    5       &         0.84  $\pm$     0.09   \\ 
0.50    &       49      &       1.8     $\pm$   0.4     &        -3.1   $\pm$     0.2    &         1.5   $\pm$     0.4   &         2.2   $\pm$     0.6   &         2.8    $\pm$     0.8   &       0.08    $\pm$   0.02    &       12      $\pm$    3       &         0.58  $\pm$     0.05   \\ 
0.25    &       213     &       0.38    $\pm$   0.08    &        -3.3   $\pm$     0.2    &         1.1   $\pm$     0.3   &         0.4   $\pm$     0.1   &         2.1    $\pm$     0.7   &       0.04    $\pm$   0.01    &        5      $\pm$    1       &         0.27  $\pm$     0.02   \\ 
\hline  
\multicolumn{9}{c}{Apertures with R$_\mathrm{Gal}>$ 7.5 kpc} \\
\hline  
1.00    &       11      &       6       $\pm$   2       &        -2.9   $\pm$     0.2    &         1.1   $\pm$     0.3   &         7     $\pm$     2     &         2.3    $\pm$     2.4   &       0.10    $\pm$   0.02    &       12      $\pm$    4       &         0.8   $\pm$     0.1    \\ 
0.50    &       42      &       1.8     $\pm$   0.5     &        -3.3   $\pm$     0.2    &         1.3   $\pm$     0.4   &         2     $\pm$     5     &         3.0    $\pm$     0.9   &       0.09    $\pm$   0.02    &        8      $\pm$    2       &         0.55  $\pm$     0.05   \\ 
0.25    &       176     &       0.36    $\pm$   0.08    &        -3.4   $\pm$     0.1    &         1.0   $\pm$     4.6   &         0.4   $\pm$     0.2   &         2.2    $\pm$     0.8   &       0.05    $\pm$   0.01    &        3      $\pm$    1       &         0.27  $\pm$     0.02   \\ 
\hline                                   %inserts single line
\end{tabular}
\tablefoot{\\
\tablefoottext{a}{Number of apertures considered.}
\tablefoottext{b}{Slope of the power-law fit.}
\tablefoottext{c}{Mean surface density within the apertures (total cloud mass divided by aperture area).}
\tablefoottext{d}{$A$ is the area of the aperture.}
\tablefoottext{e}{This is our full survey area, therefore there is only one aperture, and no standard deviation is given.}
%\textcolor{red}{Steeper N-PDFs and lower fdg, but higher SFE and SFR for smaller apertures! Why.}\\
%\tablefoottext{a}{Fractional mass of clouds for which we have YSO data; a measure of the completeness of the YSO data.}\\
%\tablefoottext{b}{SFE computed only using the clouds for which we have YSO data.}% SFE = mass(YSOs)/mass(YSOs+mol). mass(YSOs)=0.5*nYSOs.}
}
\end{table*}

\begin{figure*}
    \centering
    \includegraphics[width=0.49\textwidth]{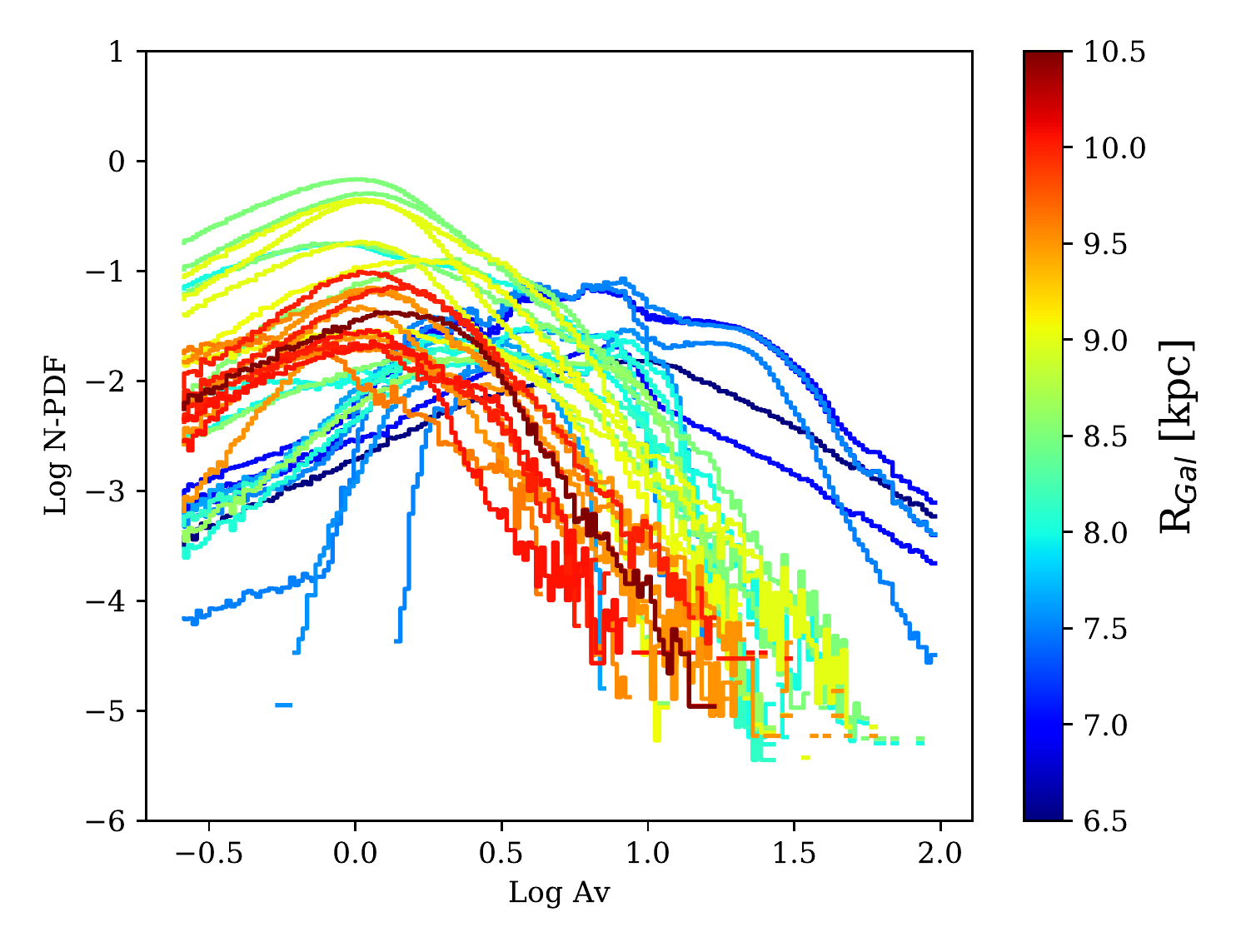}%.png}
    \includegraphics[width=0.49\textwidth]{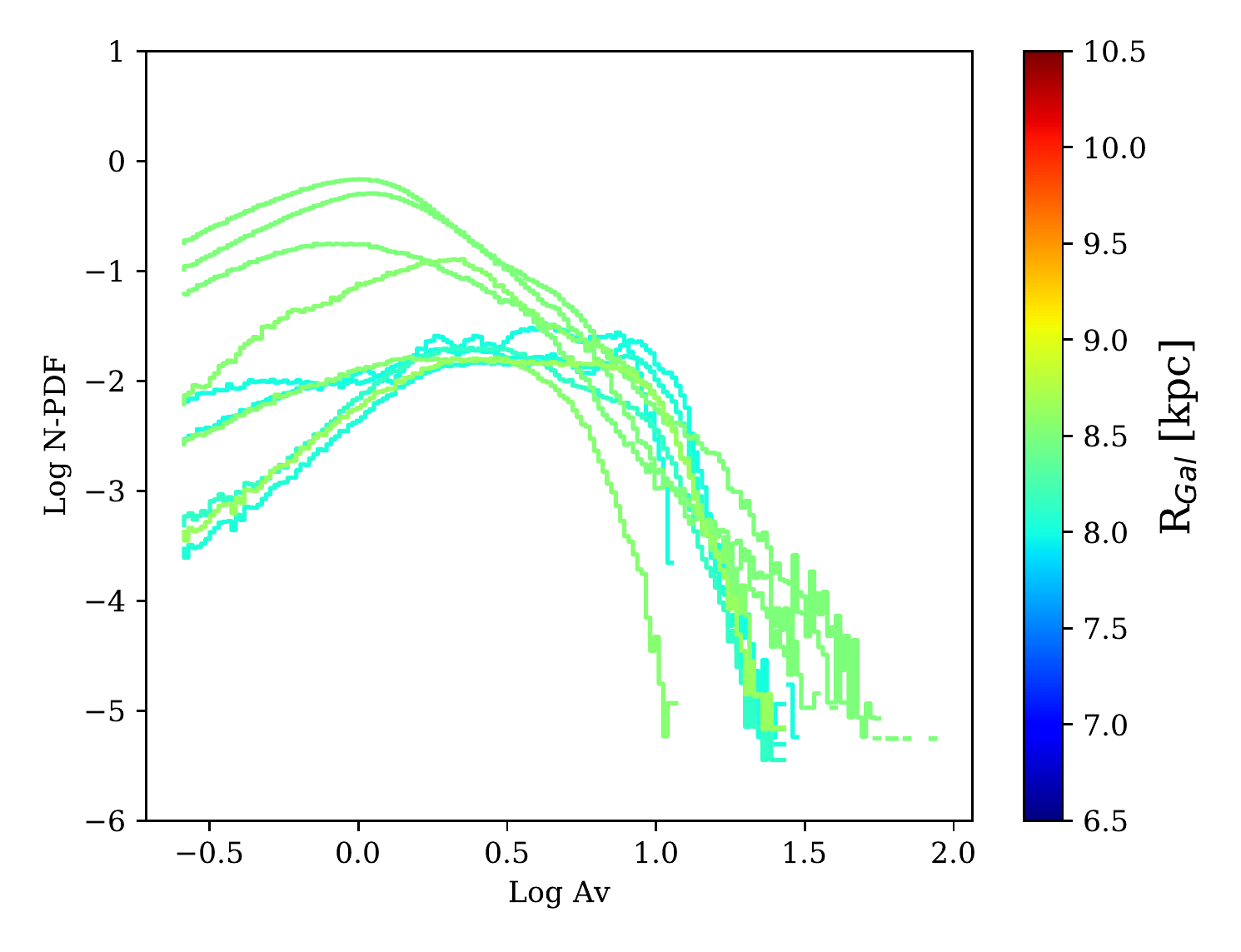}%.png}
    \caption{N-PDFs of $R=0.5$ kpc apertures. \emph{Left:} N-PDFs of $R=0.5$ kpc apertures covering the whole survey region, and \emph{right:} only apertures with centre at $R_{\mathrm{Gal}}=8-9$ kpc. The colour-coding is the same as in Fig. \ref{fig:move_cov}. }
    \label{fig:ap_moved}
\end{figure*}

The slopes of the aperture N-PDFs become flatter with increasing aperture size. This is also seen in the empirical values $f_{\mathrm{DG}}$ and $\Delta A_{\mathrm{V}}$ that have higher values for larger apertures. This trend is present even when only clouds at galactocentric radii larger than 7.5 kpc are considered (i.e. the solar inter-arm environment). It is also present even when we only fit power laws from the highest values of the last closed contour (5 mag). This suggests a scale dependence of the N-PDF shapes. This scale dependence probably arises because larger apertures are more likely to contain a larger variety of clouds, including massive clouds with very flat N-PDFs. The trend is even stronger when all apertures are considered, that is, when those are included that contain clouds in the spiral-arm environment. However, in this case, it seems clear that the deviant N-PDFs of massive clouds in the spiral-arm area contribute to the top-heaviness of the aperture N-PDFs; if these N-PDFs are strongly affected by confusion, the strength of the trend is over-estimated by our data. 

The aperture N-PDFs of a given size show significant variance. N-PDFs of $R$=0.5 kpc apertures are shown in Fig. \ref{fig:ap_moved}, revealing significant variation in the N-PDFs at this scale, and a possible dependence on galactic environment. The apertures closer to the Galactic centre reach higher column densities than those farther away. The right panel of Fig. \ref{fig:ap_moved} shows only the N-PDFs of apertures with $R_{\mathrm{Gal}}=$ 8-9 kpc, which likely represents the galactic environment of the solar neighbourhood. A significant beam-to-beam variation is seen within this environment as well. This variety is also seen in the standard deviations in Table \ref{tab:ap_shapes_new}. 

\begin{figure*}
    \centering
        \includegraphics[width=0.32\textwidth]{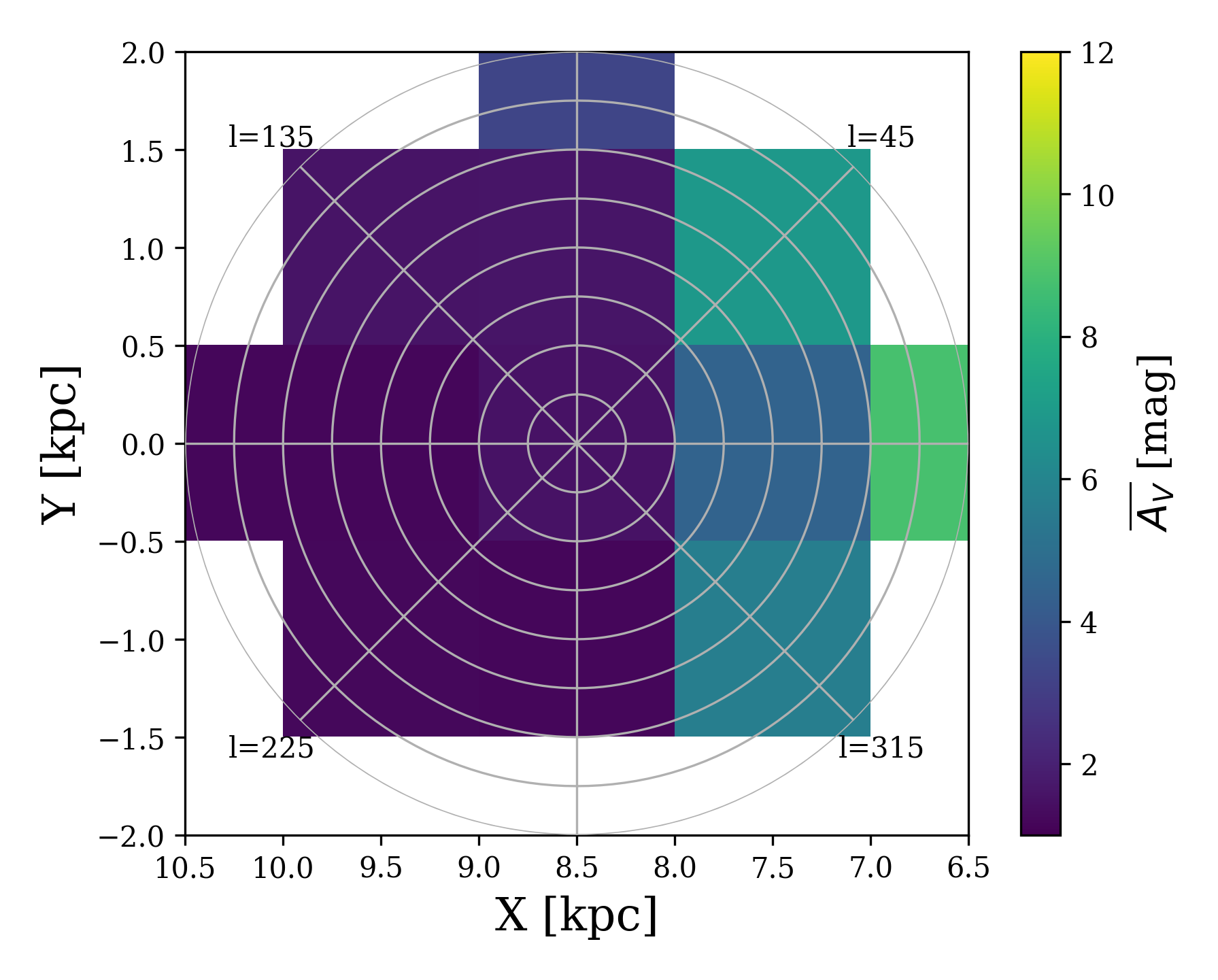}
        \includegraphics[width=0.32\textwidth]{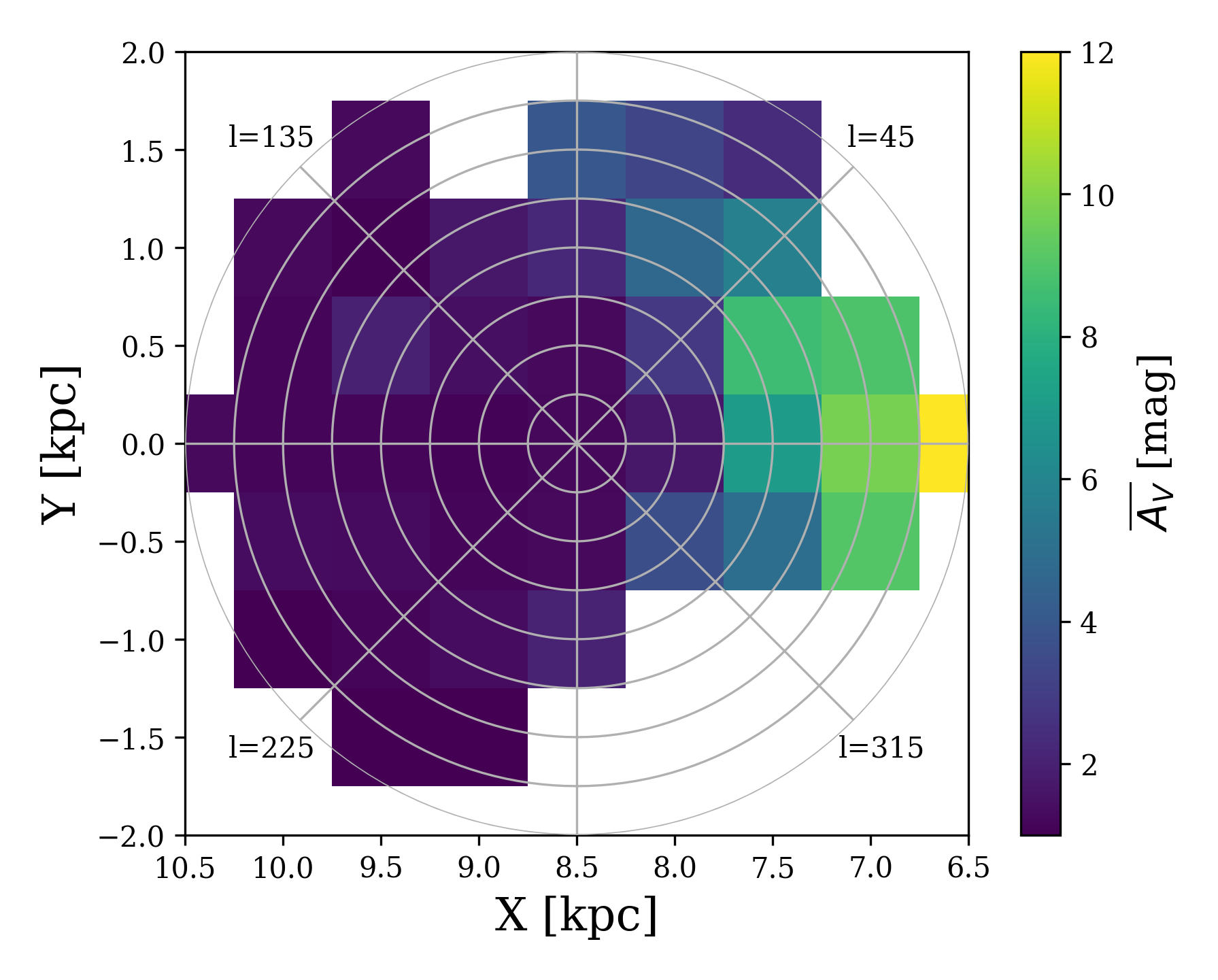}
        \includegraphics[width=0.32\textwidth]{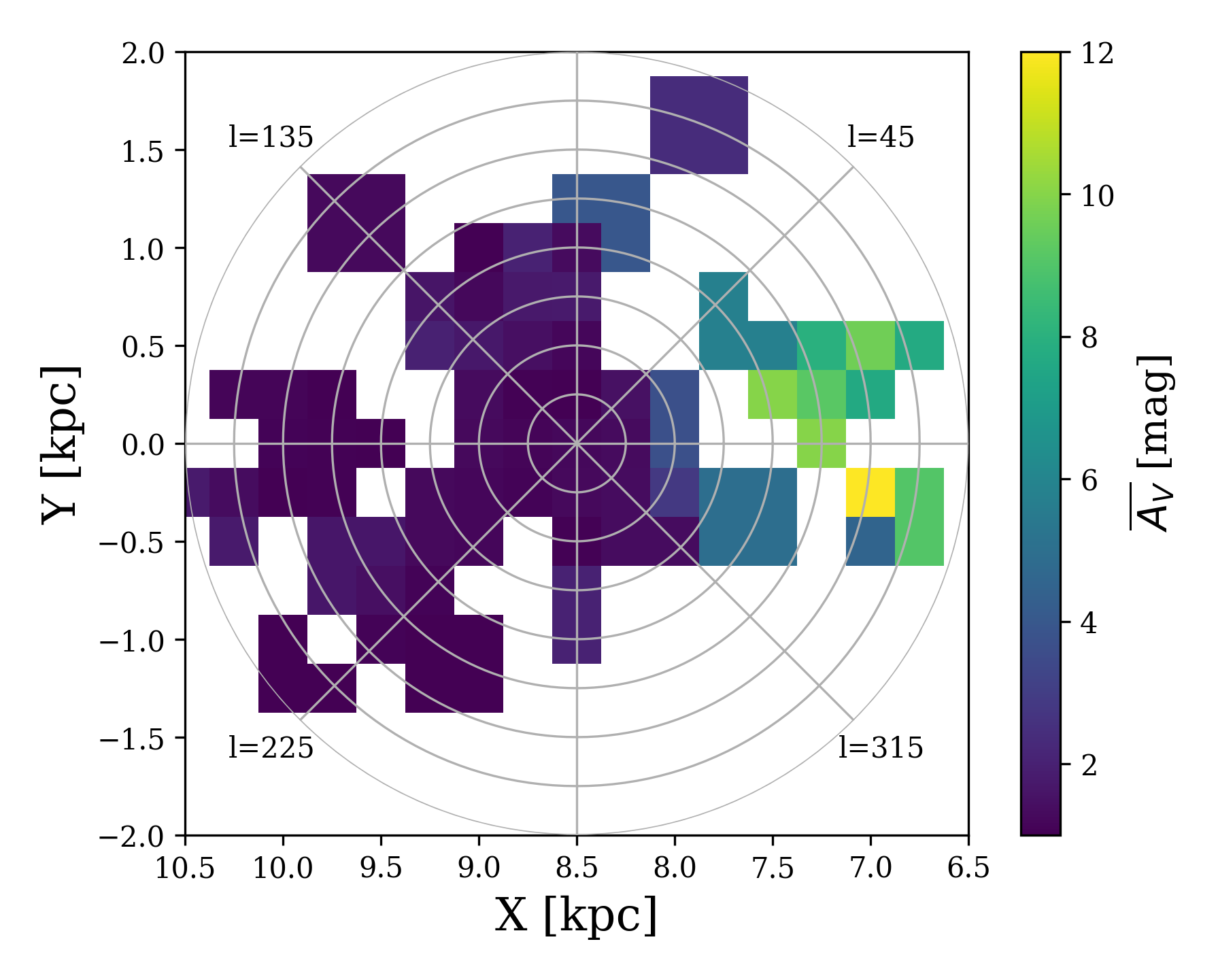}
        
        \includegraphics[width=0.32\textwidth]{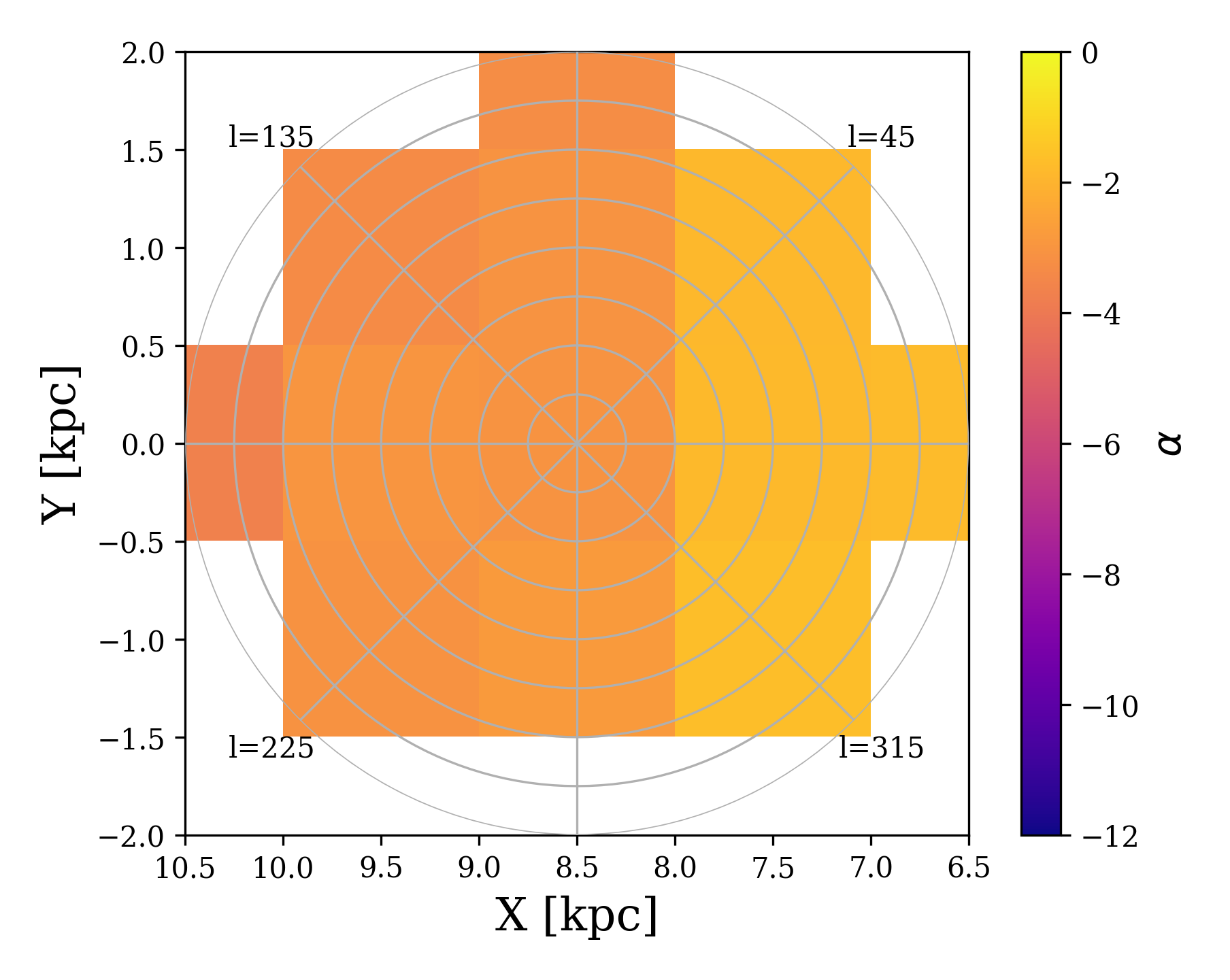}
        \includegraphics[width=0.32\textwidth]{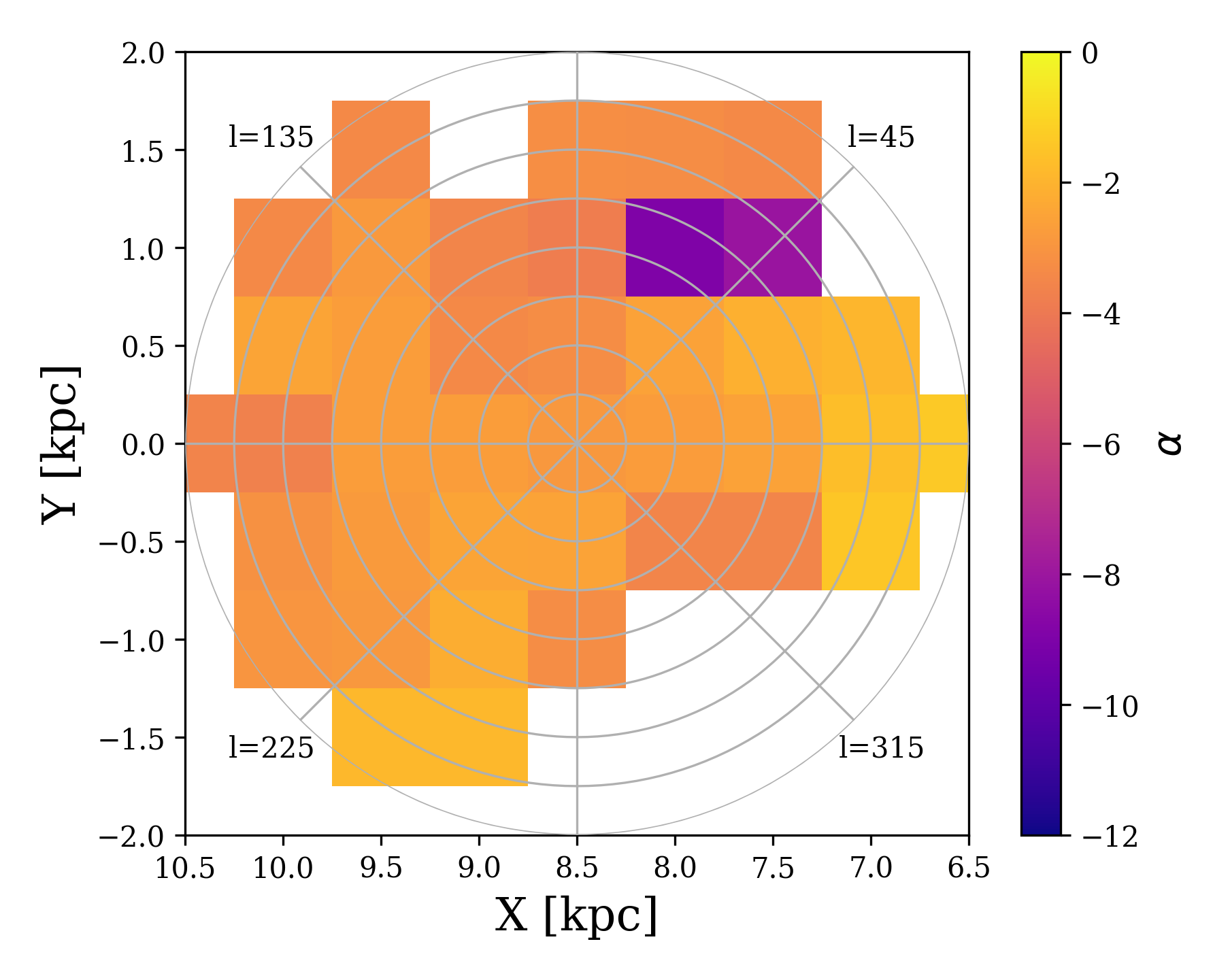}
        \includegraphics[width=0.32\textwidth]{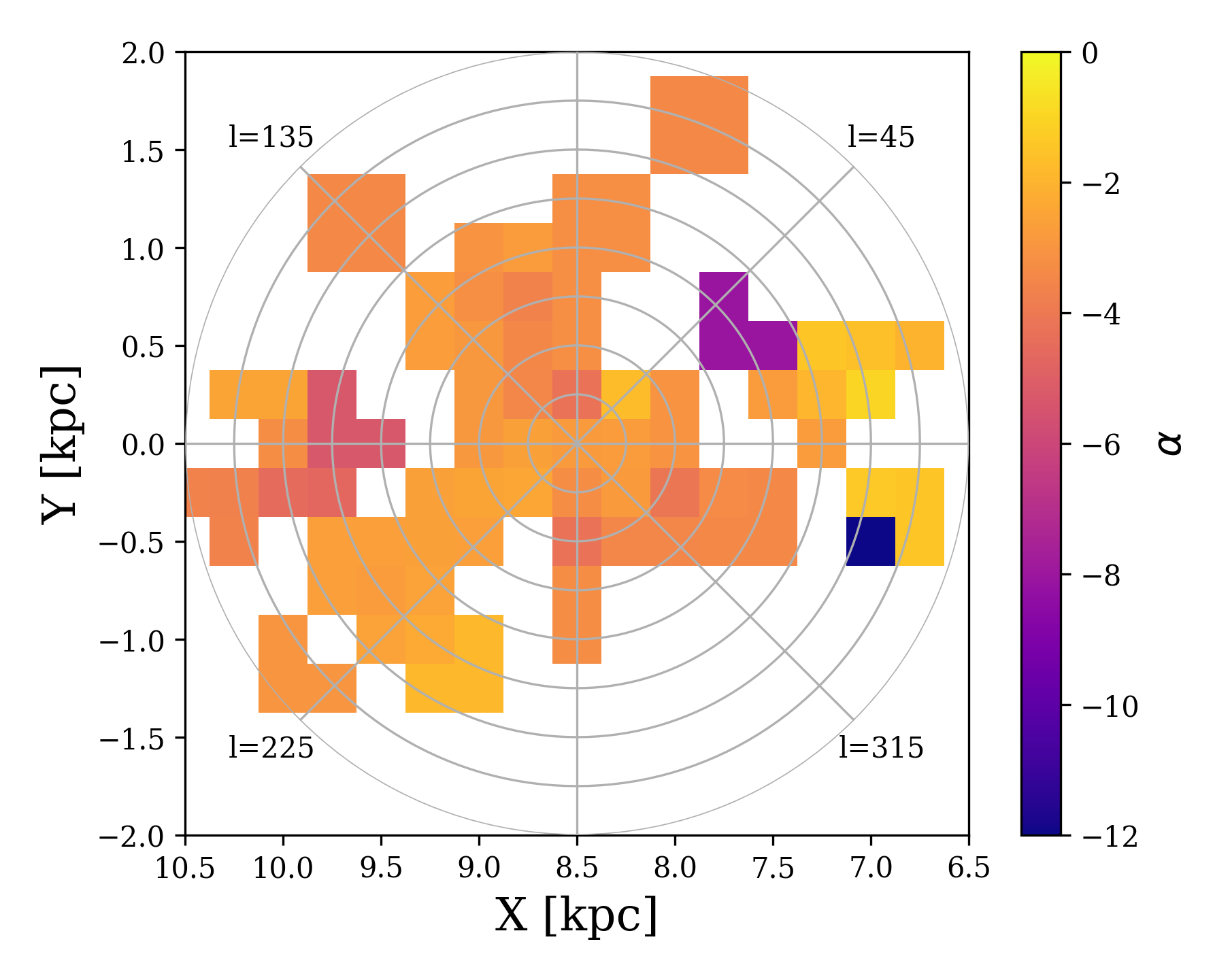}
        
        \includegraphics[width=0.32\textwidth]{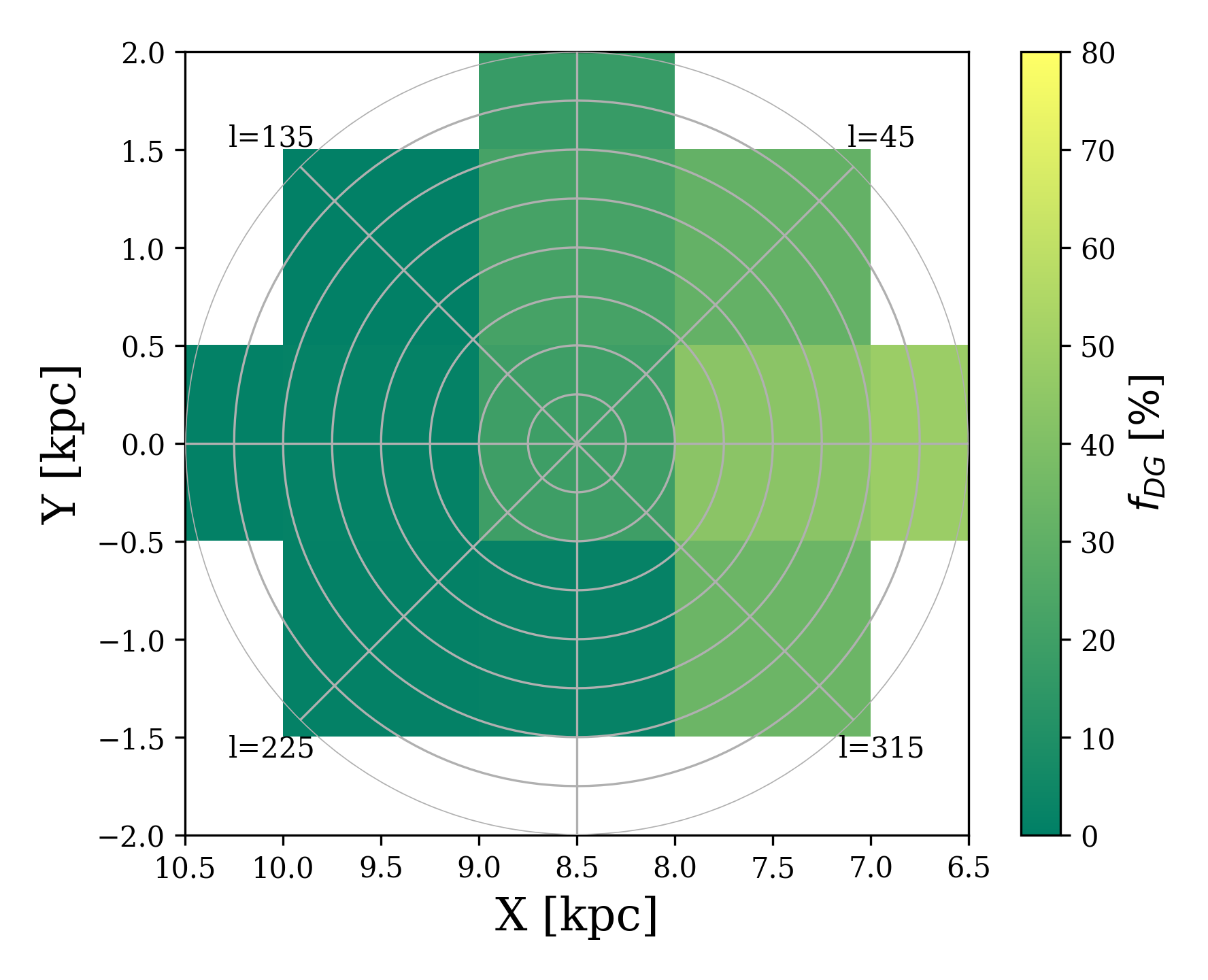}
        \includegraphics[width=0.32\textwidth]{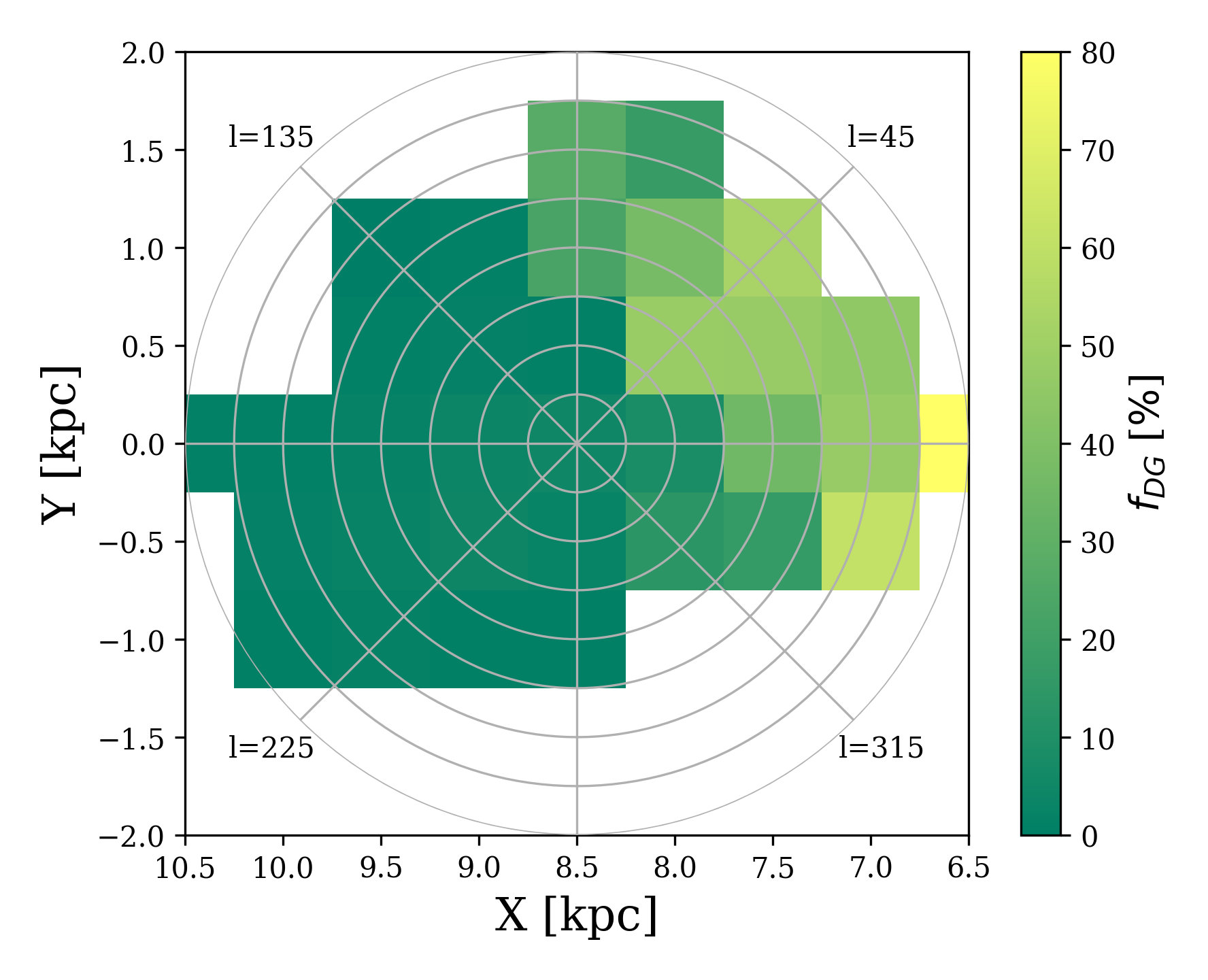}
        \includegraphics[width=0.32\textwidth]{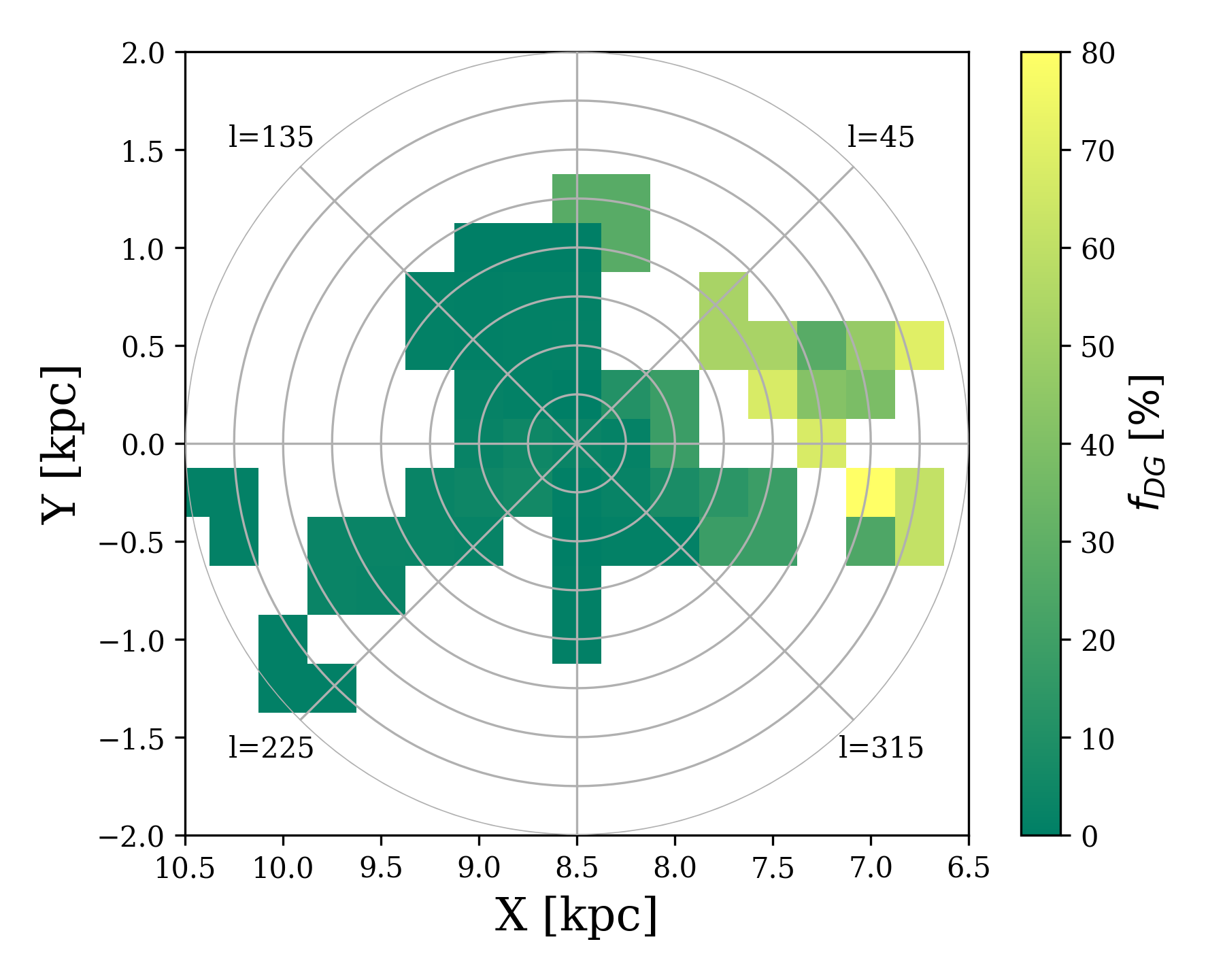}
        
        \includegraphics[width=0.32\textwidth]{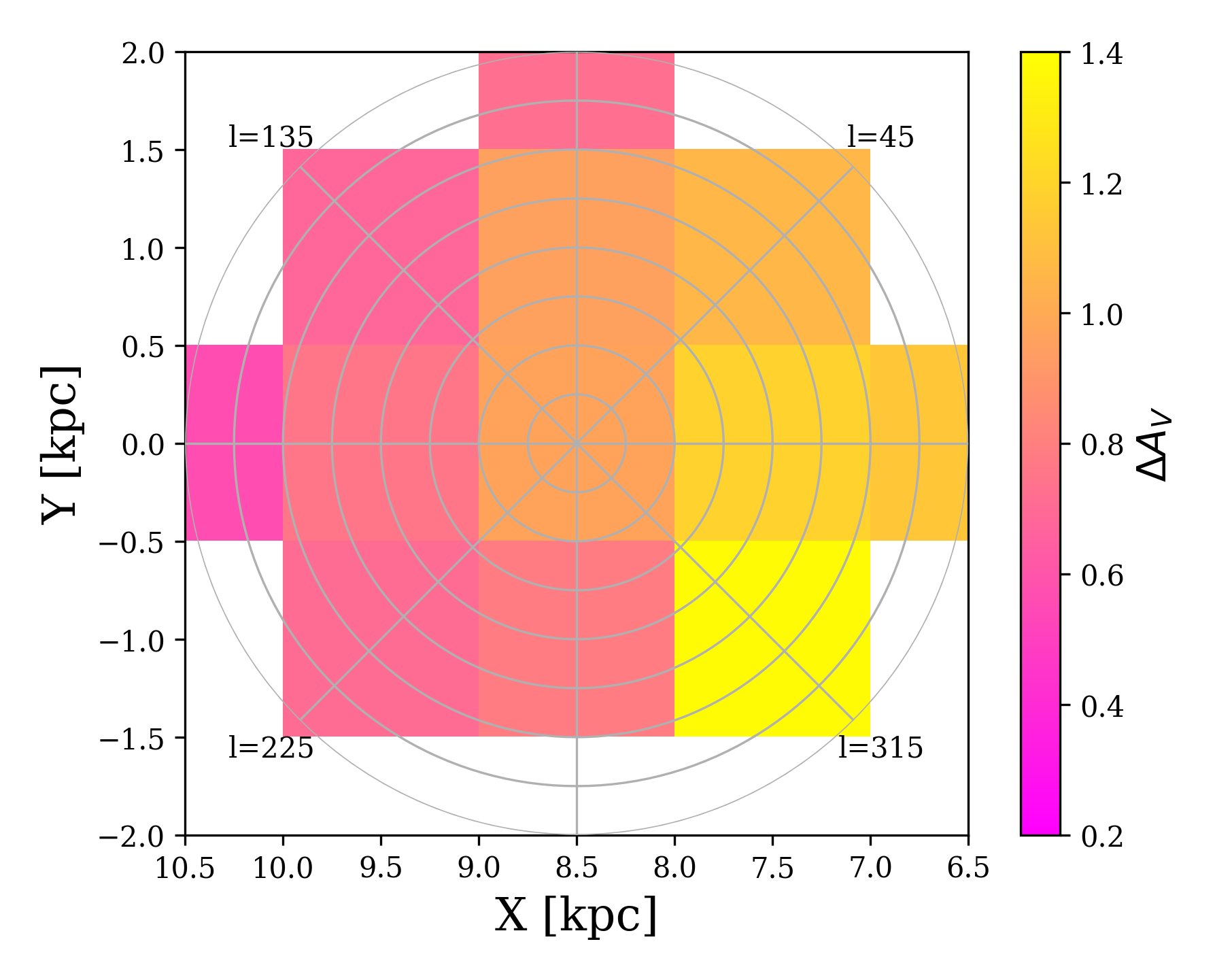}
        \includegraphics[width=0.32\textwidth]{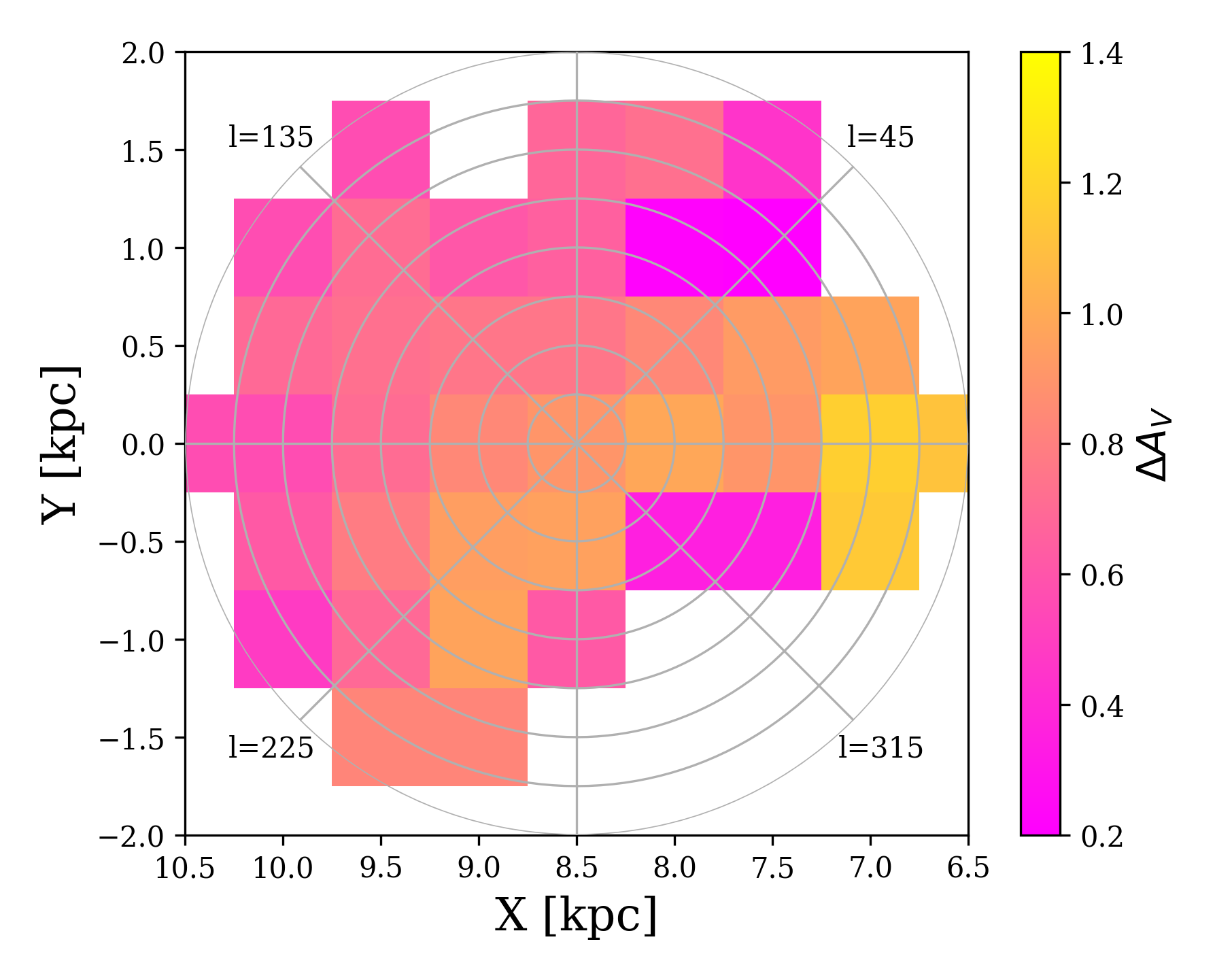}
        \includegraphics[width=0.32\textwidth]{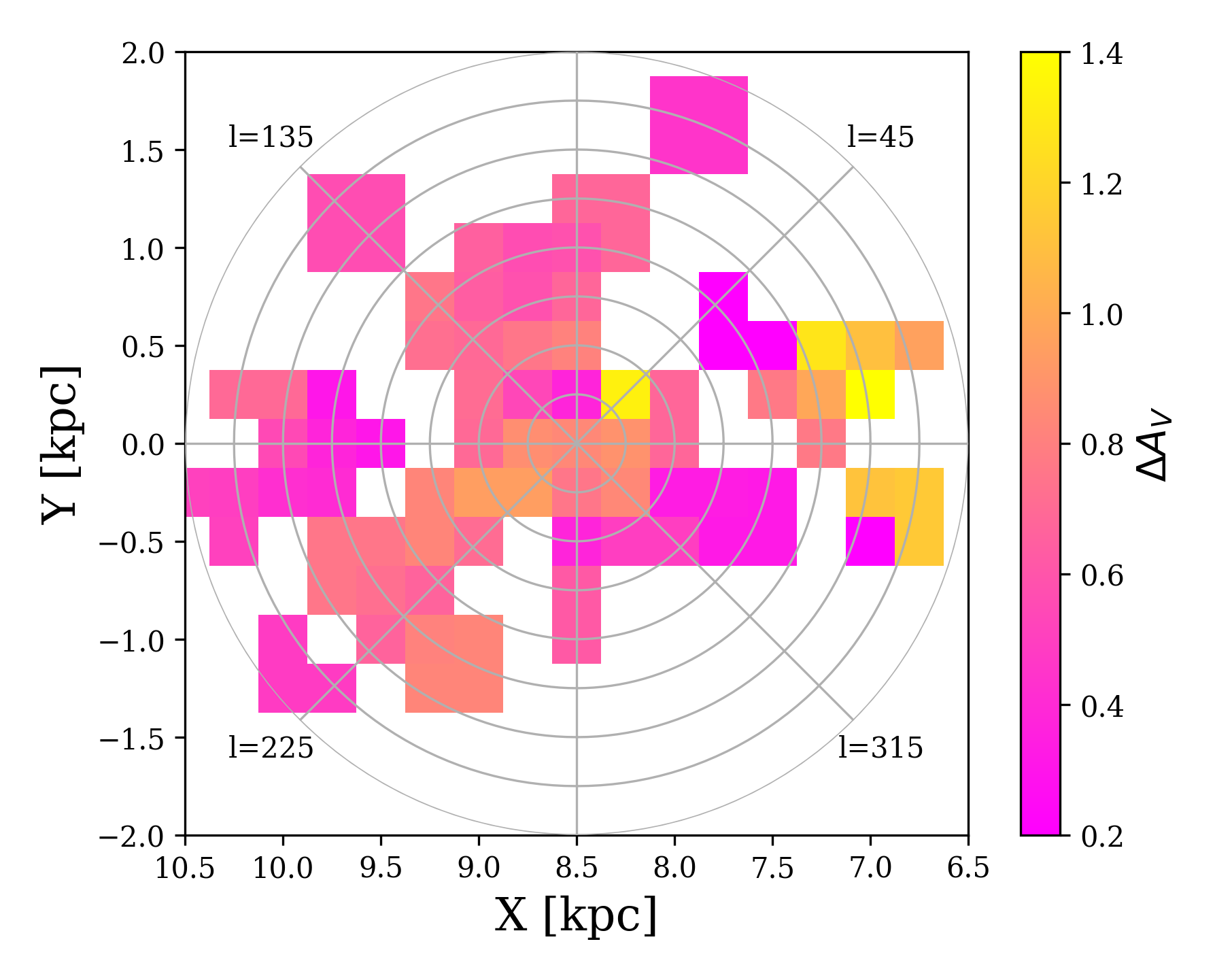}
    \caption{N-PDF characteristics within the local Galactic environment ($<2$ kpc distance). From top to bottom, the rows show the maps of mean extinction; N-PDF slope $\alpha$ (of the PL model); dense gas fraction $f_\mathrm{DG}$; and density contrast $\Delta A_\mathrm{V}$. The columns show the data for the apertures with $R=1$ kpc (left), $R=0.5$ kpc (middle), and $R=0.25$ kpc (right). The pixels correspond to Nyquist beams covering the sample, as in Fig. \ref{fig:ap_moved}. 
    }
    \label{fig:maps}
\end{figure*}

%--------------------------------------------------------
\subsubsection{Aperture N-PDFs and Galactic environment}
\label{sec:result_appos}
%--------------------------------------------------------

We already saw in the previous section that aperture N-PDFs may depend on the Galactic environment. Here, we further describe how the properties within the apertures vary across the survey area. Specifically, we investigate the variation of the mean extinction, power-law slopes, $f_{\mathrm{DG}}$, and $\Delta A_\mathrm{V}$ within the apertures. The maps of these metrics are shown in Fig. \ref{fig:maps}. The metrics clearly change for apertures closer to the Galactic centre. This is again caused by the group of clouds with high amounts of high column density gas. The maps in Fig. \ref{fig:maps} also indicate that the variance of the parameters decrease at large scales. Together, the results shown in Figs. \ref{fig:ap_moved} and \ref{fig:maps} reveal significant differences among the apertures across the survey area.

\begin{figure*}
    \centering
        \includegraphics[width=0.32\textwidth]{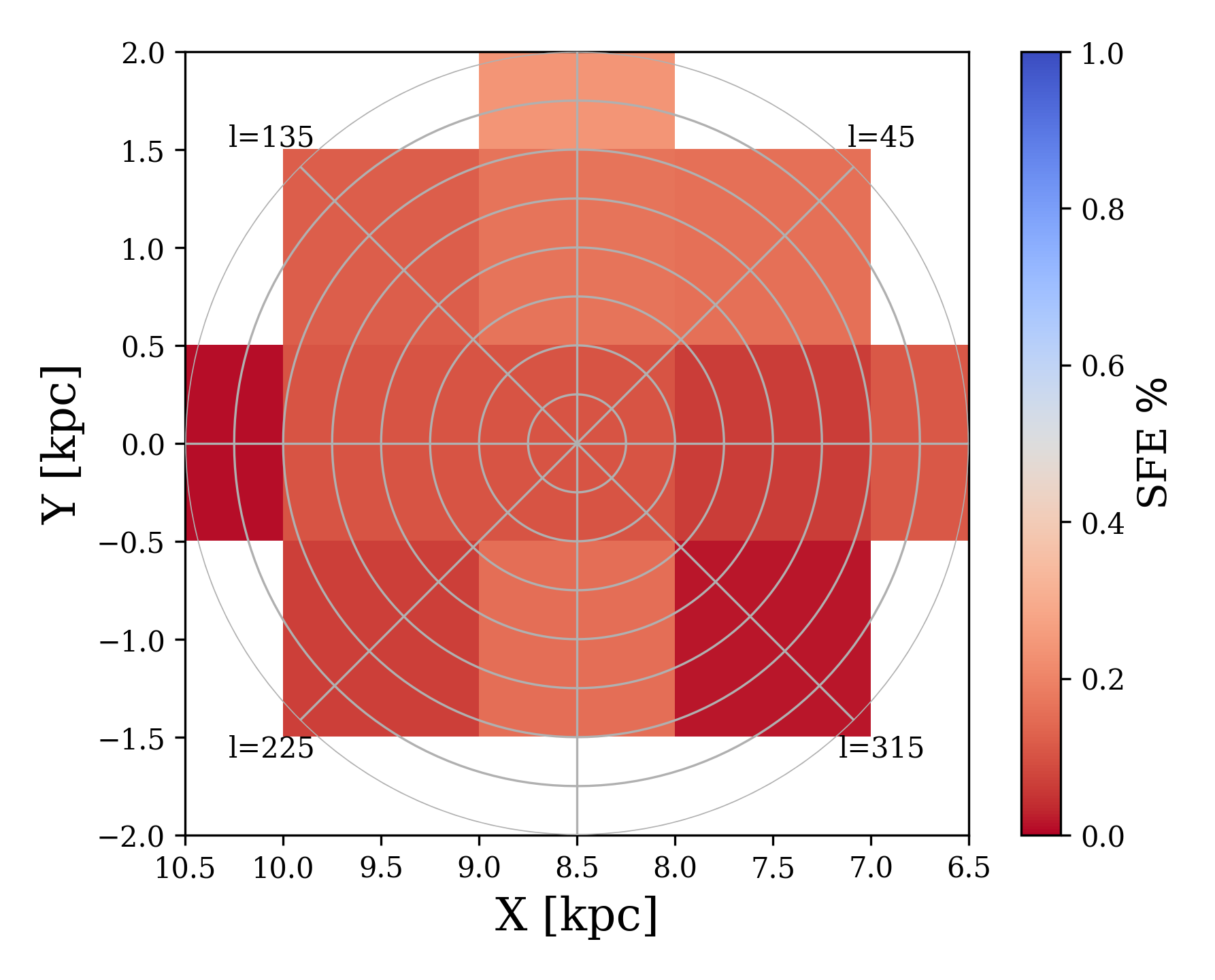}
        \includegraphics[width=0.32\textwidth]{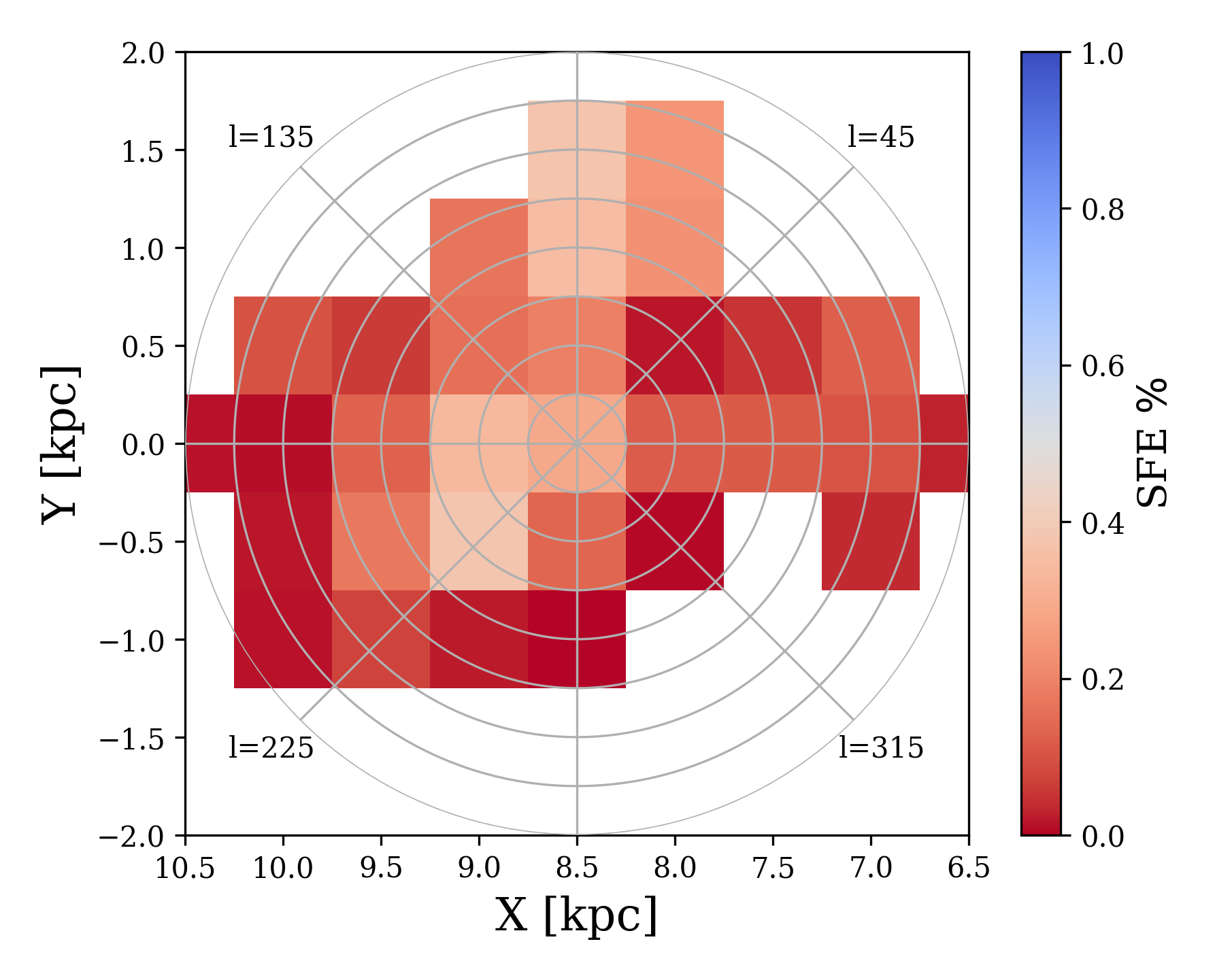}
        \includegraphics[width=0.32\textwidth]{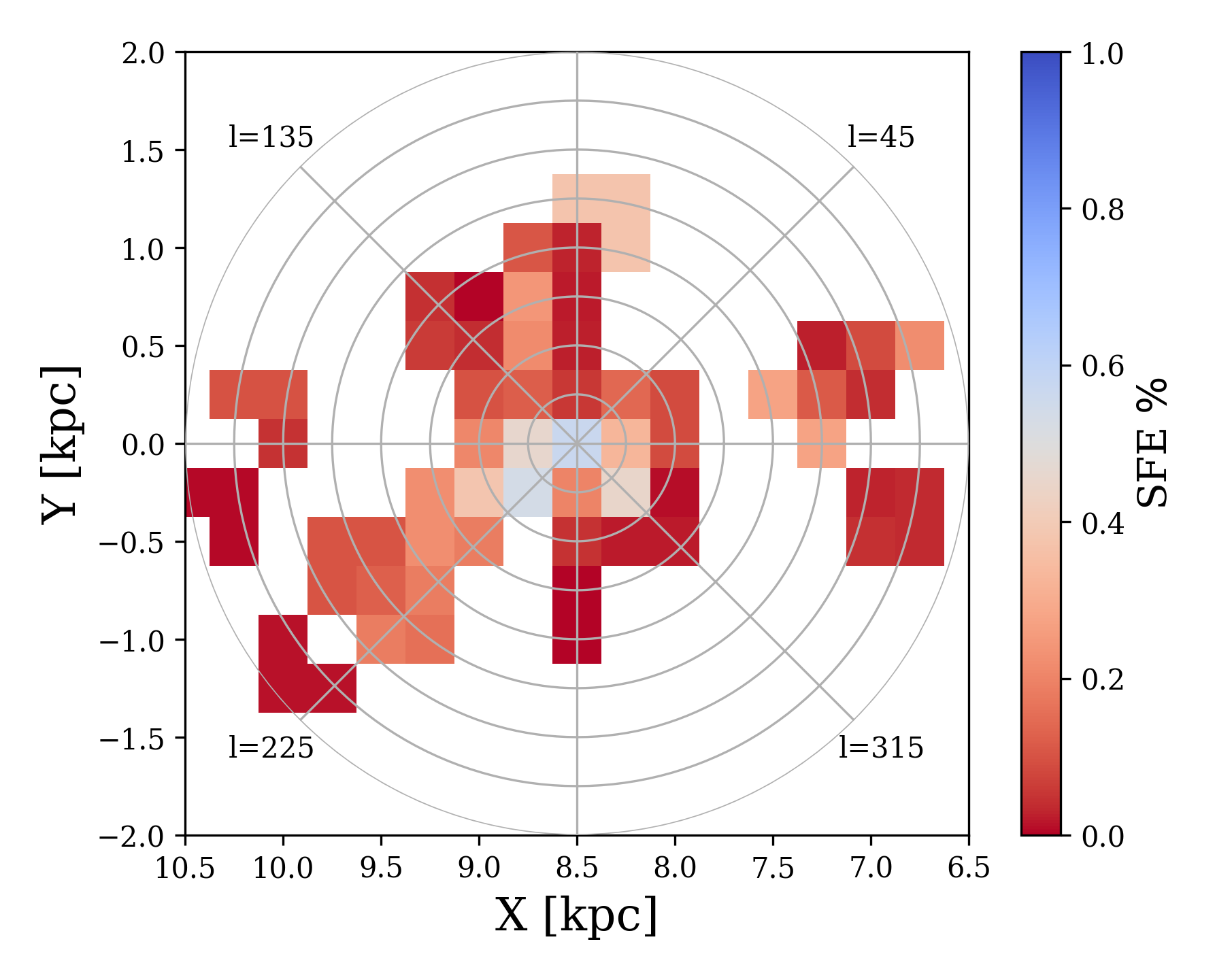}
        
        \includegraphics[width=0.32\textwidth]{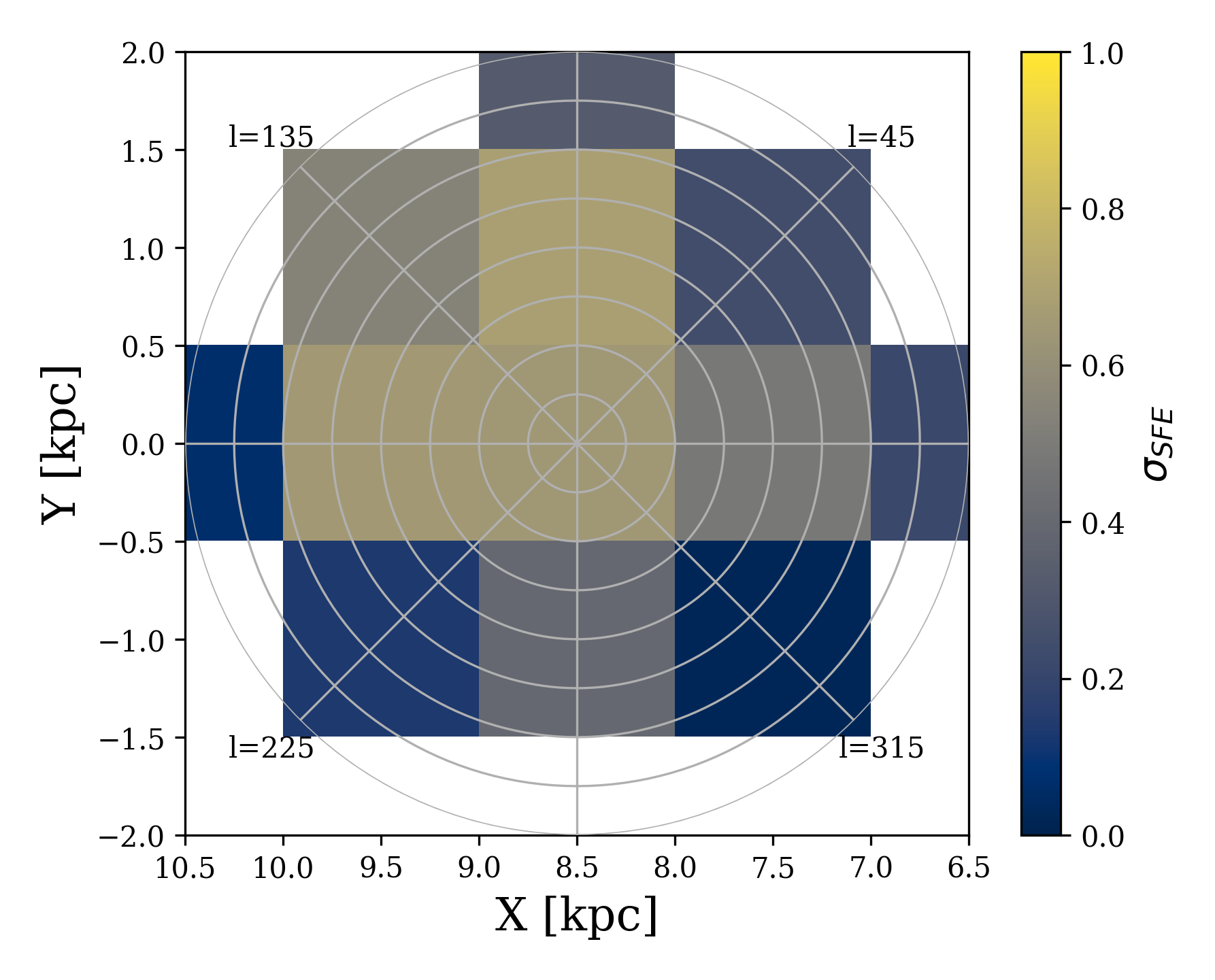}
        \includegraphics[width=0.32\textwidth]{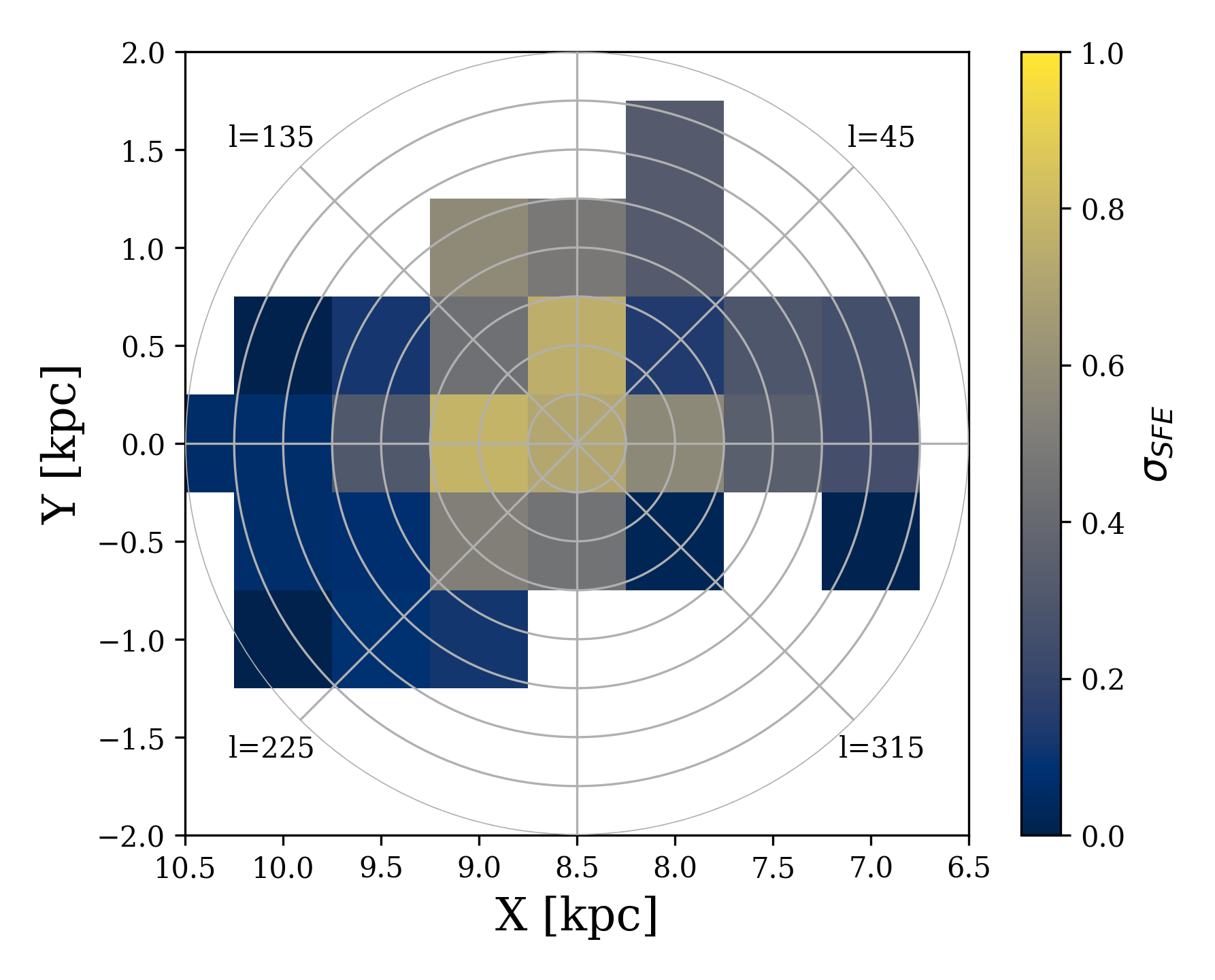}
        \includegraphics[width=0.32\textwidth]{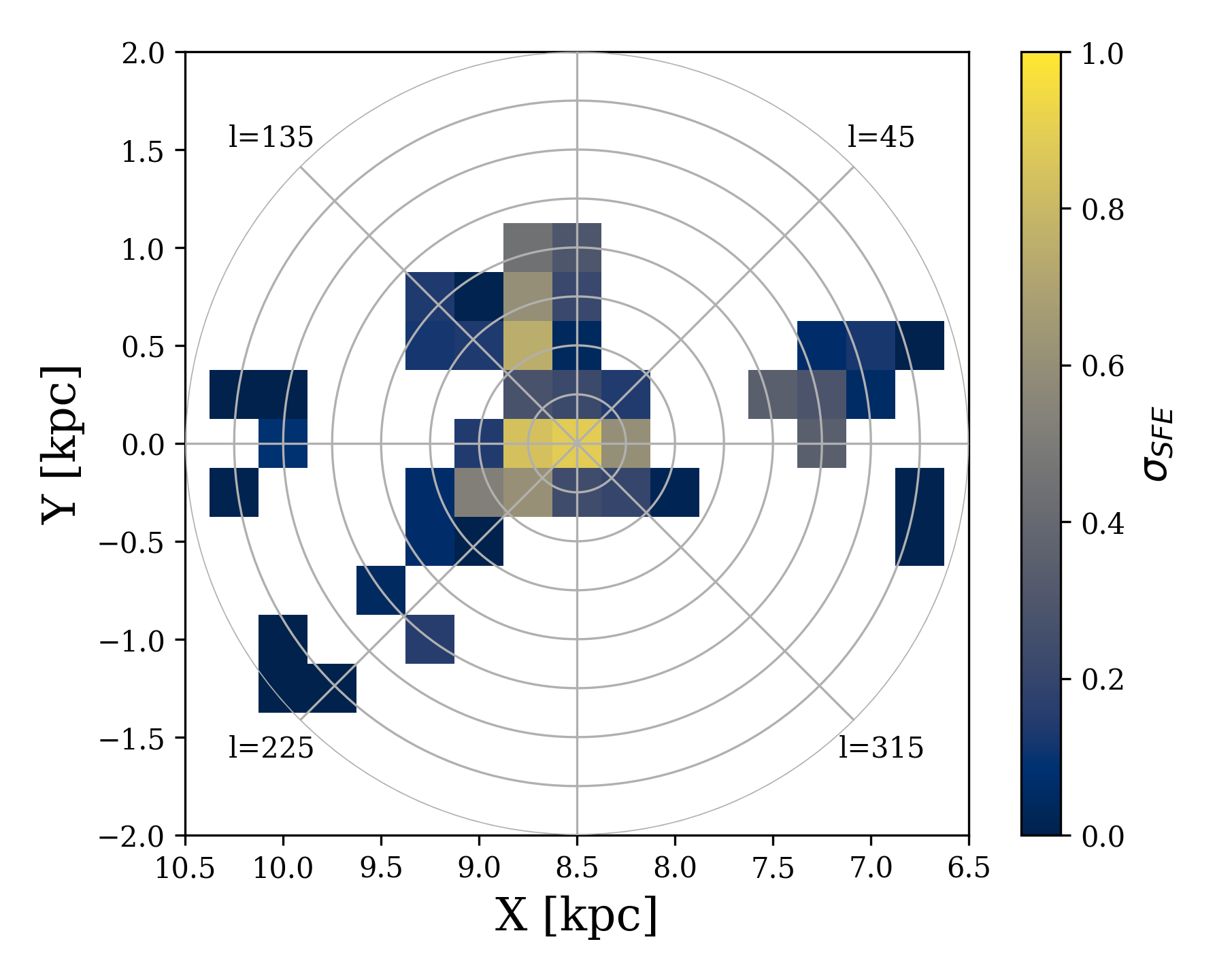}
    \caption{Maps of SFE (top) and SFE variance (bottom) for apertures with $R=1$ kpc (left), $R=0.5$ kpc (middle), and $R=0.25$ kpc (right).}
    \label{fig:maps_SFE}
\end{figure*}

\begin{figure}
    \centering
    \includegraphics[width=0.37\textwidth]{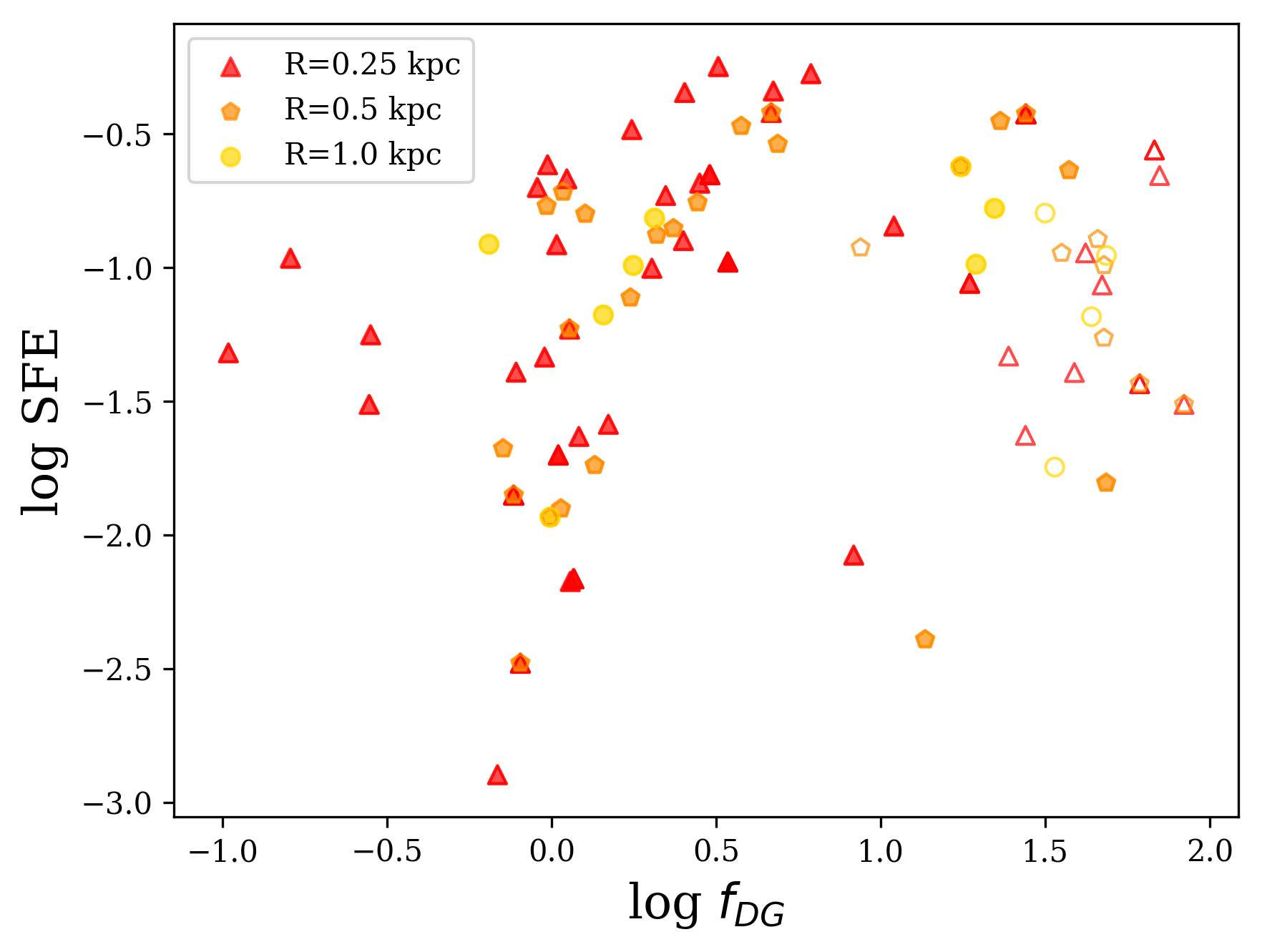}
    \includegraphics[width=0.37\textwidth]{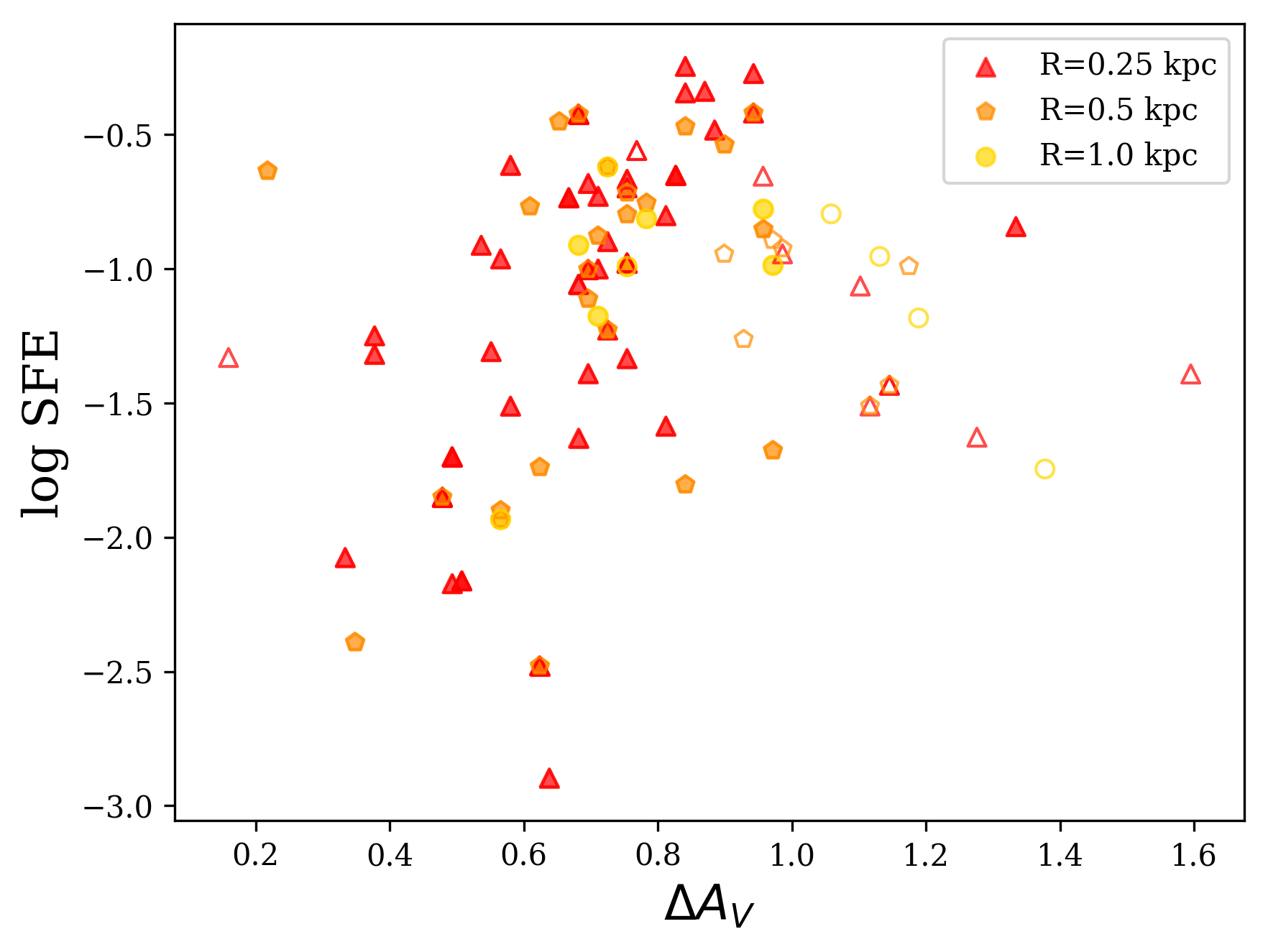}
    \caption{Relation between the aperture averaged SFE and dense gas measures. Empty symbols are apertures with $R_{\mathrm{Gal}}<8$ kpc. For the apertures with $R_{\mathrm{Gal}}>8$ kpc, a significant correlation is seen for the apertures of $R=0.25$ and $R=0.5$ kpc ($f_{\mathrm{DG}}$) and for apertures of $R=0.25$ ($\Delta A_\mathrm{V}$). }
    \label{fig:ap_SFE}
\end{figure}

%-------------------------------------------------------------------------
\subsubsection{Aperture N-PDFs and star formation}
%-------------------------------------------------------------------------
\label{sec:result_ap_SF}

 Next, we studied the relation between star formation and the shape of the aperture N-PDFs. To describe the SFE data, Fig. \ref{fig:maps_SFE} shows maps of SFE and its variance in the apertures over the survey area. We again see a larger variance for the smaller apertures. The maps show that SFE is very high for a few pixels. In addition, the SFE might be a little higher in the solar neighbourhood than in the remaining field. The star formation efficiencies of our clouds and apertures are likely underestimated especially between 1-2 kpc due to inhomogeneous YSO data, but this does not strongly affect the aperture-averaged values (see Sect. \ref{sec:discuss_inhomo} and Fig. \ref{fig:maps_ysoadjust}).
 
 In the case of individual clouds, the dense gas fraction (in clouds with $R_{Gal}>7.5$ kpc) and the density contrast correlated with SFE (Fig. \ref{fig:SF_vs_densegasfrac}). Analogously, Fig. \ref{fig:ap_SFE} shows the relation between SFE and the dense gas measures for the apertures. We find correlations for the apertures of radius 0.25 kpc ($f_{\mathrm{DG}}$ p-value: 0.01, R-value: 0.3. $\Delta A_\mathrm{V}$ p-value: 0.02, R-value: 0.3). %, and also 0.5 kpc for $f_{\mathrm{DG}}$ (p-value: 0.02, R-value: 0.4).
 When we only use the apertures with $R_{\mathrm{Gal}}>8$ kpc, the correlations are stronger (lower p and higher R), and the same is true when we correct the number of YSOs in the clouds farther than 1 kpc for incompleteness (Sect. \ref{sec:discuss_inhomo}). Thus, it seems that the relation between dense gas and SF is also present in the aperture data, at least up to $\sim$0.5 kpc.

%%%%%%%%%%%%%%%%%%%%%%%%%%%%%%%%%
%%%%%%%%%%%%%%%%%%%%%%%%%%%%%%%%%
\section{Discussion}
\label{sec:discussion}
%%%%%%%%%%%%%%%%%%%%%%%%%%%%%%%%%
%%%%%%%%%%%%%%%%%%%%%%%%%%%%%%%%%

This section is structured as follows. We first discuss the individual N-PDFs (Sect. \ref{sec:discuss_ind}) and the aperture N-PDFs (Sect. \ref{sec:discuss_aps}). Then follows a discussion of the Kennicutt-Schmidt (KS) relation (Sect. \ref{sec:discuss_ks}). Finally, we discuss the caveats and limitations of this study through investigating the difference of the N-PDFs in the Sagittarius spiral arm (Sect. \ref{sec:discuss_weirdos}) and the inhomogeneity of our YSO data (Sect. \ref{sec:discuss_inhomo}).

%--------------------------------------------------
\subsection{Individual cloud N-PDFs: New census within 2 kpc}
\label{sec:discuss_ind}
%--------------------------------------------------

% First topic: description of the spectrum
This study is the first effort to address the full spectrum of molecular cloud N-PDFs within a distance of 2 kpc from the Sun. Qualitatively, the range of N-PDF shapes we detect is similar to that in the earlier studies of smaller samples. For example, the most bottom-heavy N-PDFs in our sample resemble those commonly seen in studies of nearby clouds \citep[e.g.][]{lombardi2006pipe,kainulainen2009probing,froebrich2010}. Similarly, the most top-heavy N-PDFs resemble those of infrared dark clouds or other relatively massive clouds \citep[e.g.][]{kainulainen2013high,butler2014darkest,abreuvicente2015,schneider2016cygnus}. Our work expands on the topic by describing the complete spectrum of N-PDFs. We find suggestive evidence on the dependence of the N-PDFs on galactic environment. Previous works showed varying conclusions about this matter; some results point towards a constant mean column density and the dense gas statistics independent of it \citep[e.g.][]{battistiheyer2014}, while some claim evidence of a dependence on galactic environment \citep[e.g.][]{sawada2012structural,miville2017physical}. However, the techniques and approaches employed by these works are vastly different, and specifically, none employed N-PDFs as we do. It is therefore unclear how comparable the results are overall, and the question seems to remain open.

Previous works have established a habit of describing N-PDFs with the help of log-normal functions, power laws, or the combinations of the two. However, we have shown that only a fraction of the entire gas content is in clouds that are \emph{\textup{quantitatively}} well described by any one of the commonly adopted functional forms. Moreover, significant amount of the gas is in clouds with more complex N-PDFs than the simple shapes; the systematic deviations from the simple forms cannot be understood in terms of statistical fluctuations. A possible explanation is that molecular clouds have varying internal conditions and/or processes that affect different regions within. The total N-PDF of a cloud is then an agglomerate of the N-PDFs of these physically differing regions \citep[e.g. in terms of density and Mach number; cf.][]{elmegreen2011}. Not least, this result urges for caution about confirmation bias when cloud structure is analysed with the help of N-PDFs; using simple models may be attractive due to the connection to physical processes, but the N-PDFs at the scale of individual clouds may only rarely agree with them. It is unclear how well parameters derived from averaging over the different local PDFs reflect the physical state of the gas.

% Second topic: N-PDFs and SFRs
Our census enables us to firmly confirm the correlation between the top-heaviness of the N-PDFs and star formation activities of individual clouds in the solar galactic environment with a complete sample \citep[detected earlier by e.g.][]{kainulainen2009probing,lada2010star,schneider2013pdfs,kainulainen2014unfolding}. Because most clouds are not very well described by simple functional forms (as explained above), the relation is best recovered by the empirically determined dense gas fraction measures. We showed that a relative dense gas measure, such as our $\Delta A_\mathrm{V}$, is beneficial over an absolute measure to capture the trend (see Fig. \ref{fig:SF_vs_densegasfrac}). For a rough interpretation of $\Delta A_\mathrm{V}$, we provide the range of $\Delta A_\mathrm{V}$ also in terms of the average power-law slope of the N-PDF. If we assume the power-law shape for the N-PDF, a simple transform exists: $\alpha_\mathrm{calc} = \log{(0.05)} / \Delta A_\mathrm{V} - 2$. Thus, the range of $\Delta A_\mathrm{V} = [0.15, 1.79]$ transforms to the range of $\alpha_\mathrm{calc} = [-11, -2.7]$. This reflects an average slope over the N-PDF; some of the N-PDFs show high column density parts that are well described by flatter power laws reaching $\alpha \approx$-2 (cf. Fig. \ref{fig:PDFs_YSOcolor}).

%-------------------------------------------------------------
\subsection{Aperture N-PDFs: Relation to extragalactic works}
\label{sec:discuss_aps}
%-------------------------------------------------------------

We now discuss our aperture N-PDFs in the context of extragalactic works. Our apertures with radii of 0.25-1 kpc correspond for example to angular resolutions of $25-6\arcsec$ at the distance of M51 (8.4 Mpc), or to distances of 2.3-9.4 Mpc observed at $22\arcsec$ resolution (i.e. CO($J=1-0$) with the IRAM-30m telescope). It is important to emphasise that our aperture N-PDFs (and other aperture-averaged properties) are composed of dust-traced, relatively high column density gas that is organised in cloud-like structures. This makes our data and approach different from typical extragalactic works in two important ways. First, our apertures do not trace the \emph{\textup{total}} molecular gas content that may include a significant fraction of diffuse molecular gas not located in cloud-like structures \citep[e.g.][]{romanduval2016diffusegas}. Extragalactic works that employ molecular line emission as a gas tracer are sensitive to \emph{\textup{all}} line-emitting gas, regardless of whether it is organised in diffuse gas or cloud-like structures. While this issue is somewhat alleviated in the case of the most high-resolution works (resolution of some dozen pc), it is not overcome: it is a matter of debate where most of the molecular line emission of the usual gas tracers originate even when the parsec or tens-of-parsec scales of individual molecular clouds are studied \citep[e.g.][]{pety2017orionb}. Thus, our aperture N-PDFs describe the dense strongly concentrated gas that is more intimately connected to the star formation sites rather than the gas traced by extragalactic works.

Second, the range of column densities traced by molecular line observations is more limited than that traced by dust. Consequently, extragalactic works usually do not measure the shape of the column density distribution through N-PDFs. Instead, they commonly measure the top-heaviness of the gas distribution using a dense gas fraction constructed from observations of two molecular species \citep[e.g.][]{gallagher2018spectroscopic}. To zeroth order, this definition is analogous to our definition of $f_\mathrm{DG}$, that is, it represents the ratio of masses above two density thresholds (defined by the critical densities of the molecules). However, it is unclear where most of the emission of the molecules originates and how well, and with what calibration, line ratios trace dense gas fraction.  

When these differences are acknowledged, one-to-one comparisons with extragalactic data are not possible. Rather, we place our results in the context of those works as a first step to start understanding similarities and differences between the results. The most relevant extragalactic works to compare with are the recent interferometric observations of nearby galaxies. In particular, \citet{hughes2013probability} studied the N-PDFs of 100 pc scale CO data in different environments of the M51 galaxy. They found that the distributions change with galactic environment. The N-PDFs of the inner regions of M51 tend to be wider, and gas in inter-arm regions tends to have lower characteristic densities than gas within spiral arms. Qualitatively (and acknowledging the aforementioned caveats), these results imply more top-heavy N-PDFs and higher dense gas mass fraction in arm regions than in inter-arm regions. Overall, several recent works have studied the relation of dense gas fraction and gas surface density in nearby galaxies \citep[e.g.][]{sun2020molecular,sun2018cloud,hughes2013probability,querejeta2019dense,gallagher2018spectroscopic,leroy2017cloud}. In general, these works find a correlation between dense gas fraction and gas surface density, regardless of whether they consider clouds identified from data or integrated properties within apertures and environments. Again qualitatively, this trend is analogous to the fact that we detect higher mean surface densities and dense gas fractions ($f_\mathrm{DG}$) in the spiral arm environment than in the solar inter-arm environment.

Several relevant models that considered the relation between molecular cloud properties, star formation, physical scale, and galactic environment exist \citep[e.g.][and those more generally reviewed in \citealt{padoan2014ppvi}]{elmegreen2018appearance, meidt2020model}. Our results in the Milky Way open a door to testing these models with a large sample of data. Furthermore, the suggestive similarities in the results between our work and extragalactic studies make it interesting to link these results \emph{\textup{together}} with the models. This requires a close reconciliation of the correspondence between the details of the data and model parameters. This is beyond the current paper and an important avenue for future works.

%-------------------------------------------------------------
\subsection{Kennicutt-Schmidt relation}
\label{sec:discuss_ks}
%-------------------------------------------------------------

\begin{figure}
    \centering
        \includegraphics[width=\columnwidth]{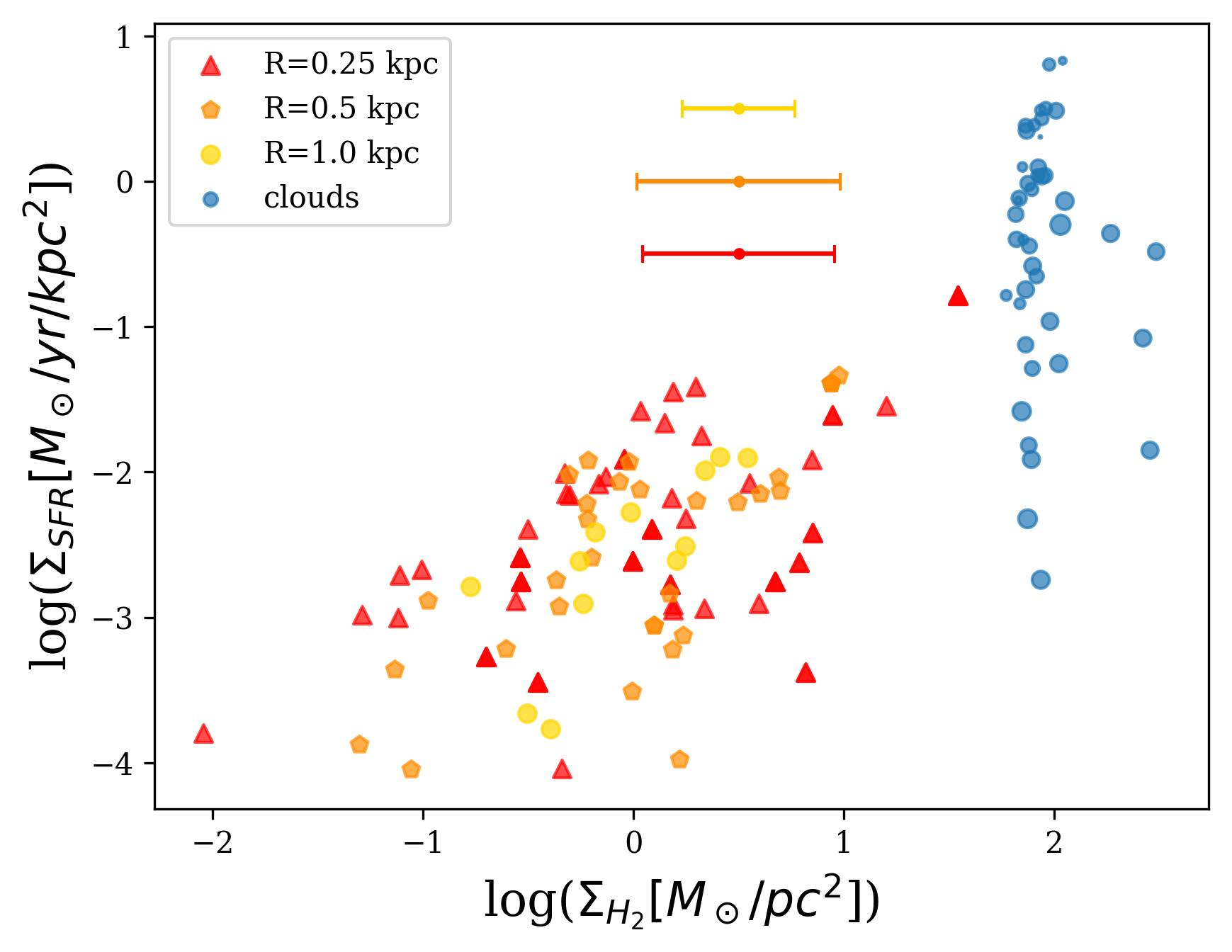}
    \caption{Kennicutt-Schmidt relation for apertures of radius 1 kpc (yellow circles), 0.5kpc (orange pentagons), 0.25kpc (red triangles), and the individual clouds (blue circles, size corresponds to cloud area). The error bars show the standard deviation of the x-axis (gas surface density) of the apertures.
    }
    \label{fig:KS}
\end{figure}

We next discuss the interpretation of our aperture experiment in the context of the Kennicutt-Schmidt (KS) relation. This relation has been extensively studied in extragalactic works \citep[see e.g.][for a review]{kennicutt2012star} and in the solar neighbourhood clouds up to distances of some $\sim$ 0.5 kpc \citep[e.g.][]{heiderman2010star,gutermuth2011,lada2013schmidt,evans2014,pokhrel2020star}. Many of these Galactic works have employed data similar to our study: extinction map data to trace gas surface density, and YSO counts to trace the star formation content of the clouds. To place our results in the context of these works, we show in Fig. \ref{fig:KS} the Kennicutt-Schmidt relation of our cloud and aperture sample. The cloud sample spans only a small range of mean surface densities above the chosen threshold level (3 mag $\approx$50 M$_\odot$ pc$^{-2}$), similarly to what earlier works have found. There is no correlation in the cloud-to-cloud data in the solar neighbourhood; this key result has been derived earlier for a smaller sample by \citet[][see also \citealt{evans2014}]{lada2013schmidt}. Our sample extends somewhat out of the Galactic environment of the solar neighbourhood. The few clouds that may be associated with the spiral arm towards the Galactic centre have considerably higher mean gas surface densities than the clouds in the solar neighbourhood (see Sect. \ref{sec:result_ind_gal_dist} and Fig. \ref{fig:galactic_densgas_profile}). A study comprising a higher number of more distant clouds would be needed to determine whether these clouds, when analysed together with the solar neighbourhood clouds, would give rise to a correlation in the KS plane. 

% FIGURE: KS relation

\begin{figure}
    \centering
        \includegraphics[width=\columnwidth]{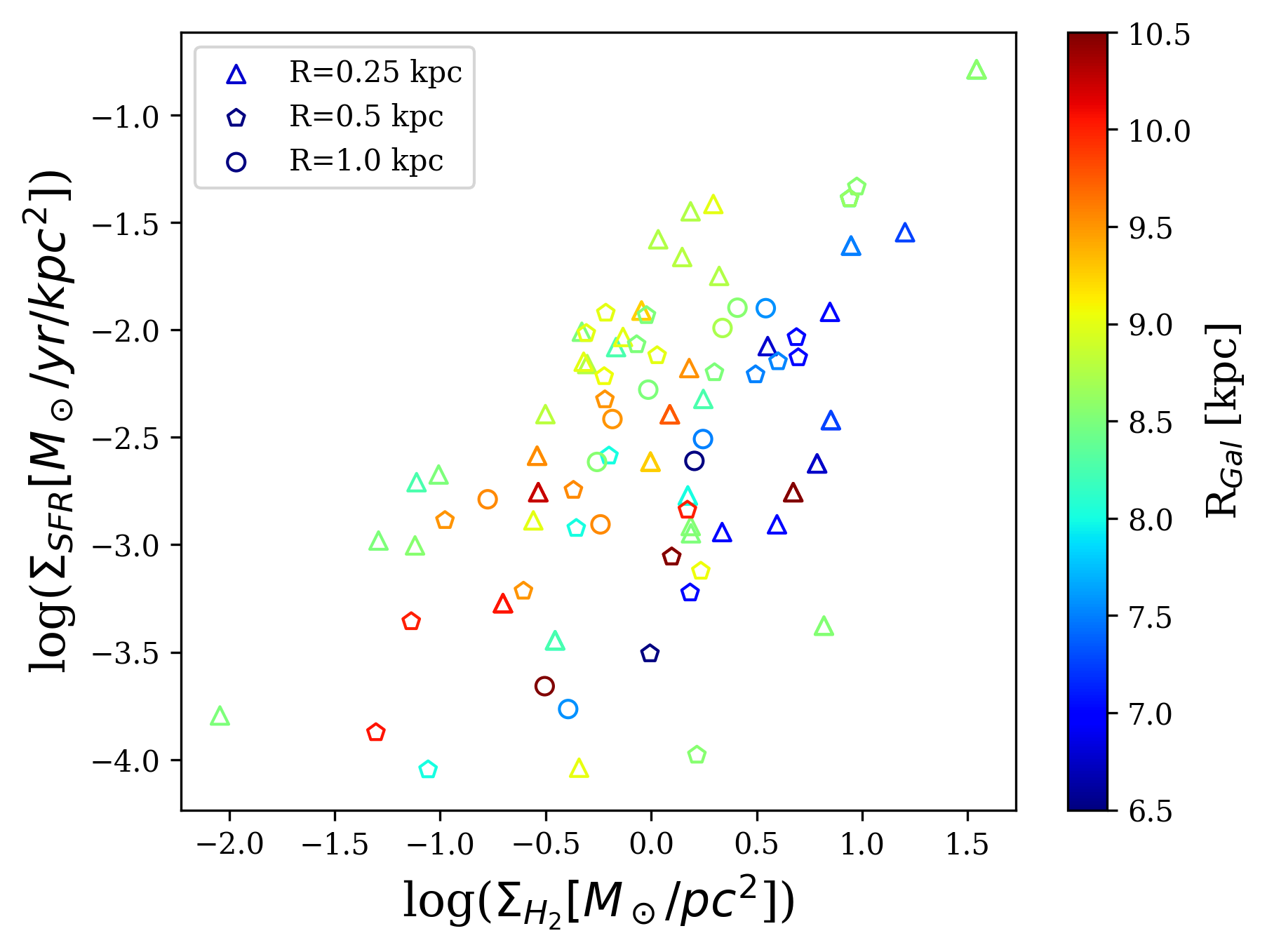}
    \caption{Kennicutt-Schmidt relation for apertures of radius 1 (circles), 0.5 (pentagons), and 0.25 kpc (triangles), colour-coded according to Galactic radius (as in Fig. \ref{fig:ap_moved}).}
    \label{fig:KS_Rgal}
\end{figure}

Figure \ref{fig:KS} also shows the KS relation for the apertures in our bird's eye experiment. The data span a wider range of column densities than the cloud-to-cloud data. We emphasise again that the aperture experiment is not directly comparable to extragalactic studies, mostly because of the lack of a diffuse molecular component. The wide span in column density originates mostly from the fact that the number density of clouds in the survey area is relatively small, and hence, the scatter in the number of clouds per beam is high. The apertures with only few clouds have very low mean gas surface density, stretching the range towards low surface densities. Importantly, the range of mean surface density decreases with increasing aperture size, as shown in the inset lines showing standard deviations. The range of mean surface densities is also slightly affected by the relation between the mean surface density and Galactocentric radius. We demonstrate this in Fig. \ref{fig:KS_Rgal}, which shows how the KS relation moves towards higher mean column densities with decreasing Galactocentric radius. The figure also indicates higher star formation rates for apertures with $R_{\mathrm{Gal}}=8-9$ kpc, which we discuss in Sect. \ref{sec:discuss_inhomo}.

These trends together help us understand the emergence of the KS relation in our aperture data as a combination of two effects. On the one hand, the near-constant mean surface densities of individual clouds cause a wide range of aperture-averaged mean column densities. If the clouds are spatially randomly distributed, the range is expected to decrease as an inverse square root of the aperture size, following from the central limit theorem. On the other hand, the mixture of galactic environments further increases the range of mean surface densities. Together, these effects give rise to a correlation in the KS plot, and hence, to the KS relation in our data. It would be interesting to further study the information about the sub-structure of the cloud distribution encoded in the relation and its scatter.

%-------------------------------------------------------------------------------
\subsection{N-PDFs towards the Sagittarius spiral arm}
\label{sec:discuss_weirdos} 
%-------------------------------------------------------------

Our results suggest that clouds and apertures towards the Galactic centre and Sagittarius spiral arm may have higher mean extinctions and column densities, higher dense gas fractions, and shallower N-PDFs than those closer to the Sun (Figs.  \ref{fig:galactic_densgas_profile}, \ref{fig:ap_moved}, and \ref{fig:maps}). These results are to a high degree driven by the N-PDFs of six clouds: M16, M17, M20, NGC6334, Cartwheel, and Cygnus. These are all massive clouds in the Galactic plane. Here we investigate whether these differences are indeed caused by the spiral arm galactic environment in which they reside, or if other explanations for the different N-PDFs of these clouds present themselves. The possibility that the different column density maps cause the difference was discussed in Sect. \ref{sec:result_ind_gal_dist}.

Two factors that might contribute to the high column density of the six spiral-arm clouds are confusion and incompleteness. The clouds are located in the plane of the Galaxy, where material that is not related to the clouds might be included in their N-PDFs, making the extinction higher (confusion). It is also possible that the location of the clouds in the Galactic plane have caused us to exclude the lower column density part of the clouds (the N-PDFs are incomplete). Accounting for incompleteness and confusion could alter the N-PDFs of the large clouds so that they appear to be more like the other N-PDFs in our sample. However, incompleteness mainly affects the low-$A_\mathrm{V}$ part of the N-PDF, and confusion is not likely either to dominate the shape of the N-PDFs at high-$A_\mathrm{V}$. The possibility of incomplete N-PDFs is explored below. 

If the N-PDFs of the clouds in the Galactic plane are incomplete, the increased mean extinctions and surface densities we see towards the Galactic centre would result from the inability of our column density maps to appropriately cover the low column density parts of massive clouds in the Galactic plane. For example, the aperture N-PDF with the size of 2 kpc is strongly affected by clouds with peaks in N-PDFs at about 10-20 mag. This might be dominantly caused by observational limitations, and the N-PDF may exclude an amount of gas. While we cannot exclude this possibility, we can study whether missing the low column density parts of the clouds appears to sensible. To do this, we performed a simple calculation in which we hypothesise that the N-PDFs of the clouds with high mean $A_\mathrm{V}$ in fact continue in a power-law-like fashion until the $A_\mathrm{V}$ of unity; our observational data have missed this gas due to the confusion in the Galactic plane. As a result, the aperture N-PDF with size of 2 kpc continues as a steeper power law until the $A_\mathrm{V}$ of unity. The total mass added in this experiment is about $2.5 \cdot 10^6 $M$_\odot$, which is  $\sim 28$\% of the total mass in the aperture. Thus, it seems entirely possible that the extinction maps have missed this component. It is possible that the apparently shallower low-$A_\mathrm{V}$ part of the aperture N-PDFs is affected by the missing mass, but missing this component would not strongly affect the shape of the remaining aperture N-PDFs. It is possible, however, that the mean surface density of clouds and apertures towards the Galactic centre would be slightly altered if this missing component were included. It is difficult to establish how strong this effect is exactly. Hence, our results regarding the relation between values such as mean column density and galactic environment remain suggestive. However, the high column density parts of the N-PDFs are not affected by these issues, and the results regarding the slopes of N-PDFs, for example, are firmer. 

We therefore argue that the N-PDFs of these clouds are indeed somewhat different from the other clouds in the sample, and that some of this difference is attributed to the cloud locations, which are closer to the Galactic centre and the Sagittarius spiral arm. Because these clouds are among the most massive in our sample, they dominate the aperture N-PDFs. Fig. \ref{fig:pie} showed that the fraction of N-PDFs that did not fit any of our functional forms decreased substantially when we excluded the spiral arm clouds. These massive clouds might better be described as cloud complexes with more complex N-PDFs as a result than the simple functional forms. These types of clouds might be the ones mostly seen in external galaxies, and conclusions about the star formation from extragalactic works would then be based on these types of clouds. It is therefore important to understand the difference between these clouds and smaller clouds.

\subsection{Inhomogeneous YSO data}
\label{sec:discuss_inhomo}

There is considerable uncertainty in the YSO counts, and hence in the star formation rates, of the clouds. Dedicated observations exploiting NIR and MIR data can reach a mass completeness of about 0.1 M$_\odot$ in clouds closer than some $\sim$1 kpc \cite[e.g.][]{pokhrel2020star}. We exploit just these data, therefore this is the mass completeness for most nearby clouds in our sample. At greater distances, and especially between 1-2 kpc, the studies from which we adopted the YSO counts are not homogeneously calibrated against each other and can have relatively large systematic differences between them. Moreover, several clouds at these distances do not have YSO censuses from dedicated NIR or MIR data, and we do not report YSO counts for them; we found YSO information for 44 of the 72 clouds. To the first degree, the mass sensitivity decreases with distance squared. A rough estimate therefore is that on average, the mass sensitivity of the YSO data for our farthest clouds (2 kpc) is a factor of some $\sim$4 lower than for the most nearby clouds. However, the shape of the system initial mass function (IMF) likely flattens at low masses \citep{chabrier2003review}, meaning that loss of mass sensitivity does not translate directly into a similar loss in the number of sources. We can approximately estimate the effect of this on the YSO count censuses by examining the integral over the IMF above a distance-dependent mass sensitivity limit, M$_\mathrm{lim}$. If we adopt a conservative mass sensitivity limit of 0.2 M$_\odot$ at 1 kpc, and hence 0.8 M$_\odot$ at 2 kpc, we find that that at 2 kpc distance, we would detect roughly 30\% of the YSOs detected in nearby clouds. This means that the star formation rates and efficiencies may be systematically lower by a factor of roughly three at the farthest clouds of our sample compared to the nearest ones. 

We therefore performed a test to determine how increasing the number of YSOs at far distances affects our results. We increased the number linearly from $1-2$kpc so that we reached a factor of three at 2kpc. The result that more star-forming clouds have more top-heavy N-PDFs (Sect. \ref{sec:results_ind_npdfs_and_sf} and Fig. \ref{fig:SF_vs_densegasfrac}) is likely not significantly affected by the YSO incompleteness because the correlations remain when only clouds closer than 1.2 kpc are considered (see Fig. \ref{fig:SF_vs_densegasfrac_12}). The YSO incompleteness does affect the aperture SFRs and SFEs, however. The effects are shown in Appendix \ref{app:ysoadjust}. Only a small change is seen in the SFE maps (Fig.\ref{fig:maps_ysoadjust}) because the mass exceeds the number of YSOs for most of the far clouds. The correlation between star formation properties and dense gas measures in Fig. \ref{fig:SF_vs_densegasfrac} was seen for apertures of $R = 0.25$ kpc, and the correlation is stronger when the YSO numbers are adjusted for incompleteness (Fig. \ref{fig:SFE_ysoadjust}). Our results on aperture SFE therefore appear to hold with the adjusted number of YSOs, and the correlation between dense gas properties and aperture SFE is clearer after the adjustment. We cannot adjust the number of YSOs for the clouds without YSO information, however, and this is difficult to correct for. We therefore caution that there might still be some bias in our data towards lower SFR and SFE of distant clouds and apertures.

%%%%%%%%%%%%%%%%%%%%%%%%%%%%%%%%%%%%%
%%%%%%%%%%%%%%%%%%%%%%%%%%%%%%%%%%%%%
\section{Conclusions}
\label{sec:conclusion}
%%%%%%%%%%%%%%%%%%%%%%%%%%%%%%%%%%%%%
%%%%%%%%%%%%%%%%%%%%%%%%%%%%%%%%%%%%%

We performed a study of the column density distributions of a complete sample of major molecular clouds within 2 kpc distance from the Sun. We employed dust extinction and dust emission data and YSO counts from the literature. We studied the column density distributions of the clouds individually, enabling us to quantify the full spectrum of molecular cloud N-PDFs and their relation with star formation. We also performed an experiment in which we simulated the column density distributions as they would be viewed from outside the Milky Way, within apertures of radius 0.25-2 kpc. The experiment enabled us to determine links between unresolved and resolved properties of the density distributions, and from there, start developing a link between Galactic resolved observations of molecular clouds and extragalactic unresolved observations. Our main results and conclusions are listed below.

   \begin{enumerate}
      \item We present a census of molecular clouds within 2 kpc distance by collecting all large structures prominent in CO and/or dust extinction. The census consists of 72 objects, for 44 of which we find YSO count information in the literature (see Table \ref{tab:mastertable}). The data set represents the most complete study to date of molecular cloud structures in the local Galactic environment. Within the uncertainties, our sample recovers the mass of CO traced clouds in the survey area, indicating high completeness.
      
      \item We describe the spectrum of molecular cloud N-PDF shapes within 2 kpc of the Sun. The N-PDFs show a broad spectrum of shapes and are generally poorly described by any single simple model (LN, PL, or LN+PL, see Sect. \ref{sec:results_npdfs}). Furthermore, using any single model to study a cloud population can lead to biases in how the N-PDF shapes are recovered. Most of the clouds are best fit with the LN+PL model, but a significant amount of the mass is in clouds whose N-PDFs are poorly described by any of the forms. This mass is dominated by the large clouds towards the Sagittarius spiral arm. When these are excluded, $\sim26\%$ of the mass is in clouds with LN+PL shapes. The most top-heavy N-PDFs decrease at high column densities roughly with a power-law slope of -2. 
      
      \item We describe the relationship between the N-PDF shapes and star formation in our sample. For the clouds with $R_{\mathrm{Gal}}>7.5$ kpc, the top-heaviness of the N-PDFs correlates with SFR and SFE. This correlation is present in both the dense gas mass fraction computed using two column density thresholds ($f_\mathrm{DG}$) and the relative density contrast of the clouds ($\Delta A_\mathrm{V}$, see Sect. \ref{sec:results_ind_npdfs_and_sf}). However, the correlation between SFE and $f_\mathrm{DG}$ breaks down when different galactic environments are considered together. This suggests that density contrast may be more fundamentally linked to star formation than $f_\mathrm{DG}$. 
      
      \item We find indications of a dependence on Galactic environment for the N-PDF shapes and dense gas measures. The clouds and apertures associated with smaller galactic radii (likely associated with the Sagittarius spiral arm) have higher mean column densities and are in relative terms more top-heavy N-PDFs than those outside it. While this conclusion is susceptible to uncertainties related to mapping the column density of clouds in the Galactic plane, we find the evidence suggestive.
      \item Our aperture N-PDF experiment describes how the internal cloud N-PDFs give rise to averaged low-resolution N-PDFs. The aperture N-PDFs are in general more shallow and top-heavy than those of the individual clouds because they are dominated by the gas in the massive star-forming clouds. A scale dependence on the shape and variance of the aperture N-PDFs is seen: larger apertures give rise to more shallow and top-heavy density distributions. We also describe the decrease in variance of the aperture properties with scale. 
 
      \item We approximately quantify the shapes of aperture N-PDFs using power-law functions. For our entire $R=2$ kpc survey area, we obtain the relation $P(A_\mathrm{V})\propto A_\mathrm{V}^{-1.9}$. If we only consider the Galactic environment of the Sun, and/or decrease the aperture size, the slope of the relation becomes steeper. For apertures with $R=0.25$ kpc, the relation is $P(A_\mathrm{V})\propto A_\mathrm{V}^{-3.3}$. 

      \item Analogously to individual clouds, we find a correlation between the SFE and the shape of aperture N-PDFs for apertures with $R=0.25$ kpc, at least in the solar environment ($R_{\mathrm{Gal}}>8$ kpc). The relation is not seen for apertures with $R=0.5$ or $R=1$ kpc, suggesting a possible break-down of the relation around that scale.
      
      \item We show how a Kennicutt-Schmidt-like relation emerges from the data of individual clouds in our aperture experiment. It emerges as a combination of two effects: the number of clouds that have a narrow range of mean column densities varies strongly within the apertures, causing a wide range of aperture-averaged column densities. This range is further widened by the effect of Galactic environment; in other words, simultaneously analysing apertures from different environments contributes to the emergence of the relation. While this result arises from a dust-based analysis of cloud-like structures, it remains to be studied how exactly it applies to CO-based data before direct comparisons with extragalactic works can be attempted. 
   \end{enumerate}

Our work takes important steps towards using our own galaxy, and the detailed information of its ISM, as a point of comparison to understand information decoded in extragalactic scaling and star formation relations. While several issues remain to be considered before a direct comparison is possible (e.g. the role of diffuse gas, use of different tracers, or completeness issues), our first results demonstrate the feasibility and advantage of this. In the continuation of this work, we aim at developing the approach to this direction and integrating velocity information into the analysis. This will enable us to examine scaling relations that include kinematic information, such as Larson's relations, and take yet another step towards connecting Galactic and extragalactic studies of the topic.

\begin{acknowledgements}
The authors thank the anonymous referees for constructive comments. This project has received funding from the European Union's Horizon 2020 research and innovation programme under grant agreement No 639459 (PROMISE). Jan Orkisz acknowledges funding from the Swedish Research Council, grant No. 2017-03864.
\end{acknowledgements}

\bibliographystyle{aa}
\bibliography{Birds-eye_final.bib}

%%%%%%%%%%%%%%%%%%%%%%%%%
\begin{appendix}

%\let\part=\chapter\appendix
%\input{Z_appendix}

%%%%%%%%%%%%%%%%%%%%%%%%%%%%%%%%%%%%%%%%%%%%%%%%%%%%%%%%
\section{Dust extinction maps of Cygnus and Cartwheel}
\label{app:cygnus_and_cartwheel}
%%%%%%%%%%%%%%%%%%%%%%%%%%%%%%%%%%%%%%%%%%%%%%%%%%%%%%%%

Two major clouds in our sample have no available column density maps: Cygnus and Cartwheel. The \emph{Herschel} PPMAP-based maps would ideally be adapted for these clouds, together with other Galactic plane clouds for which no extinction-based maps are available. However, both Cygnus and Cartwheel extend over a considerably wider latitude range than what is covered by the \emph{Herschel} HiGal survey; using these data would make the fields prohibitively incomplete. To include these clouds in our analyses, we derived new extinction-based column density maps for them. We employed an adaptation of the colour-excess mapping technique presented in \cite{kainulainen2011alves}. In short, this technique is based on the NICER technique \citep{lombardi2001alves}, with modifications that make the technique better applicable to clouds in the Galactic plane where a large fraction of stars are located between the observer and the cloud. We refer to \cite{kainulainen2011alves} for the full description of the technique. We applied this mapping technique to Cygnus and Carthwheel. For Cygnus, we used the technique in conjunction with near-infrared JHK data from the UKIDSS Galactic plane survey \citep{lucas2008ukidss, lawrence2007ukidss}, specifically, the data release DR11PLUS. For Cartwheel, we used near-infrared JHK data from the VISTA/VVV survey \citep{minniti2010vvv} and specifically the photometric data published by \citet{zhang2019deep}. The resulting maps are presented with all other clouds in Appendix \ref{app:extinction_maps}. In general, they recover the column density structure of the two clouds between about $A_\mathrm{V}=1-25$ mag. While the extinction maps reveal the general structure of the clouds well, they do contain some artefacts caused by imperfect foreground star removal and the high column density regions where no or very few stars are detected. We note that while extinction mapping may not be optimal for the Galactic plane regions, currently, no higher-fidelity data are available that would cover the wide areas spanned by the clouds.

\onecolumn
\section{Extinction maps of all clouds}
\label{app:extinction_maps}

%\iffalse
\begin{figure*}[h!]%[H]
    \centering
    \begin{minipage}[b]{0.85\textwidth}
    \includegraphics[width=\textwidth]{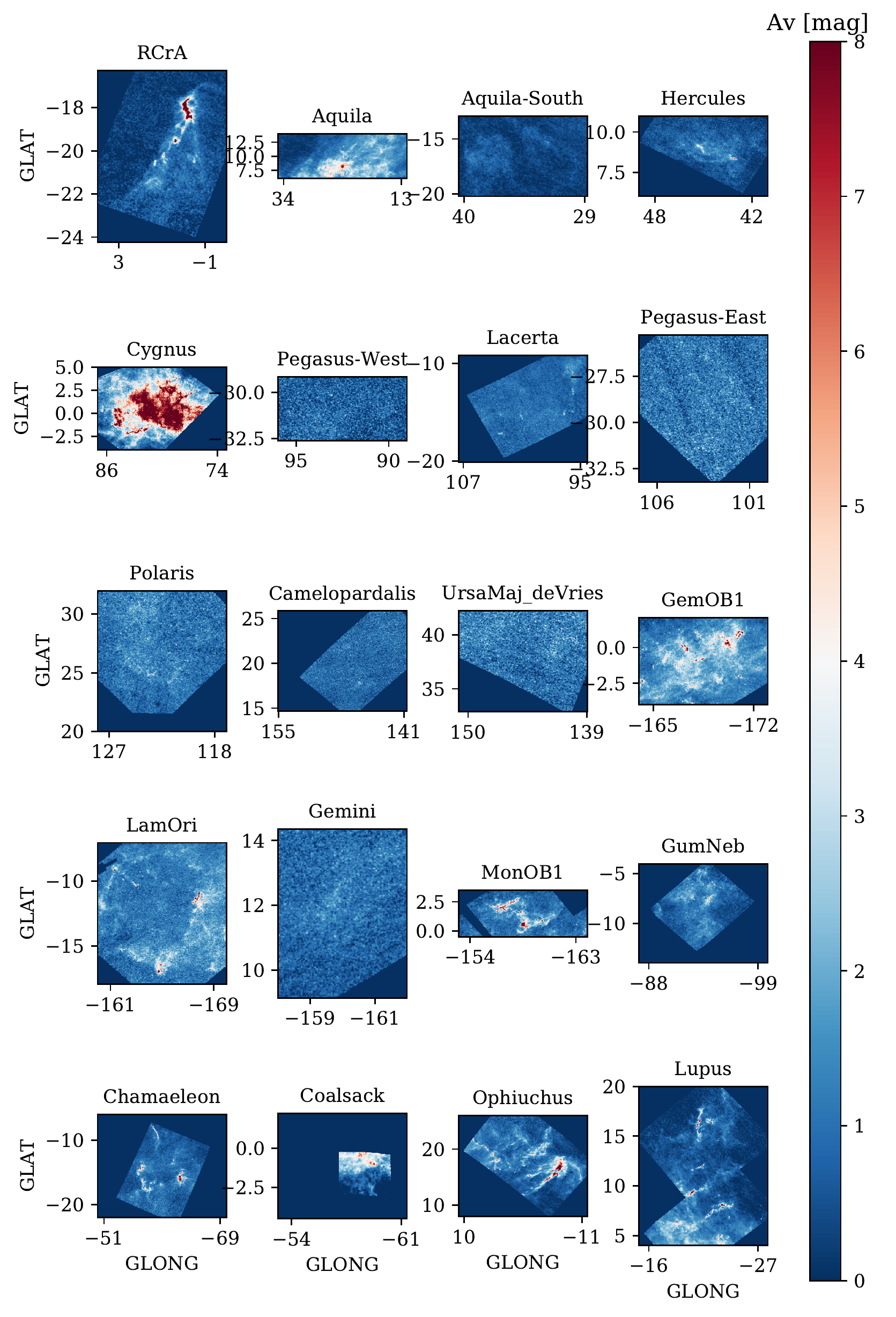}
    %Maps.png}
    \end{minipage}
    \caption{Extinction maps of the clouds within 2kpc from us (Figs. \ref{fig:cloudext}-\ref{fig:cloudext4}).} 
    \label{fig:cloudext}
\end{figure*}

\begin{figure*}[h!]%[H]
    \centering
    \begin{minipage}[b]{0.9\textwidth}
    \includegraphics[width=\textwidth]{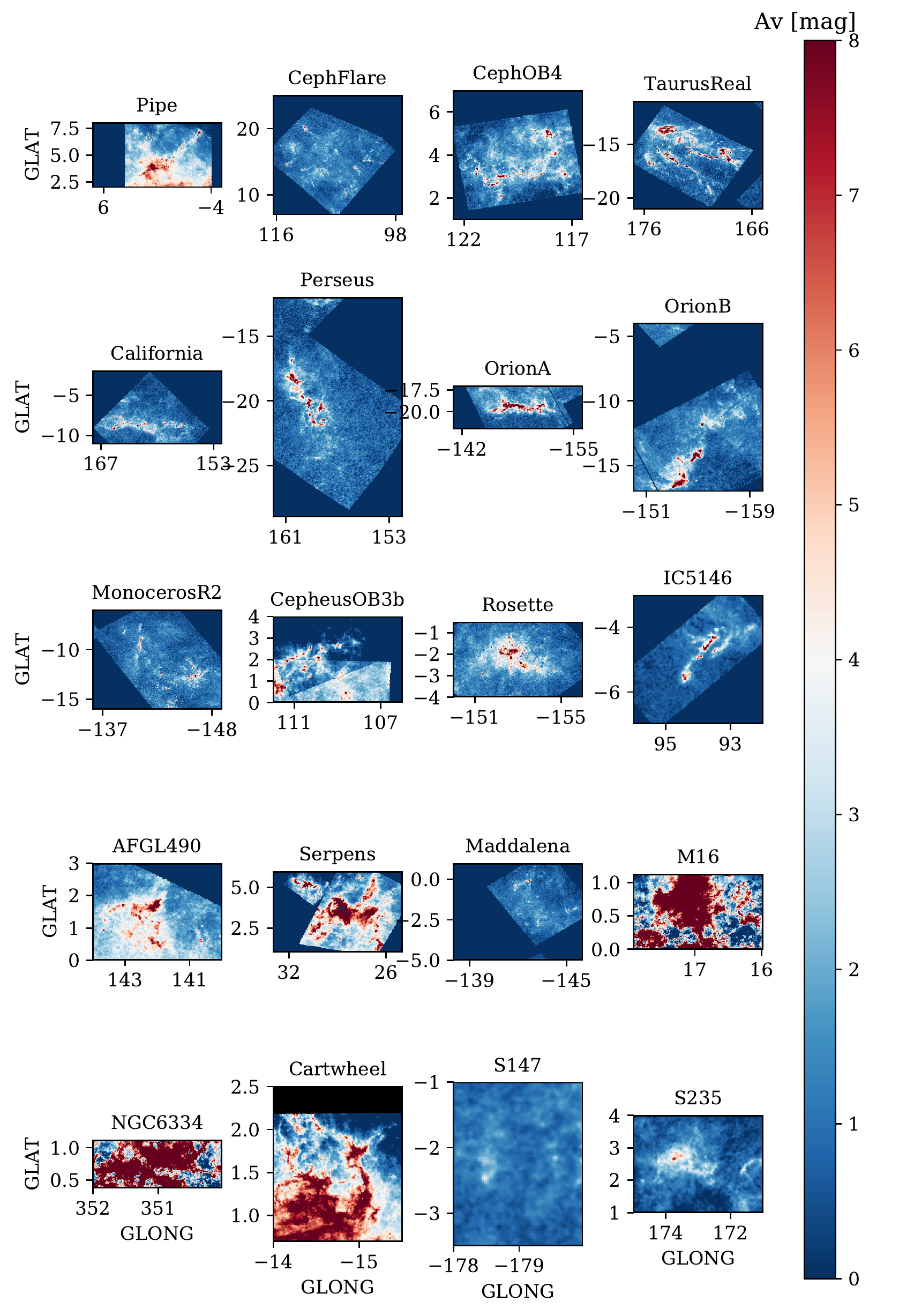}
    %Maps.png}
    \end{minipage}
    \caption{Extinction maps of all clouds, continued. }
    \label{fig:cloudext2}
\end{figure*}

\begin{figure*}[h!]%[H]
    \centering
    \begin{minipage}[b]{0.99\textwidth}
    \includegraphics[width=\textwidth]{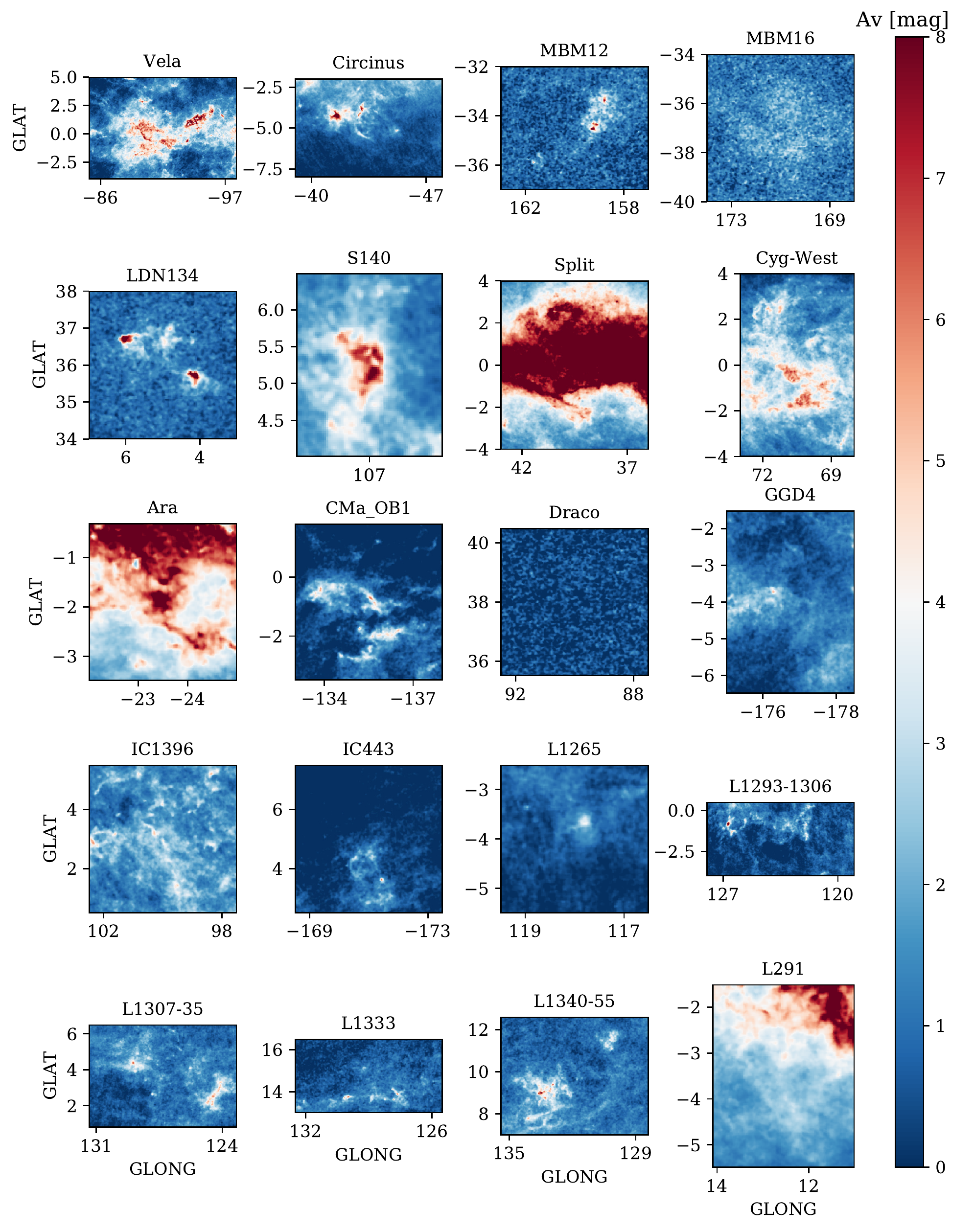}
    %Maps.png}
    \end{minipage}
    \caption{Extinction maps of all clouds, continued. }
    \label{fig:cloudext3}
\end{figure*}

\begin{figure*}[h!]%[H]
    \centering
    \begin{minipage}[b]{0.99\textwidth}
    \includegraphics[width=\textwidth]{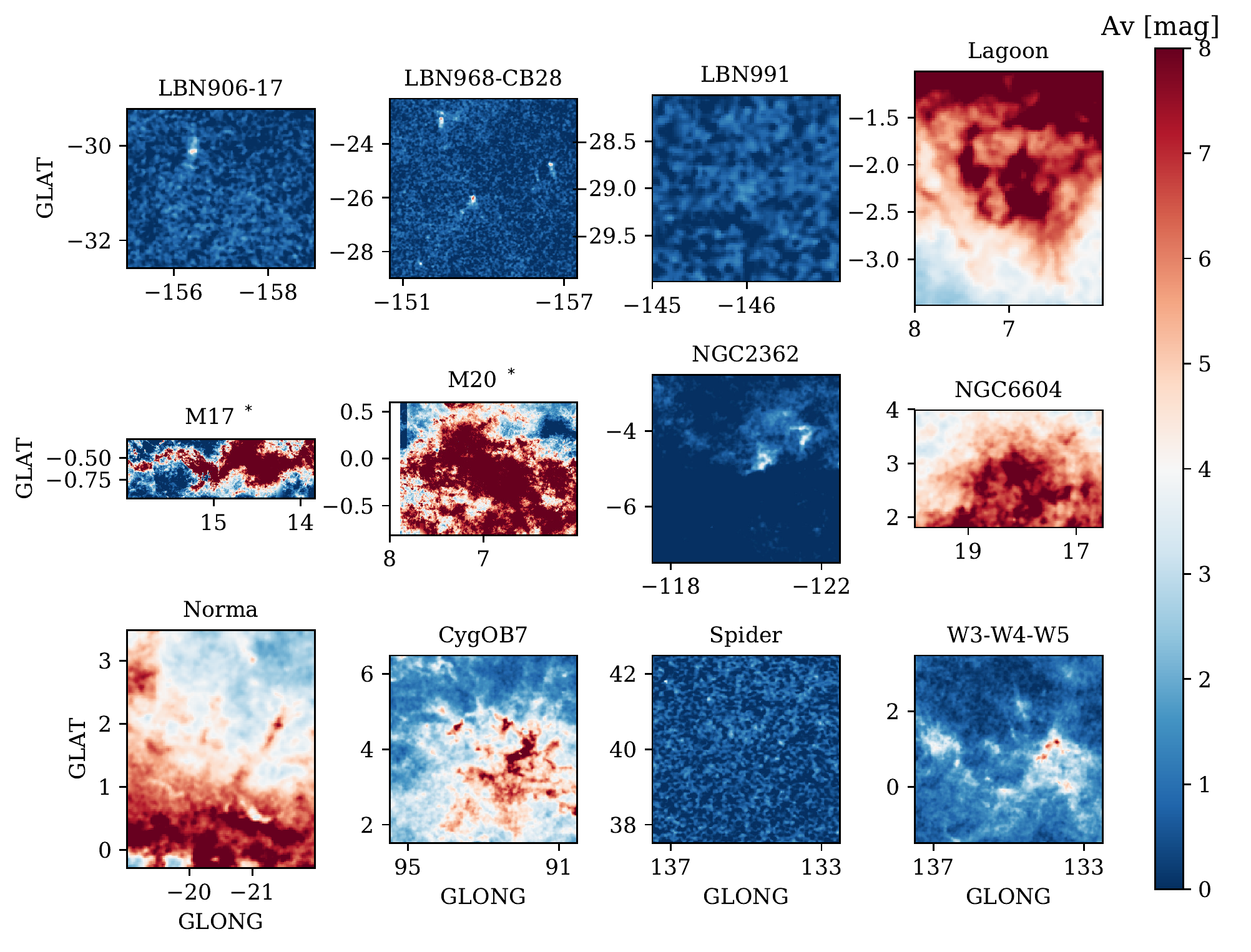}
    %Maps.png}
    \end{minipage}
    \caption{Extinction maps of all clouds, continued. M17 and M20 are marked with stars as their colour bar extends to higher extinctions than the other clouds, to 18 mag instead of 8.}
    \label{fig:cloudext4}
\end{figure*}
%\fi

% ALL THE PDFS
\clearpage
\section{N-PDFs of all clouds}

\begin{figure*}[h!]
    \centering
    \begin{minipage}[b]{0.24\textwidth}
        \includegraphics[width=\textwidth]{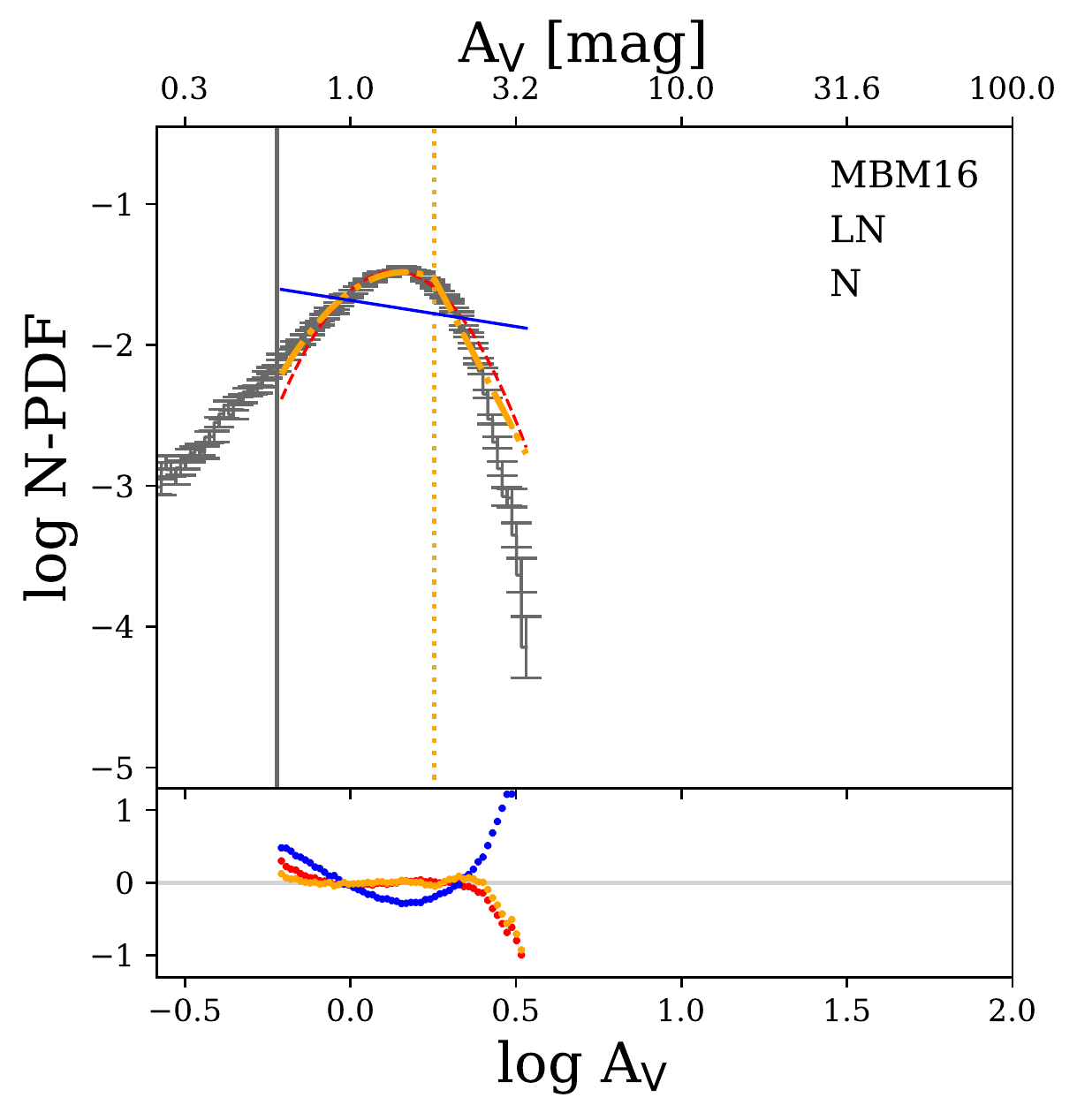}
        %RCrA_area_loglog_logbins_fit_resid_flag_wjuvela_noleg-eps-converted-to.pdf}
    \end{minipage}
    \begin{minipage}[b]{0.24\textwidth}
        \includegraphics[width=\textwidth]{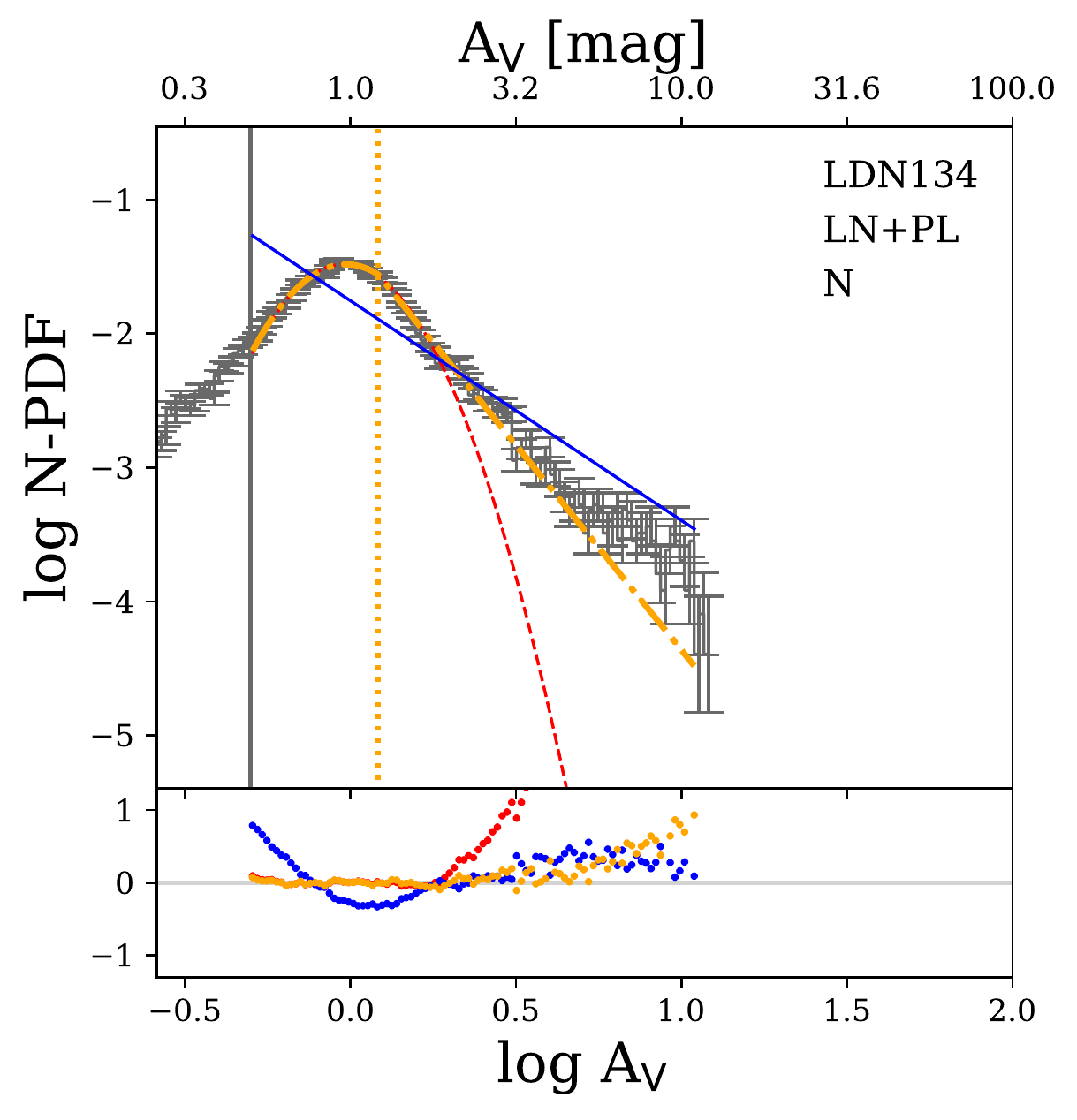}
    \end{minipage}
    \begin{minipage}[b]{0.24\textwidth}
        \includegraphics[width=\textwidth]{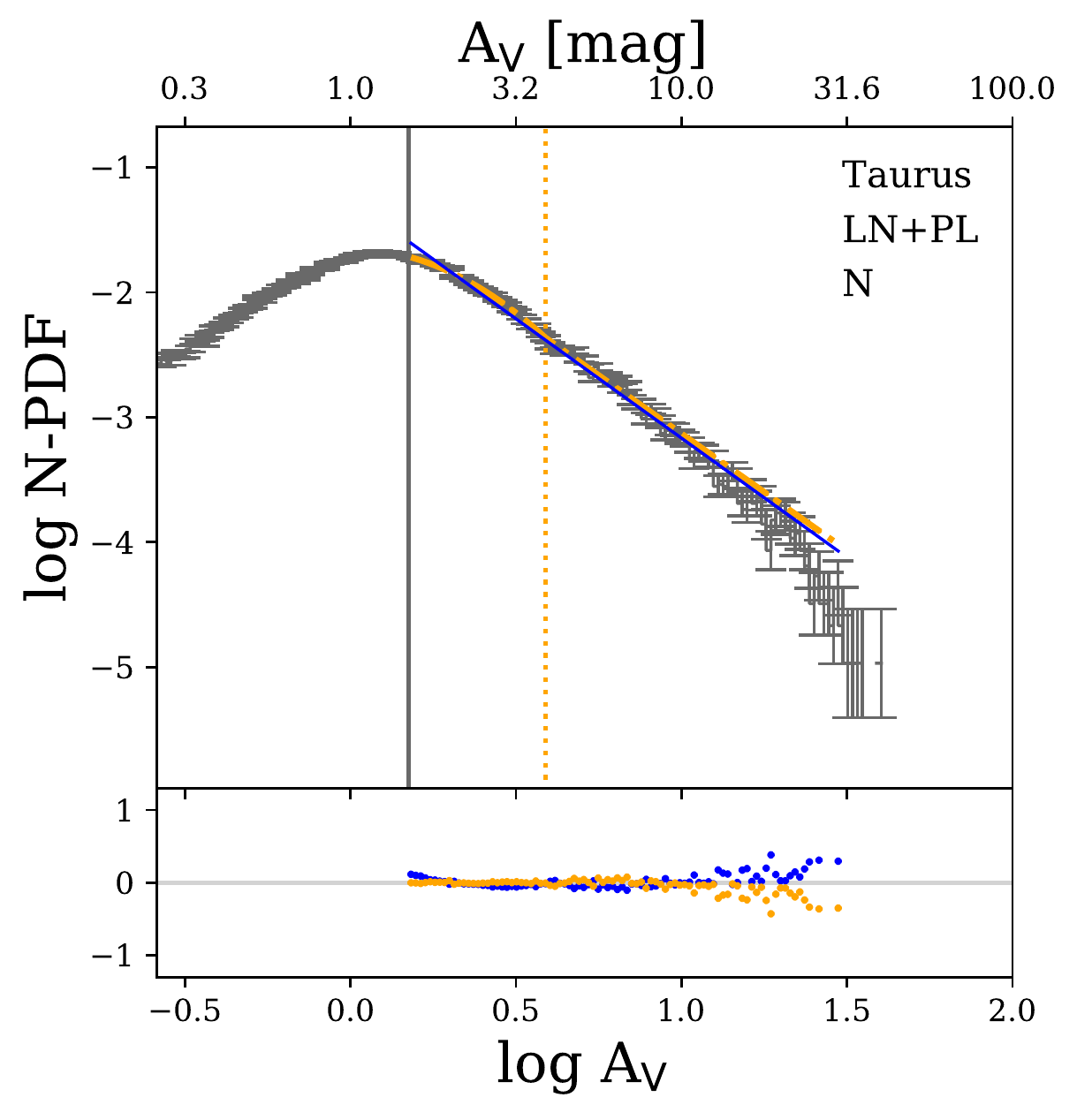}
    \end{minipage}
    \begin{minipage}[b]{0.24\textwidth}
        \includegraphics[width=\textwidth]{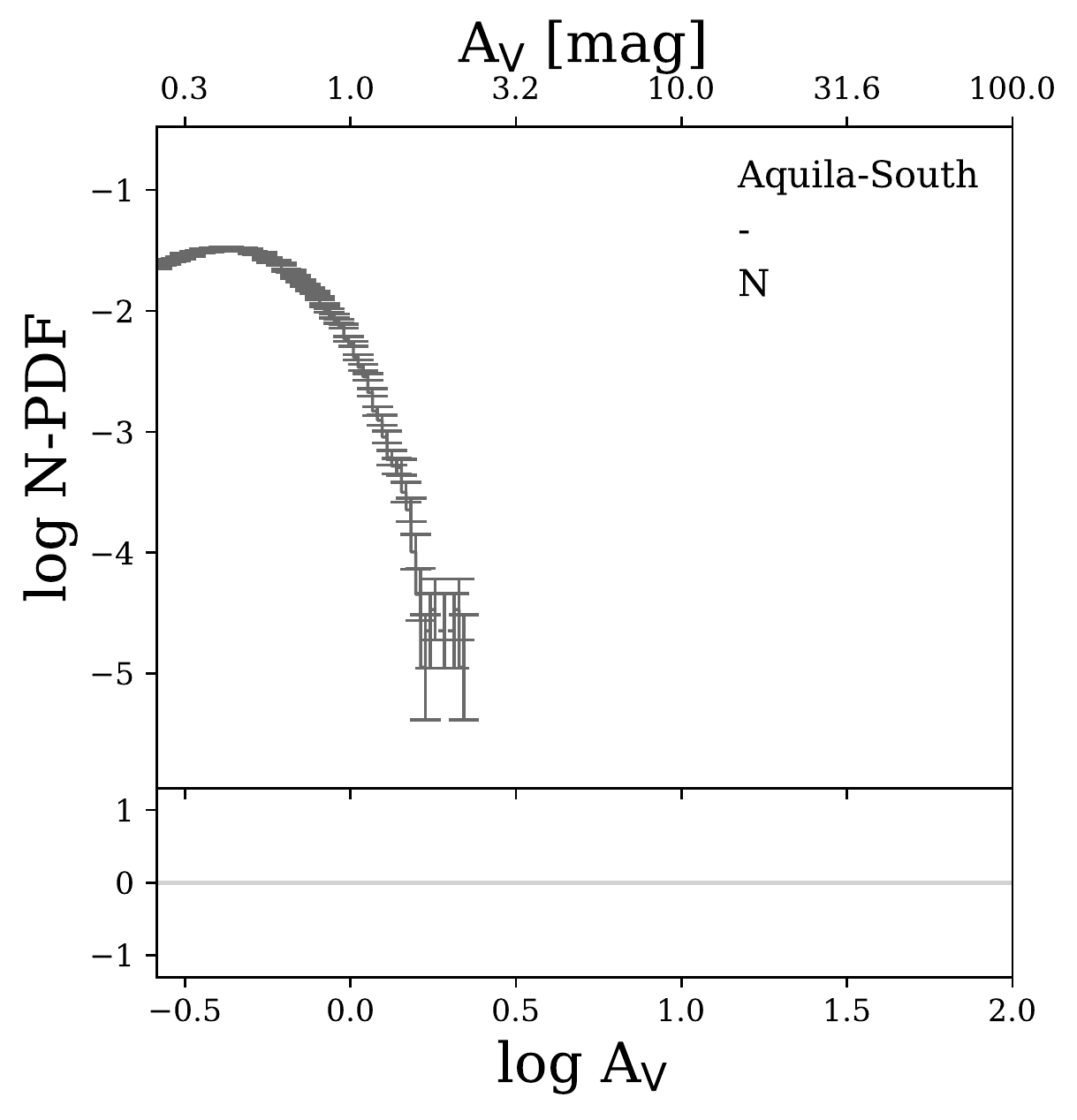}
    \end{minipage}
    
    \begin{minipage}[b]{0.24\textwidth}
        \includegraphics[width=\textwidth]{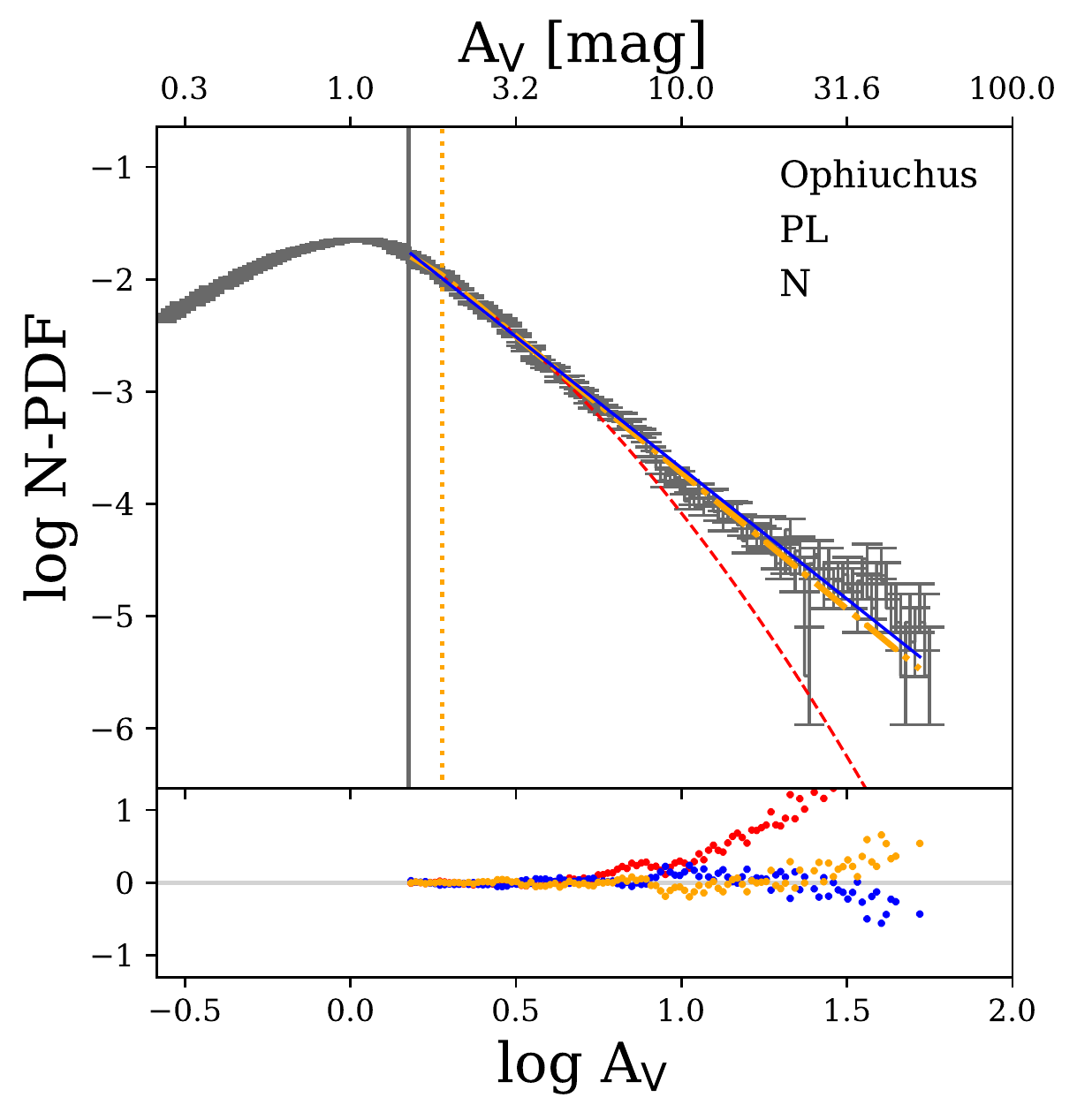}
    \end{minipage}
    \begin{minipage}[b]{0.24\textwidth}
        \includegraphics[width=\textwidth]{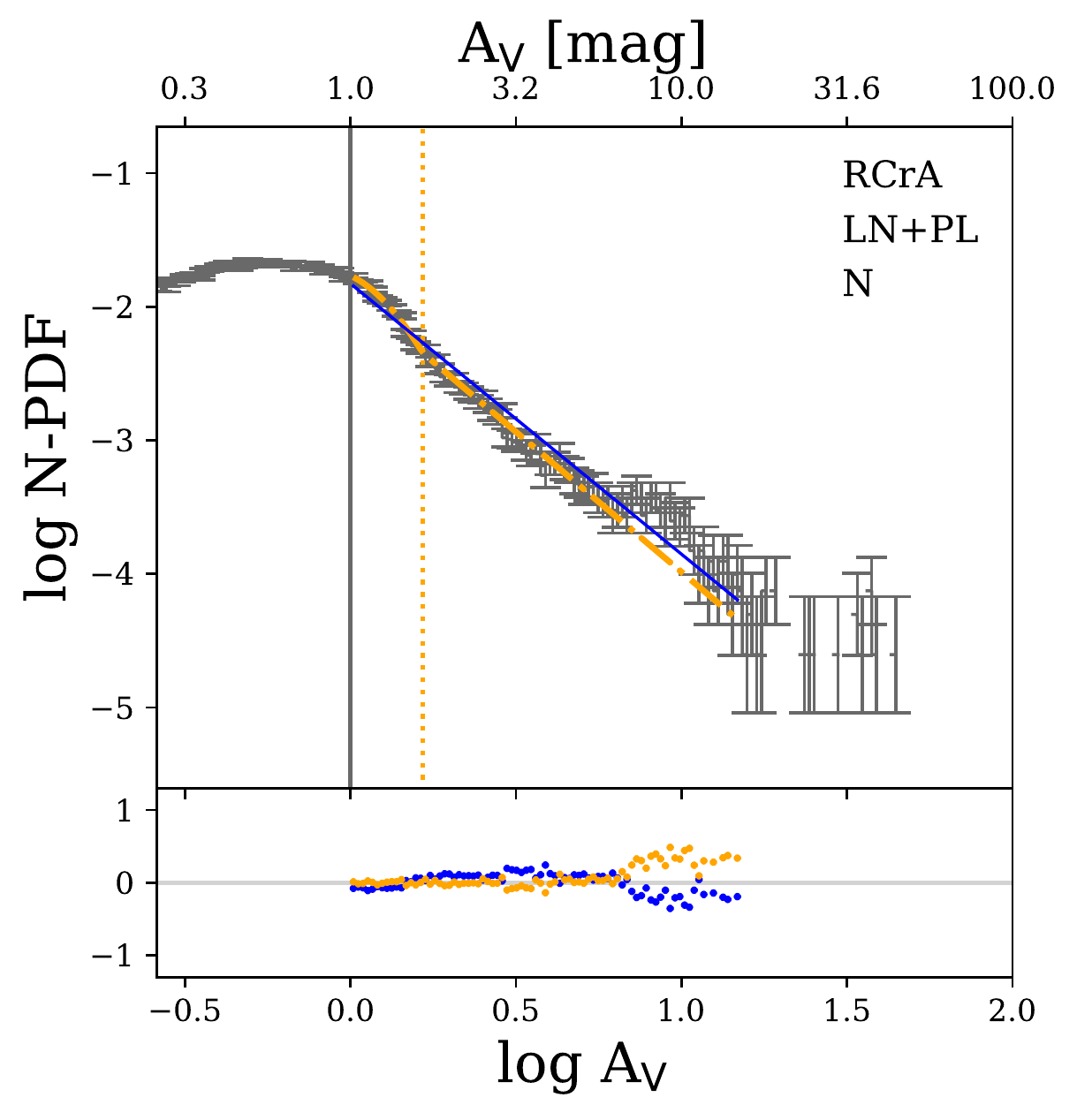}
    \end{minipage}
    \begin{minipage}[b]{0.24\textwidth}
        \includegraphics[width=\textwidth]{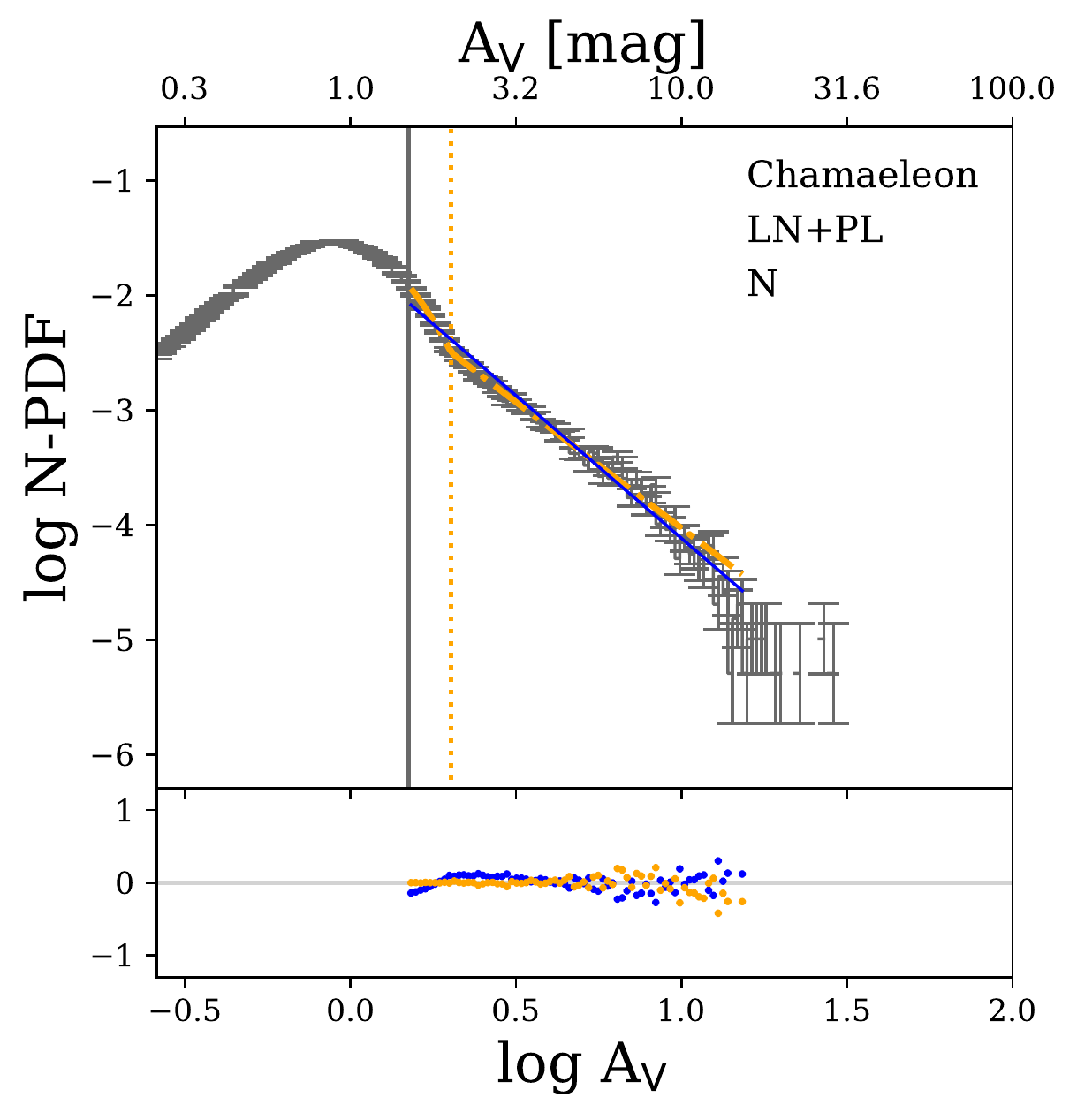}
    \end{minipage}
    \begin{minipage}[b]{0.24\textwidth}
        \includegraphics[width=\textwidth]{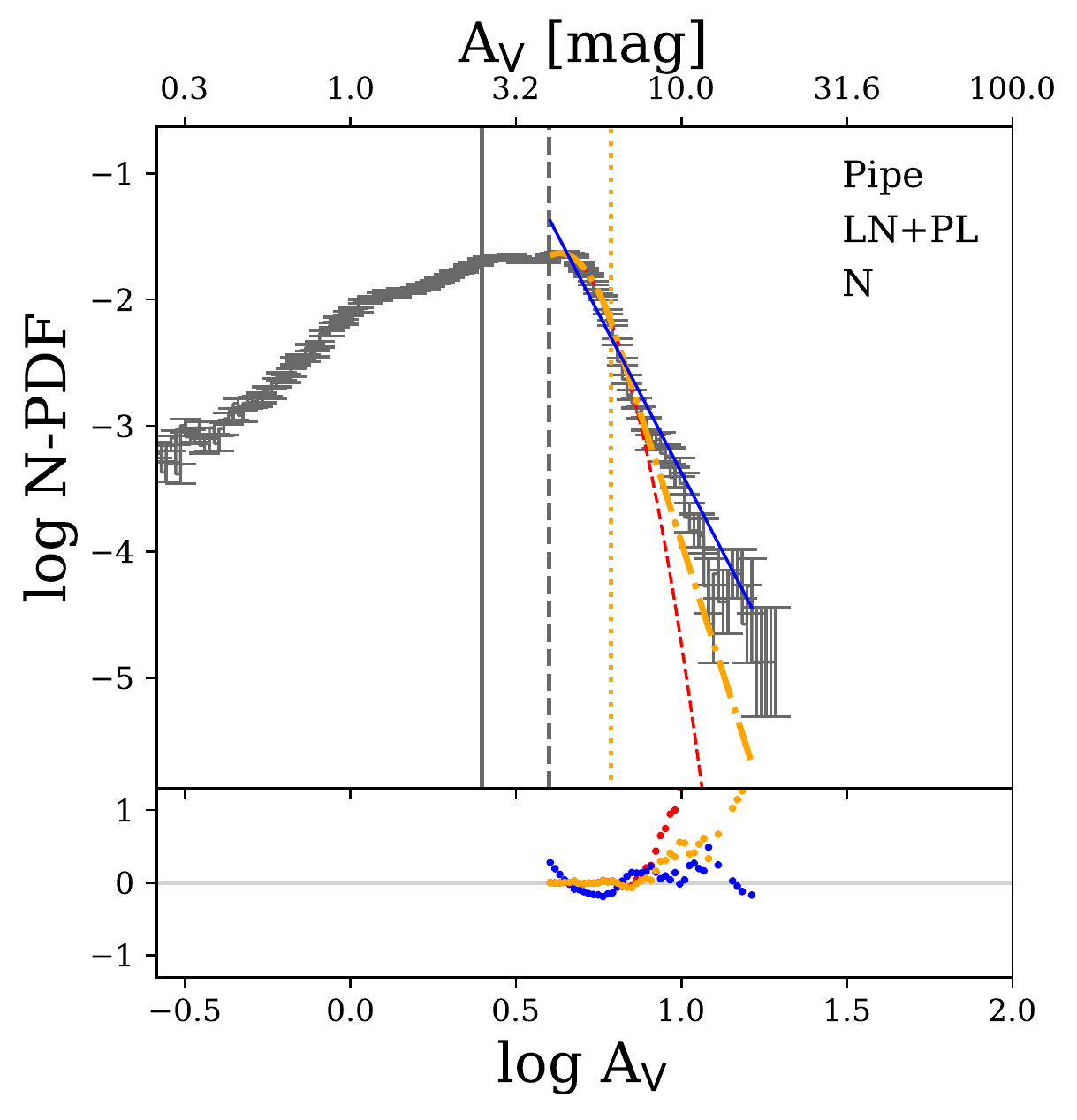}
    \end{minipage}
    
    \begin{minipage}[b]{0.24\textwidth}
        \includegraphics[width=\textwidth]{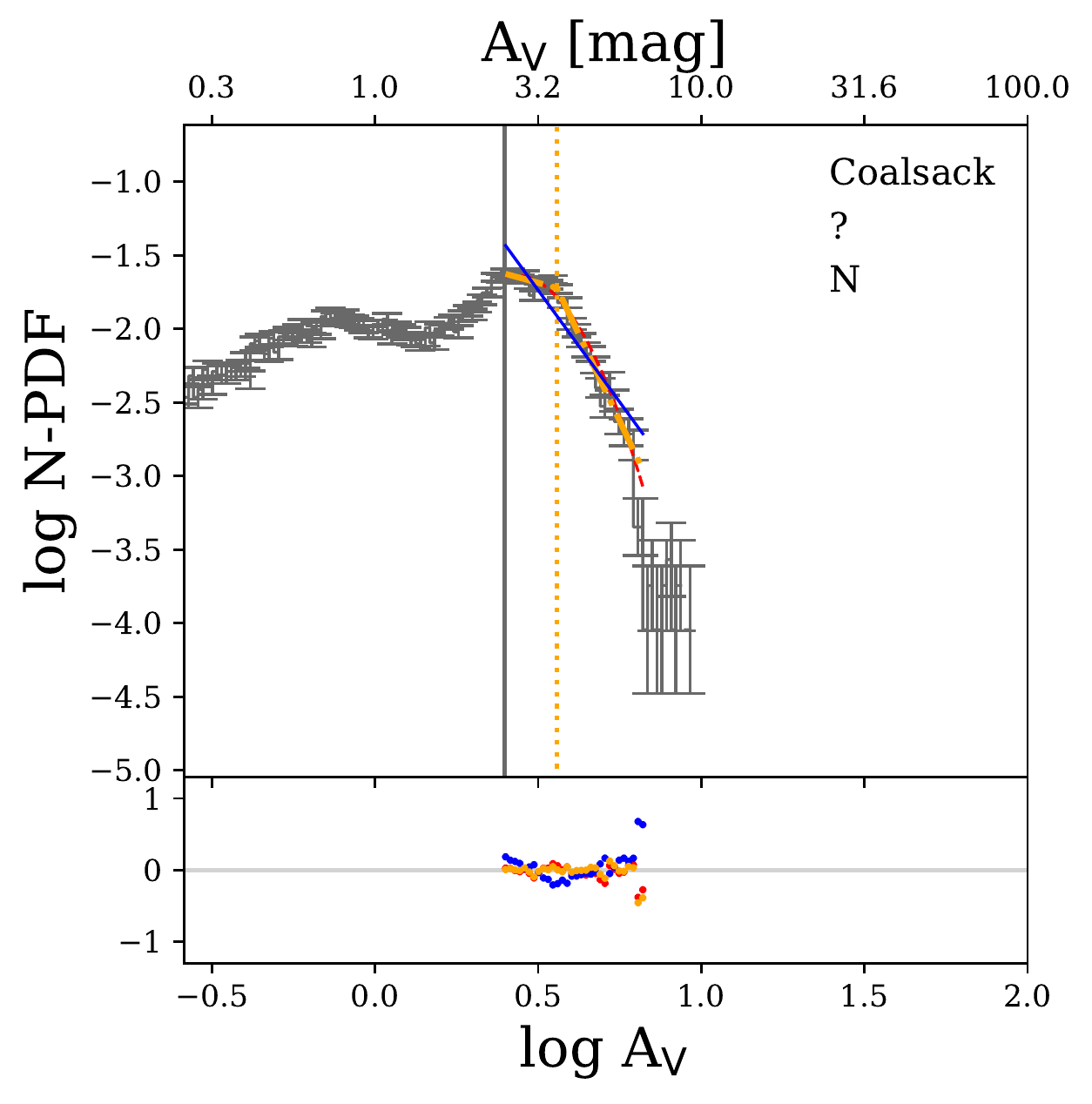}
    \end{minipage}
    \begin{minipage}[b]{0.24\textwidth}
        \includegraphics[width=\textwidth]{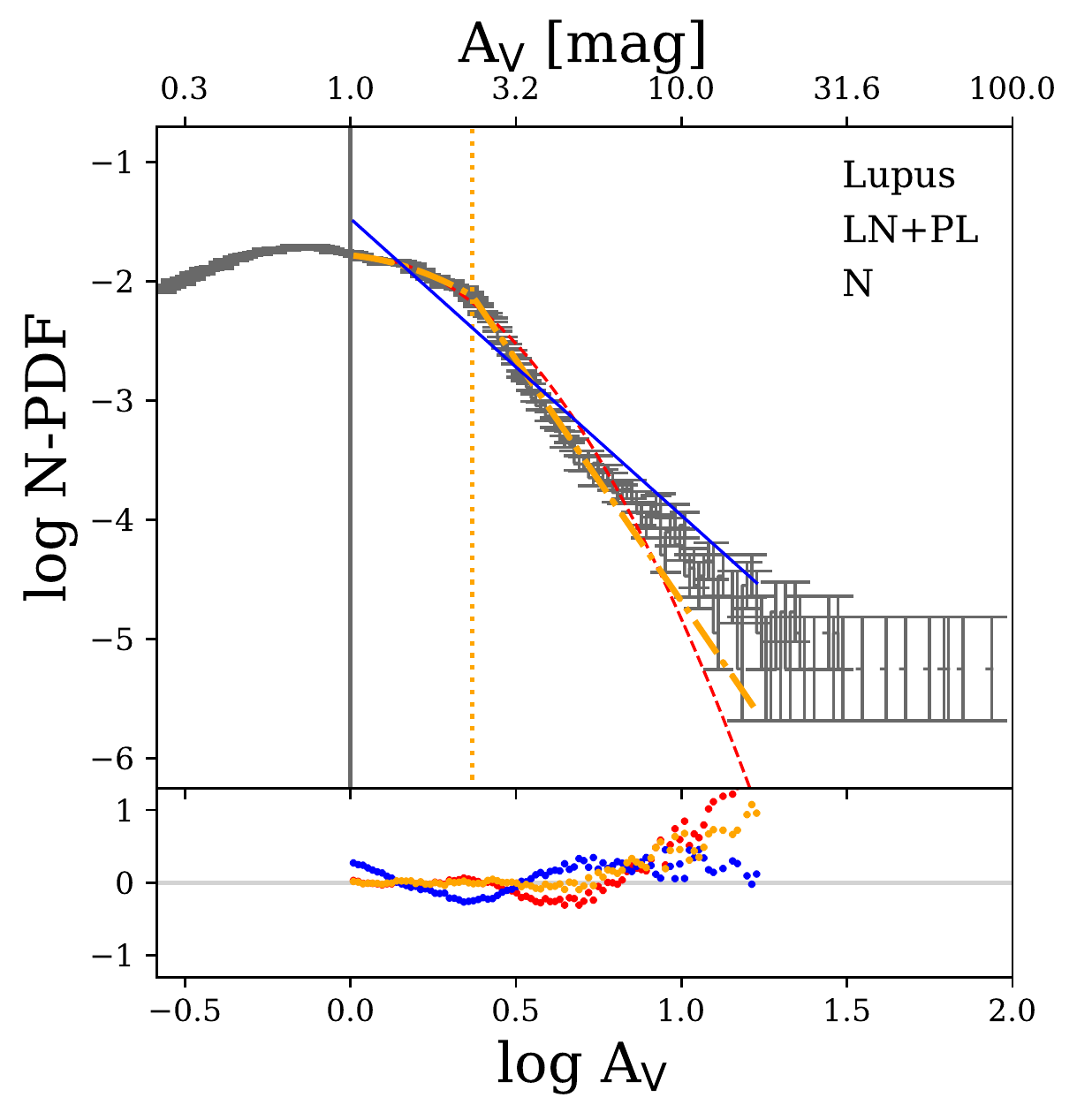}
    \end{minipage}
    \begin{minipage}[b]{0.24\textwidth}
        \includegraphics[width=\textwidth]{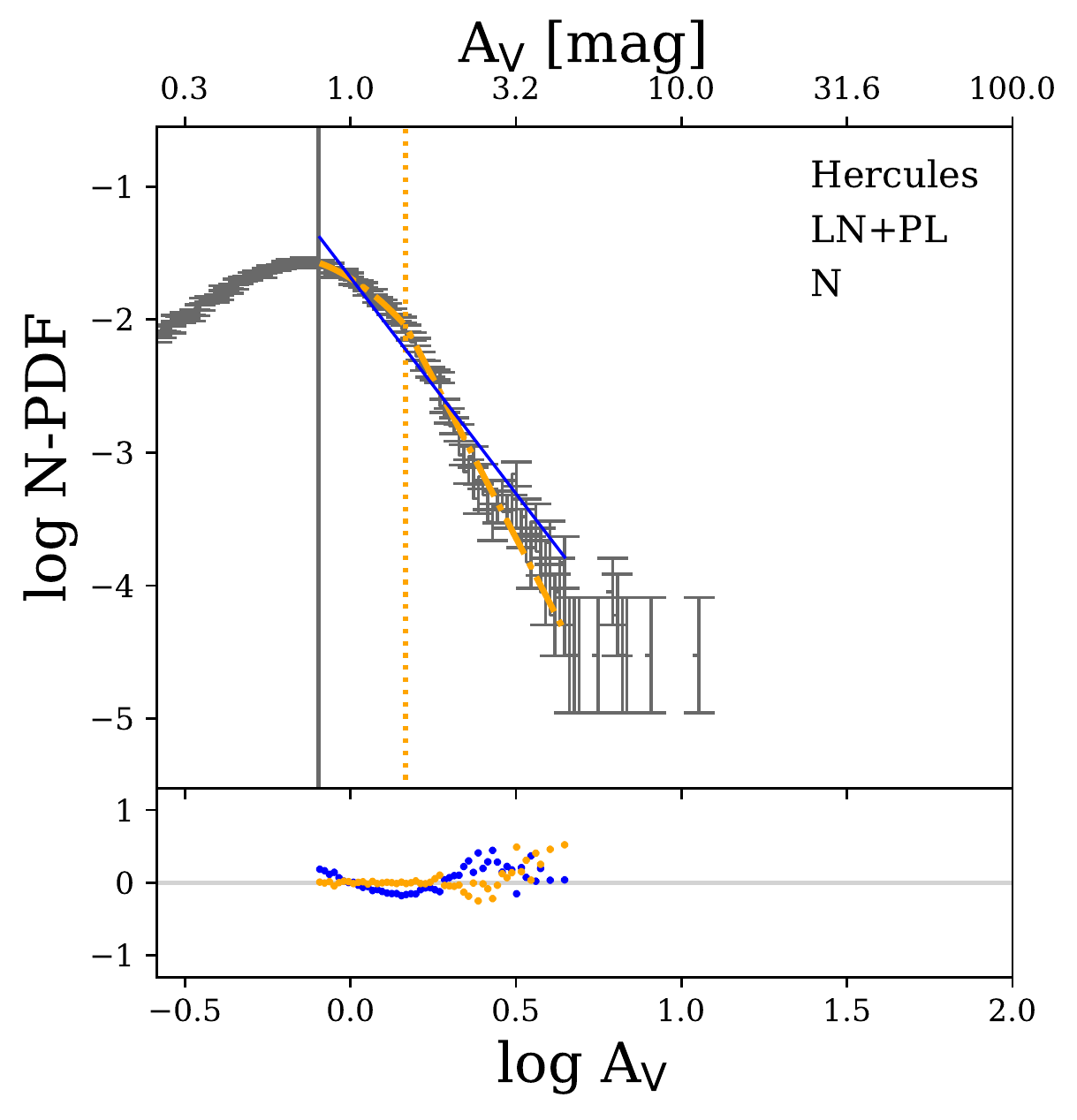}
    \end{minipage}
    \begin{minipage}[b]{0.24\textwidth}
        \includegraphics[width=\textwidth]{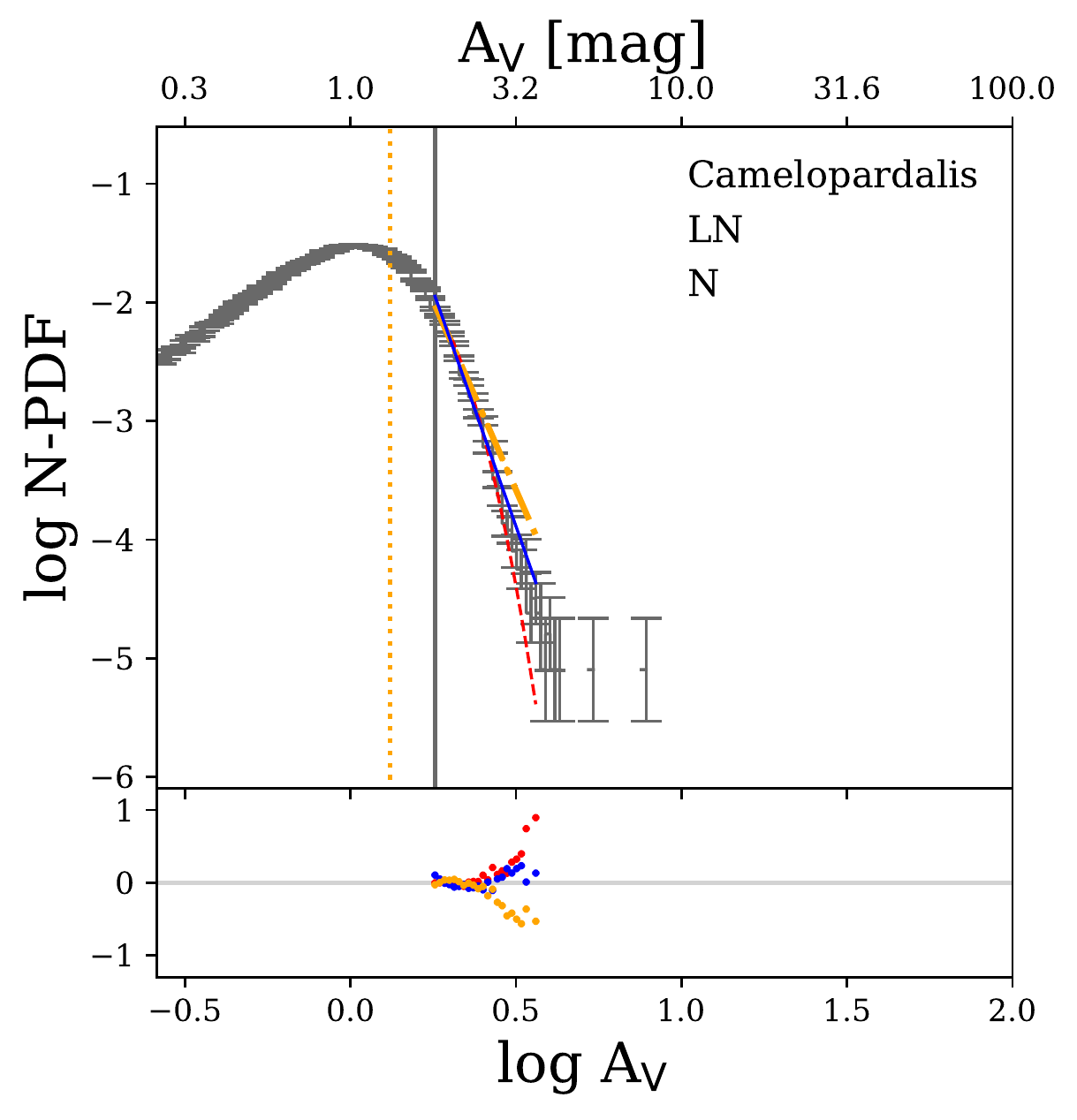}
    \end{minipage}
    
    \begin{minipage}[b]{0.24\textwidth}
        \includegraphics[width=\textwidth]{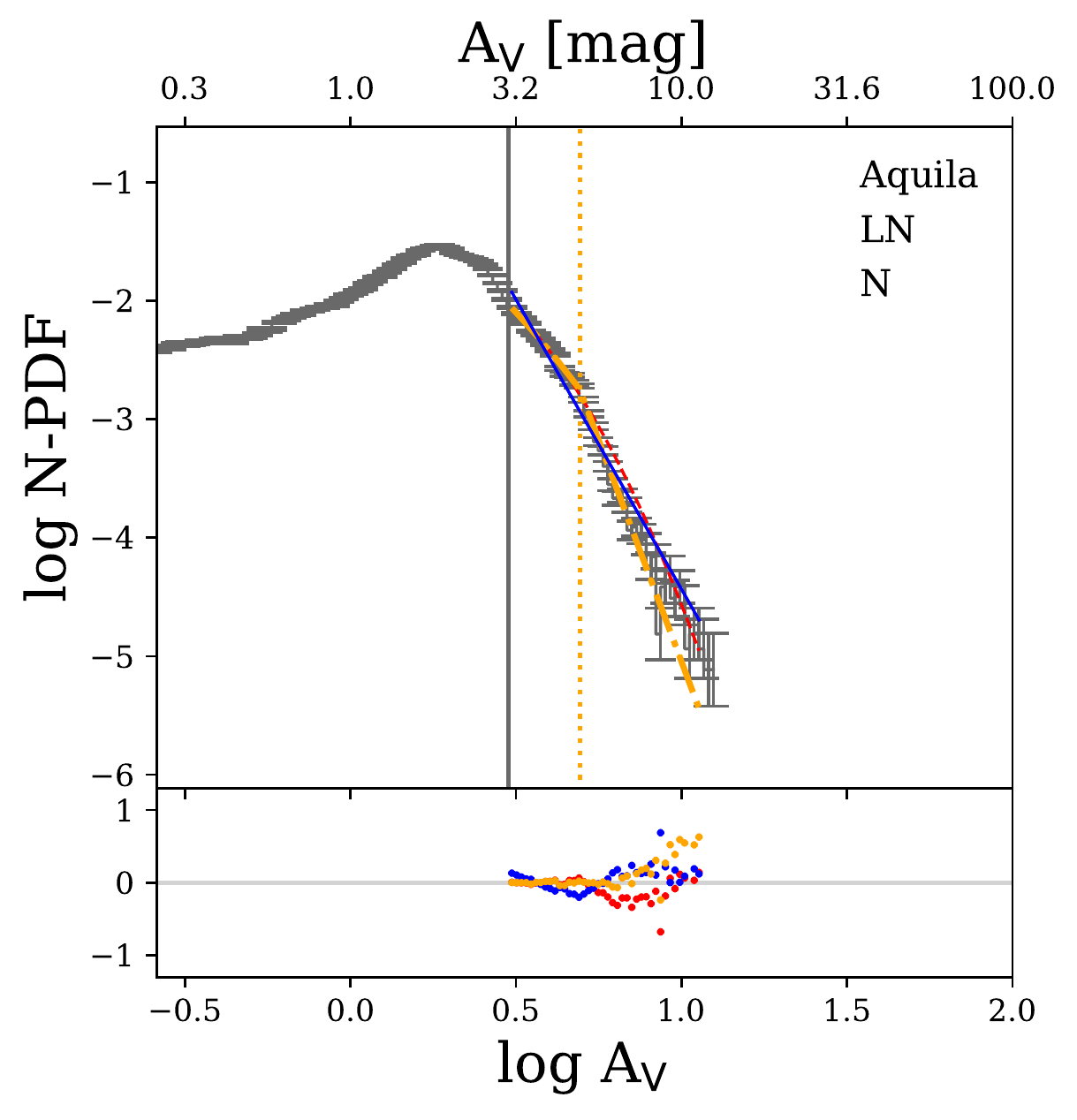}
    \end{minipage}
    \begin{minipage}[b]{0.24\textwidth}
        \includegraphics[width=\textwidth]{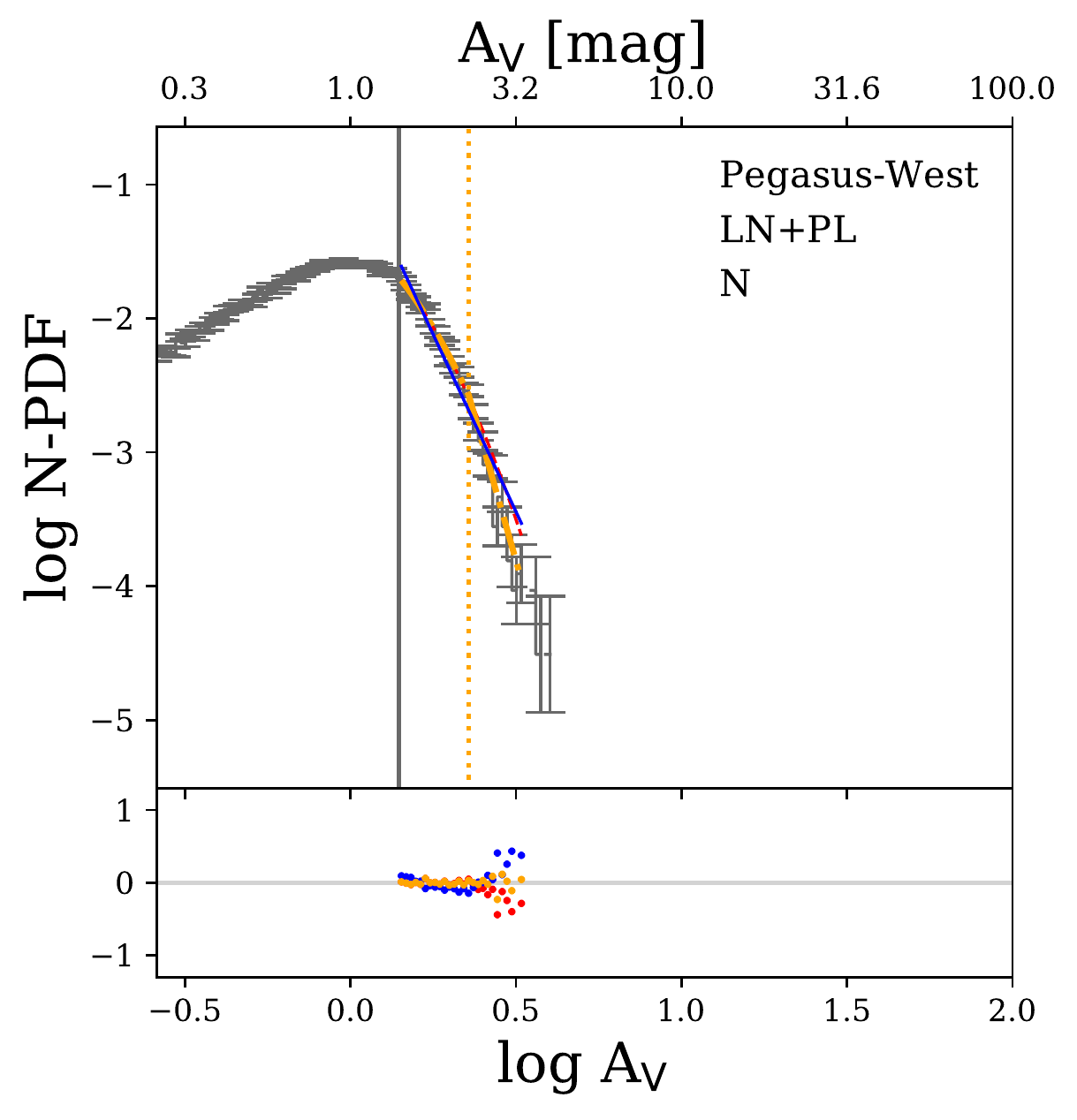}
    \end{minipage}
    \begin{minipage}[b]{0.24\textwidth}
        \includegraphics[width=\textwidth]{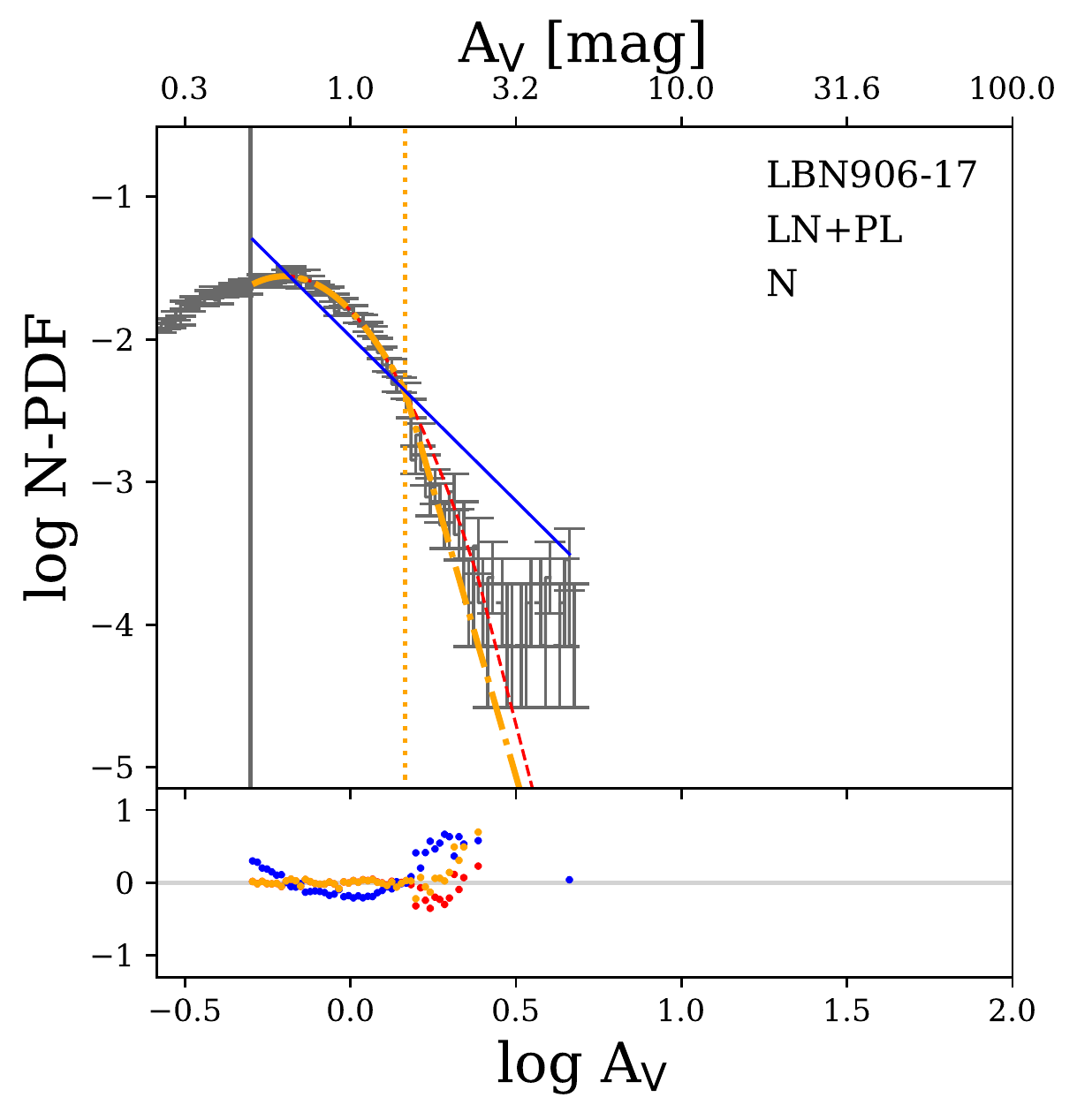}
    \end{minipage}
    \begin{minipage}[b]{0.24\textwidth}
        \includegraphics[width=\textwidth]{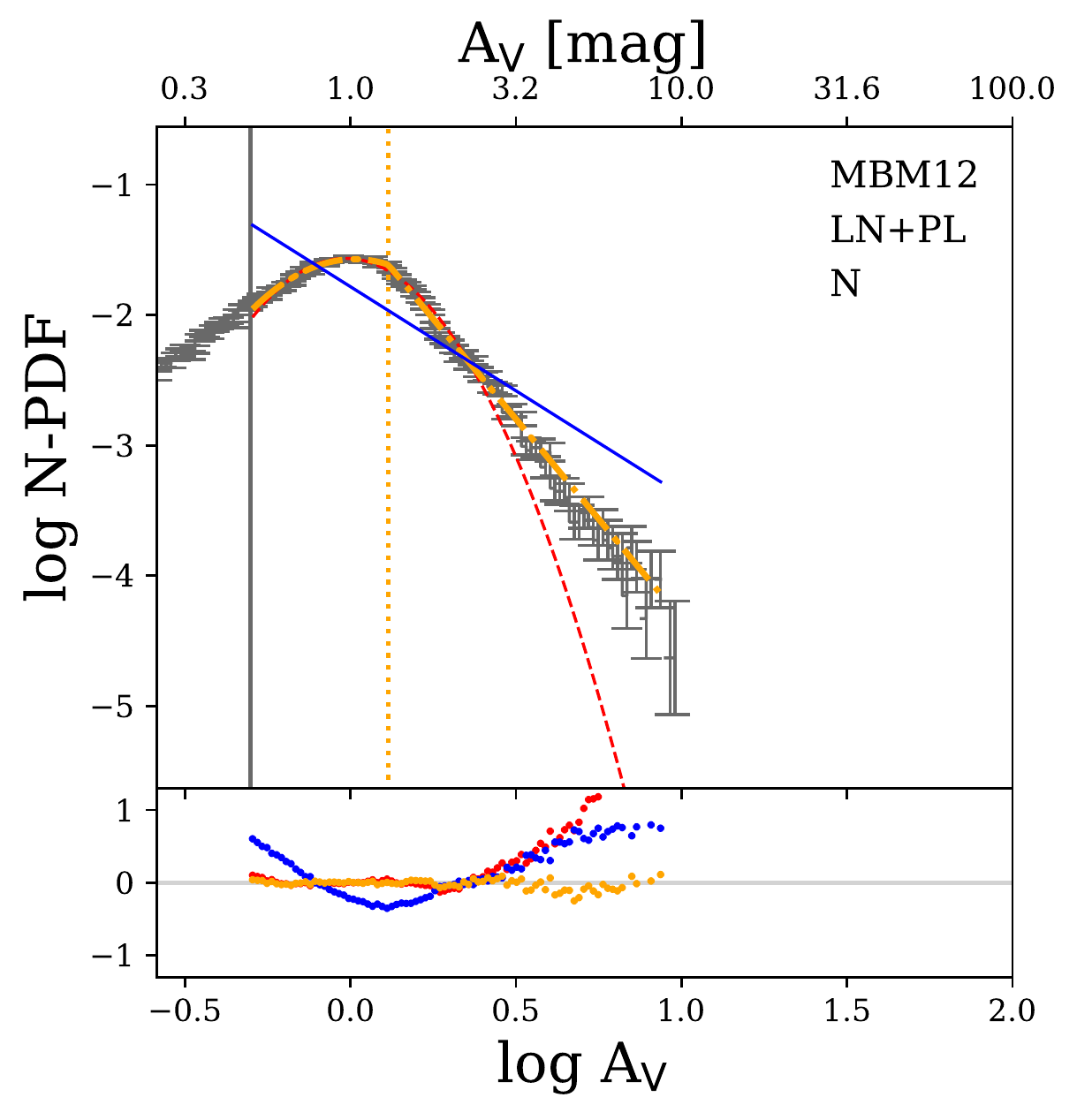}
    \end{minipage}

    \caption{N-PDFs of all clouds (Figs. \ref{fig:Allfits1}-\ref{fig:Allfits4}). The N-PDF used in the analysis is shown in dark grey. The extinction map used is noted in the legend, N for NICEST, pp for PPMAP by \citet{marsh2015temperature}, and A for the extinction maps derived in Appendix \ref{app:cygnus_and_cartwheel}. For the N-PDFs not derived from NICEST maps, the N-PDF derived from \citet{juvela2016allsky} is shown in light grey for comparison. The blue, red, and yellow lines show the fits of PL, LN, and LN+PL models, respectively. The transition point of the LN+PL model is shown with a vertical dotted yellow line. All fits were made above the extinction at the last closed contour, which is marked with the vertical grey line. The residuals of the fitted models are shown below the N-PDFs, and the best-fit shape is noted in the legend. The N-PDFs are ordered according to distance, as in Table \ref{tab:mastertable}.}
    \label{fig:Allfits1}
\end{figure*}

\begin{figure*}[h!]
    \centering
    \begin{minipage}[b]{0.24\textwidth}
        \includegraphics[width=\textwidth]{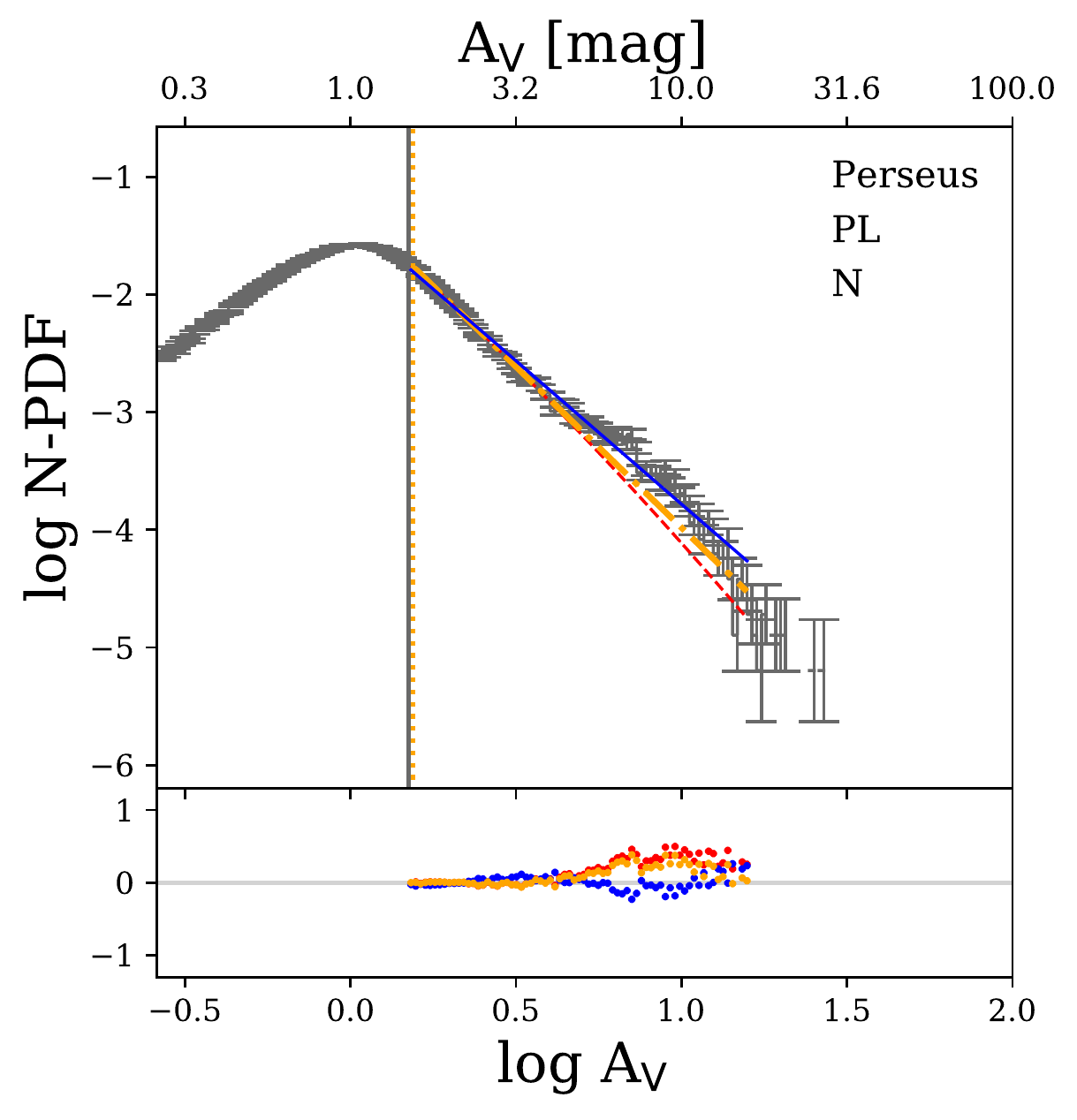}
    \end{minipage}
    \begin{minipage}[b]{0.24\textwidth}
        \includegraphics[width=\textwidth]{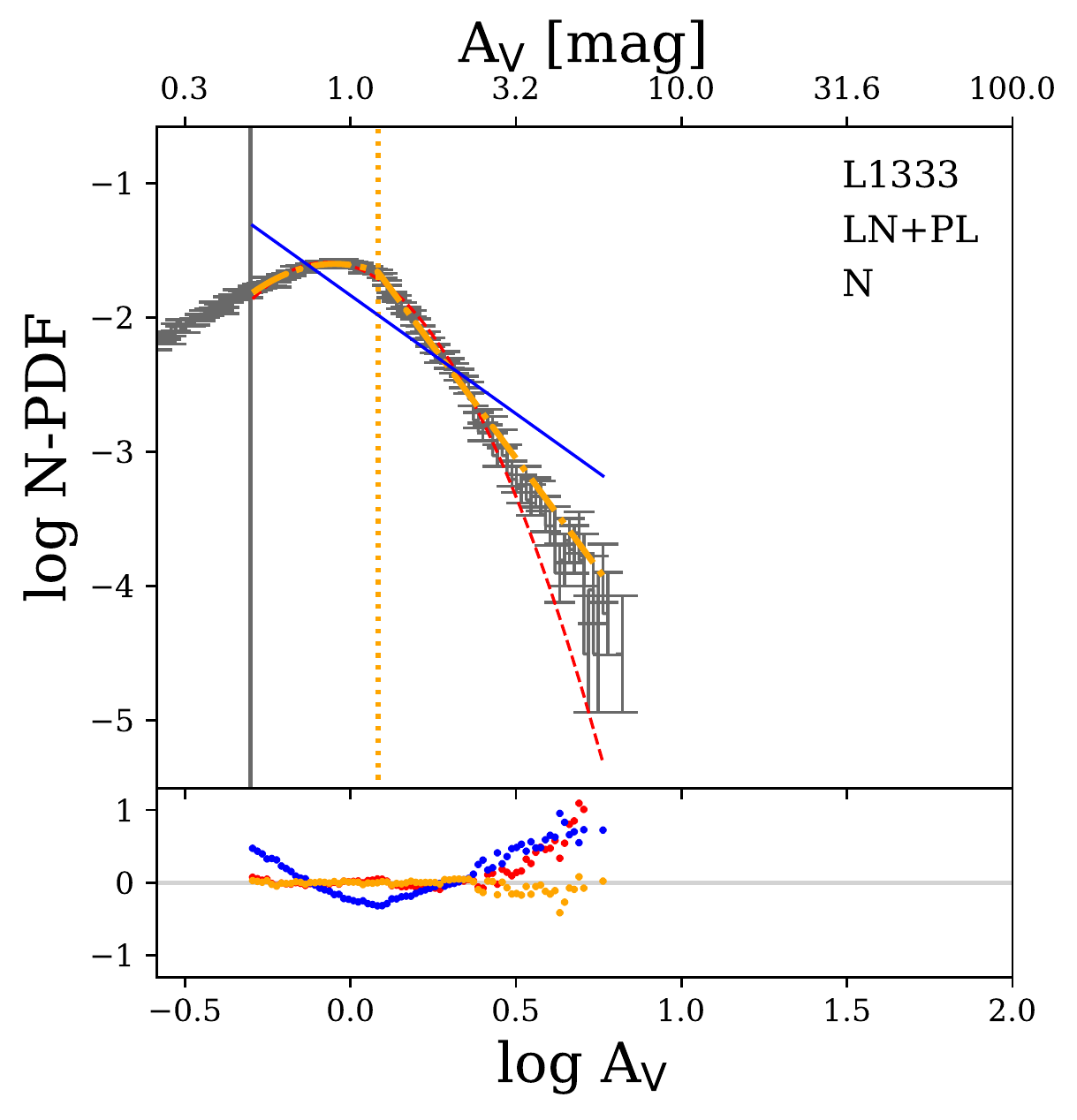}
    \end{minipage}
    \begin{minipage}[b]{0.24\textwidth}
        \includegraphics[width=\textwidth]{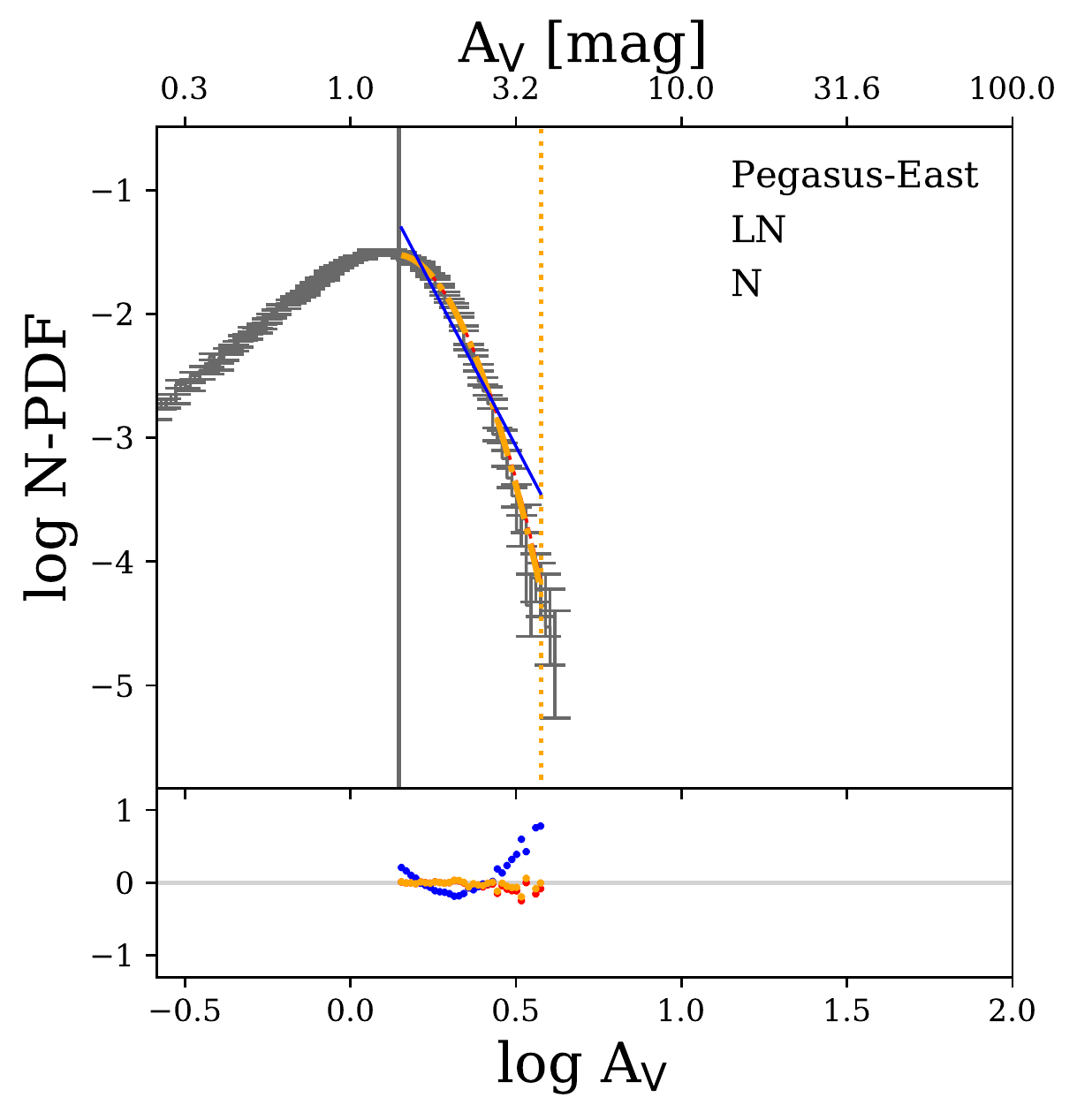}
    \end{minipage}
    \begin{minipage}[b]{0.24\textwidth}
        \includegraphics[width=\textwidth]{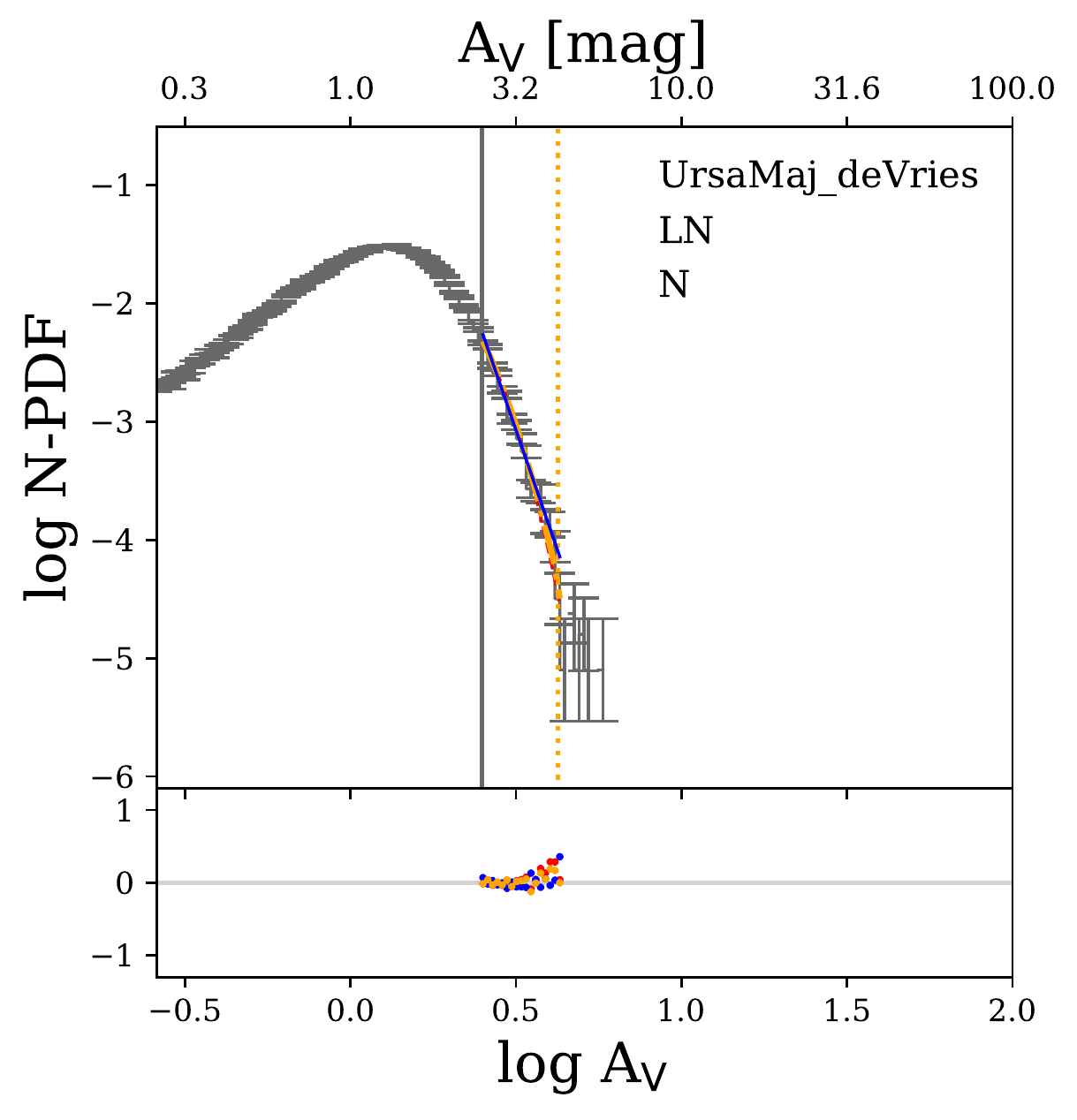}
    \end{minipage}
    
    \begin{minipage}[b]{0.24\textwidth}
        \includegraphics[width=\textwidth]{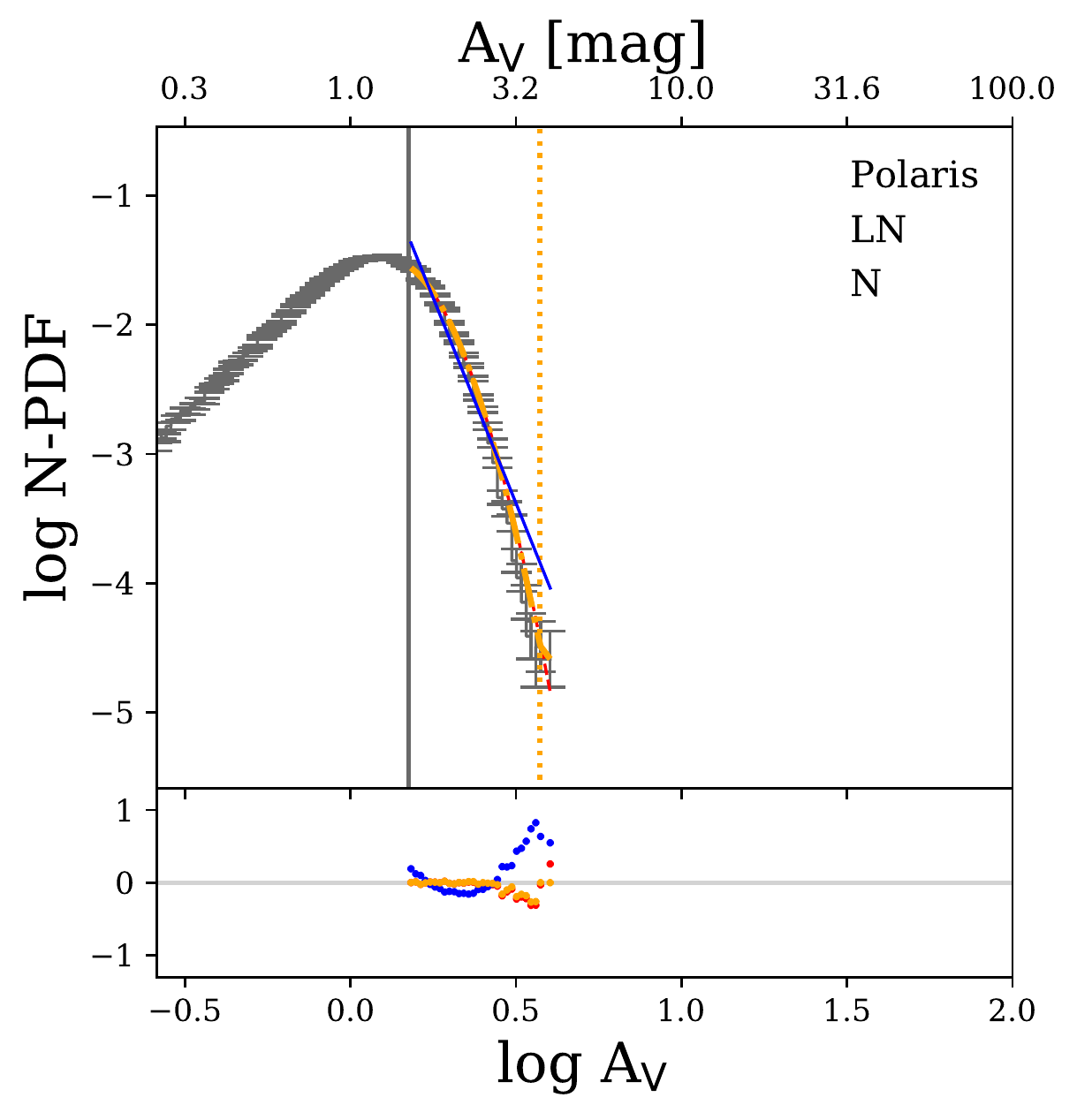}
    \end{minipage}
    \begin{minipage}[b]{0.24\textwidth}
        \includegraphics[width=\textwidth]{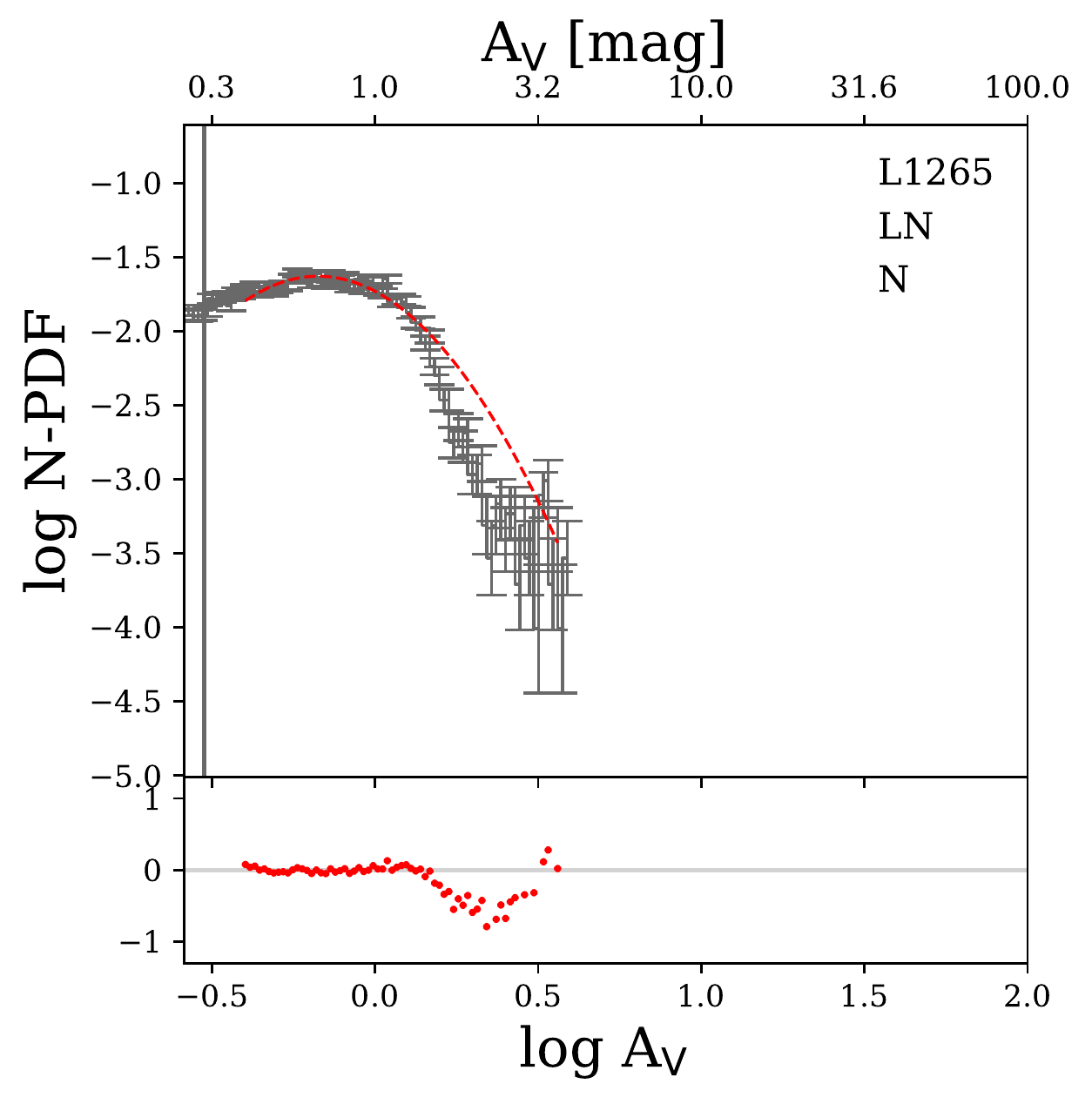}
    \end{minipage}
    \begin{minipage}[b]{0.24\textwidth}
        \includegraphics[width=\textwidth]{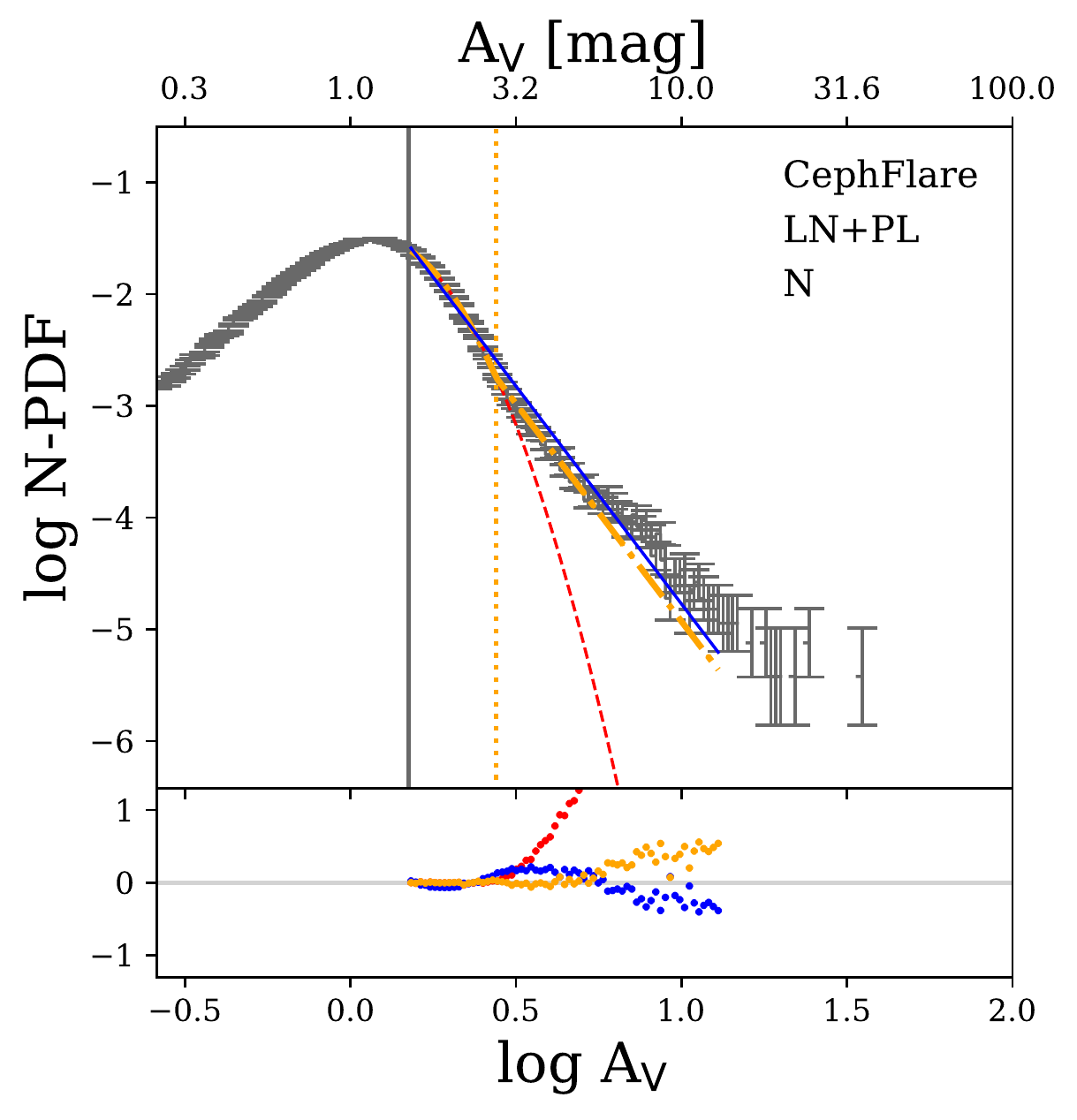}
    \end{minipage}
    \begin{minipage}[b]{0.24\textwidth}
        \includegraphics[width=\textwidth]{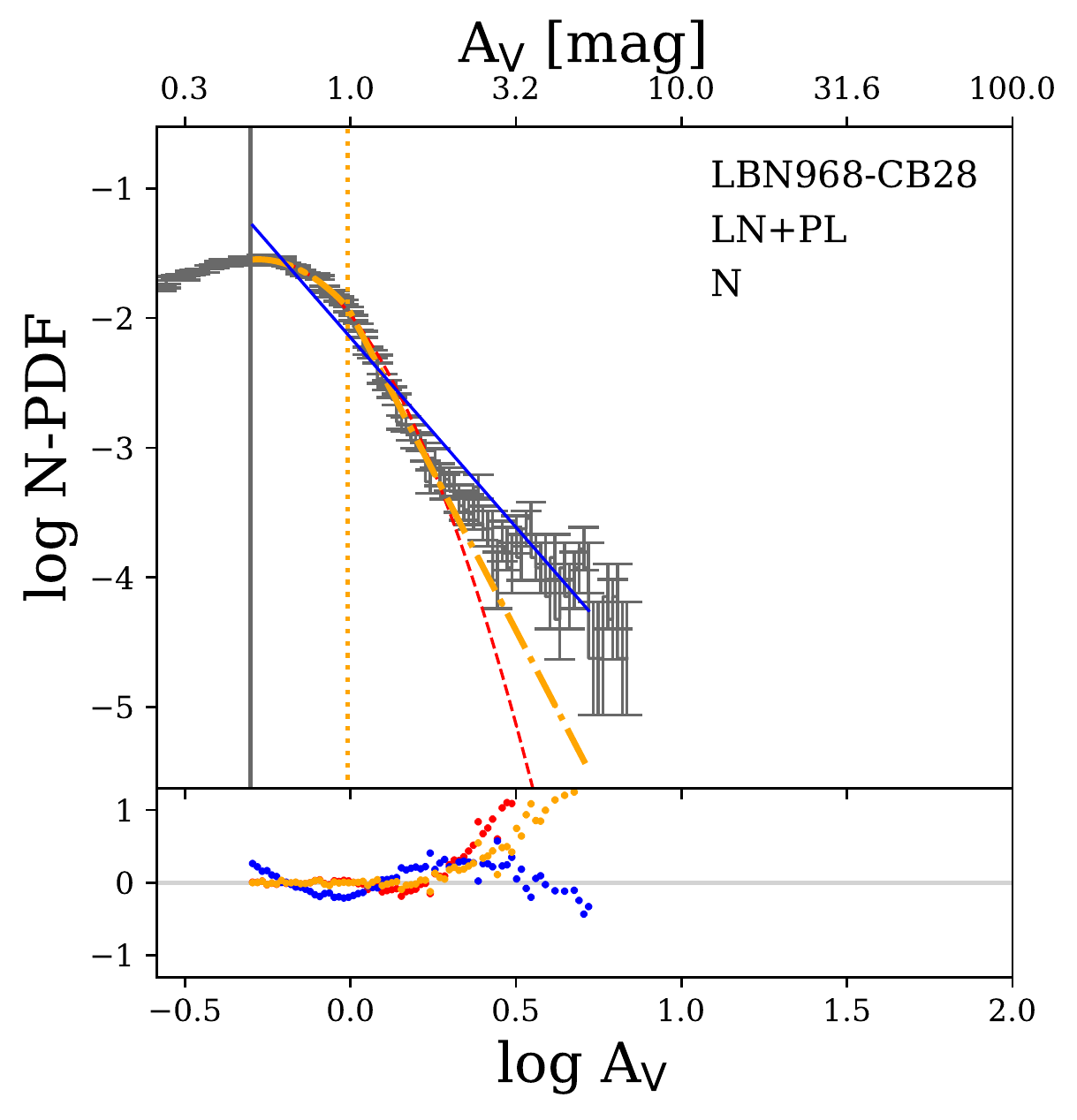}
    \end{minipage}
    
    \begin{minipage}[b]{0.24\textwidth}
        \includegraphics[width=\textwidth]{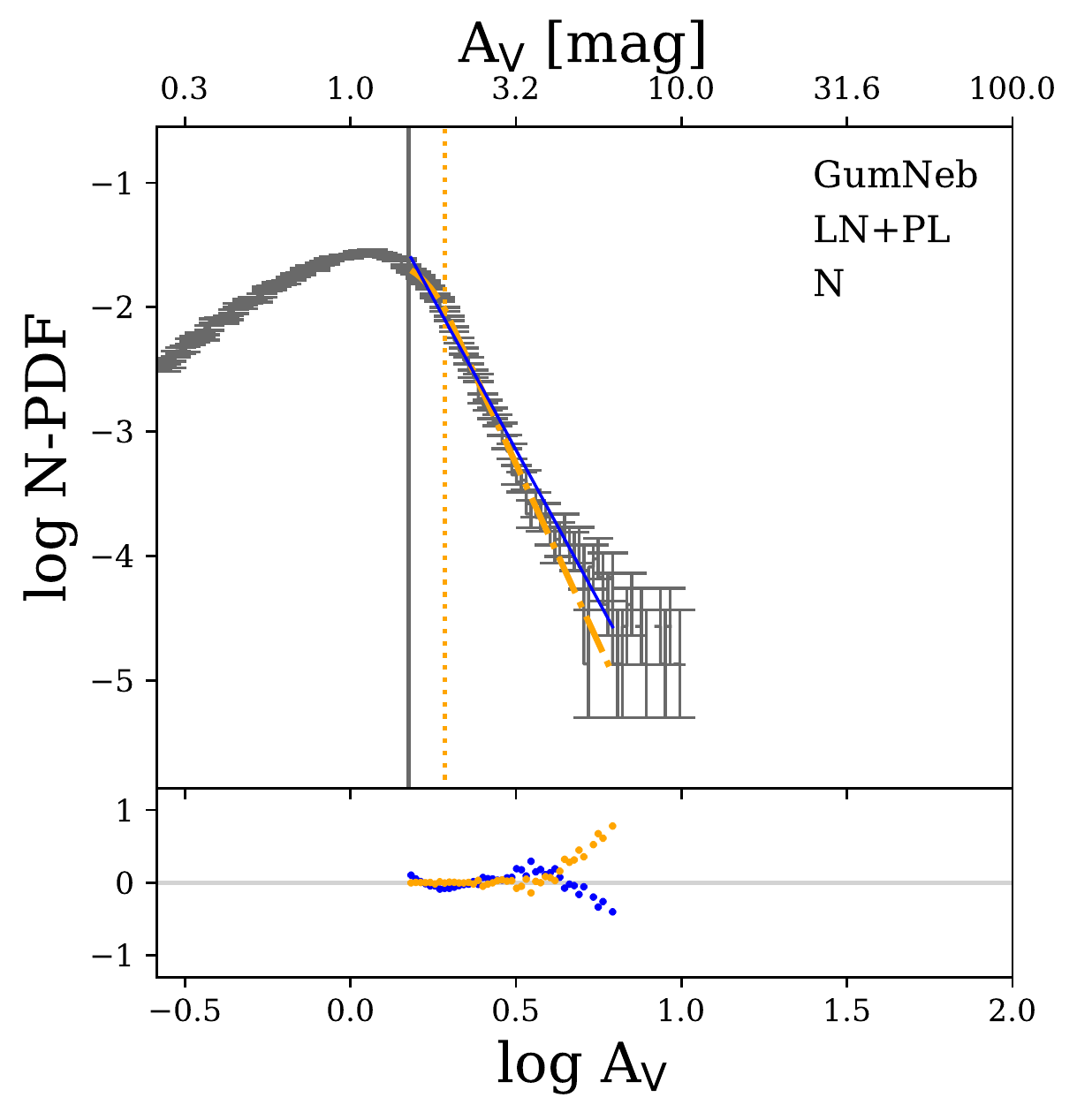}
    \end{minipage}
    \begin{minipage}[b]{0.24\textwidth}
        \includegraphics[width=\textwidth]{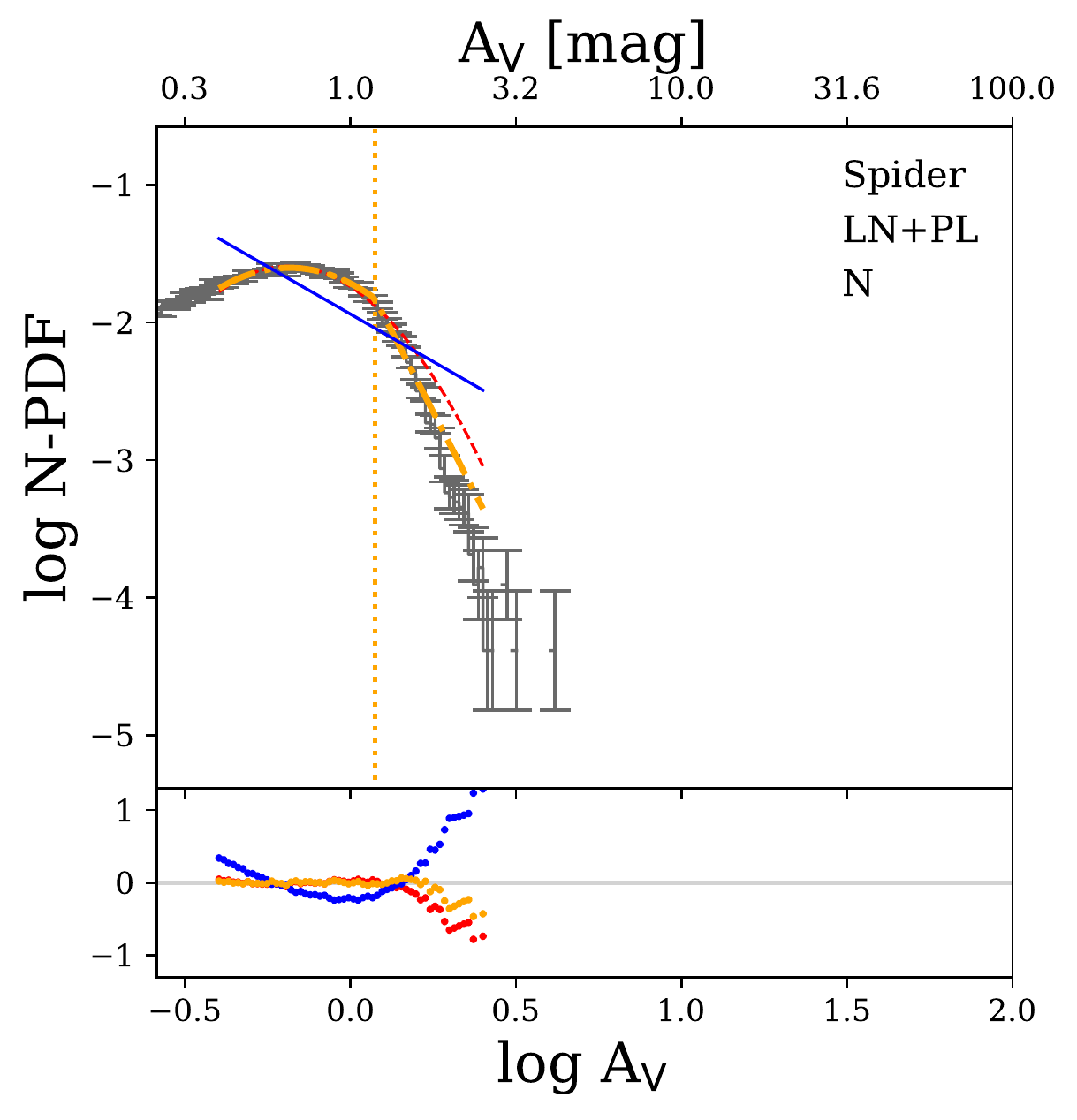}
    \end{minipage}
    \begin{minipage}[b]{0.24\textwidth}
        \includegraphics[width=\textwidth]{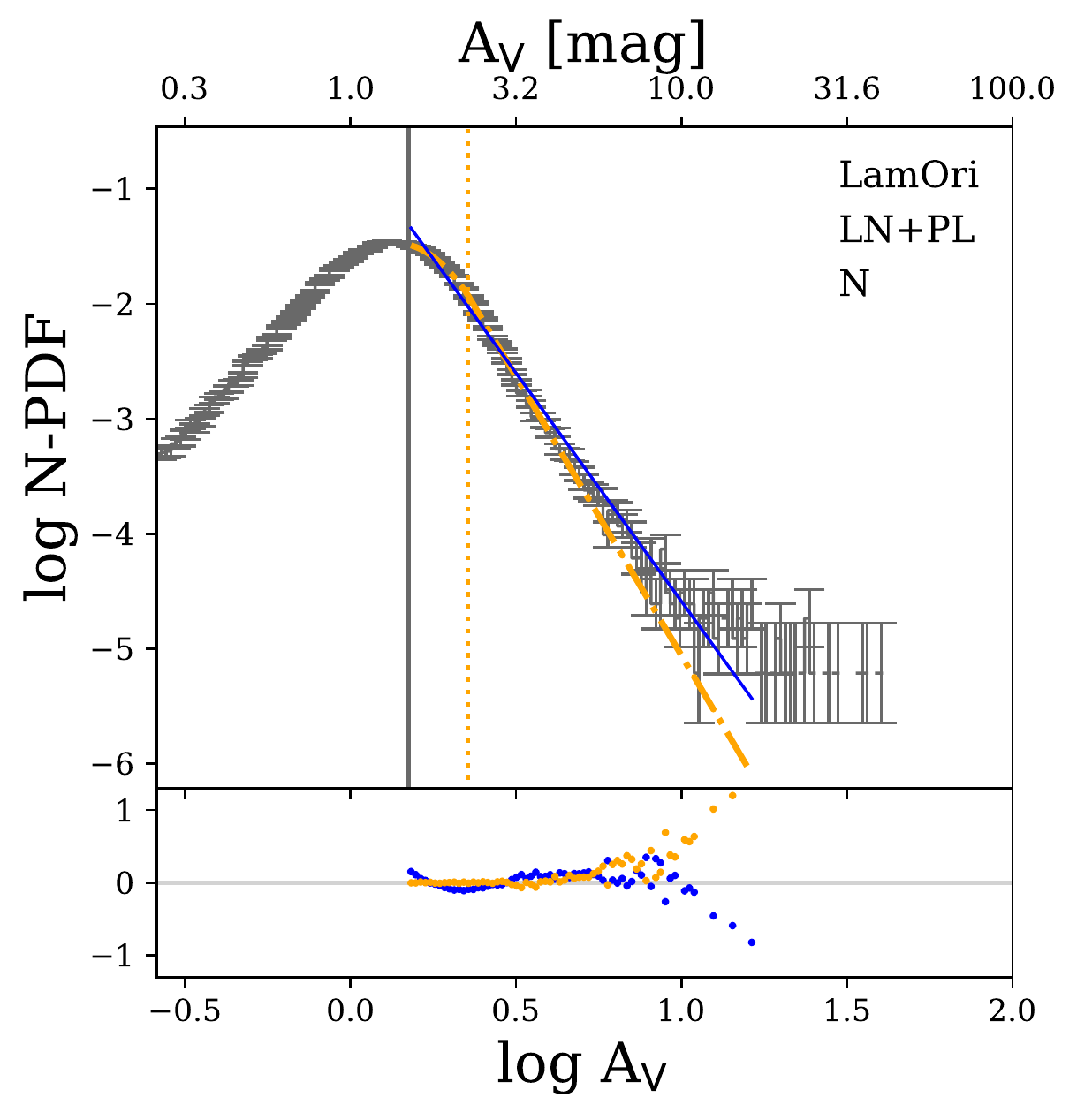}
    \end{minipage}
    \begin{minipage}[b]{0.24\textwidth}
        \includegraphics[width=\textwidth]{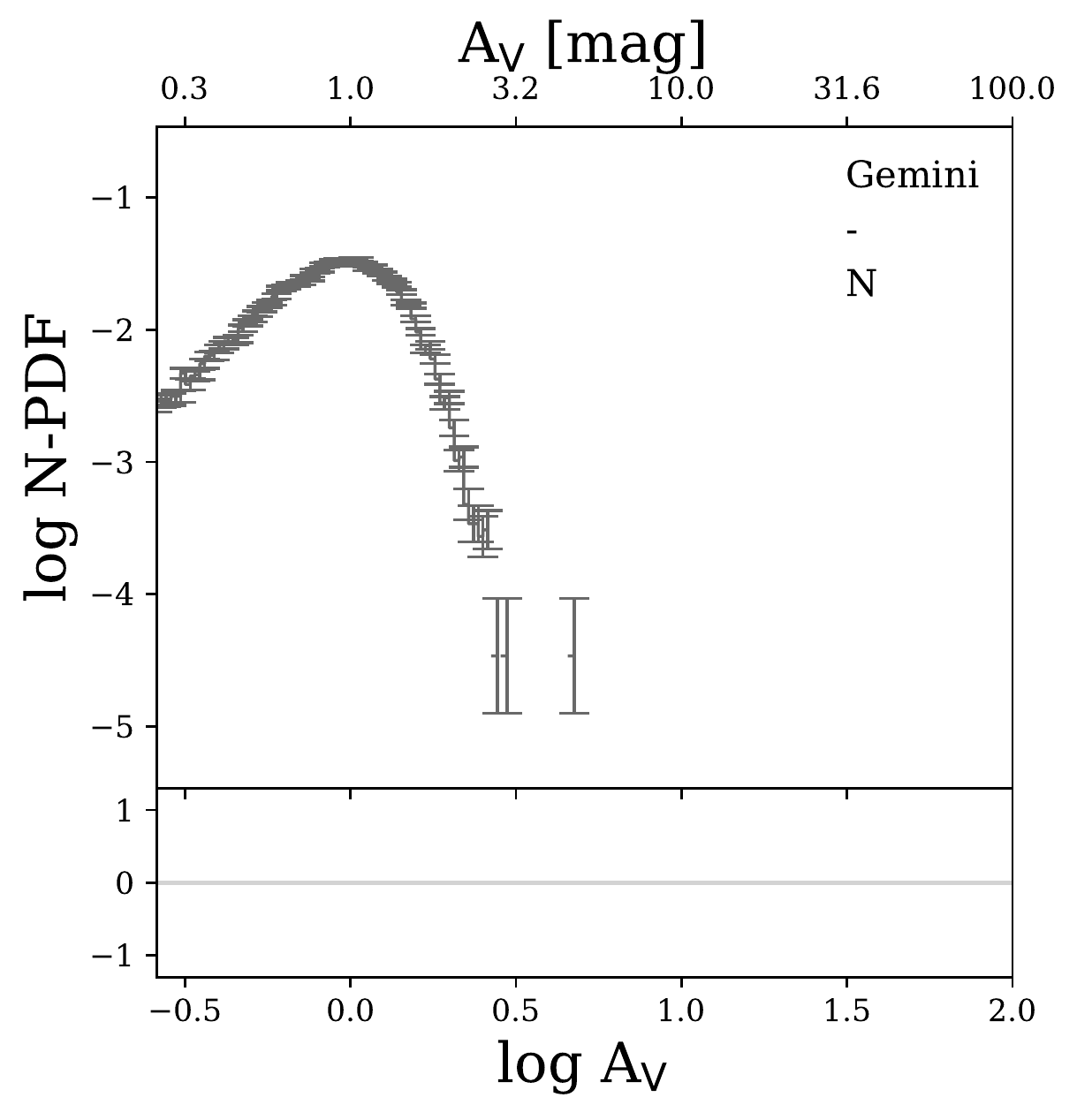}
    \end{minipage}
    
    \begin{minipage}[b]{0.24\textwidth}
        \includegraphics[width=\textwidth]{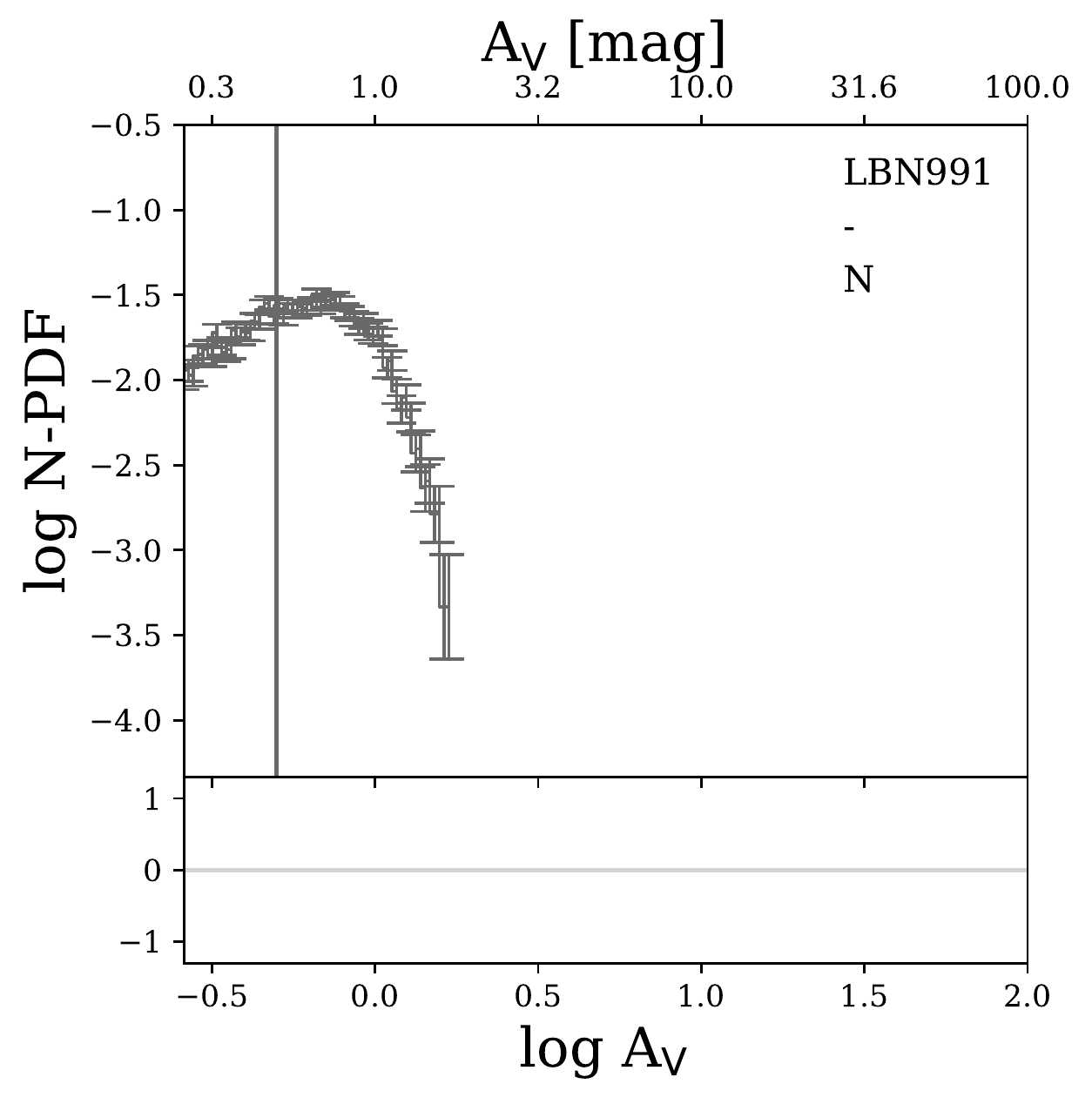}
    \end{minipage}
    \begin{minipage}[b]{0.24\textwidth}
        \includegraphics[width=\textwidth]{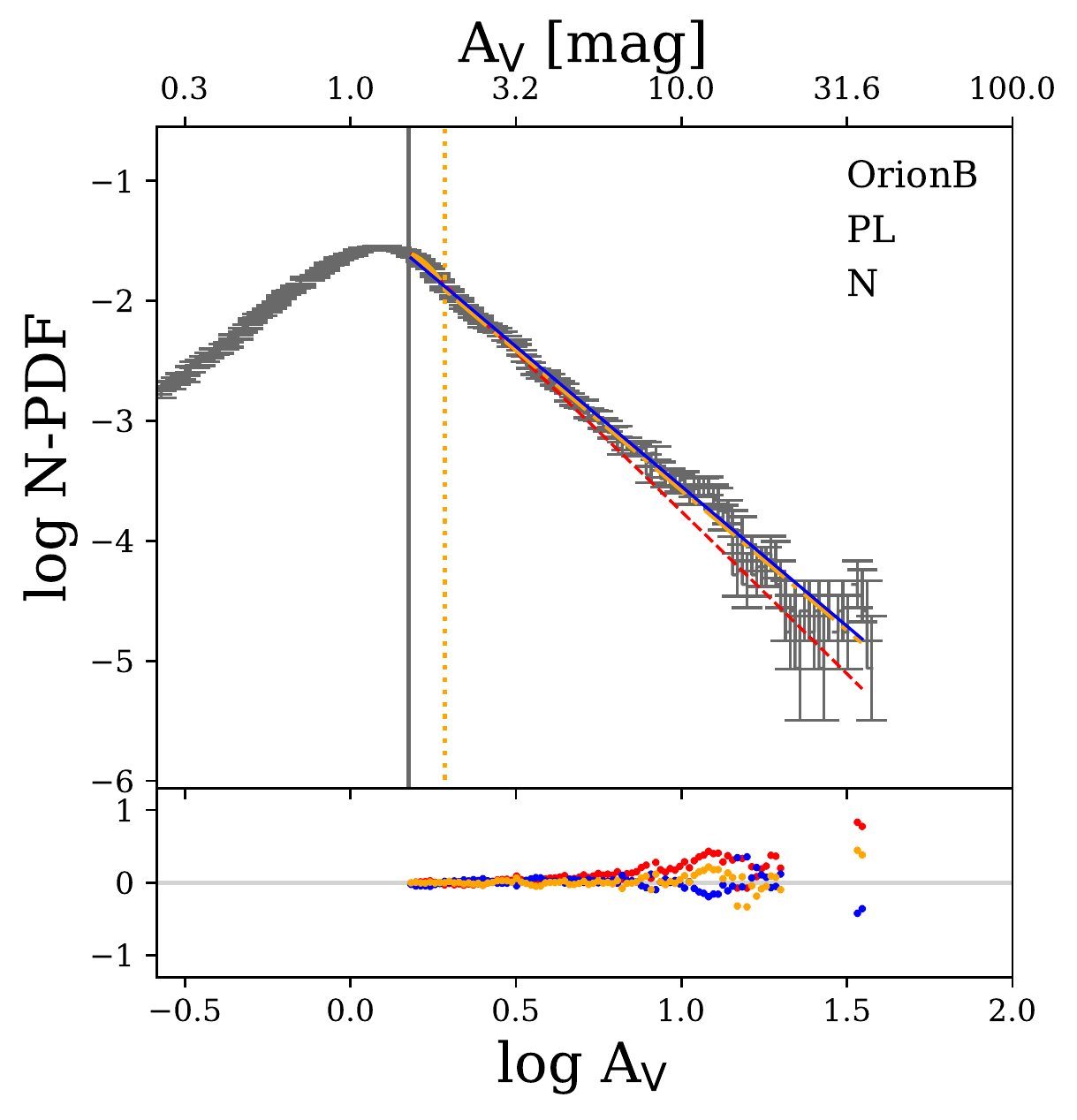}
    \end{minipage}
    \begin{minipage}[b]{0.24\textwidth}
        \includegraphics[width=\textwidth]{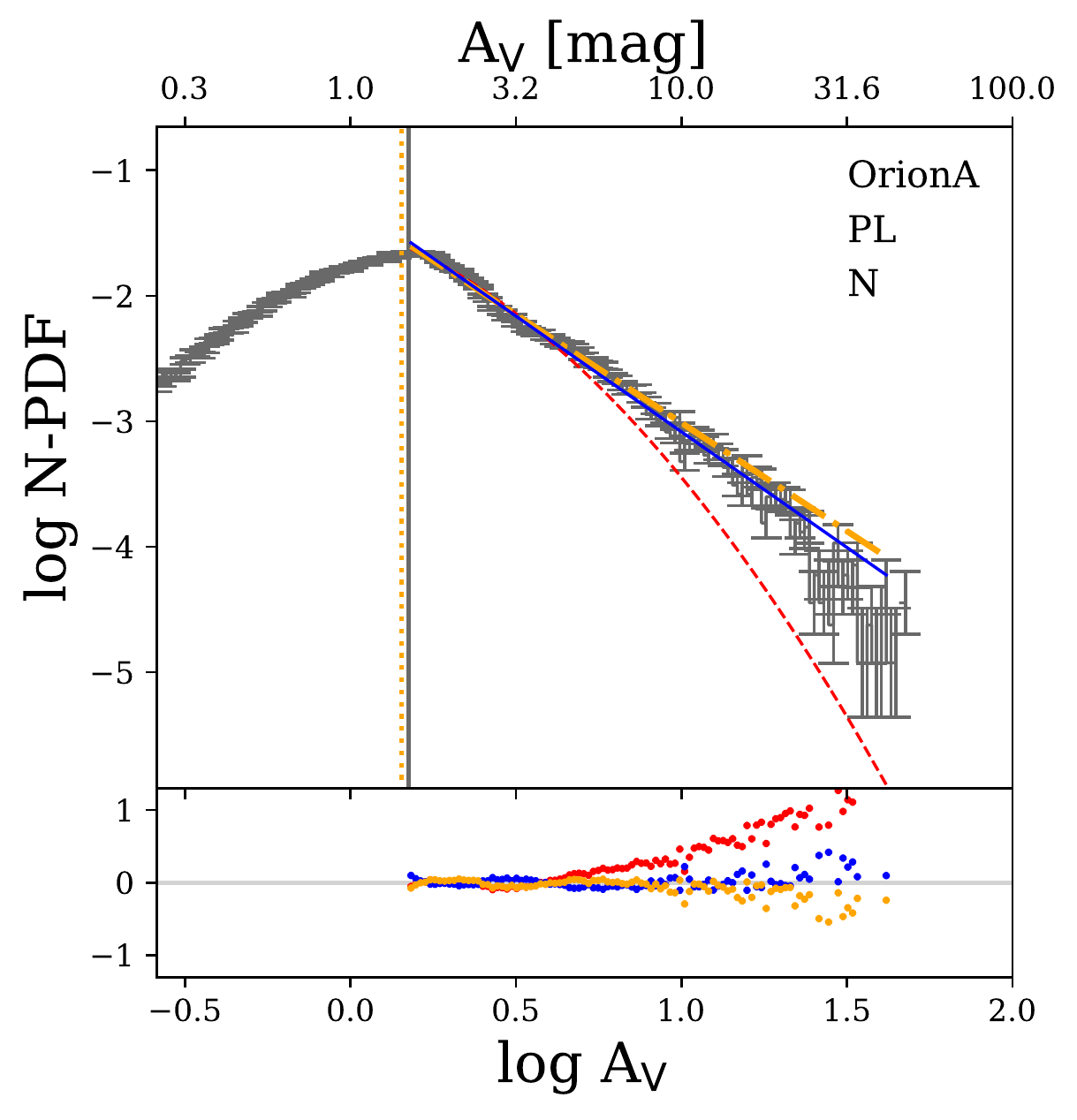}
    \end{minipage}
    \begin{minipage}[b]{0.24\textwidth}
        \includegraphics[width=\textwidth]{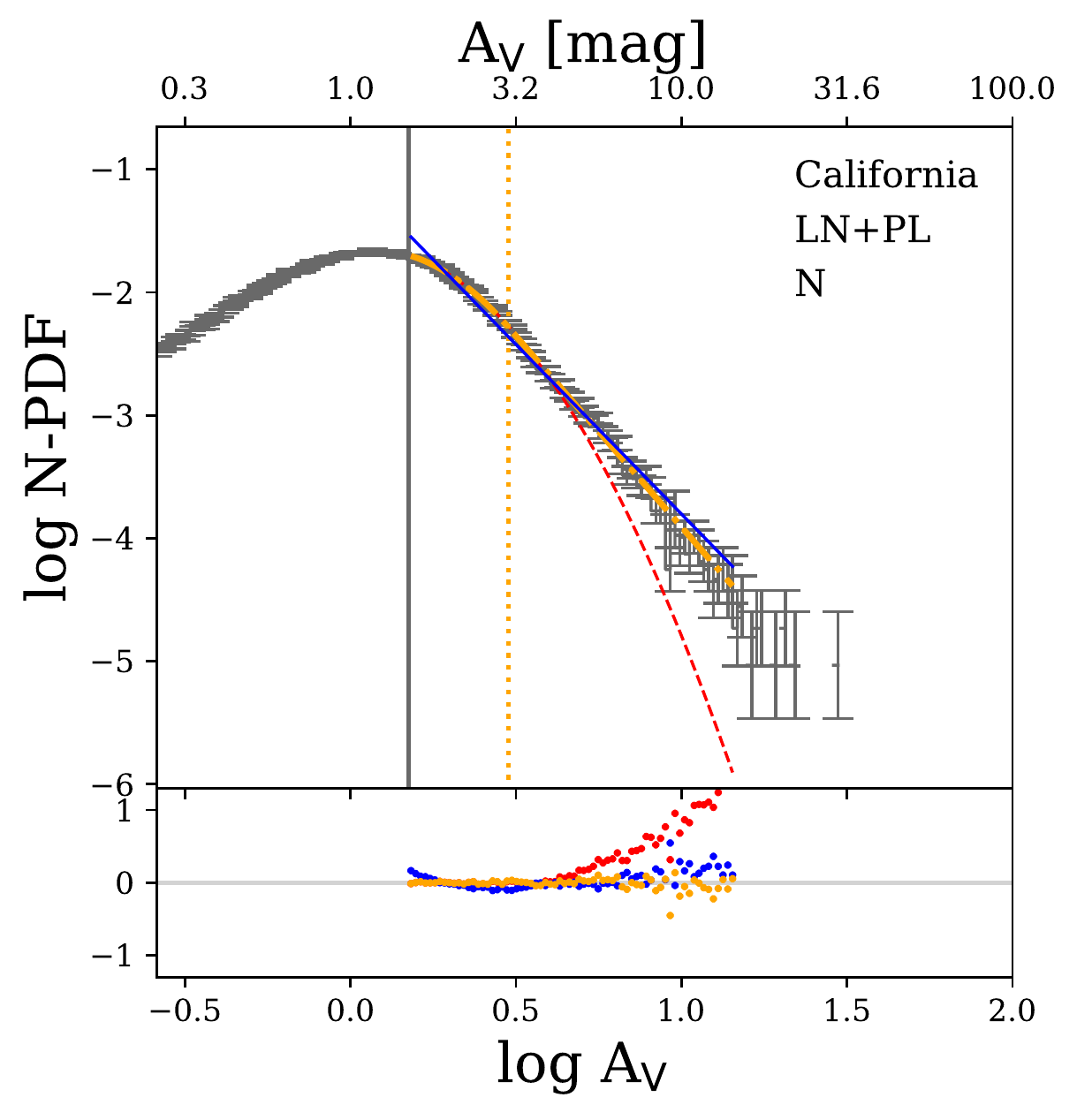}
    \end{minipage}
    
    \begin{minipage}[b]{0.24\textwidth}
        \includegraphics[width=\textwidth]{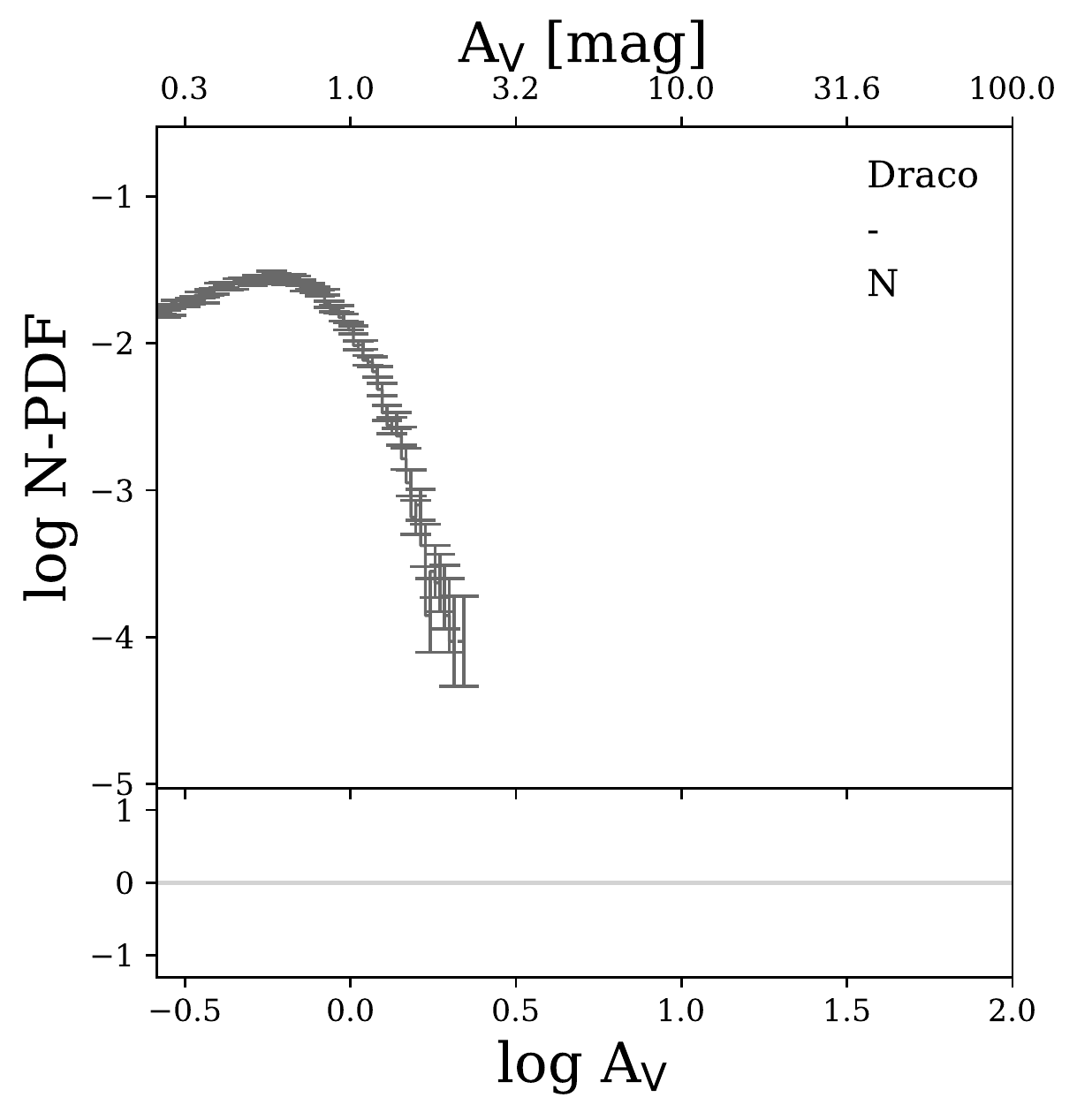}
    \end{minipage}
    \begin{minipage}[b]{0.24\textwidth}
        \includegraphics[width=\textwidth]{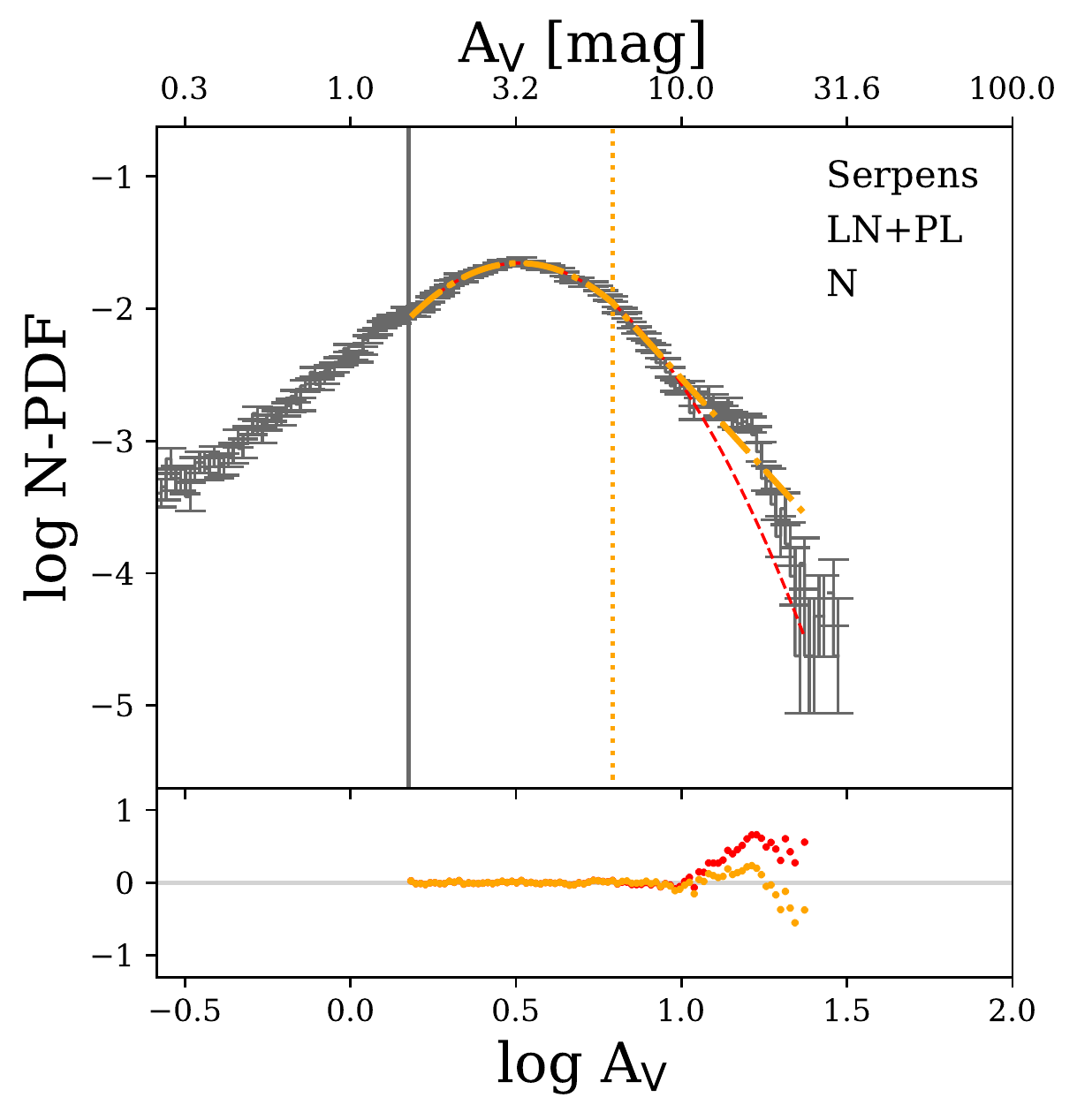}
    \end{minipage}
    \begin{minipage}[b]{0.24\textwidth}
        \includegraphics[width=\textwidth]{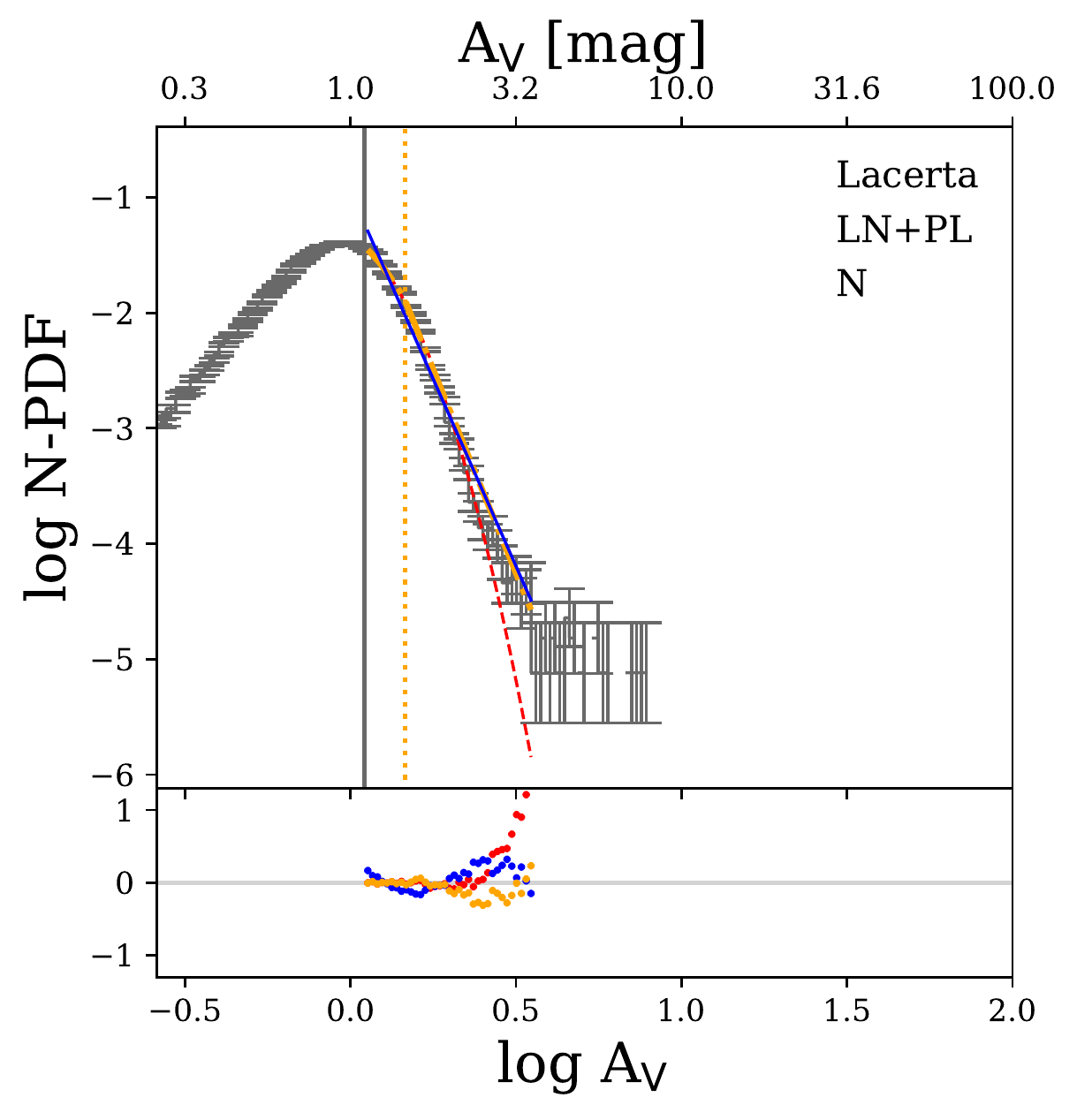}
    \end{minipage}
    \begin{minipage}[b]{0.24\textwidth}
        \includegraphics[width=\textwidth]{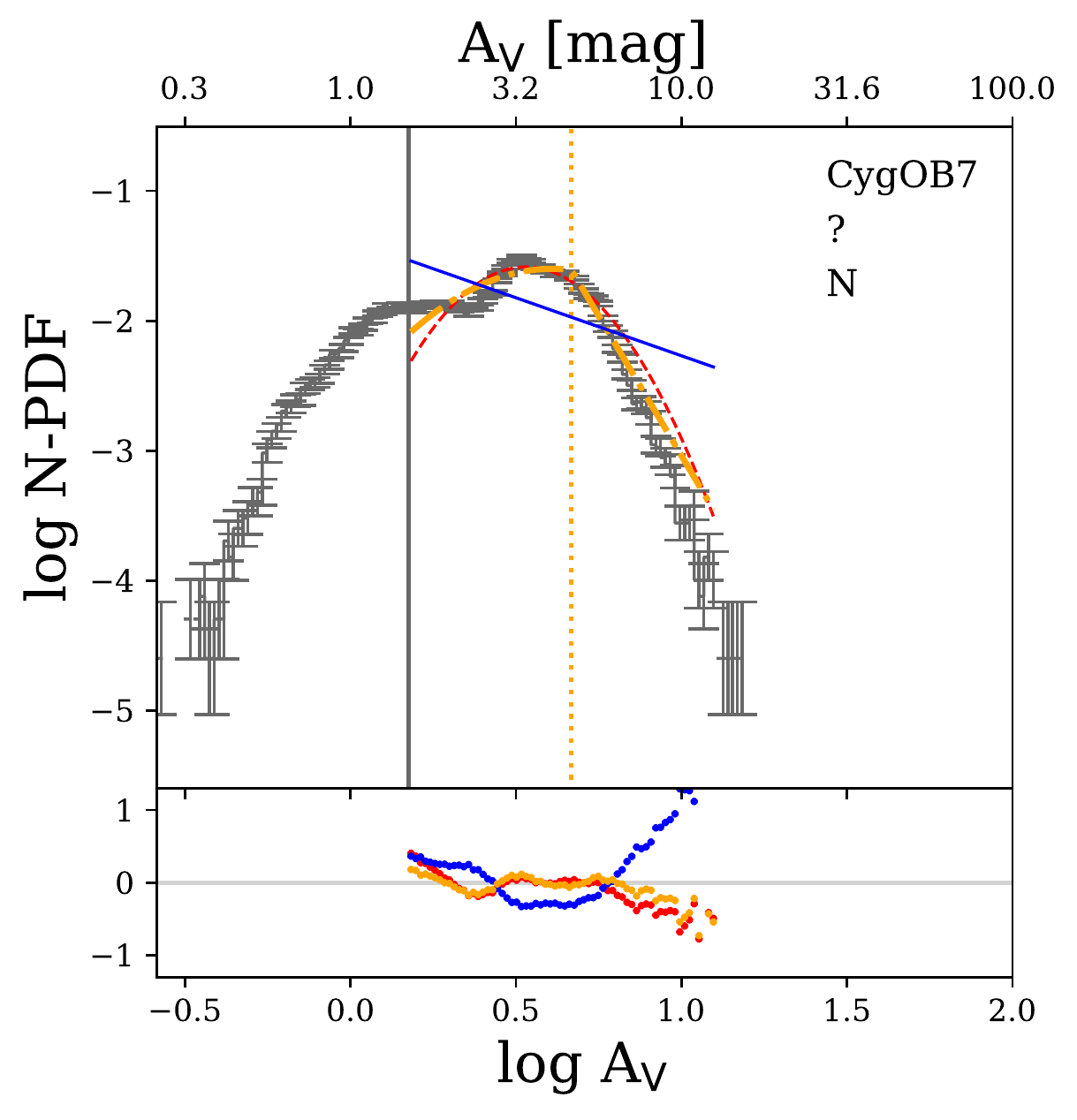}
    \end{minipage}
    \caption{N-PDFs of all clouds, continued.}
    \label{fig:Allfits2}
\end{figure*}

\begin{figure*}[h!]
    \centering
    \begin{minipage}[b]{0.24\textwidth}
        \includegraphics[width=\textwidth]{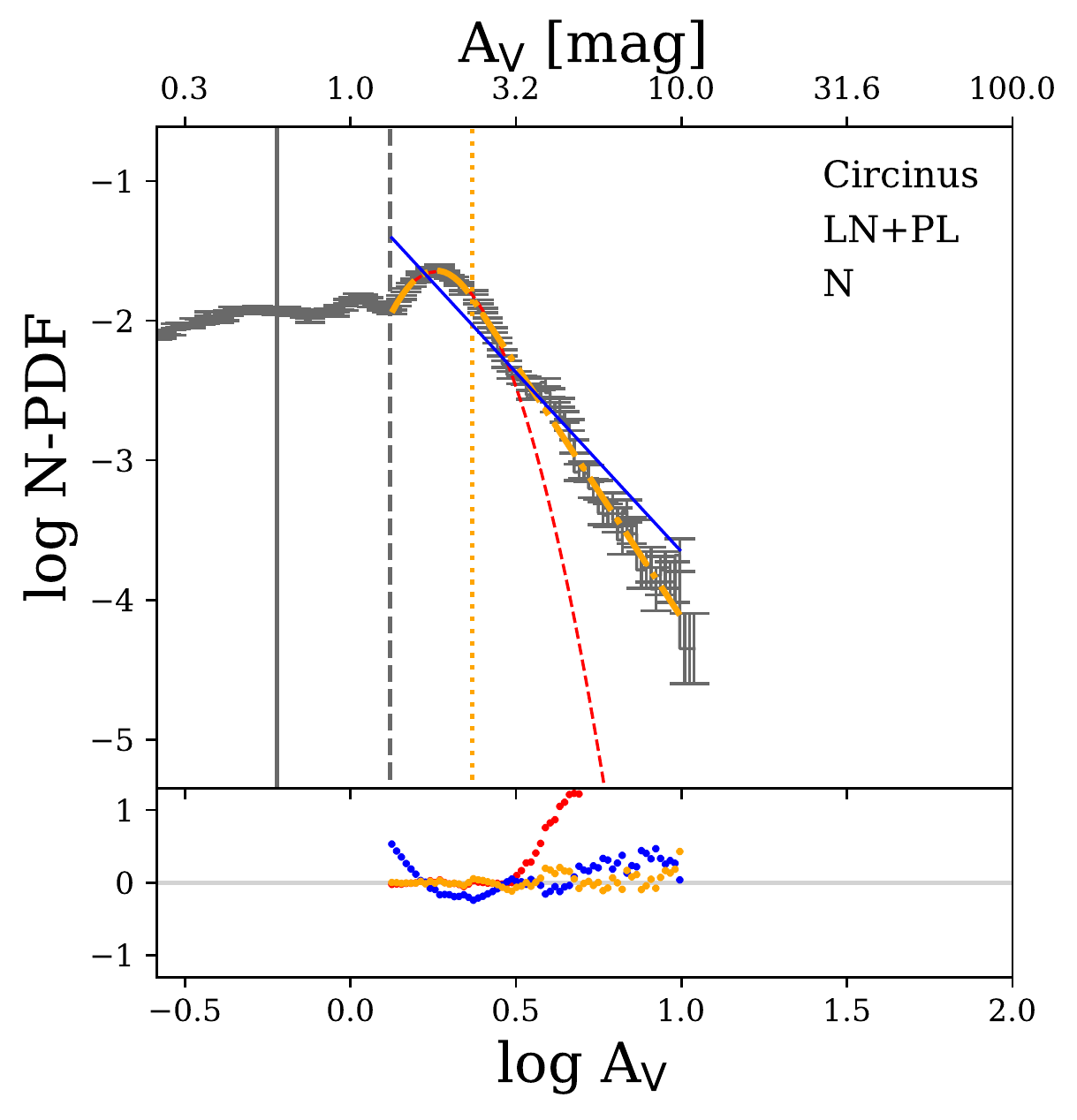}
    \end{minipage}
    \begin{minipage}[b]{0.24\textwidth}
        \includegraphics[width=\textwidth]{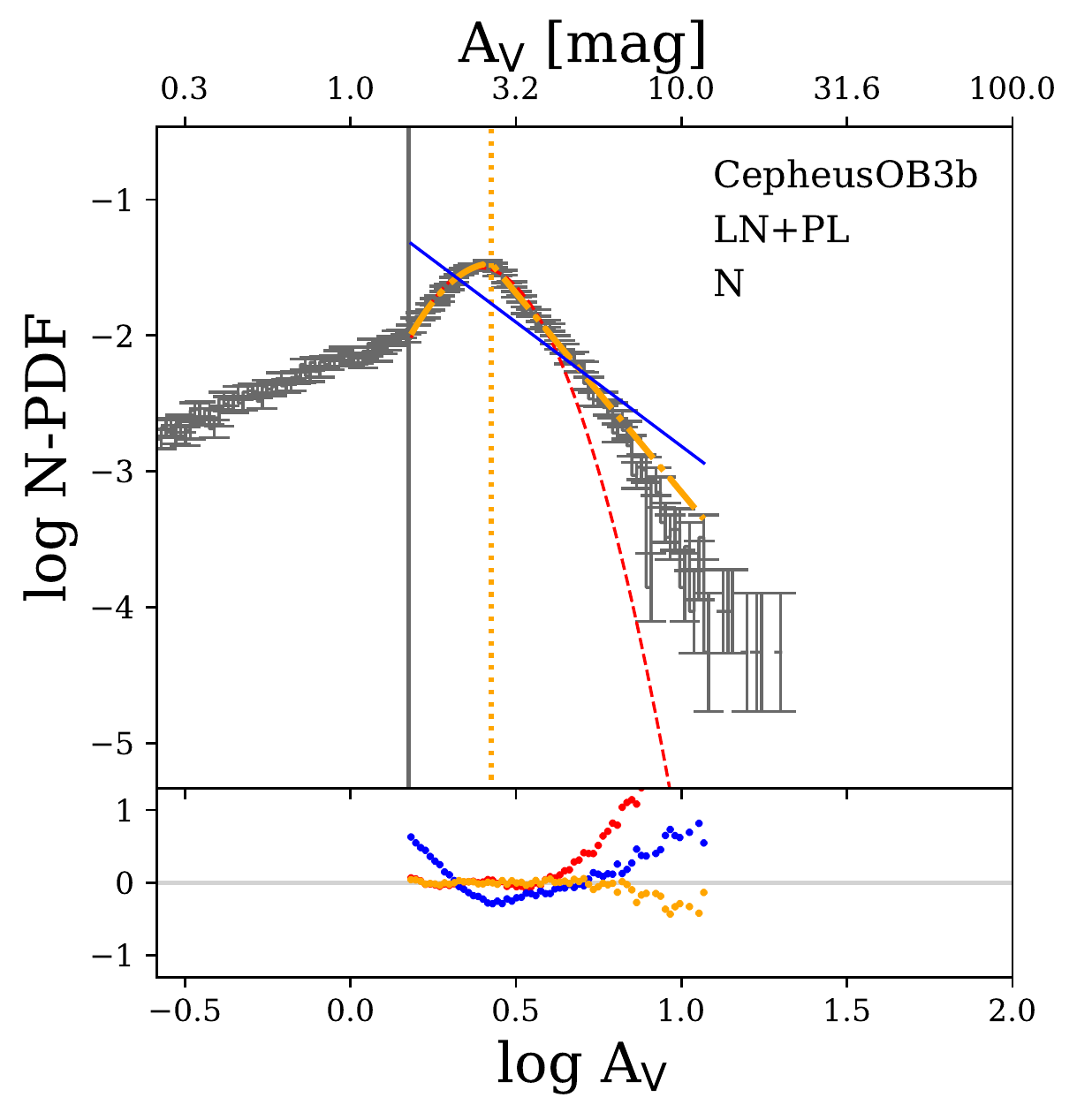}
    \end{minipage}
    \begin{minipage}[b]{0.24\textwidth}
        \includegraphics[width=\textwidth]{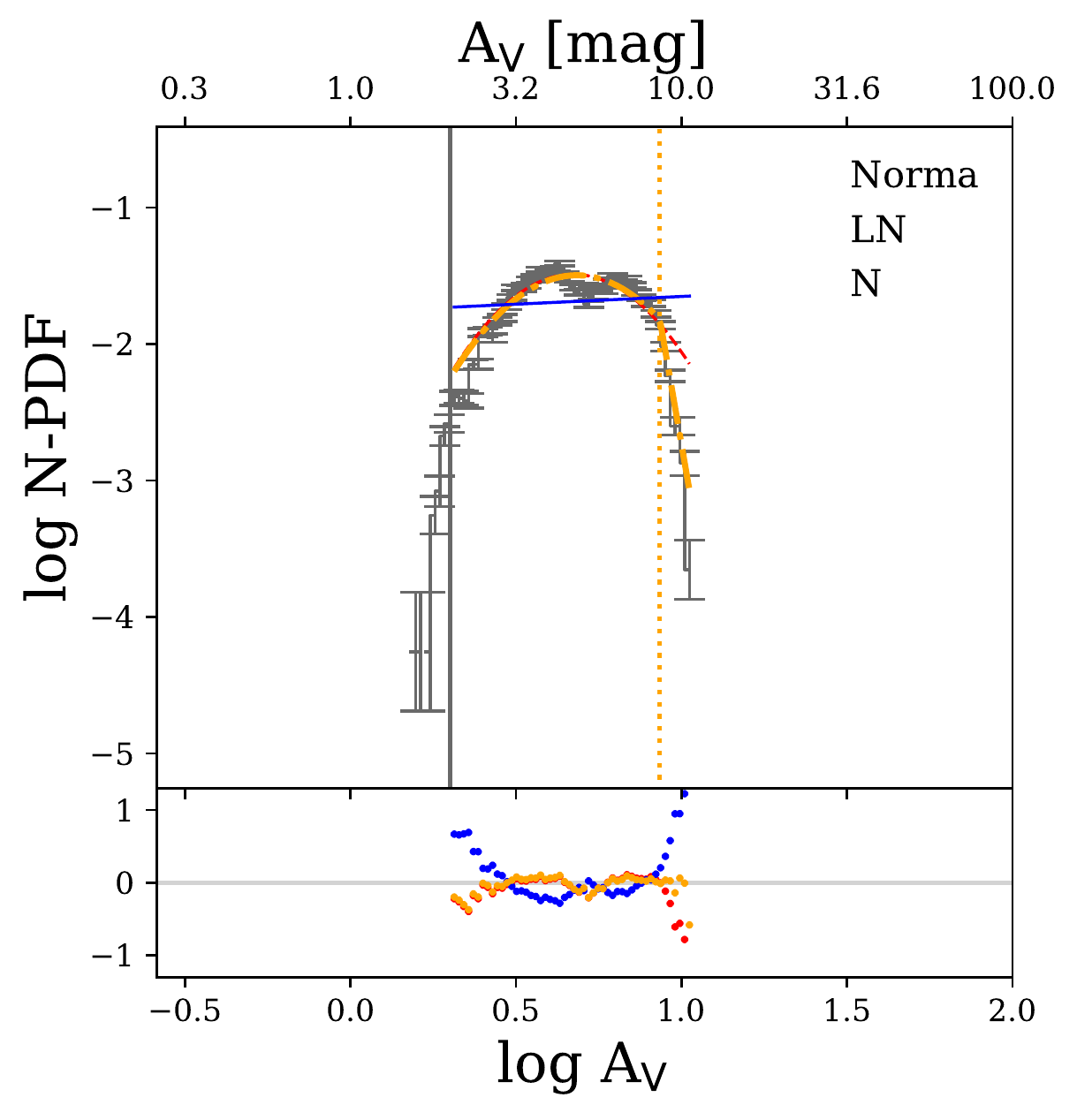}
    \end{minipage}
        \begin{minipage}[b]{0.24\textwidth}
        \includegraphics[width=\textwidth]{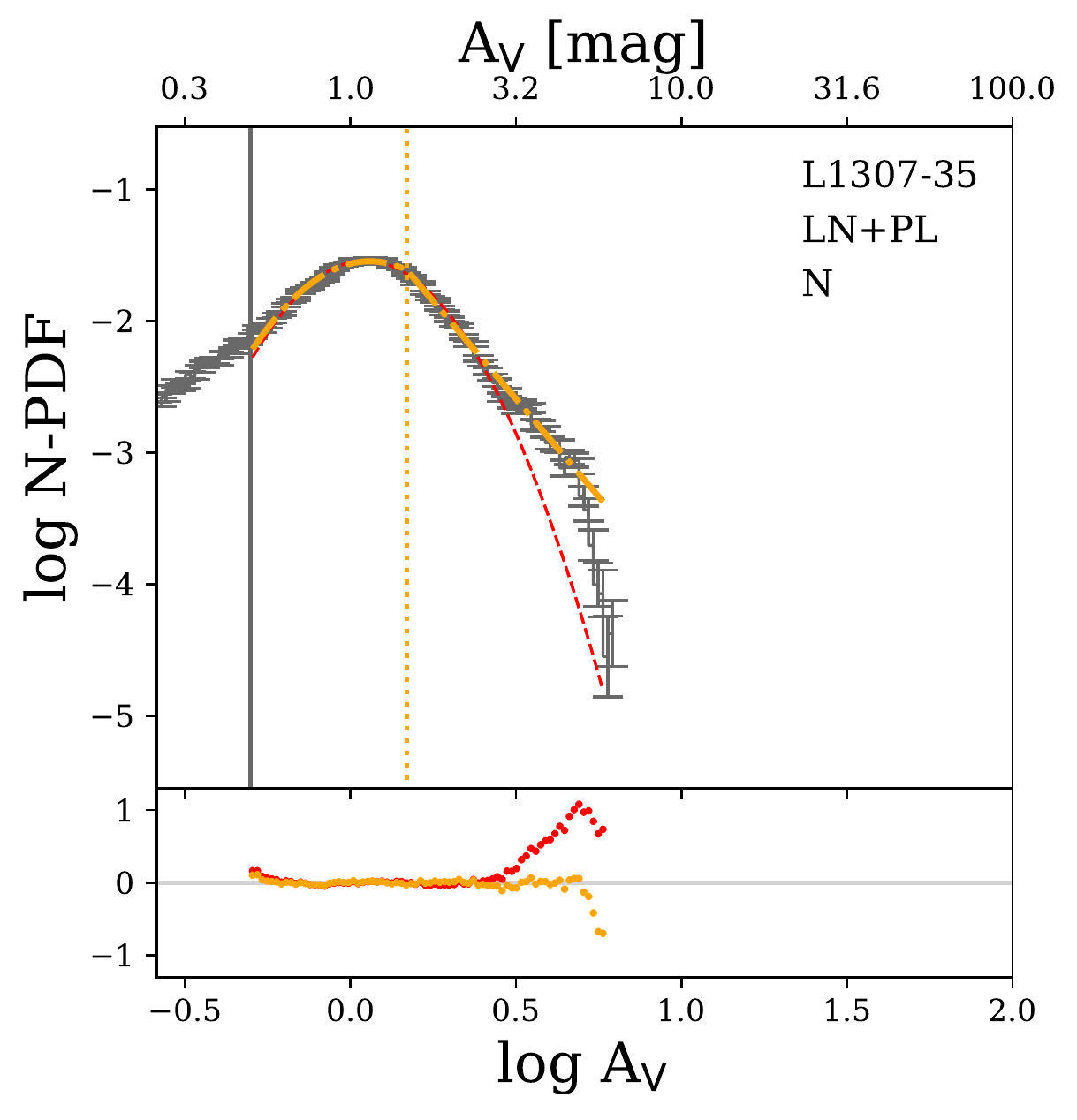}
    \end{minipage}
    
    \begin{minipage}[b]{0.24\textwidth}
        \includegraphics[width=\textwidth]{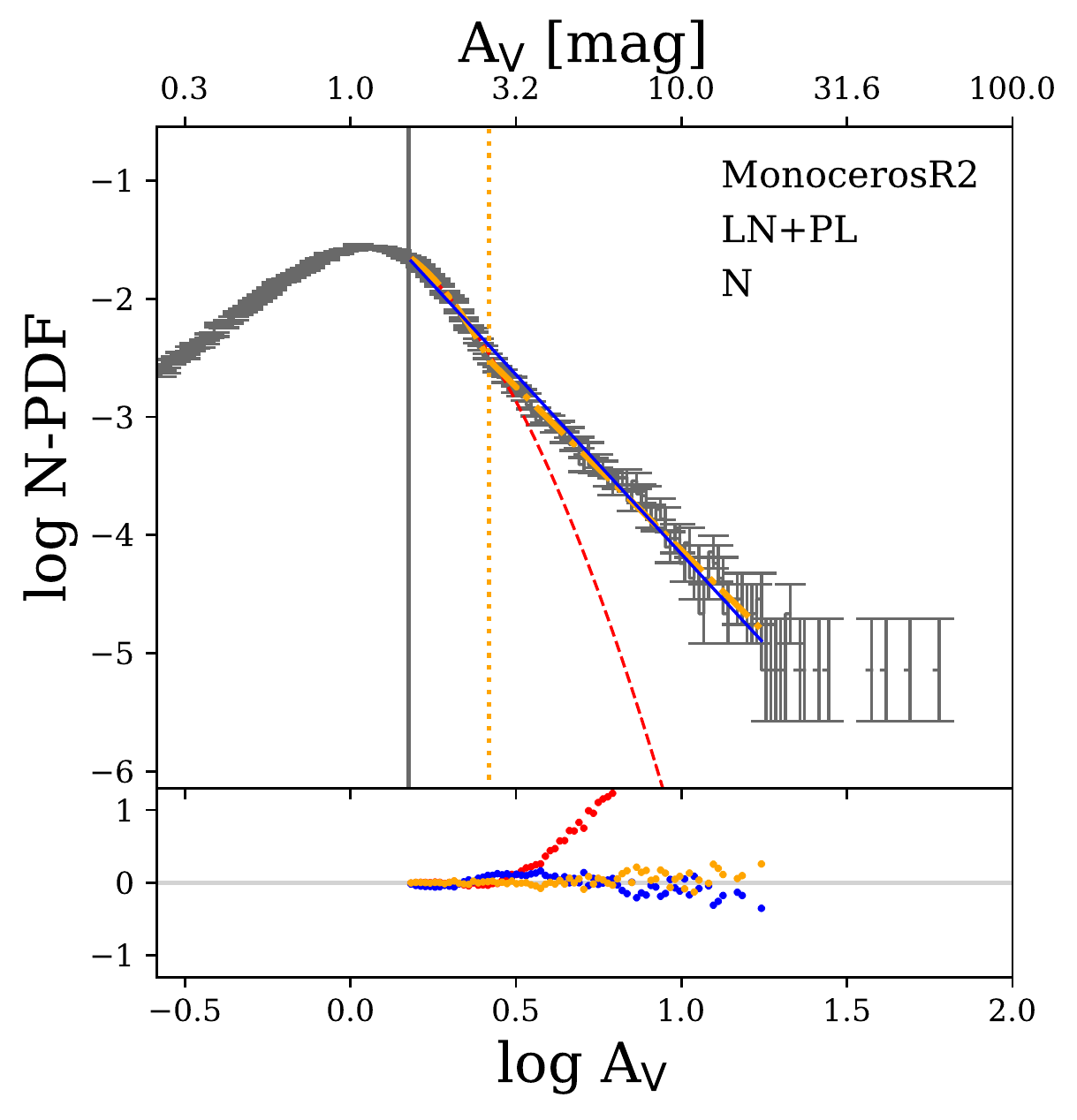}
    \end{minipage}
    \begin{minipage}[b]{0.24\textwidth}
        \includegraphics[width=\textwidth]{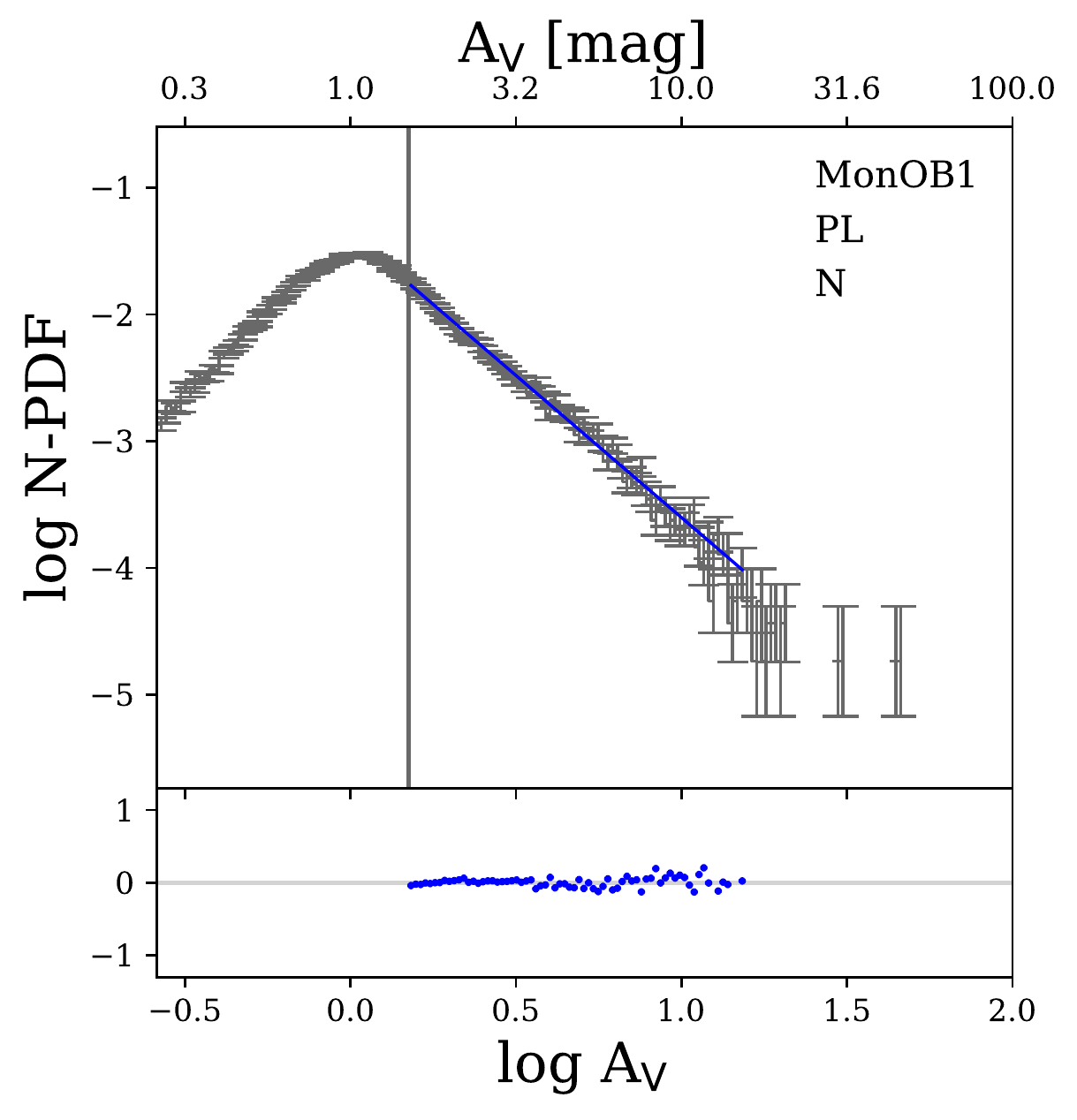}
    \end{minipage}
    \begin{minipage}[b]{0.24\textwidth}
        \includegraphics[width=\textwidth]{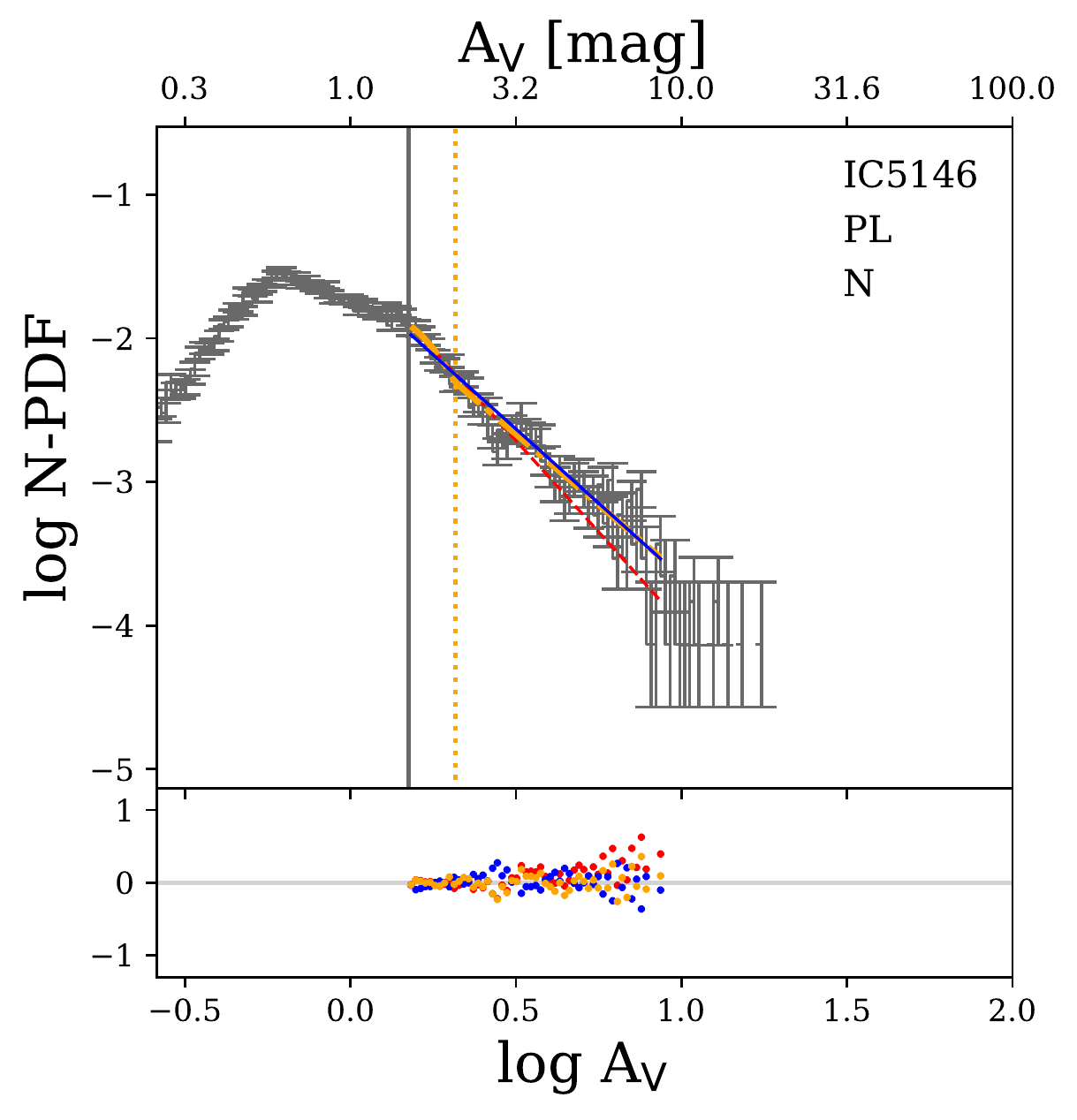}
    \end{minipage}
    \begin{minipage}[b]{0.24\textwidth}
        \includegraphics[width=\textwidth]{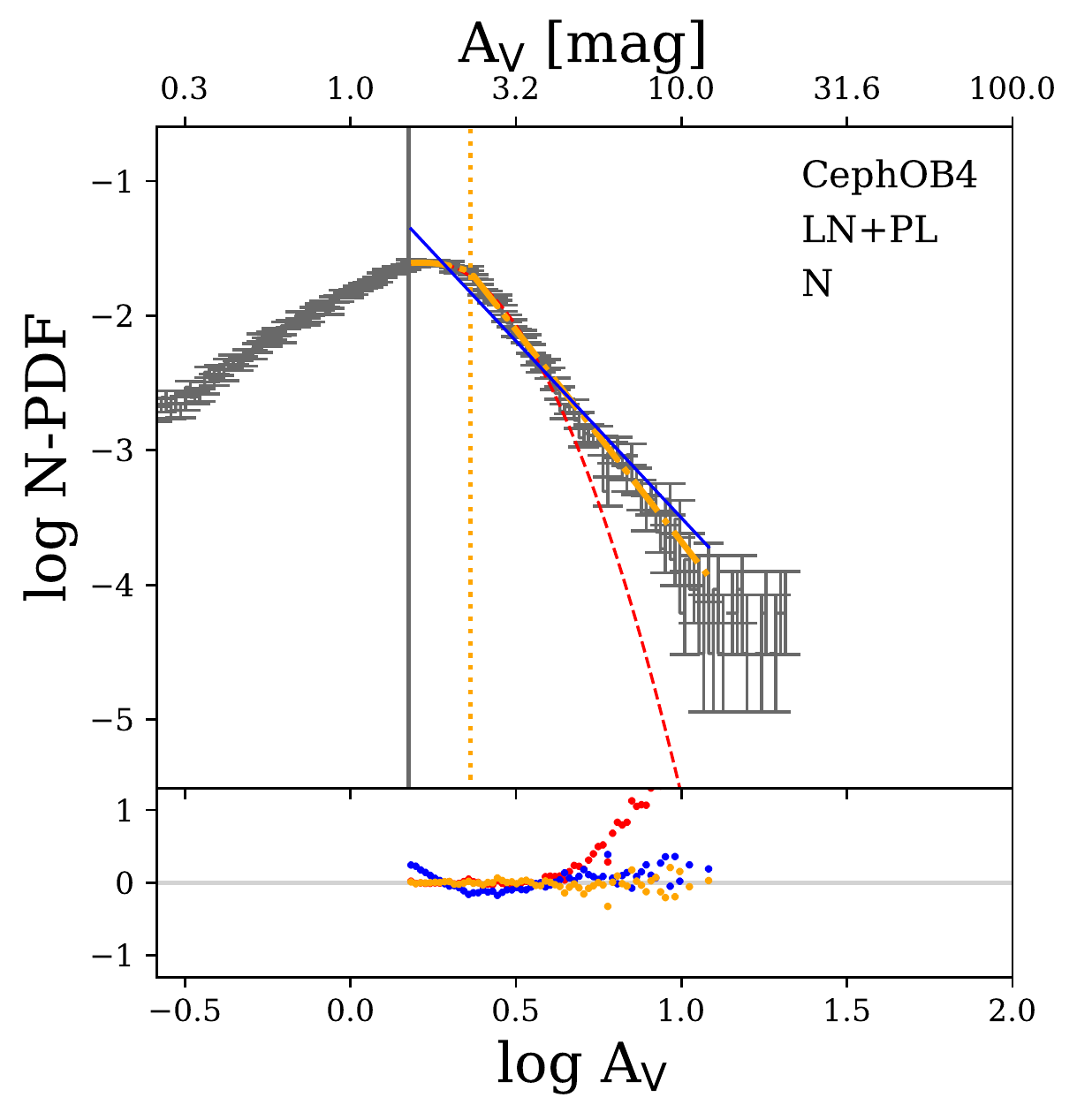}
    \end{minipage}
    
    \begin{minipage}[b]{0.24\textwidth}
        \includegraphics[width=\textwidth]{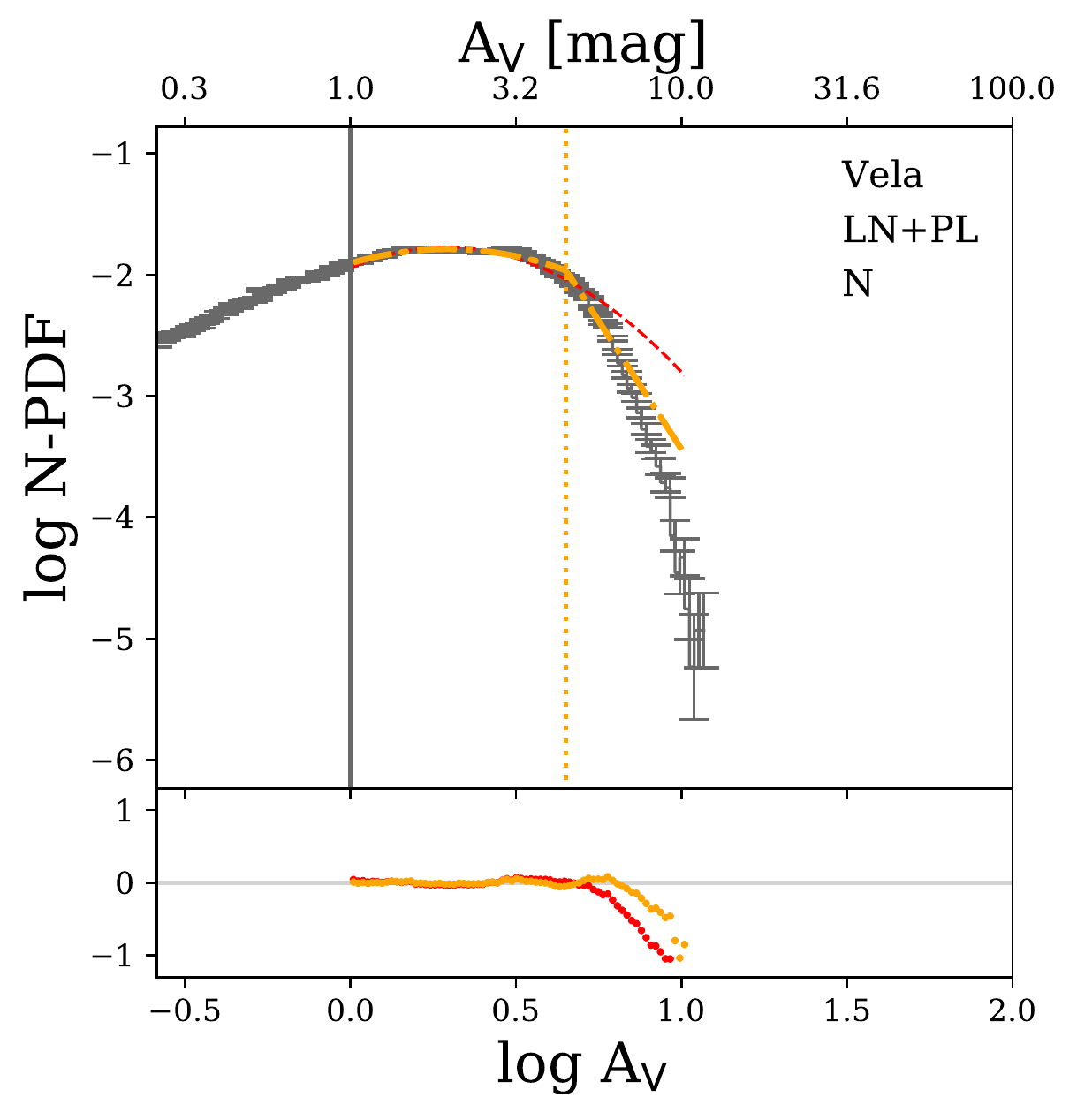}
    \end{minipage}
    \begin{minipage}[b]{0.24\textwidth}
        \includegraphics[width=\textwidth]{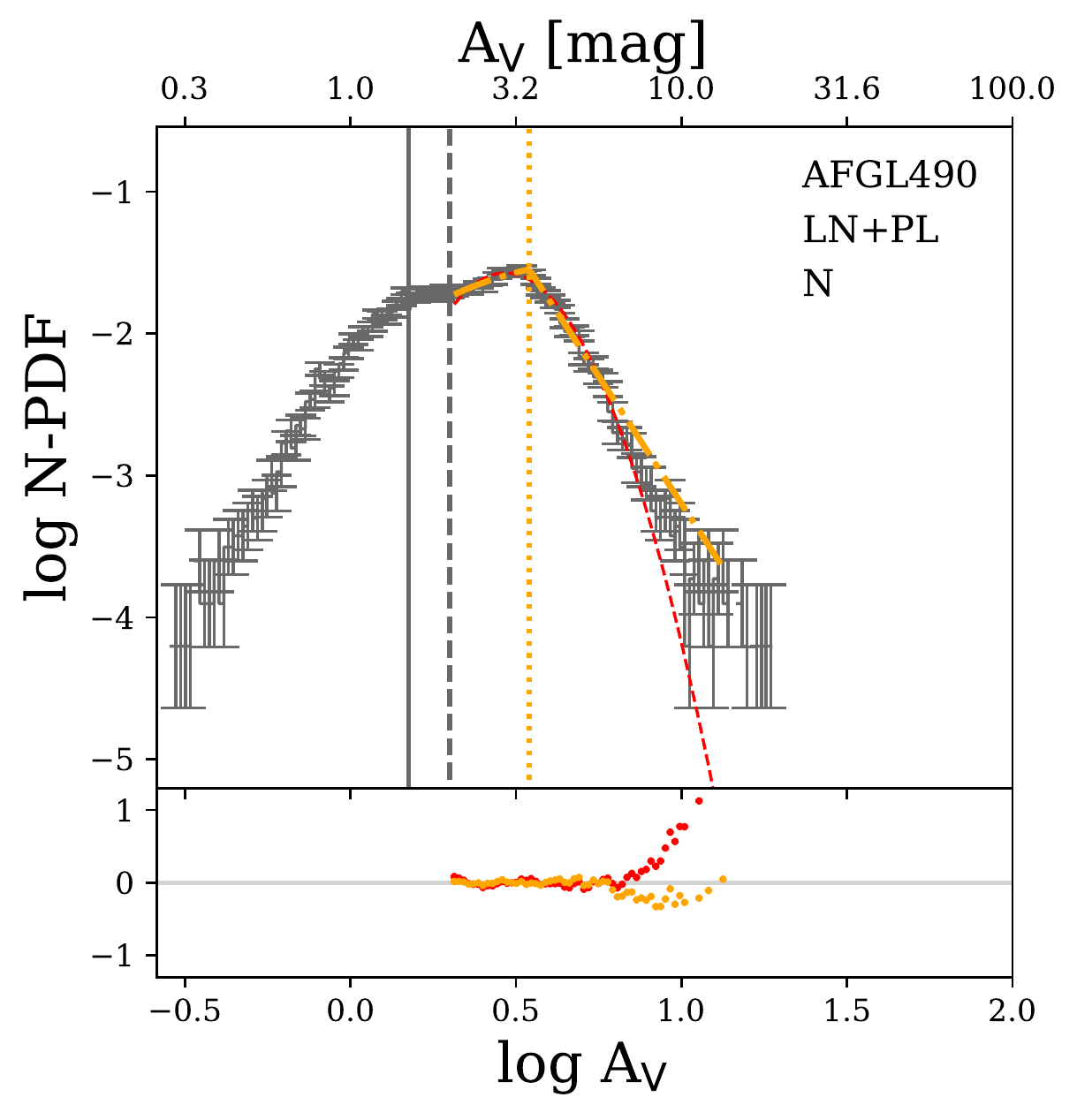}
    \end{minipage}
    \begin{minipage}[b]{0.24\textwidth}
        \includegraphics[width=\textwidth]{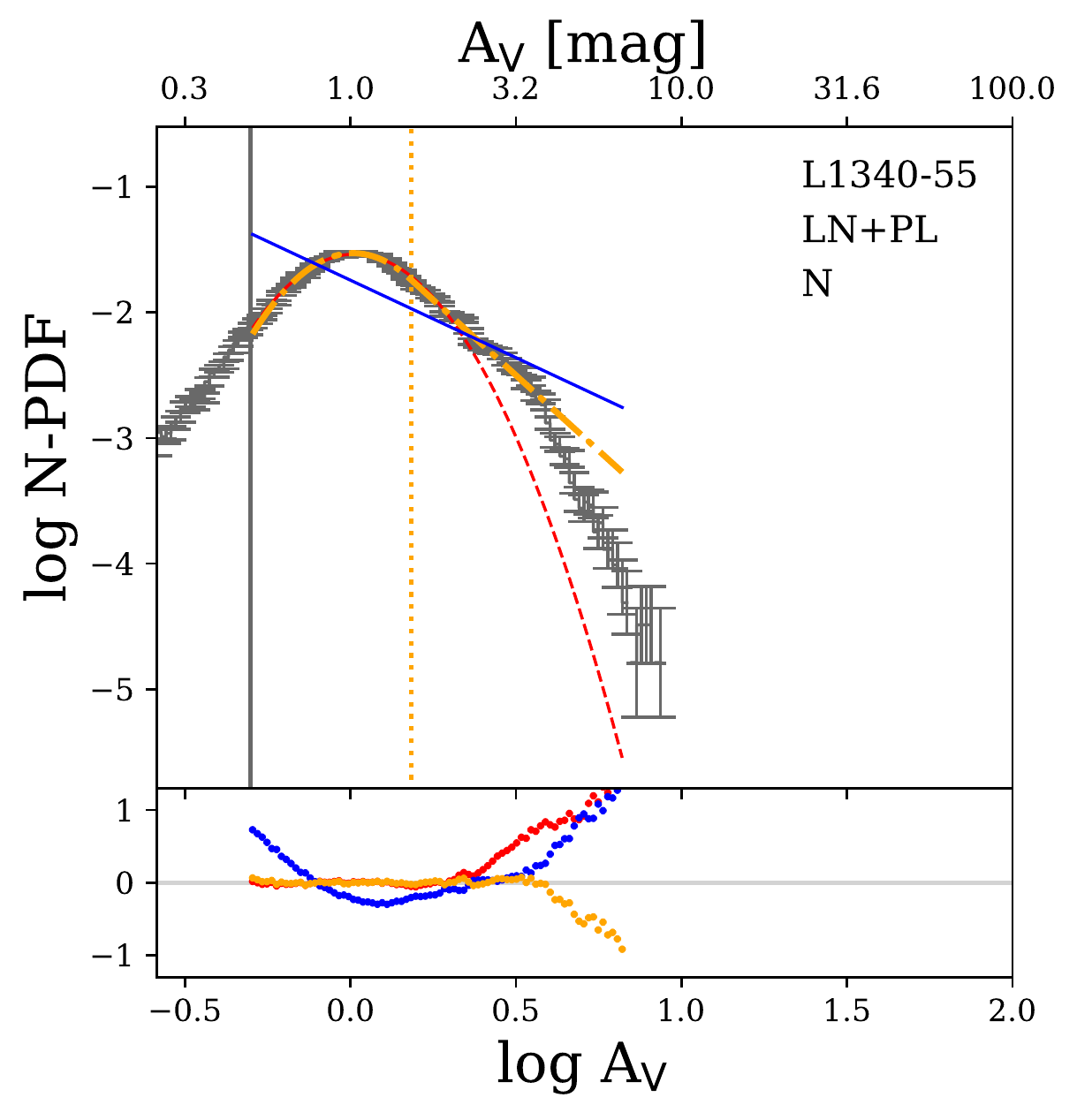}
    \end{minipage}
    \begin{minipage}[b]{0.24\textwidth}
        \includegraphics[width=\textwidth]{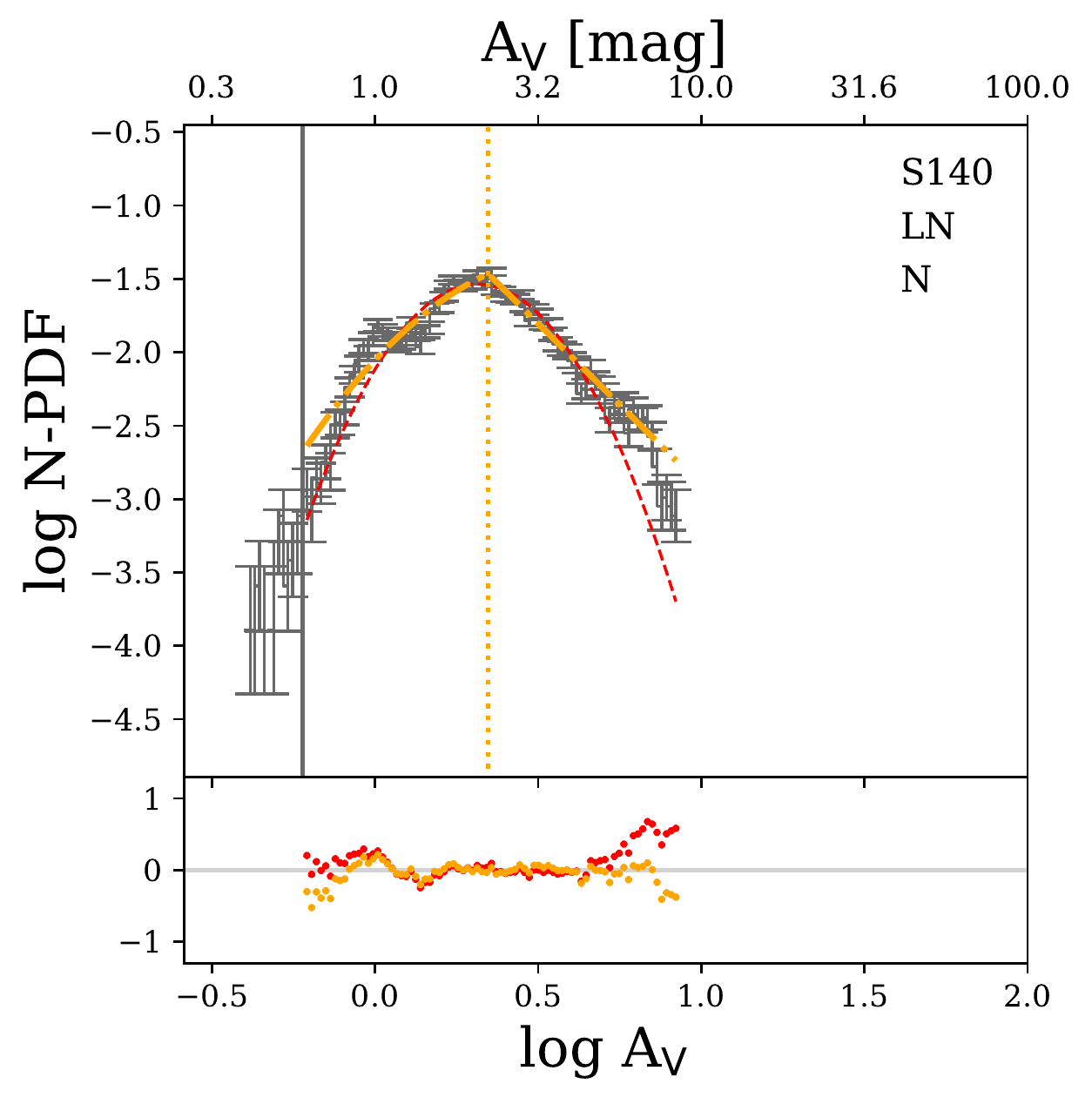}
    \end{minipage}
    
    \begin{minipage}[b]{0.24\textwidth}
        \includegraphics[width=\textwidth]{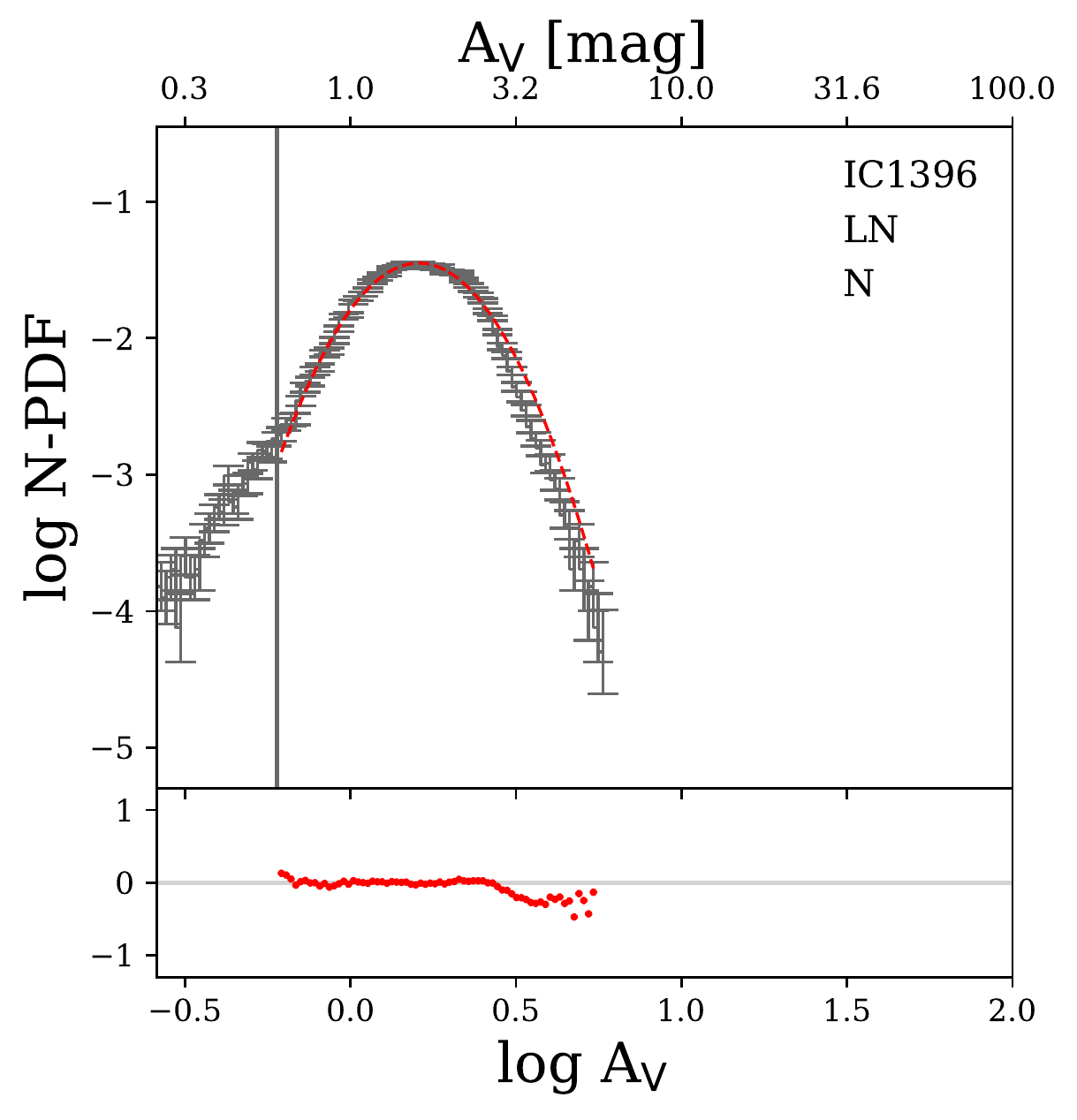}
    \end{minipage}%J
    \begin{minipage}[b]{0.24\textwidth}
        \includegraphics[width=\textwidth]{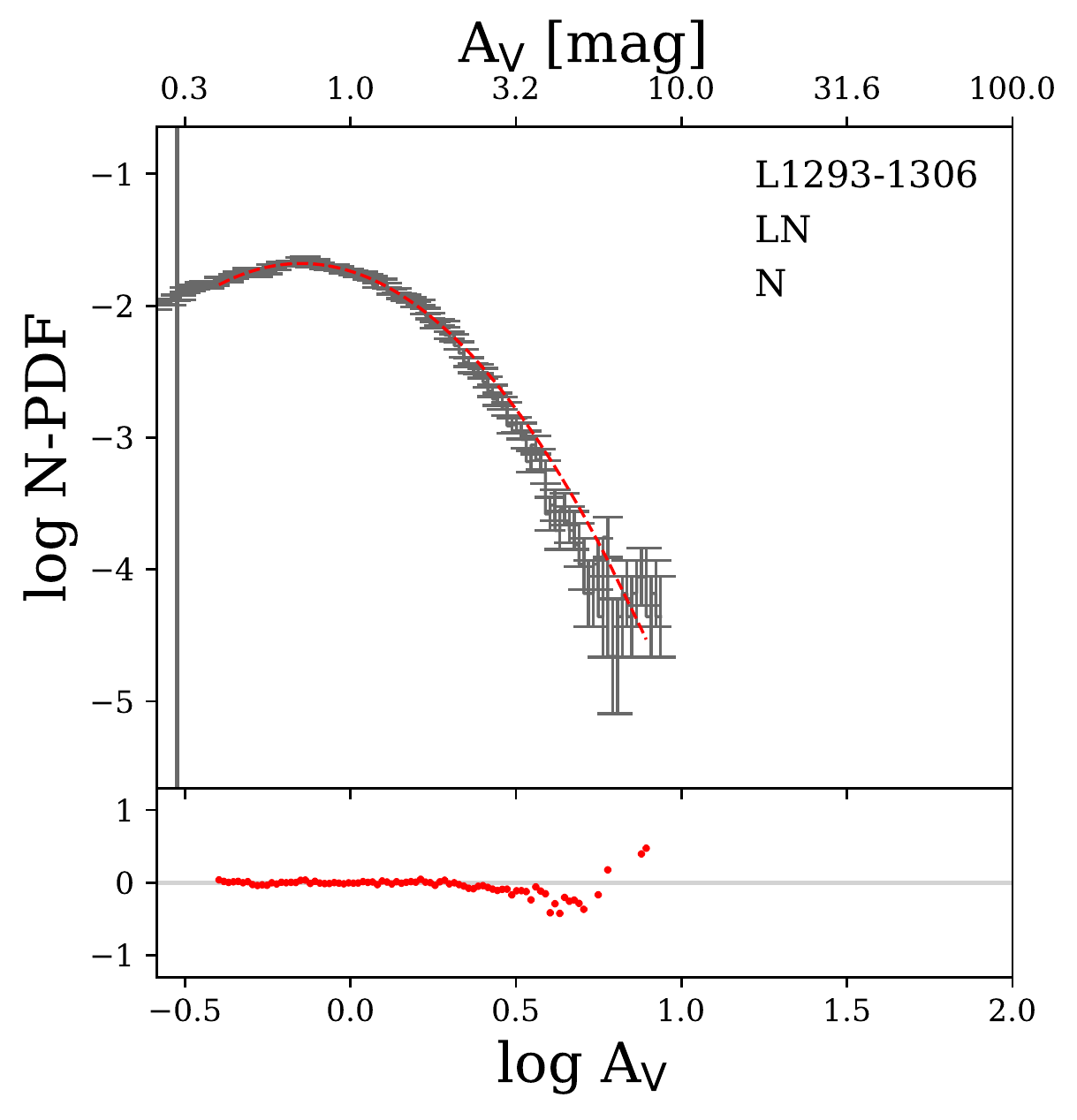}
    \end{minipage}
    \begin{minipage}[b]{0.24\textwidth}
        \includegraphics[width=\textwidth]{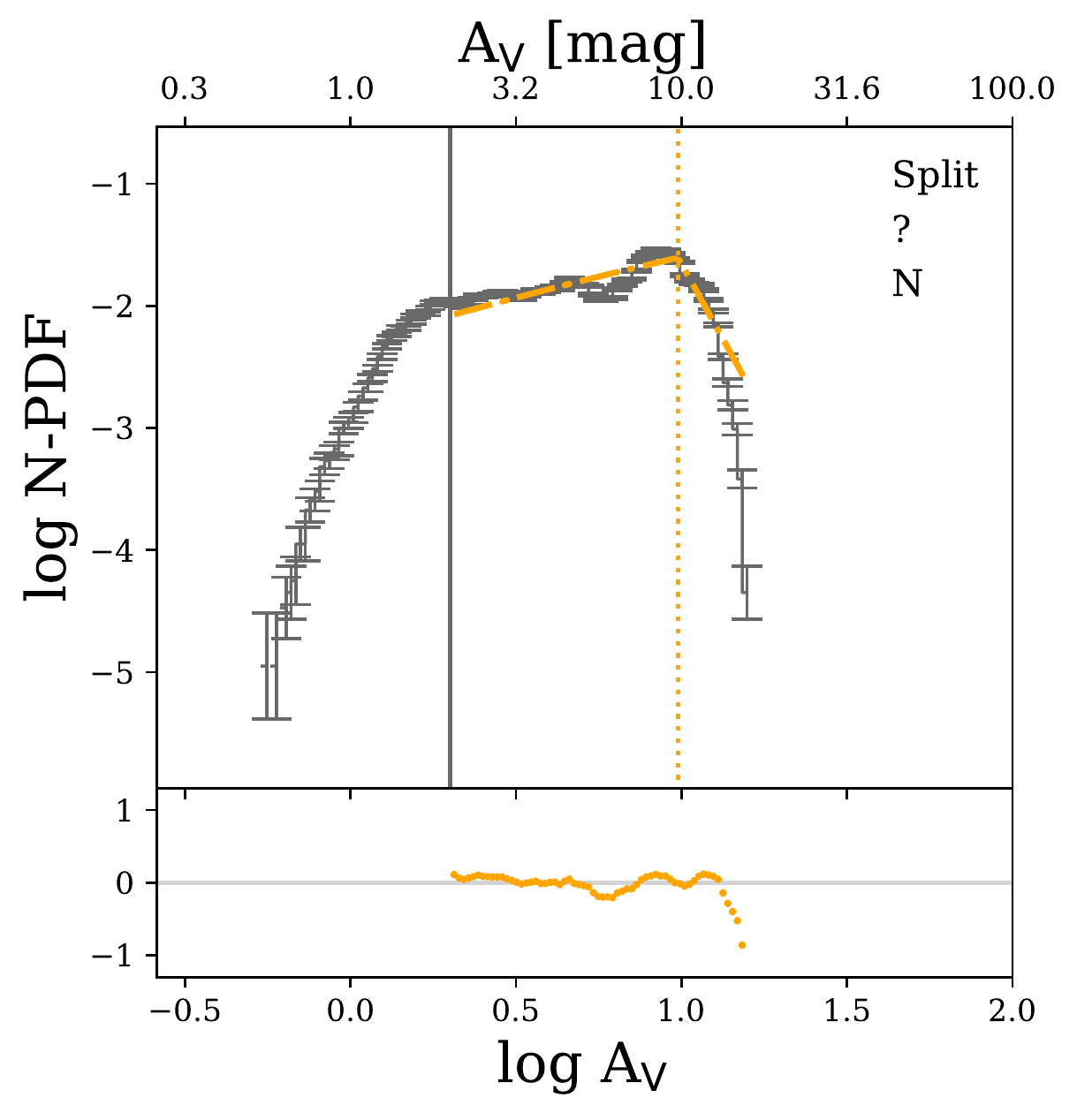}
    \end{minipage}%J
    %%% S140. LDN, MBM
    \begin{minipage}[b]{0.24\textwidth}
        \includegraphics[width=\textwidth]{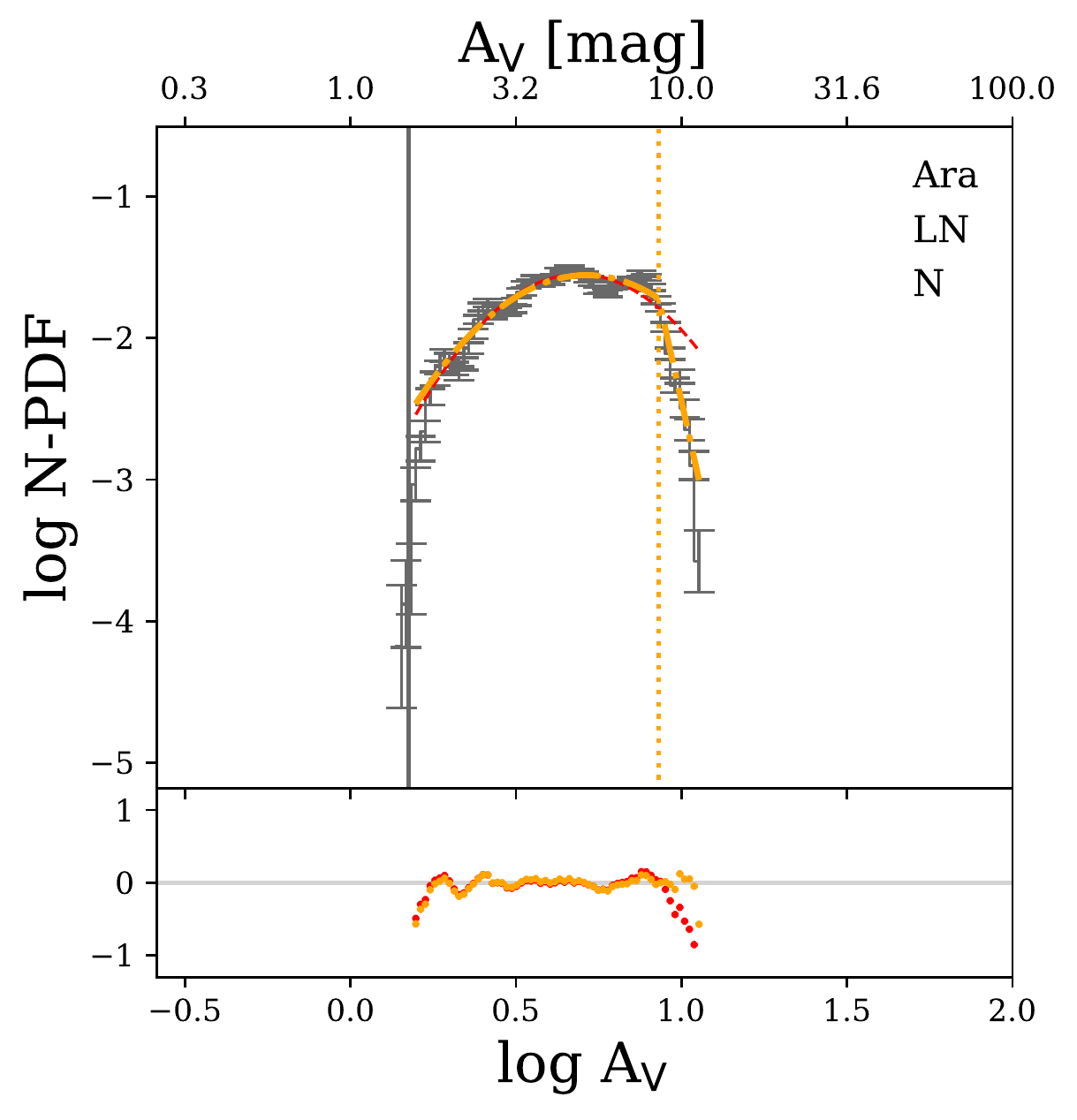}
    \end{minipage}%J
    
    \begin{minipage}[b]{0.24\textwidth}
        \includegraphics[width=\textwidth]{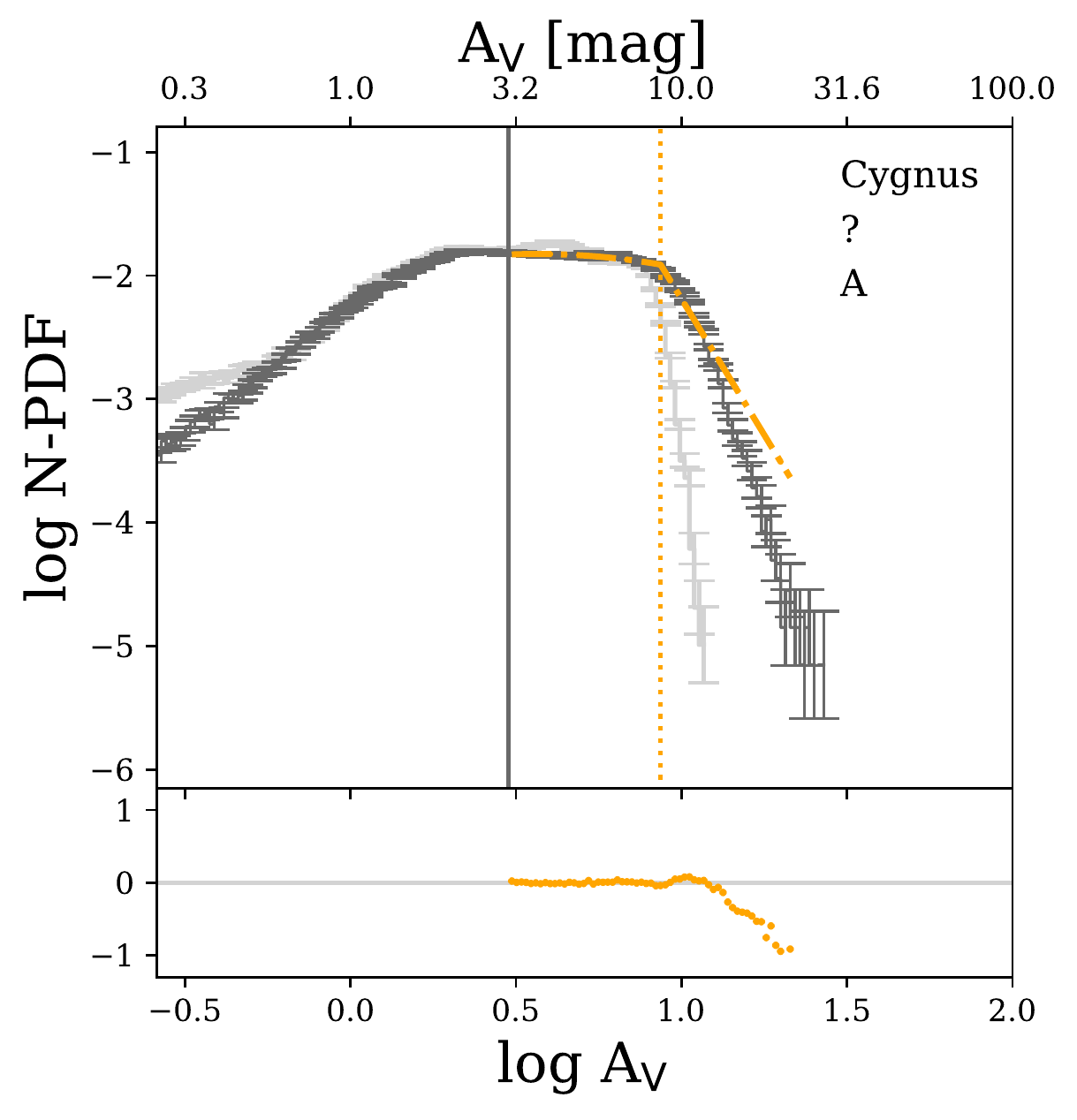}
    \end{minipage}
    \begin{minipage}[b]{0.24\textwidth}
        \includegraphics[width=\textwidth]{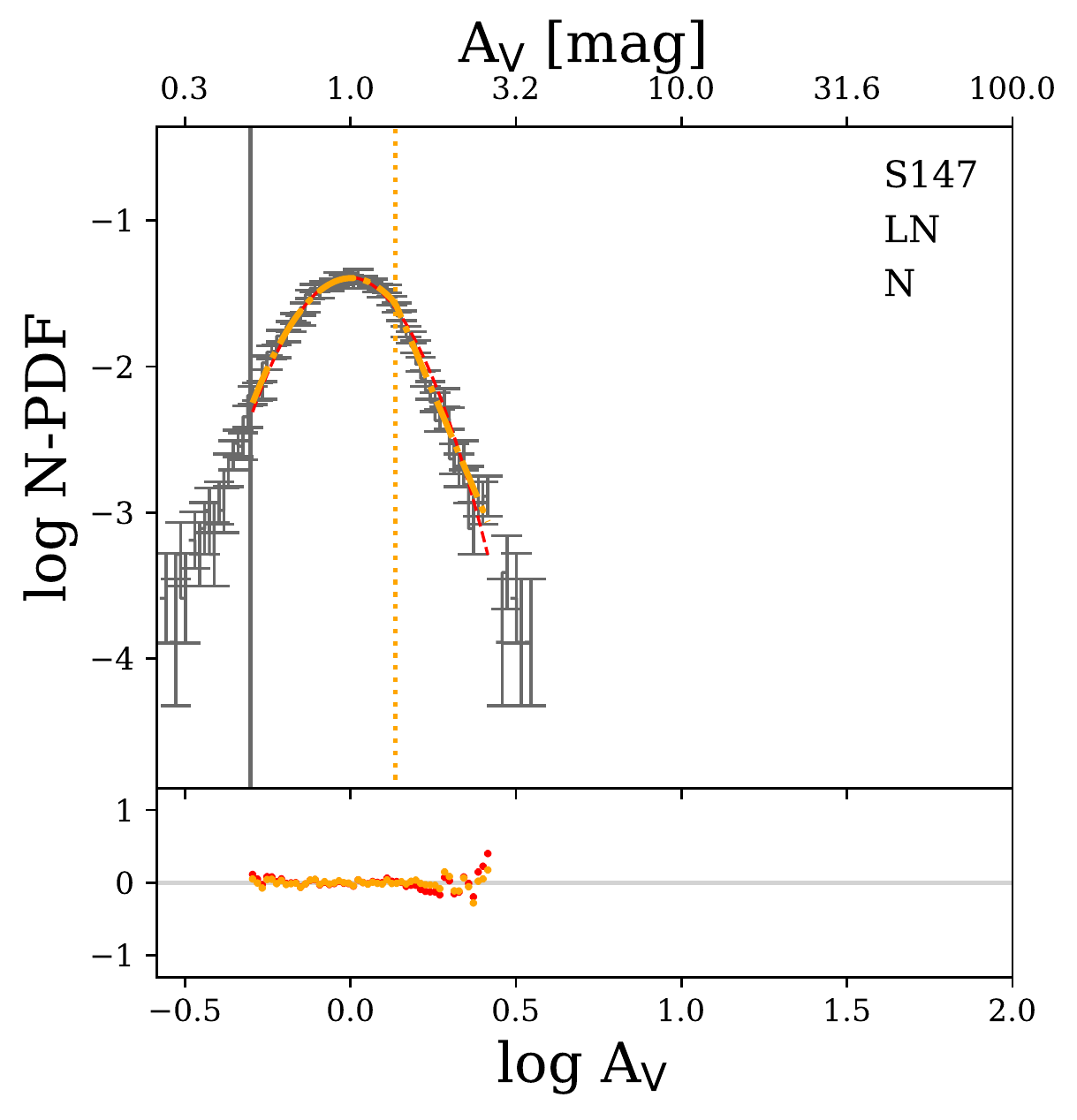}
    \end{minipage}
    \begin{minipage}[b]{0.24\textwidth}
        \includegraphics[width=\textwidth]{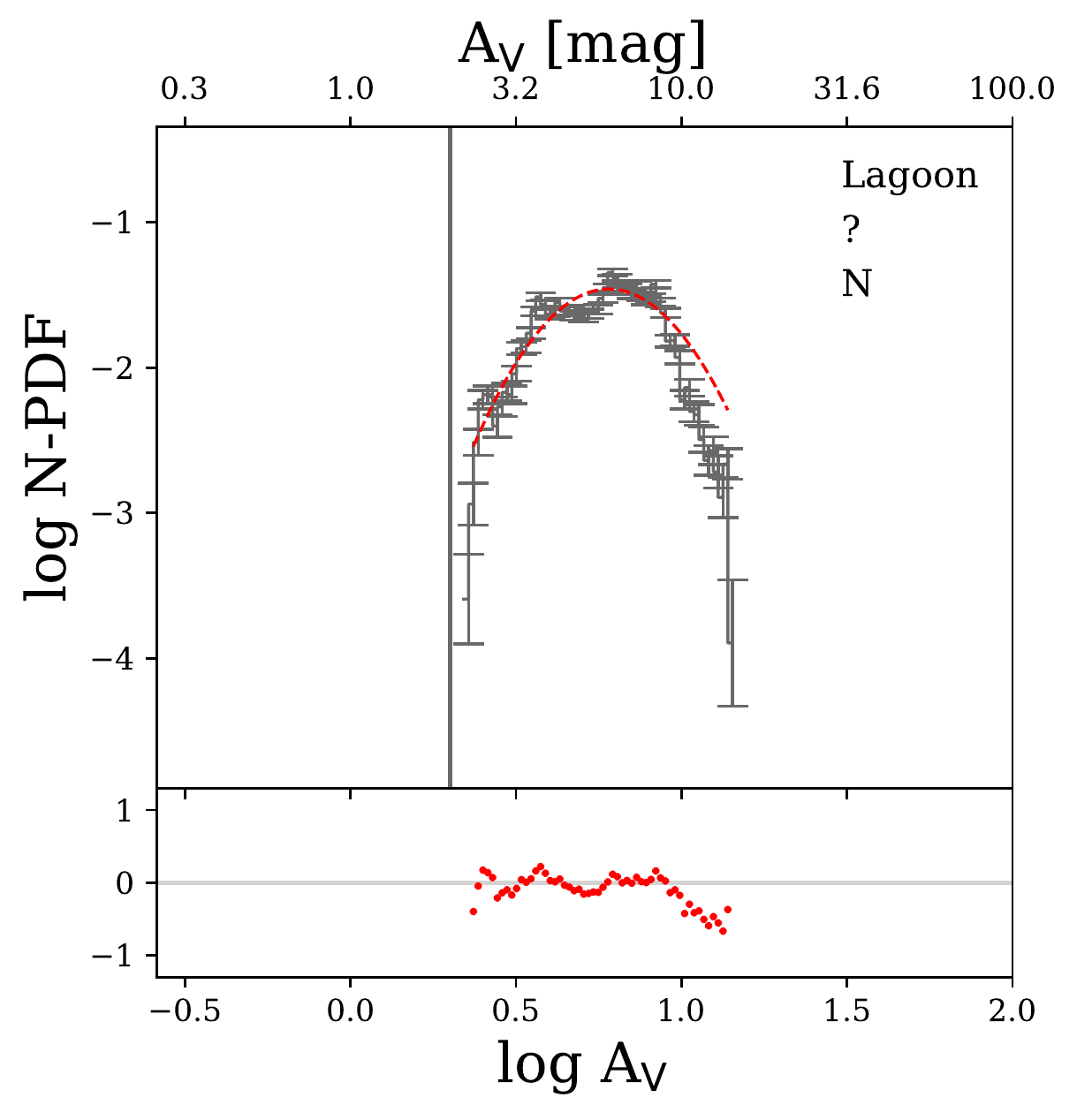}
    \end{minipage}
    \begin{minipage}[b]{0.24\textwidth}
        \includegraphics[width=\textwidth]{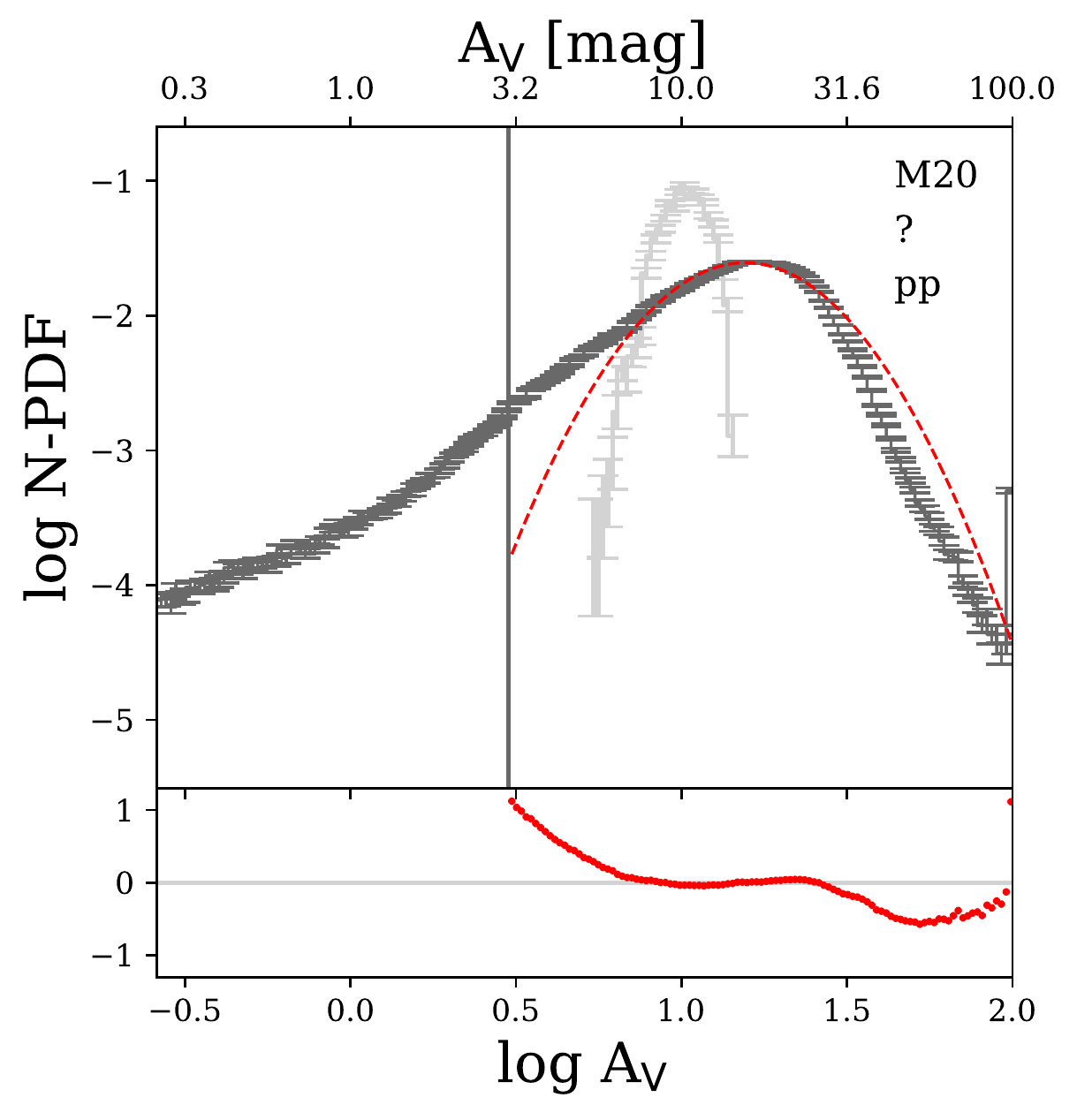}
    \end{minipage}
    \caption{N-PDFs of all clouds, continued.}
    \label{fig:Allfits3}
\end{figure*}

\begin{figure*}[h!]
    \centering
    \begin{minipage}[b]{0.24\textwidth}
        \includegraphics[width=\textwidth]{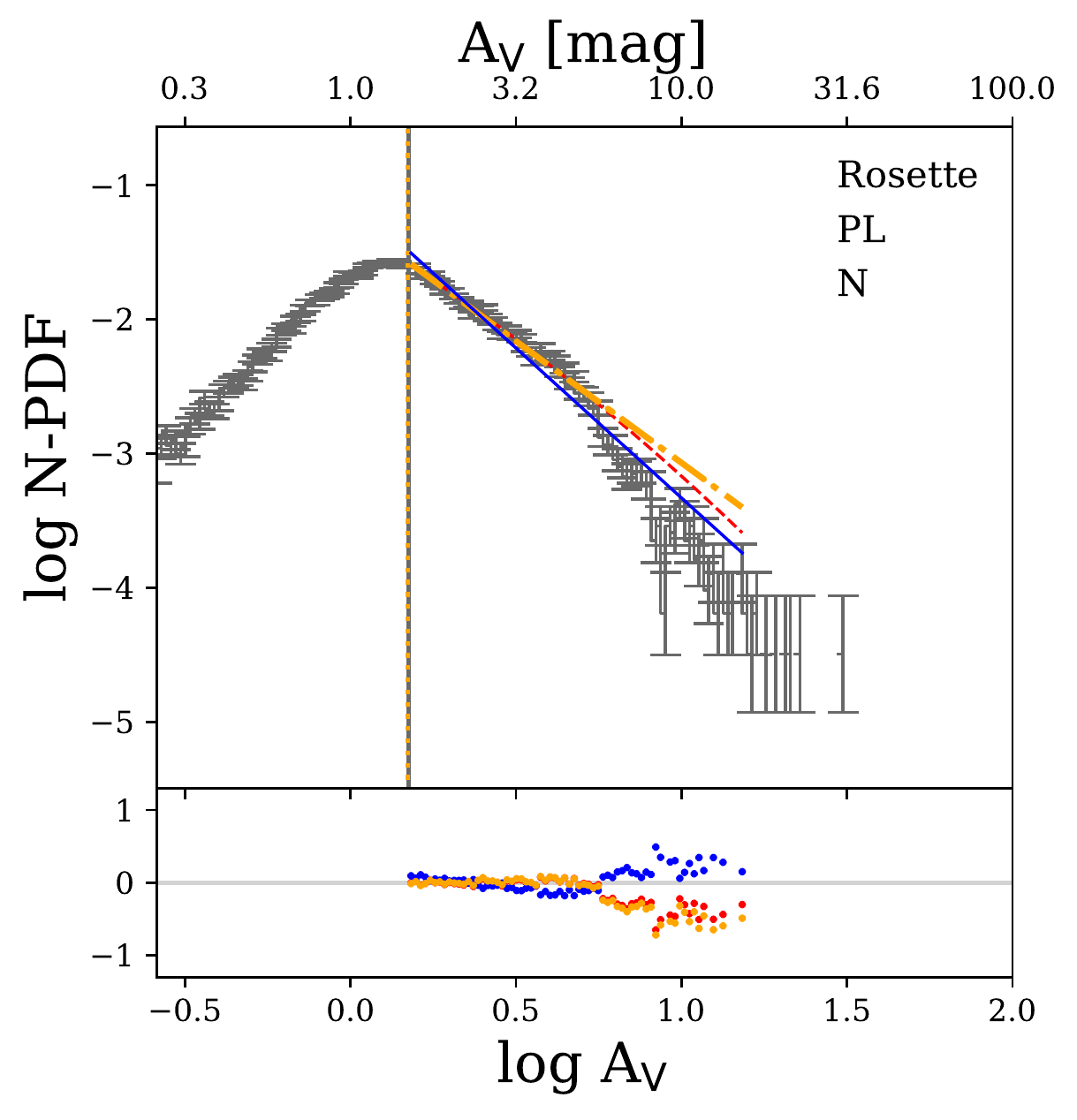}
    \end{minipage}
    \begin{minipage}[b]{0.24\textwidth}
        \includegraphics[width=\textwidth]{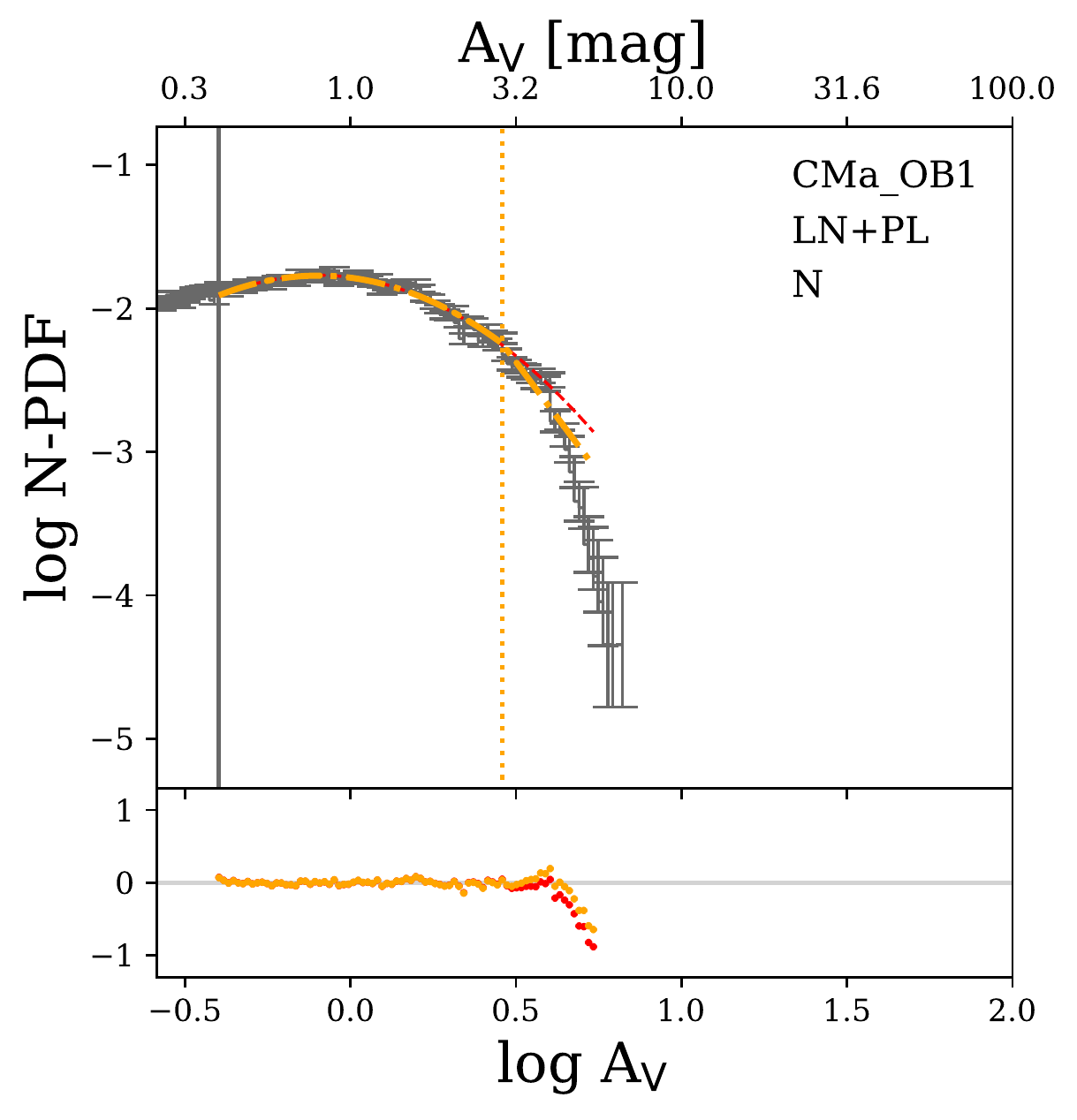}
    \end{minipage}
    \begin{minipage}[b]{0.24\textwidth}
        \includegraphics[width=\textwidth]{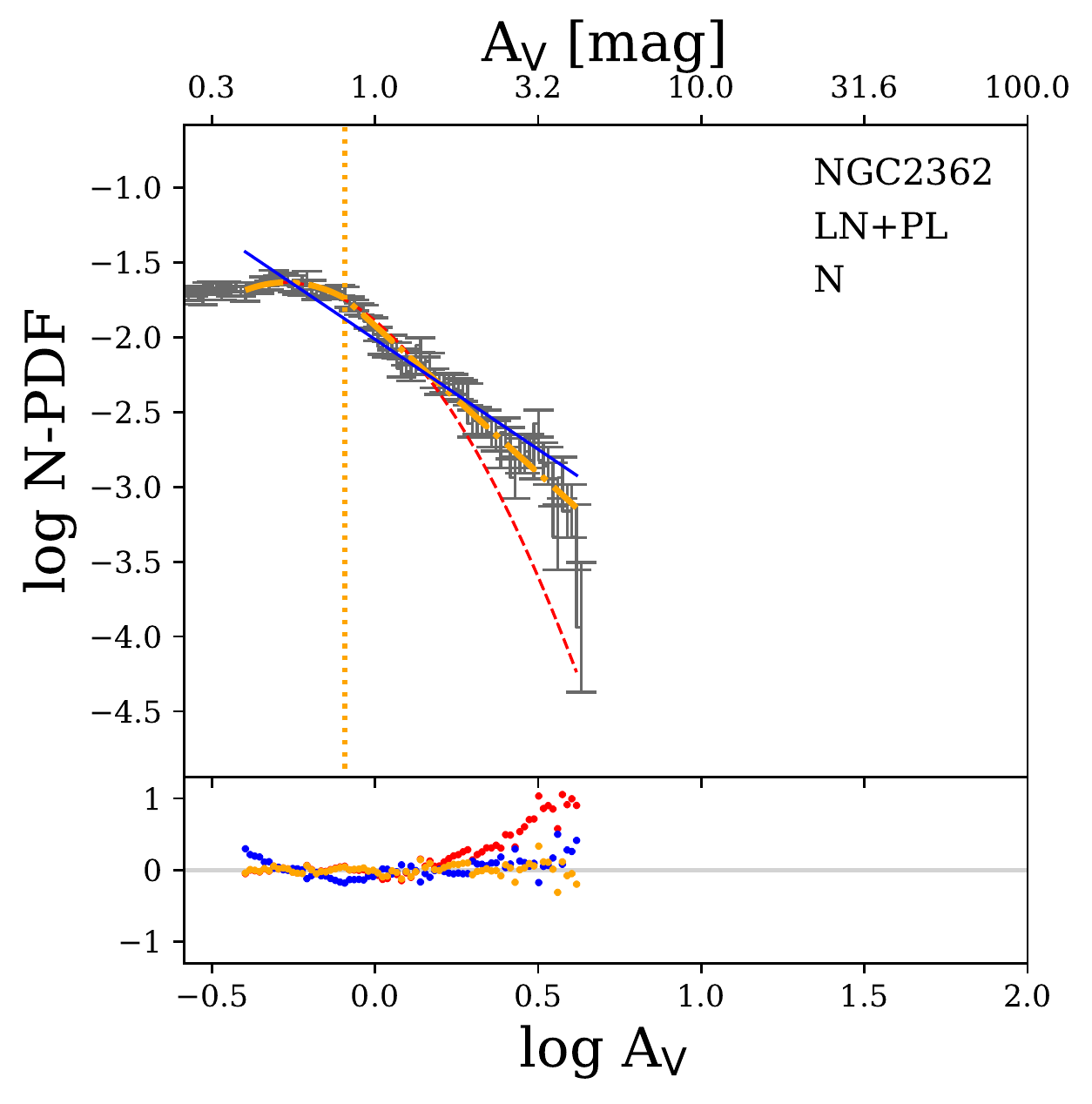}
    \end{minipage}
    \begin{minipage}[b]{0.24\textwidth}
        \includegraphics[width=\textwidth]{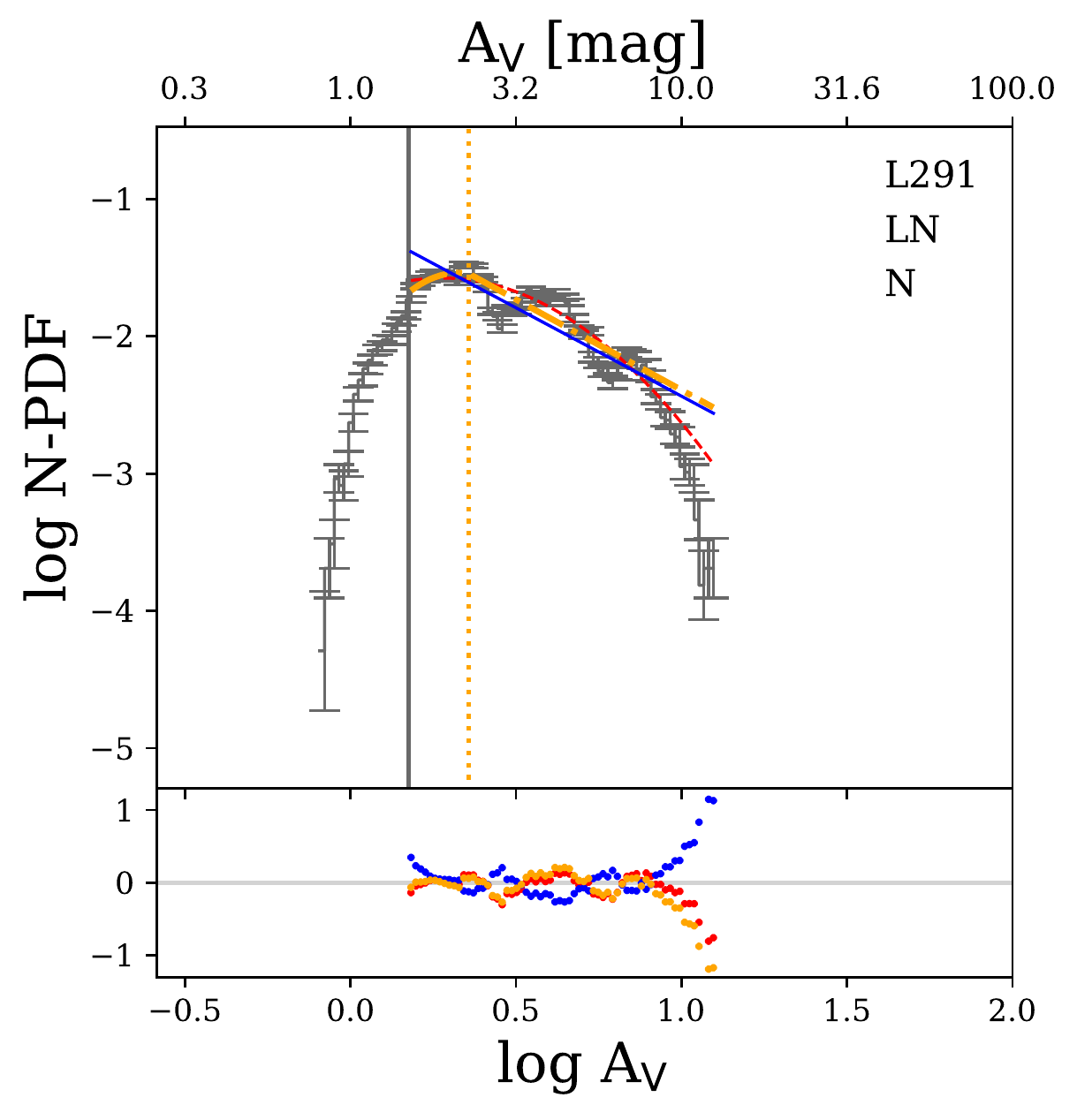}
    \end{minipage}
    
    \begin{minipage}[b]{0.24\textwidth}
        \includegraphics[width=\textwidth]{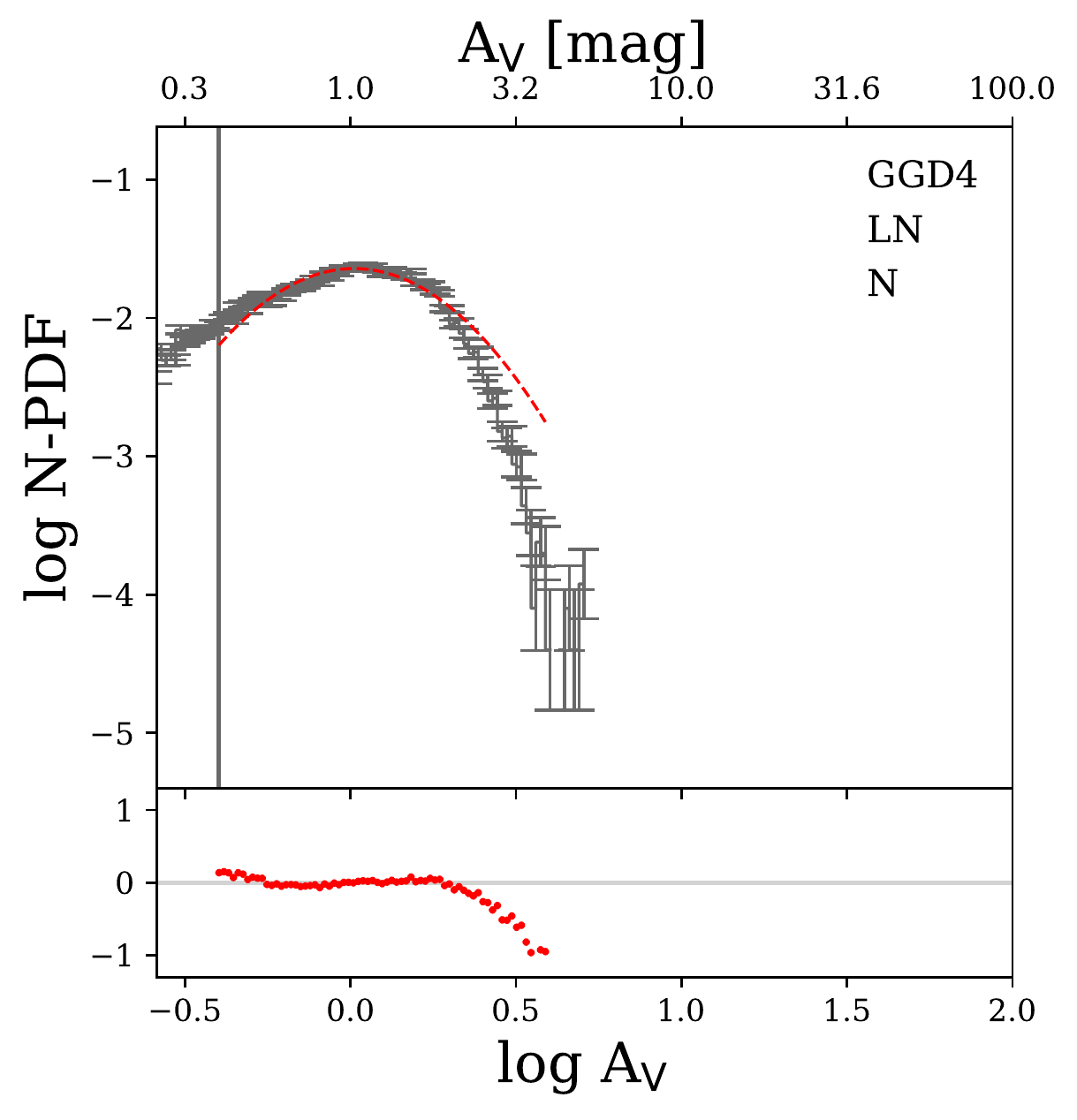}
    \end{minipage}
    \begin{minipage}[b]{0.24\textwidth}
        \includegraphics[width=\textwidth]{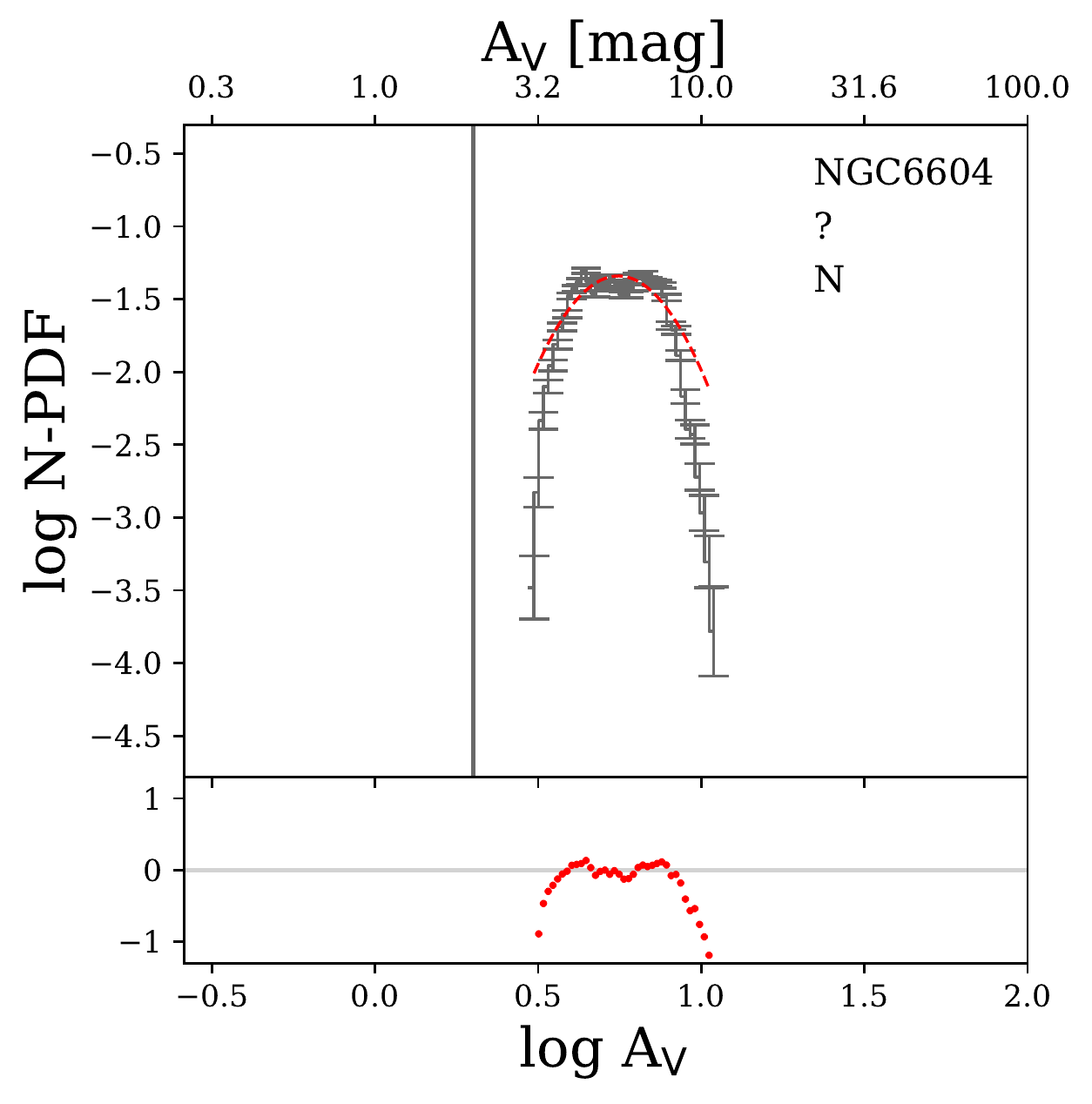}
    \end{minipage}
    %\begin{minipage}[b]{0.24\textwidth}
    %    \includegraphics[width=\textwidth]{newfigs/converted/Sag_Car_area_loglog_logbins_fit_resid_flag_noleg-eps-converted-to.pdf}
    %\end{minipage}
    \begin{minipage}[b]{0.24\textwidth}
        \includegraphics[width=\textwidth]{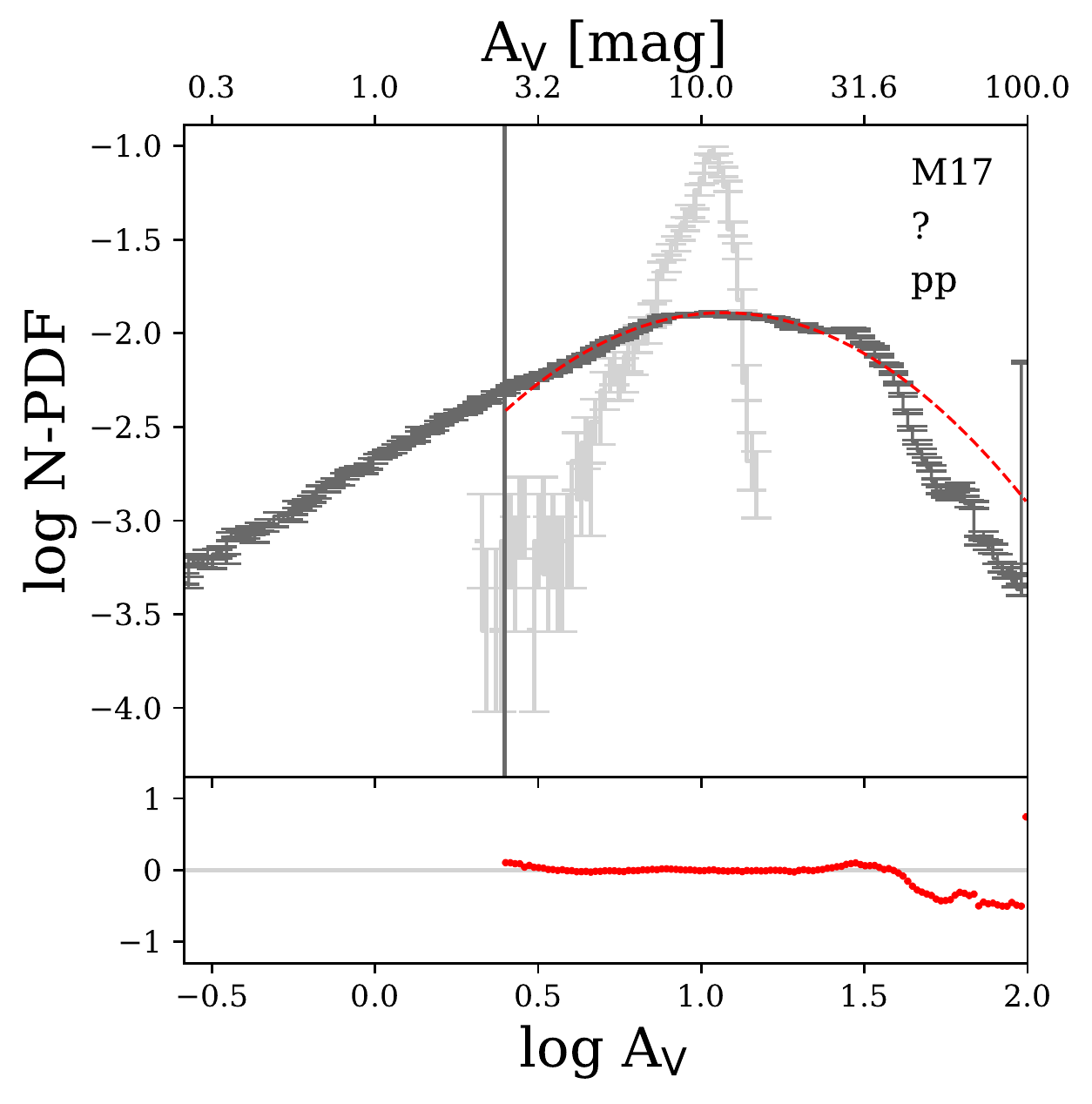}
    \end{minipage}
    \begin{minipage}[b]{0.24\textwidth}
        \includegraphics[width=\textwidth]{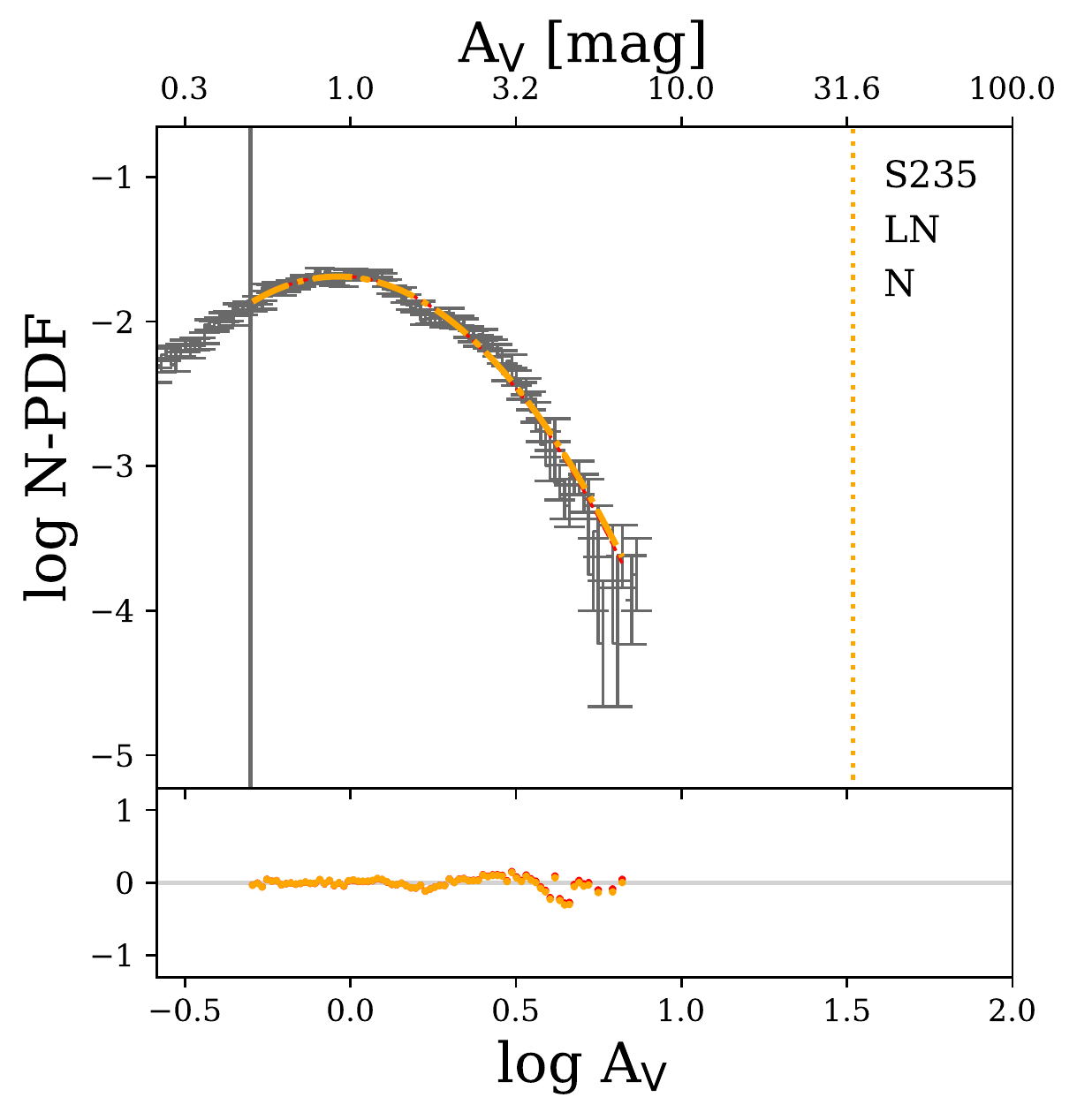}
    \end{minipage}
    
    \begin{minipage}[b]{0.24\textwidth}
        \includegraphics[width=\textwidth]{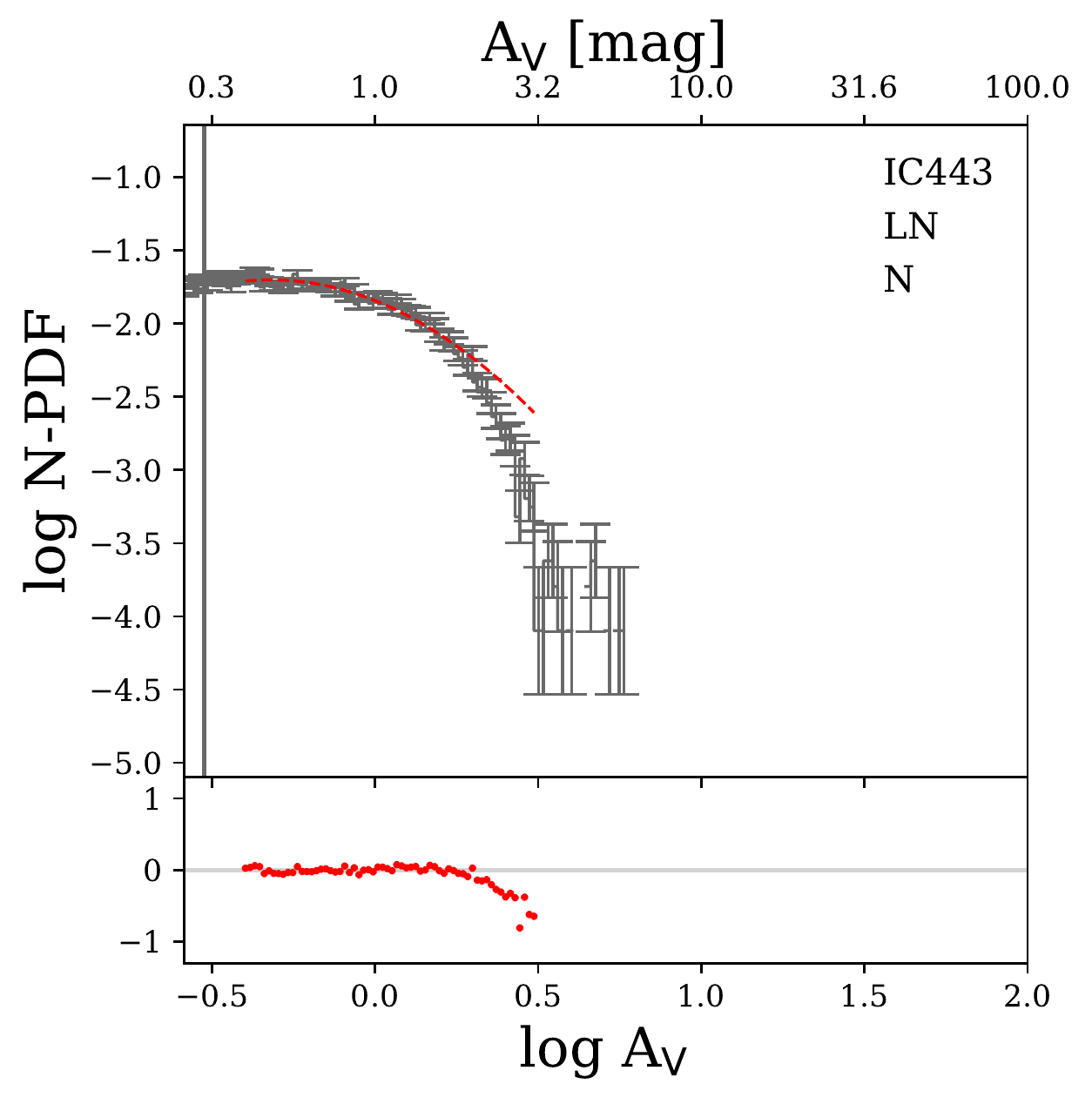}
    \end{minipage}%J
    \begin{minipage}[b]{0.24\textwidth}
        \includegraphics[width=\textwidth]{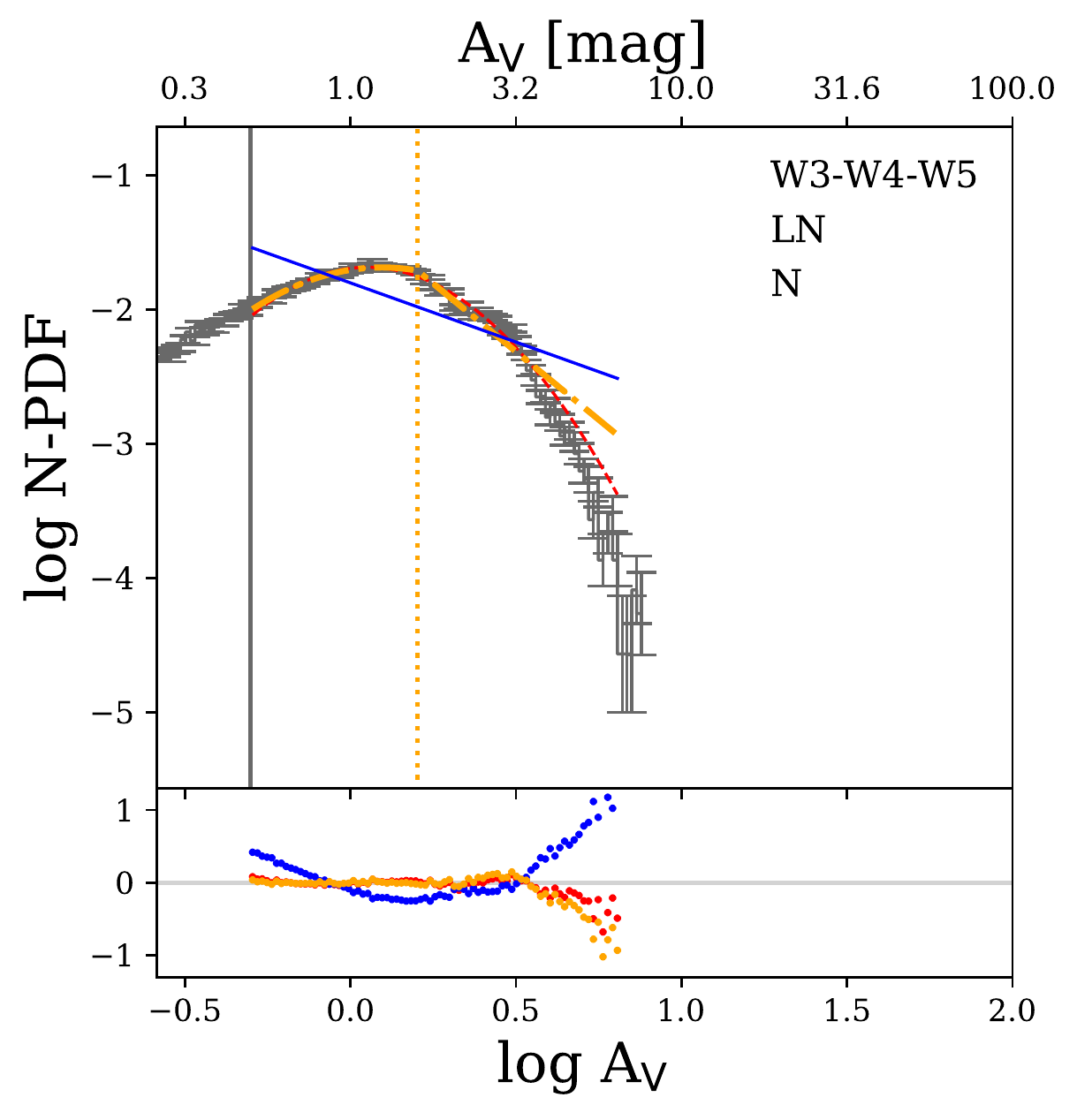}
    \end{minipage}
    \begin{minipage}[b]{0.24\textwidth}
        \includegraphics[width=\textwidth]{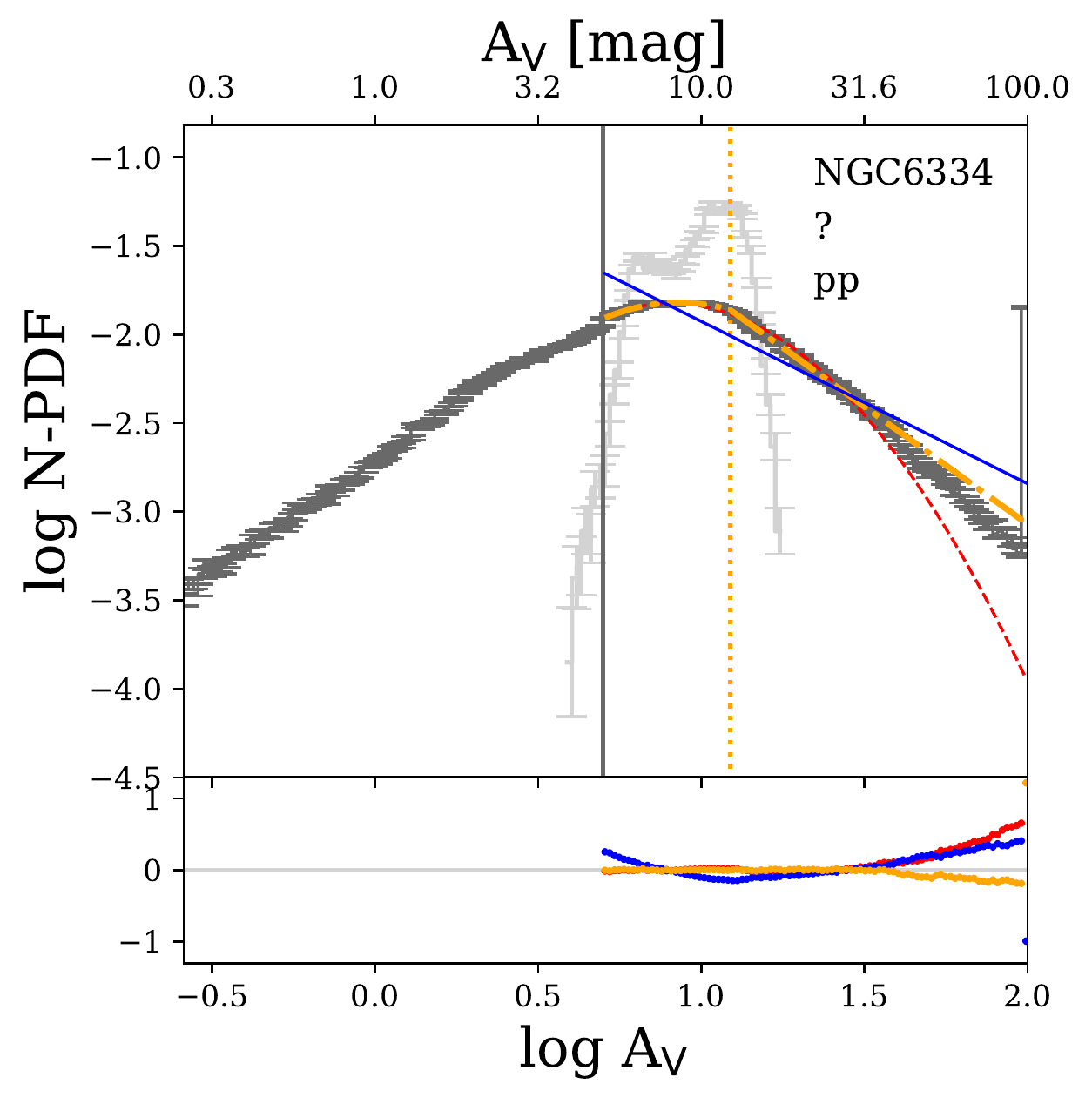}
    \end{minipage}%J
    \begin{minipage}[b]{0.24\textwidth}
        \includegraphics[width=\textwidth]{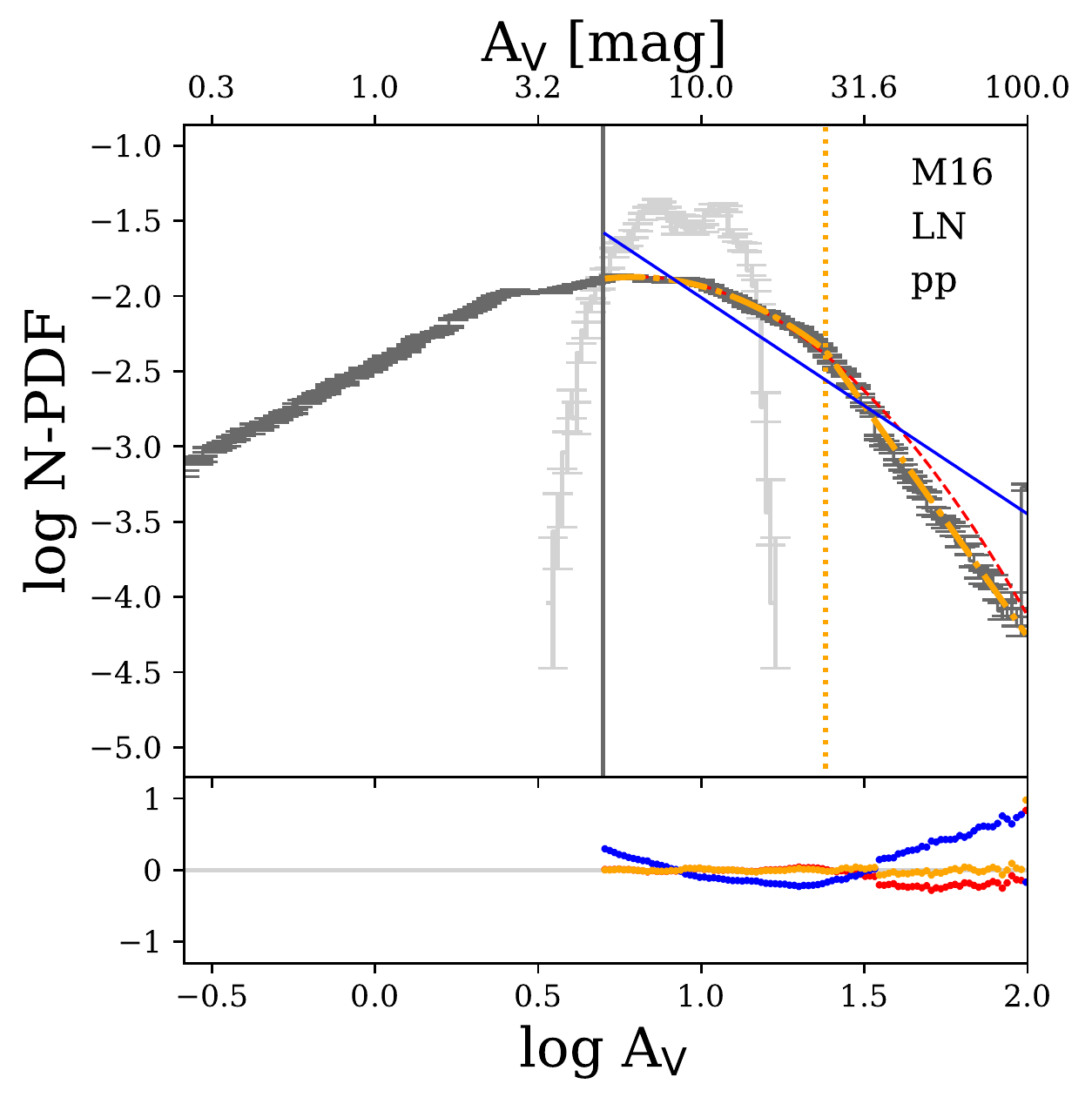}
    \end{minipage}%J
    
    \begin{minipage}[b]{0.24\textwidth}
        \includegraphics[width=\textwidth]{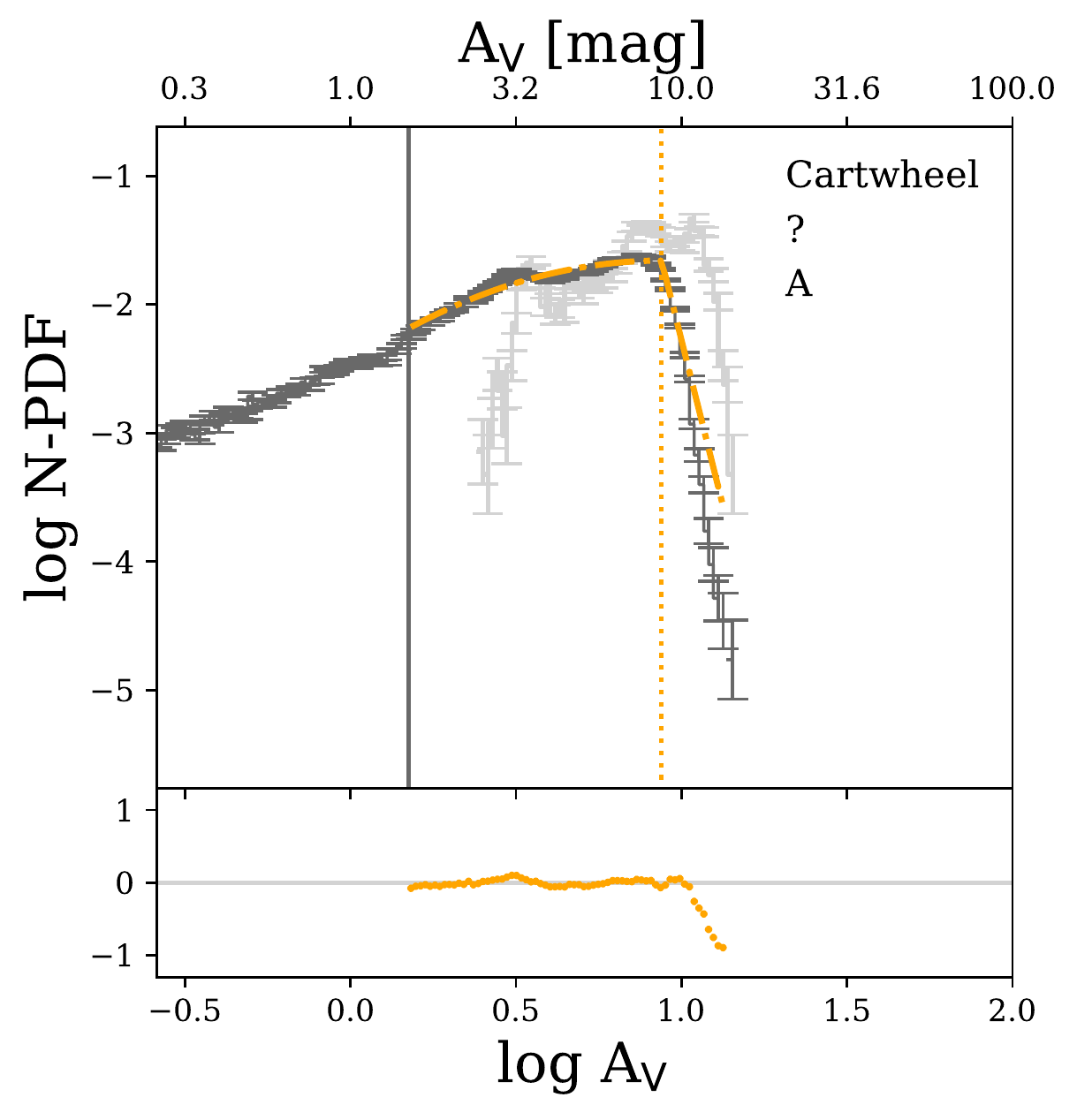}
    \end{minipage}%J
    \begin{minipage}[b]{0.24\textwidth}
        \includegraphics[width=\textwidth]{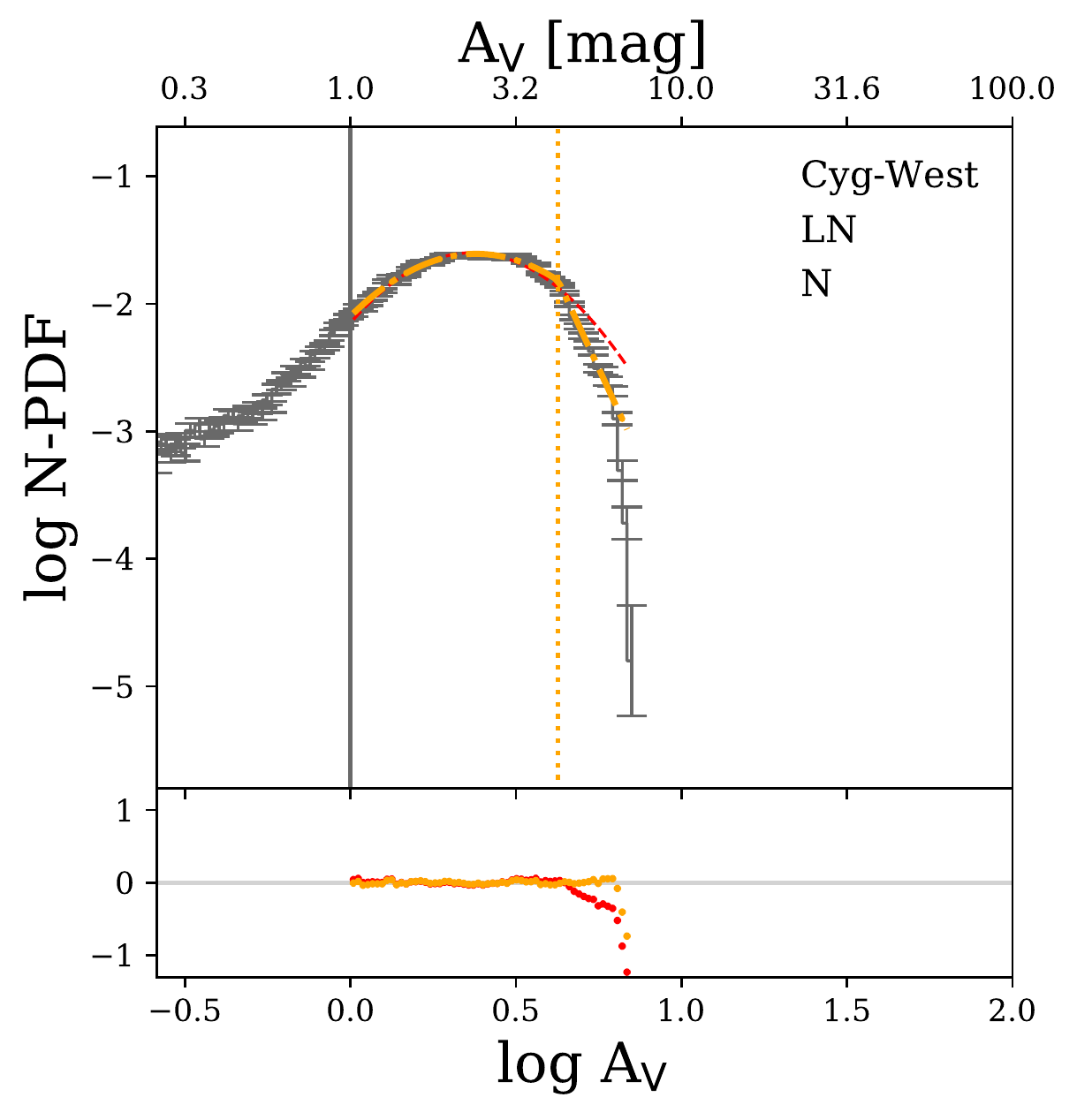}
    \end{minipage}%J
    \begin{minipage}[b]{0.24\textwidth}
        \includegraphics[width=\textwidth]{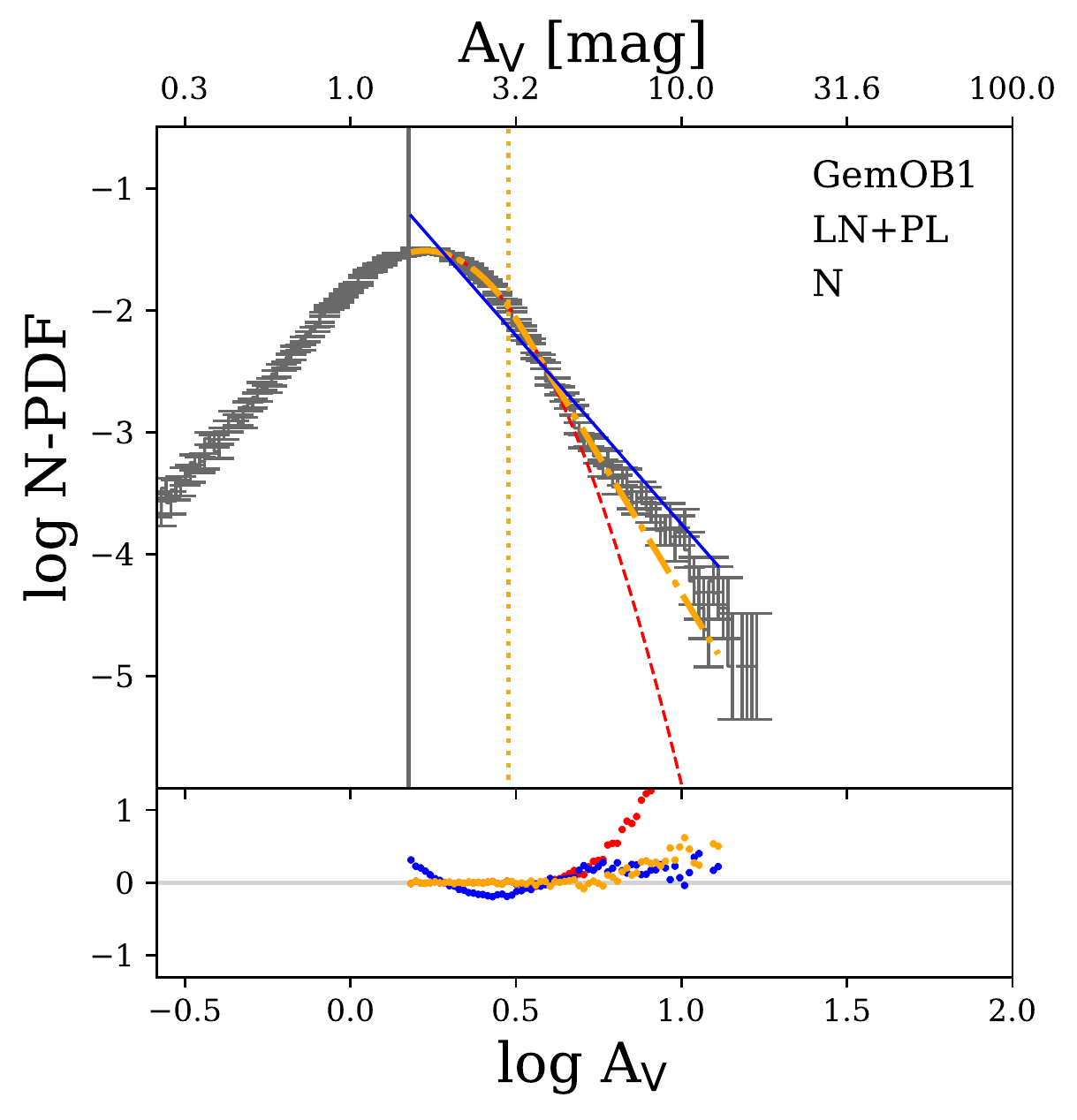}
    \end{minipage}
    \begin{minipage}[b]{0.24\textwidth}
        \includegraphics[width=\textwidth]{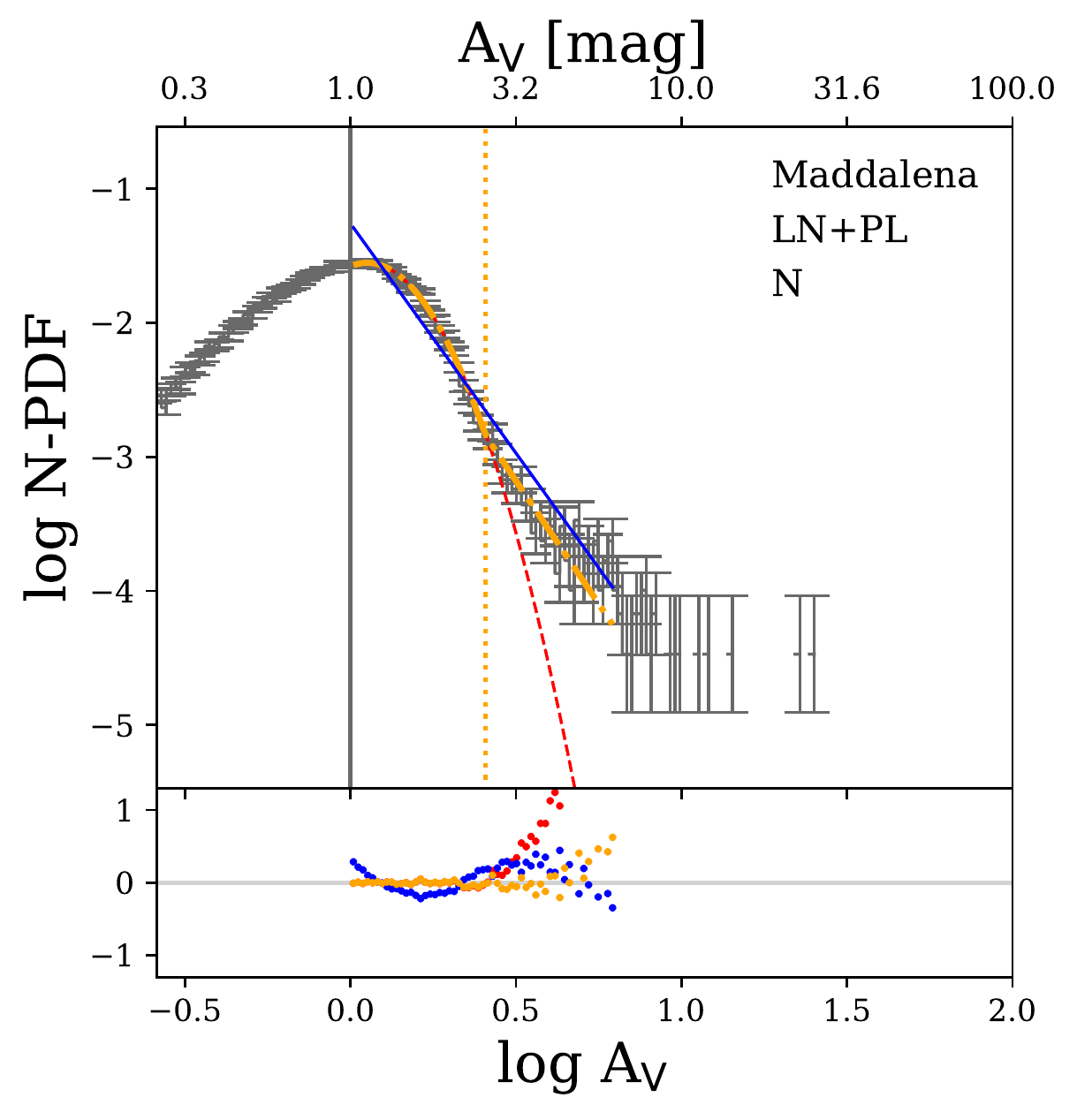}
    \end{minipage}
    \caption{N-PDFs of all clouds, continued.}
    \label{fig:Allfits4}
\end{figure*}

\clearpage

\section{Resolution study}
\label{app:resolution}

Here, we study the effect of resolution on our column density maps and N-PDFs. Because the difference in distance between the clouds is a factor of $\sim10$, the physical resolution can also be different by a similar factor. However, we note that this is the maximum\emph{} resolution difference; some of the distant clouds, for example all those with Herschel-based data, have a resolution similar to nearby clouds (nearby clouds at 200 pc away have a resolution of about 150\arcsec, while clouds 2 kpc away with Herschel data have the resolution of 12\arcsec, yielding a similar physical resolution). Thus, the resolution issue is relevant only for a subset of clouds.

\begin{figure}[h!]
    \centering
        \includegraphics[width=0.32\textwidth]{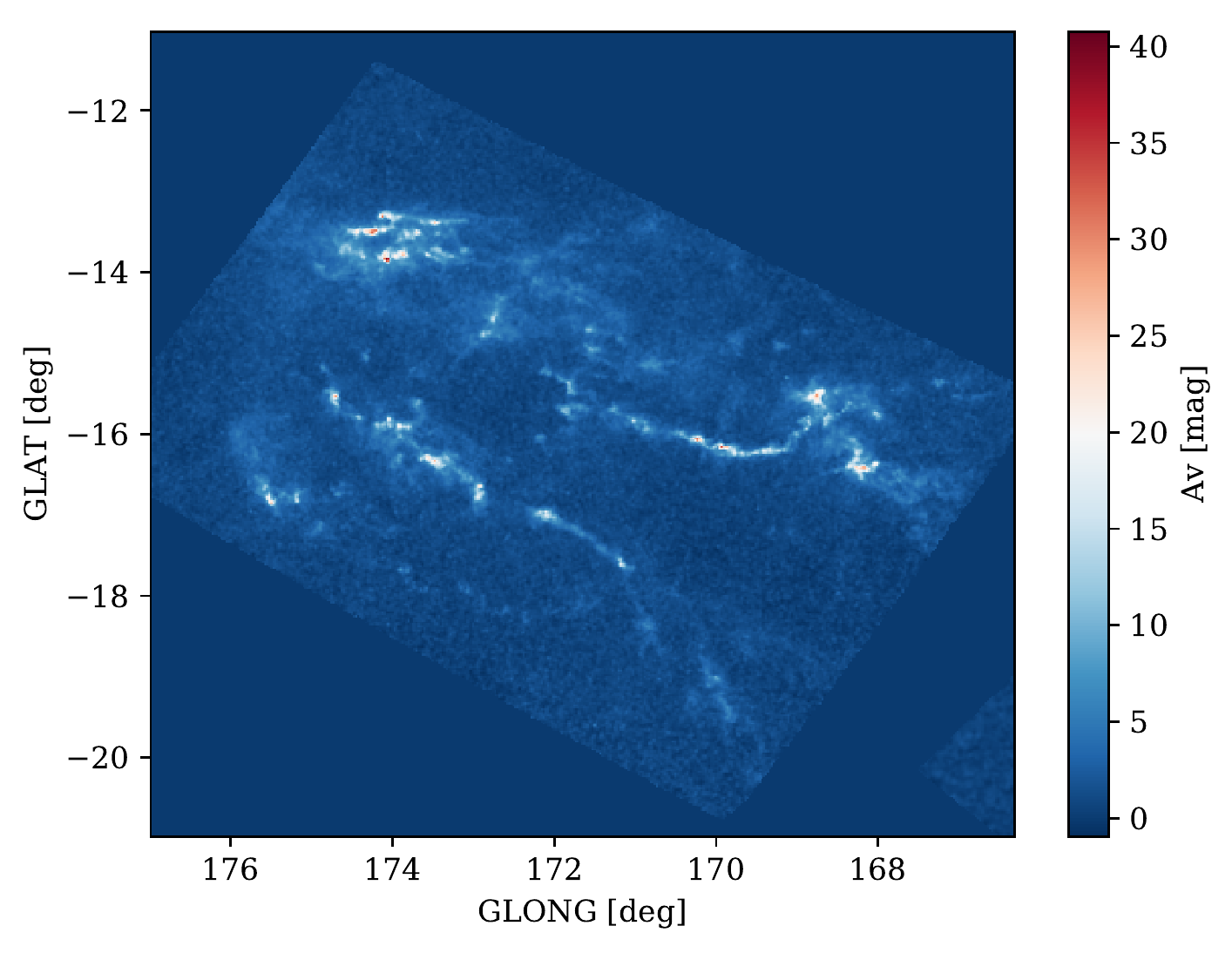}
        \includegraphics[width=0.32\textwidth]{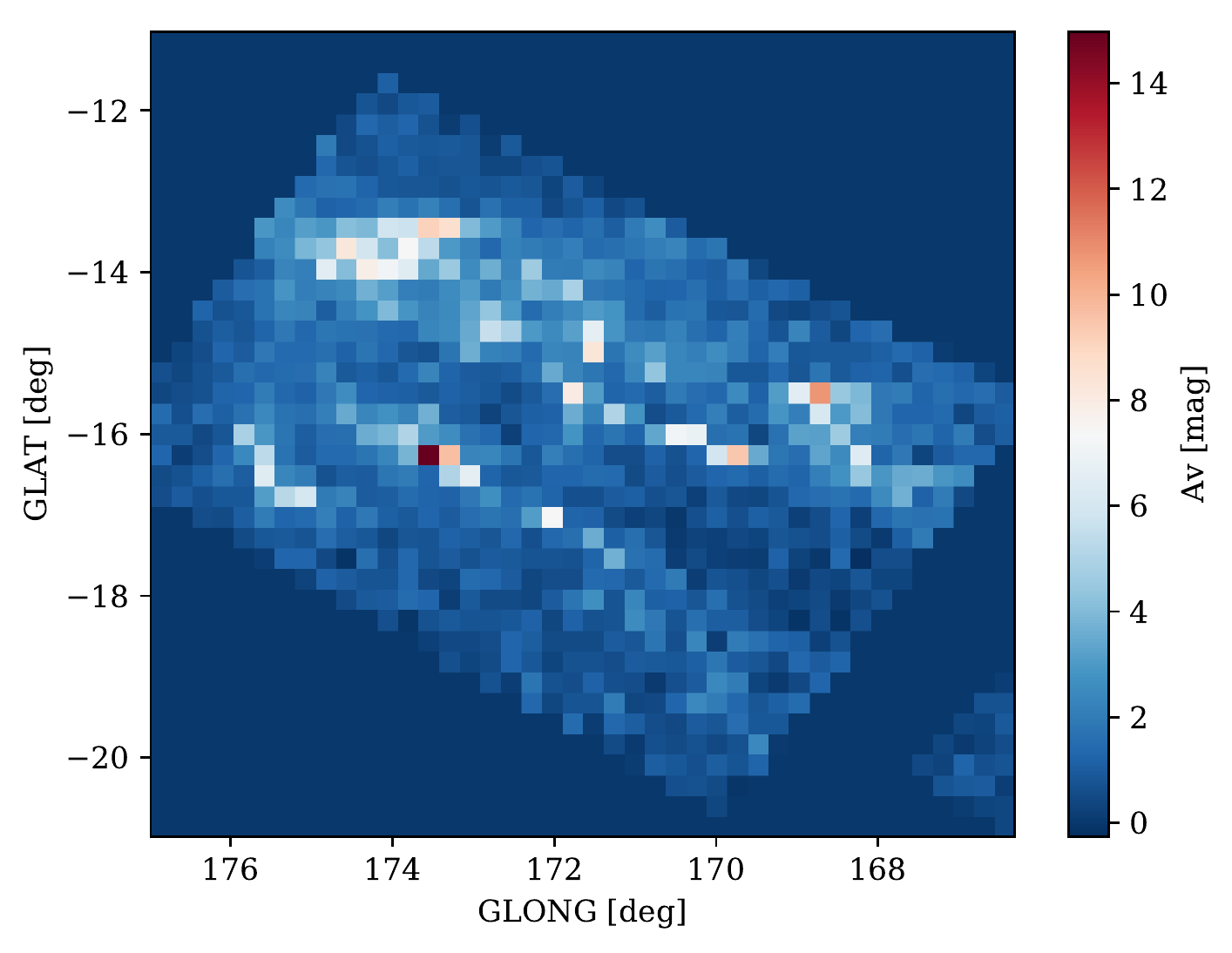}
        \includegraphics[width=0.32\textwidth]{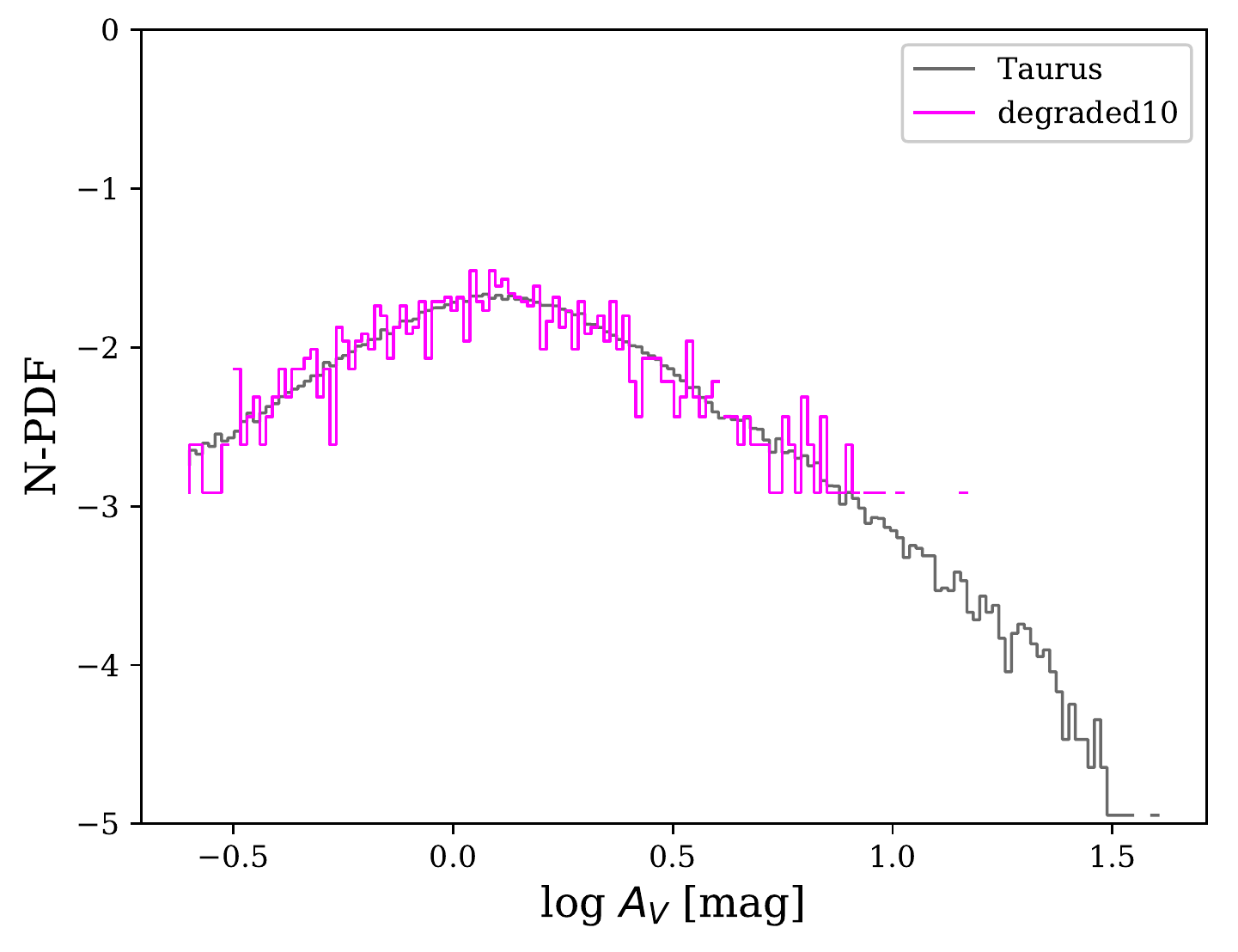}
    \caption{Column density maps of Taurus and resulting N-PDFs. \emph{Left:} Original column density map. \emph{Middle:} Degraded by a factor of 10. \emph{Right:} N-PDFs computed from the maps. The original N-PDF is shown in black, and the degraded N-PDF is shown in magenta.}
    \label{fig:resolution}
\end{figure}

To examine this, we degraded the resolution of one of the nearby clouds (Taurus) by a factor of ten to see how this would affect the shape of the derived N-PDF. To degrade the resolution, we used the Python package \texttt{scipy.interpolate.RegularGridInterpolator}. Figure \ref{fig:resolution} shows the column density map of Taurus at the two different resolutions and the N-PDFs derived from these maps. It appears that the shape of the N-PDF is relatively well preserved with the lower resolution, but some information in the high-density part of the N-PDF is lost at low resolution due to poorer sampling. This loss of the high-density part can affect the derived mass and relative dense gas measures for the clouds.

\begin{figure}[h!]
    \centering
        \includegraphics[width=0.4\textwidth]{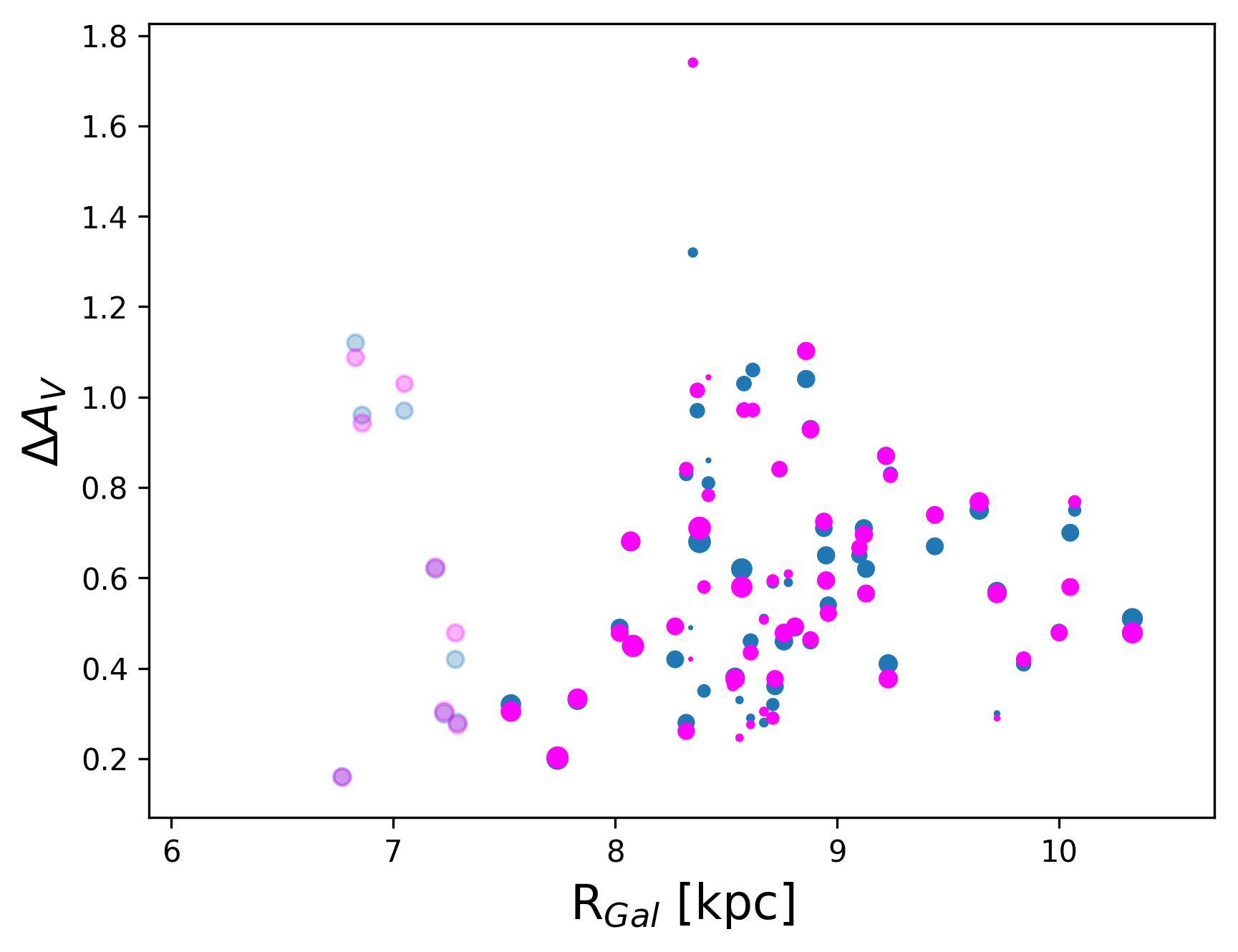}%Cloud_delAv_Rgal_pc-eps-converted-to.pdf}
        \includegraphics[width=0.4\textwidth]{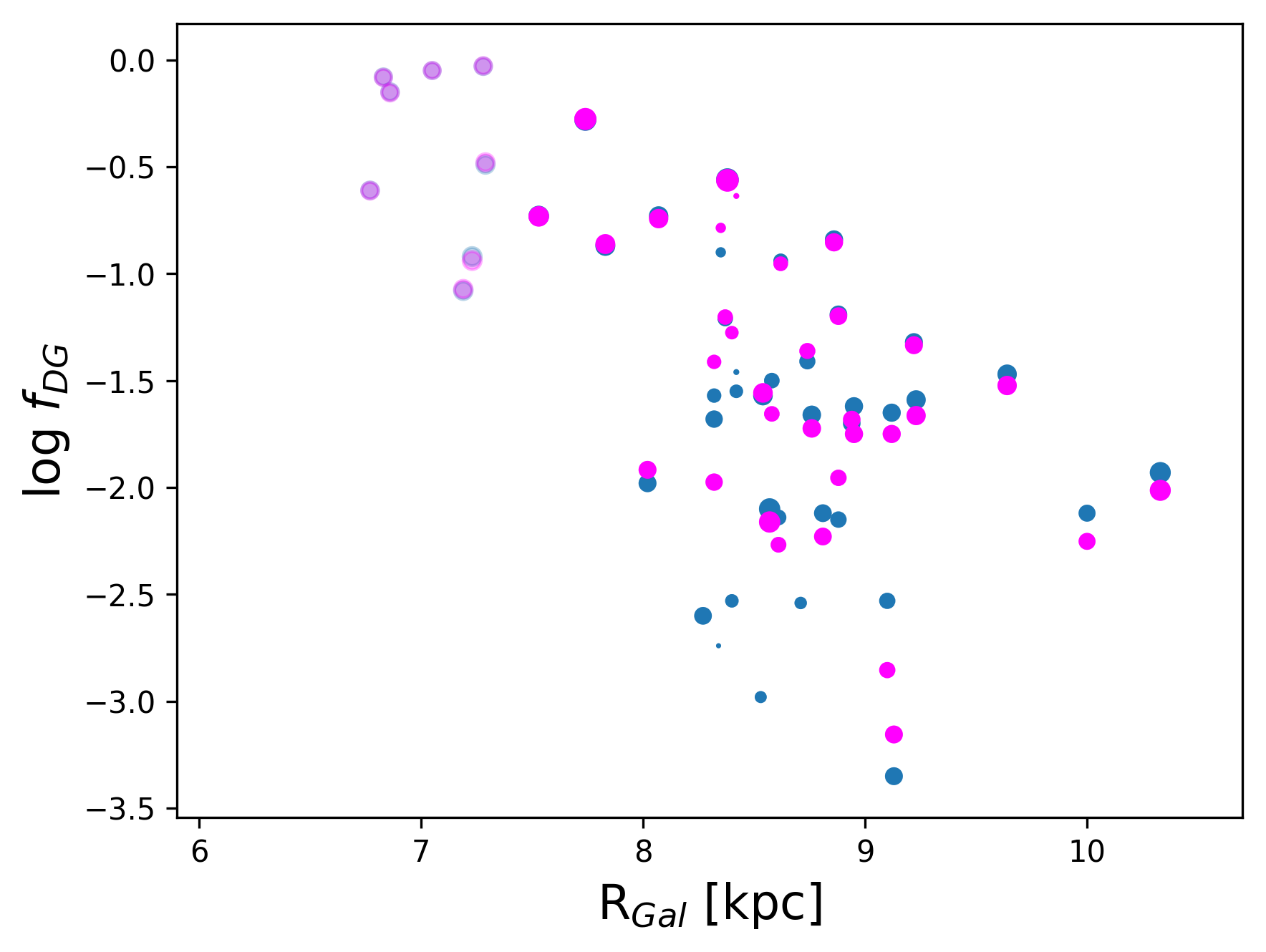}%Cloud_fdg_Rgal_pc-eps-converted-to.pdf}
    \caption{Relation between dense gas measures and galactocentric radius. The original data points for the clouds with native resolution are shown in blue (same as Fig. \ref{fig:galactic_densgas_profile}), and the same clouds with degraded resolution are shown in magenta.}
    \label{fig:resolution_dense}
\end{figure}

\begin{figure}[h!]
    \centering
        \includegraphics[width=0.4\textwidth]{newfigs/converted/Dens_in_r2000_newcuts_res-eps-converted-to.pdf}
        \includegraphics[width=0.4\textwidth]{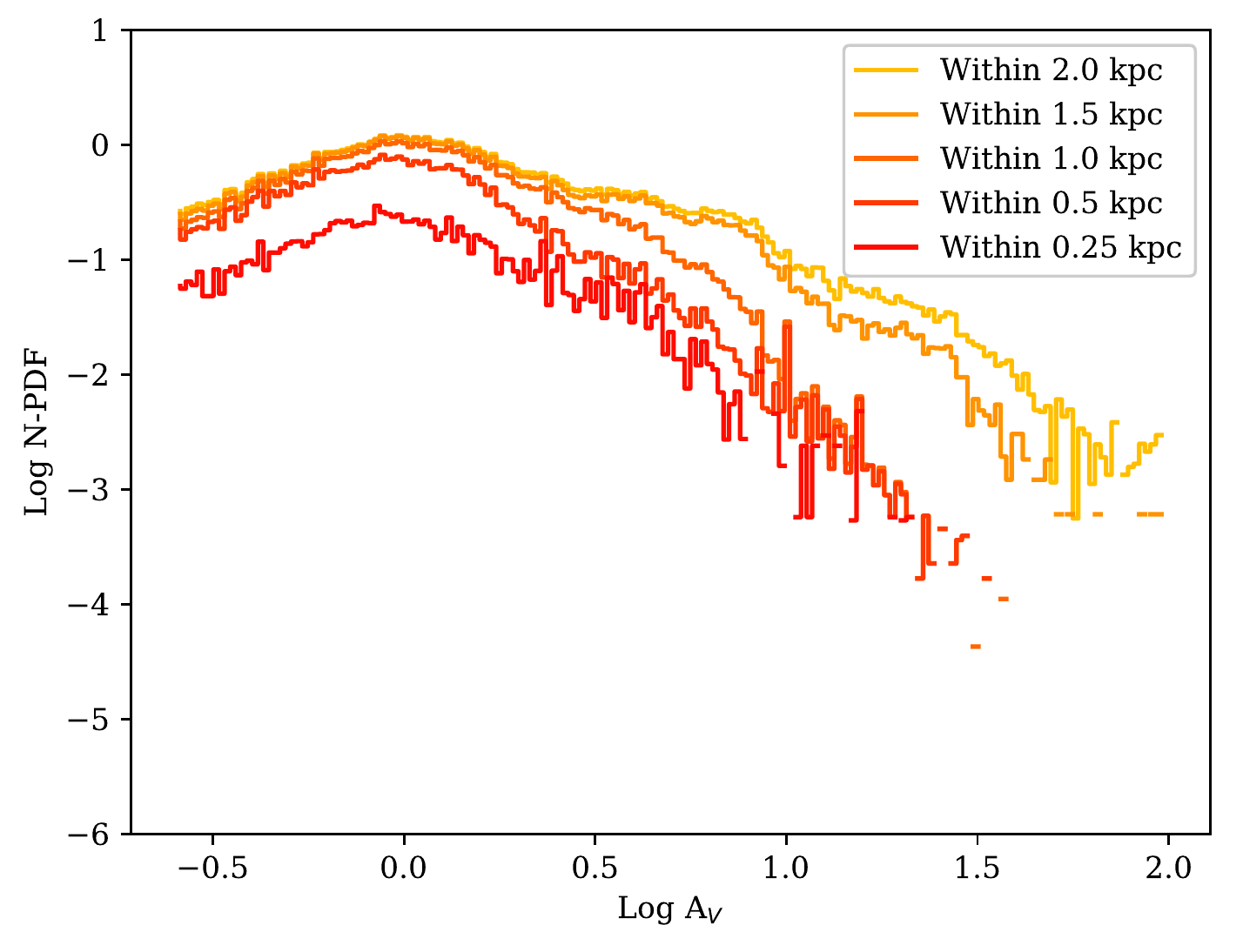}
    \caption{Solar centred apertures with native resolution (\emph{left}) and with all cloud maps degraded to a resolution of 0.824 pc (\emph{right}).}
    \label{fig:resolution_aps}
\end{figure}

To study the effect of resolution differences on our full sample, we degraded all clouds to the lowest resolution in our sample. The cloud with the lowest resolution is Maddalena, with a physical resolution of 0.824 pc. Most clouds have similar masses and dense gas measures when degraded; a few clouds have significantly different values. The effect of the degraded resolution on the dense gas measures is shown in Fig. \ref{fig:resolution_dense}, which shows that the behaviour with respect to the galactocentric radius is preserved for the degraded resolution. Figure \ref{fig:resolution_aps} shows N-PDFs for solar centred apertures before and after degrading. The shape is relatively well preserved here as well, except for the truncation of the highest-density part. Table \ref{tab:ap_shapes_res} shows the effects of degrading the resolution on the full sample of apertures covering the survey area. The degraded apertures appear to have slightly more mass, shallower slopes, and higher dense gas fractions than the original apertures, but the numbers generally agree within the errors. Hence only minor differences are seen, and we can conclude that the varying resolution of the column density maps of the clouds in this study does not significantly affect the results.

% Table of the aperture N-PDF properties
\begin{table}
\caption{Properties of the aperture N-PDFs. The top of the table is the same as Table \ref{tab:ap_shapes_new}, and the bottom shows the same quantities when all clouds are degraded to a resolution of 0.824 pc.} % title of Table
\label{tab:ap_shapes_res}      % is used to refer this table in the text
\centering                          % used for centering table
\begin{tabular}{c c c c c c c c c c}        % centered columns (4 columns)
\hline\hline                 % inserts double horizontal lines
R  & N\tablefootmark{a} & M$_{A_\mathrm{V} > 3 \ \mathrm{mag}}$   &  $\alpha$\tablefootmark{b} & $\overline{\Sigma}$\tablefootmark{c} & SFR & $\frac{\mathrm{SFR}}{A}$\tablefootmark{d} & SFE &  $f_\mathrm{DG}$ & $\Delta A_\mathrm{V}$\\   
\hline
kpc & & $10^6$ M$_\odot$ & & M$_\odot$ pc$^{-2}$ & $10^3$M$_\odot$ Myr$^{-1}$ & $10^3$M$_\odot$ Myr$^{-1}$kpc$^{-2}$ & \% & \% &  \\
\hline      
2.00\tablefootmark{e} & 1 & 8.8 & -1.9 & 0.7  & 9.7 & 0.8  & 0.13 &  23  & 1.45 \\ 

1.00    &       13      &       6       $\pm$   2       &        -2.7   $\pm$     0.2    &         1.2   $\pm$     0.3   &         7     $\pm$     2     &         2.2    $\pm$     0.6   &       0.10    $\pm$   0.02    &       17      $\pm$    5       &         0.84  $\pm$     0.09   \\ 
0.50    &       49      &       1.8     $\pm$   0.4     &        -3.1   $\pm$     0.2    &         1.5   $\pm$     0.4   &         2.2   $\pm$     0.6   &         2.8    $\pm$     0.8   &       0.08    $\pm$   0.02    &       12      $\pm$    3       &         0.58  $\pm$     0.05   \\ 
0.25    &       213     &       0.38    $\pm$   0.08    &        -3.3   $\pm$     0.2    &         1.1   $\pm$     0.3   &         0.4   $\pm$     0.1   &         2.1    $\pm$     0.7   &       0.04    $\pm$   0.01    &        5      $\pm$    1       &         0.27  $\pm$     0.02   \\ 
\hline  
\multicolumn{9}{c}{Apertures with R$_\mathrm{Gal}>$ 7.5 kpc} \\
\hline  
1.00    &       11      &       6       $\pm$   2       &        -2.9   $\pm$     0.2    &         1.1   $\pm$     0.3   &         7     $\pm$     2     &         2      $\pm$     2     &       0.10    $\pm$   0.02    &       12      $\pm$    4       &         0.8   $\pm$     0.1    \\ 
0.50    &       42      &       1.8     $\pm$   0.5     &        -3.3   $\pm$     0.2    &         1.3   $\pm$     0.4   &         2     $\pm$     5     &         3.0    $\pm$     0.9   &       0.09    $\pm$   0.02    &        8      $\pm$    2       &         0.55  $\pm$     0.05   \\ 
0.25    &       176     &       0.36    $\pm$   0.08    &        -3.4   $\pm$     0.1    &         1     $\pm$     5     &         0.4   $\pm$     0.2   &         2.3    $\pm$     0.8   &       0.05    $\pm$   0.01    &        3      $\pm$    1       &         0.27  $\pm$     0.02   \\ 
\hline  
\multicolumn{9}{c}{All apertures, now with all clouds at same resolution.} \\
\hline
1.00    &       13      &       8       $\pm$   2       &       -2.4    $\pm$    0.2     &        1.7    $\pm$    0.4    &        7      $\pm$    2      &        2.2     $\pm$    0.6    &       0.09    $\pm$   0.02    &       22      $\pm$    7       &        0.9    $\pm$    0.1     \\ 
0.50    &       49      &       2.2     $\pm$   0.5     &       -2.9    $\pm$    0.2     &        1.9    $\pm$    0.5    &        2.2    $\pm$    0.6    &        2.8     $\pm$    0.8    &       0.08    $\pm$   0.02    &       13      $\pm$    3       &        0.58   $\pm$    0.05    \\ 
0.25    &       213     &       0.47    $\pm$   0.09    &       -3.0    $\pm$    0.2     &        1.6    $\pm$    0.4    &        0.4    $\pm$    0.1    &        2.1     $\pm$    0.7    &       0.04    $\pm$   0.01    &        5      $\pm$    1       &        0.26   $\pm$    0.02    \\ 

\hline  
\multicolumn{9}{c}{Apertures with R$_\mathrm{Gal}>$ 7.5 kpc, now with all clouds at same resolution.} \\
\hline
1.00    &       11      &       6       $\pm$   2       &       -2.6    $\pm$    0.2     &        1.3    $\pm$    0.4    &        7      $\pm$    2      &        2       $\pm$    2      &       0.10    $\pm$   0.02    &       15      $\pm$    6       &        0.78   $\pm$    0.09    \\ 
0.50    &       42      &       1.8     $\pm$   0.5     &       -3.1    $\pm$    0.2     &        1.4    $\pm$    0.4    &        2      $\pm$    5      &        3.0     $\pm$    0.9    &       0.09    $\pm$   0.02    &        8      $\pm$    2       &        0.55   $\pm$    0.05    \\ 
0.25    &       176     &       0.36    $\pm$   0.09    &       -3.1    $\pm$    0.1     &        1      $\pm$    5      &        0.4    $\pm$    0.2    &        2.3     $\pm$    0.8    &       0.05    $\pm$   0.01    &        3      $\pm$    1       &        0.26   $\pm$    0.02    \\ 
\hline                                   %inserts single line
\end{tabular}
\tablefoot{\\
\tablefoottext{a}{Number of apertures considered.}
\tablefoottext{b}{Slope of the linear fit.}
\tablefoottext{c}{Mean surface density within the apertures (total cloud mass divided by aperture area).}
\tablefoottext{d}{$A$ is the area of the aperture.}
\tablefoottext{e}{This is our full survey area. Therefore there is only one aperture, and no standard deviation is given.}
}
\end{table}

\clearpage

\section{YSO completeness}
\label{app:ysoadjust}
\begin{figure}[h!]
    \centering
        \includegraphics[width=0.24\textwidth]{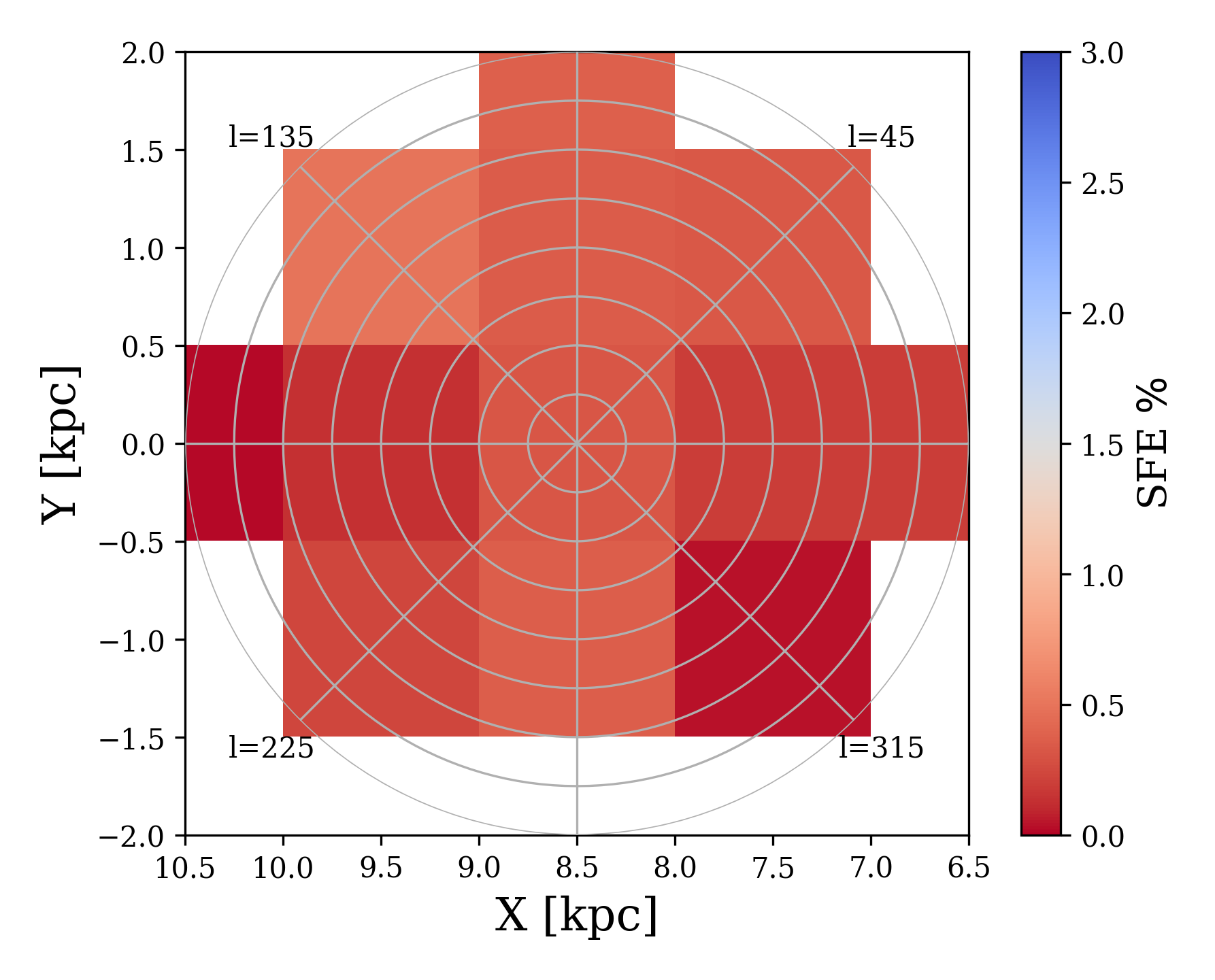}
        \includegraphics[width=0.24\textwidth]{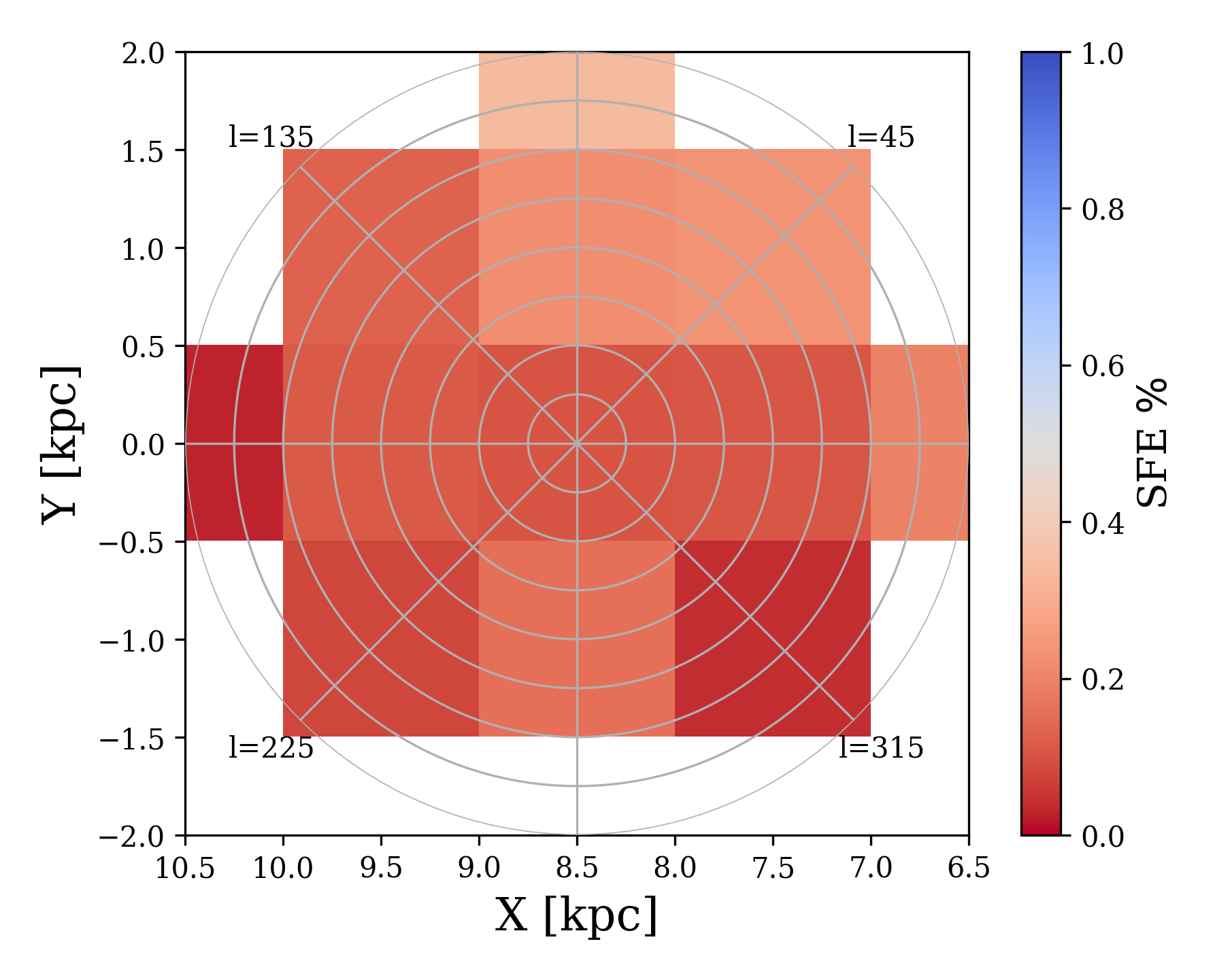}
        \includegraphics[width=0.24\textwidth]{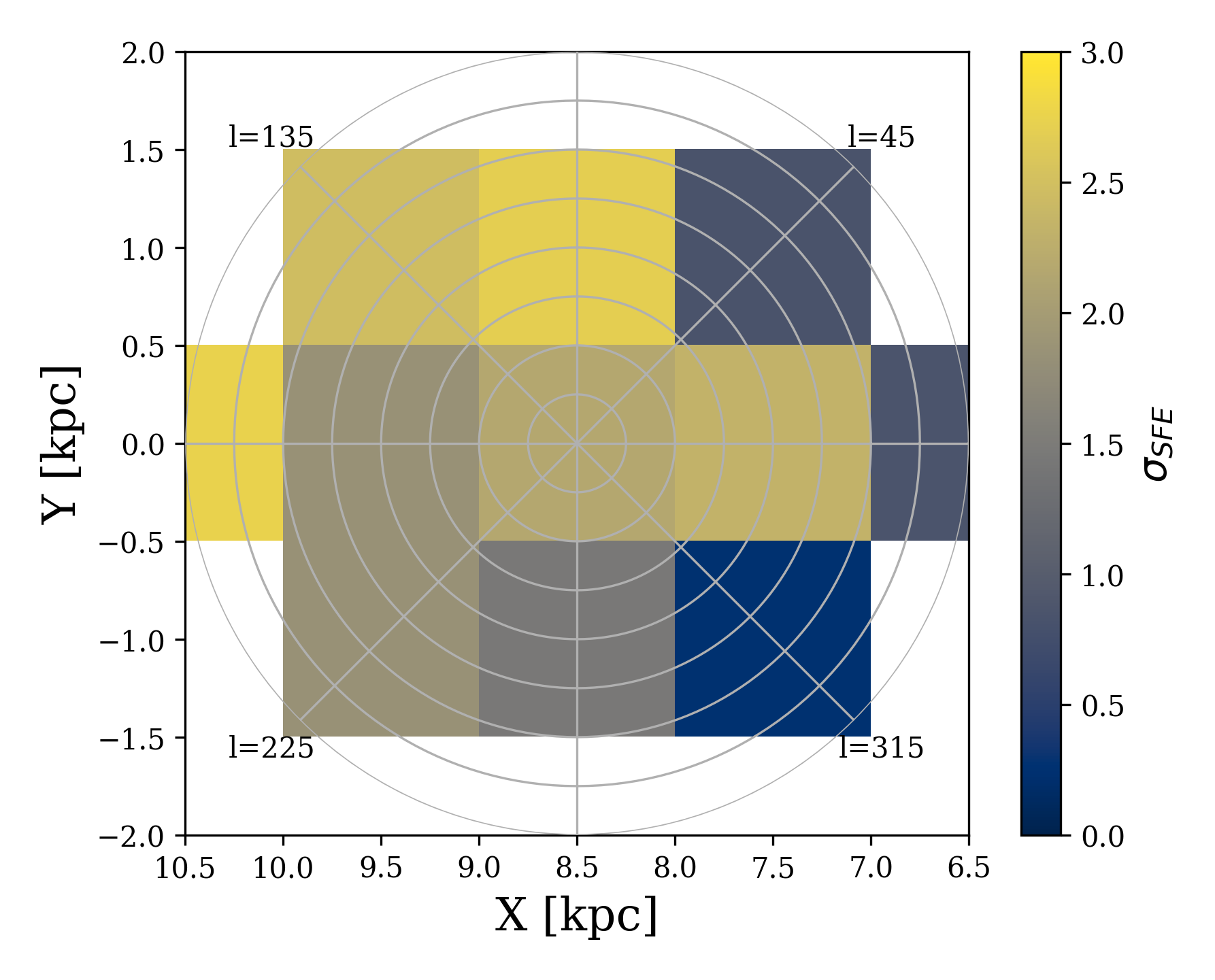}
        \includegraphics[width=0.24\textwidth]{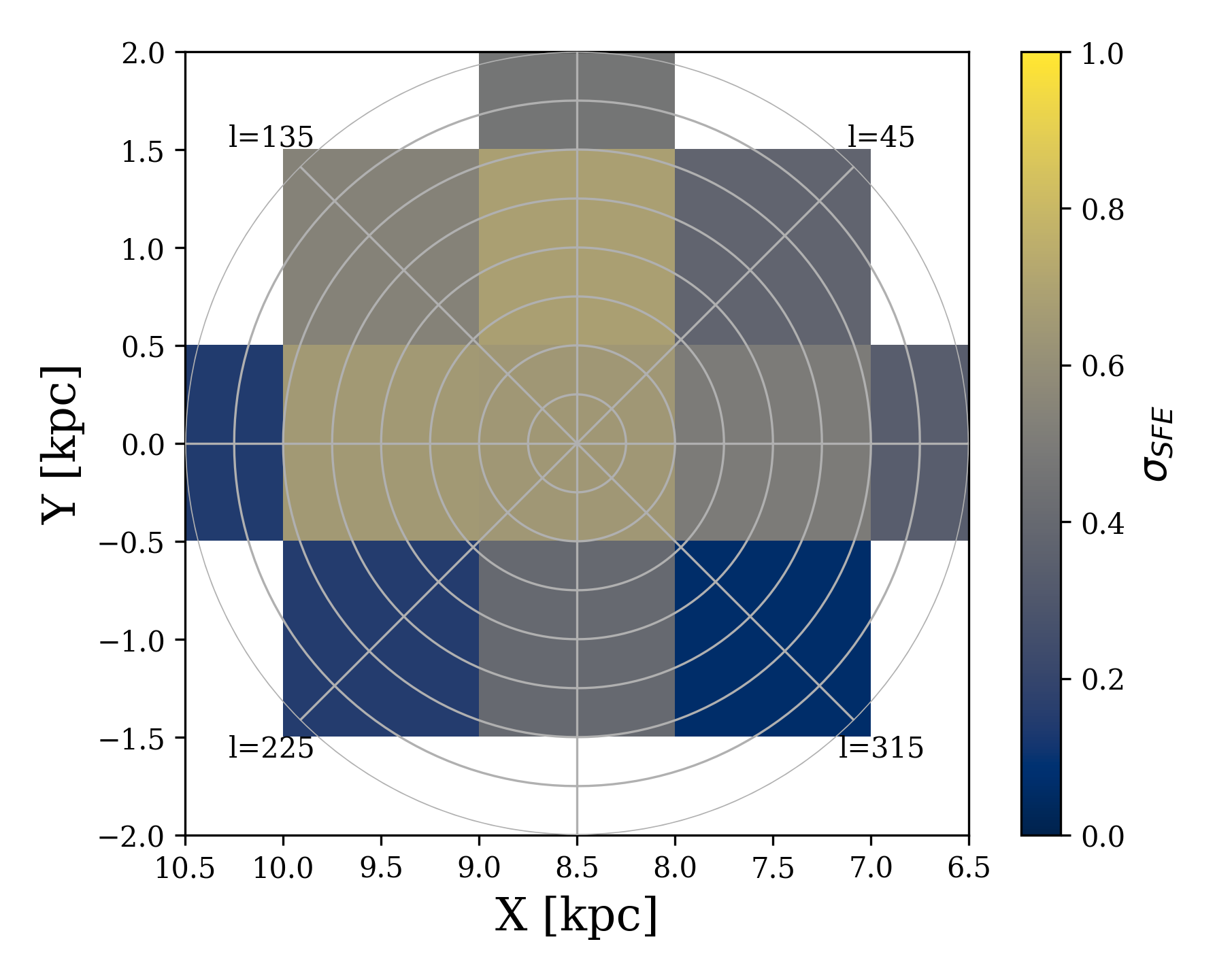}
         
        \includegraphics[width=0.24\textwidth]{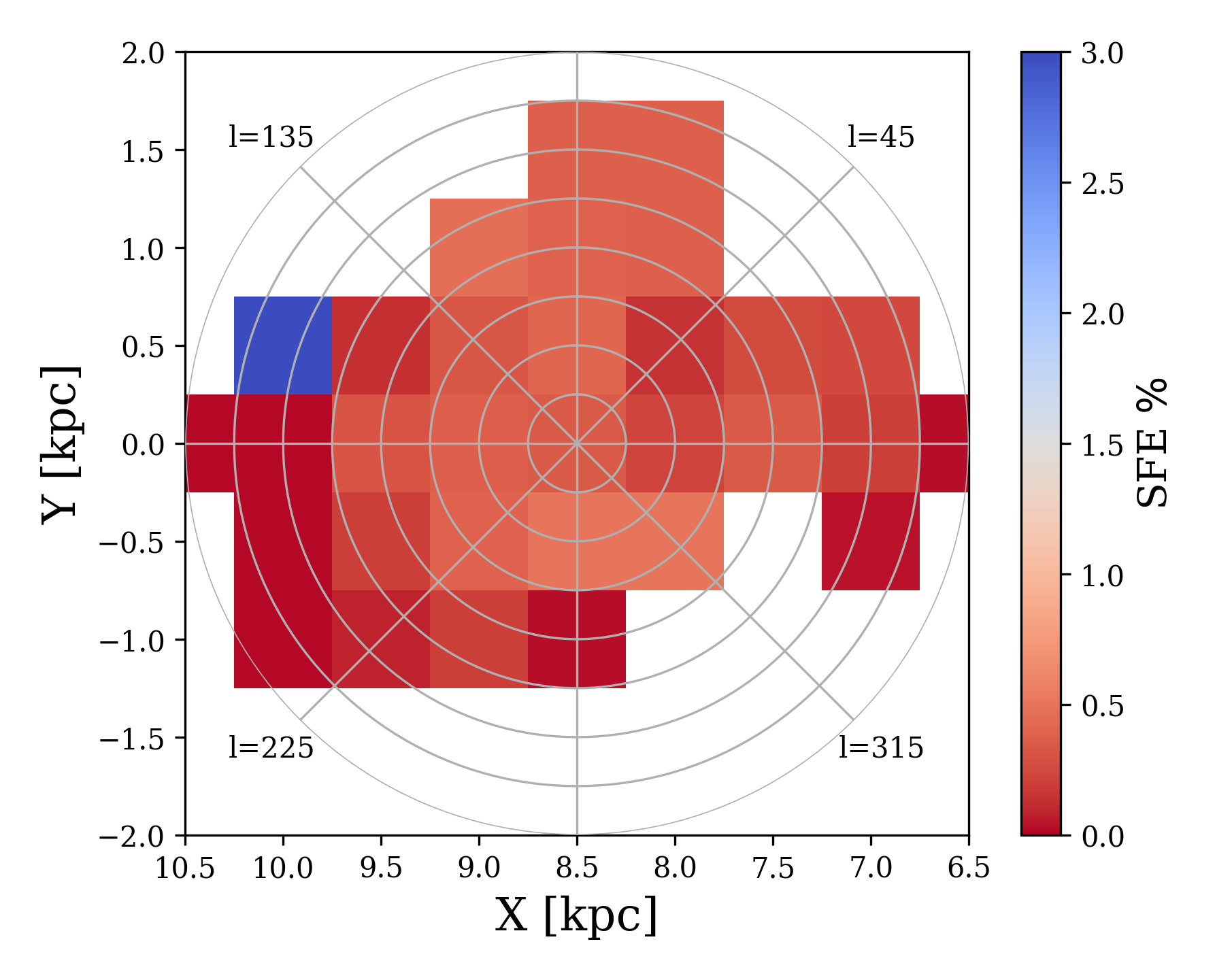}     
        \includegraphics[width=0.24\textwidth]{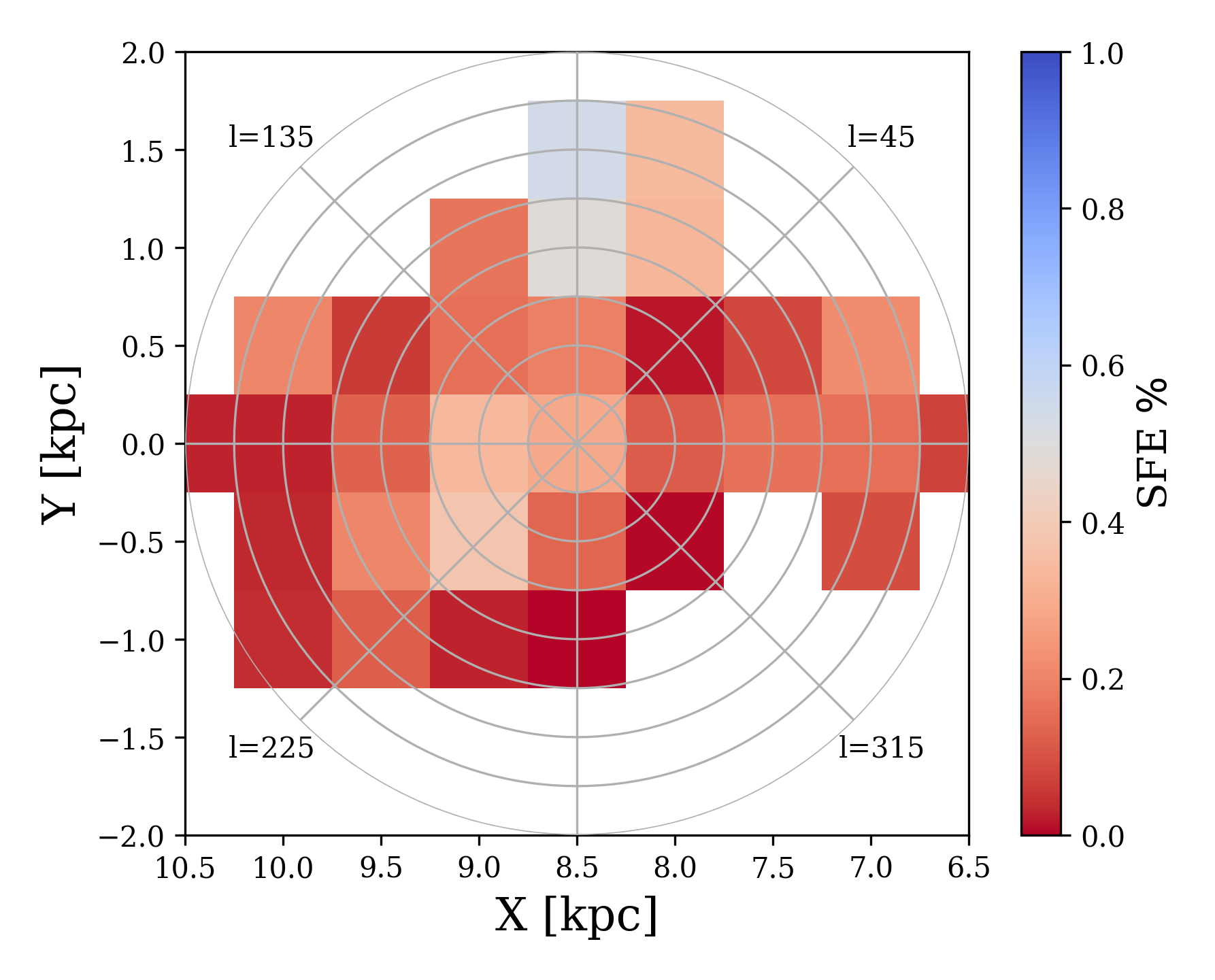}
        \includegraphics[width=0.24\textwidth]{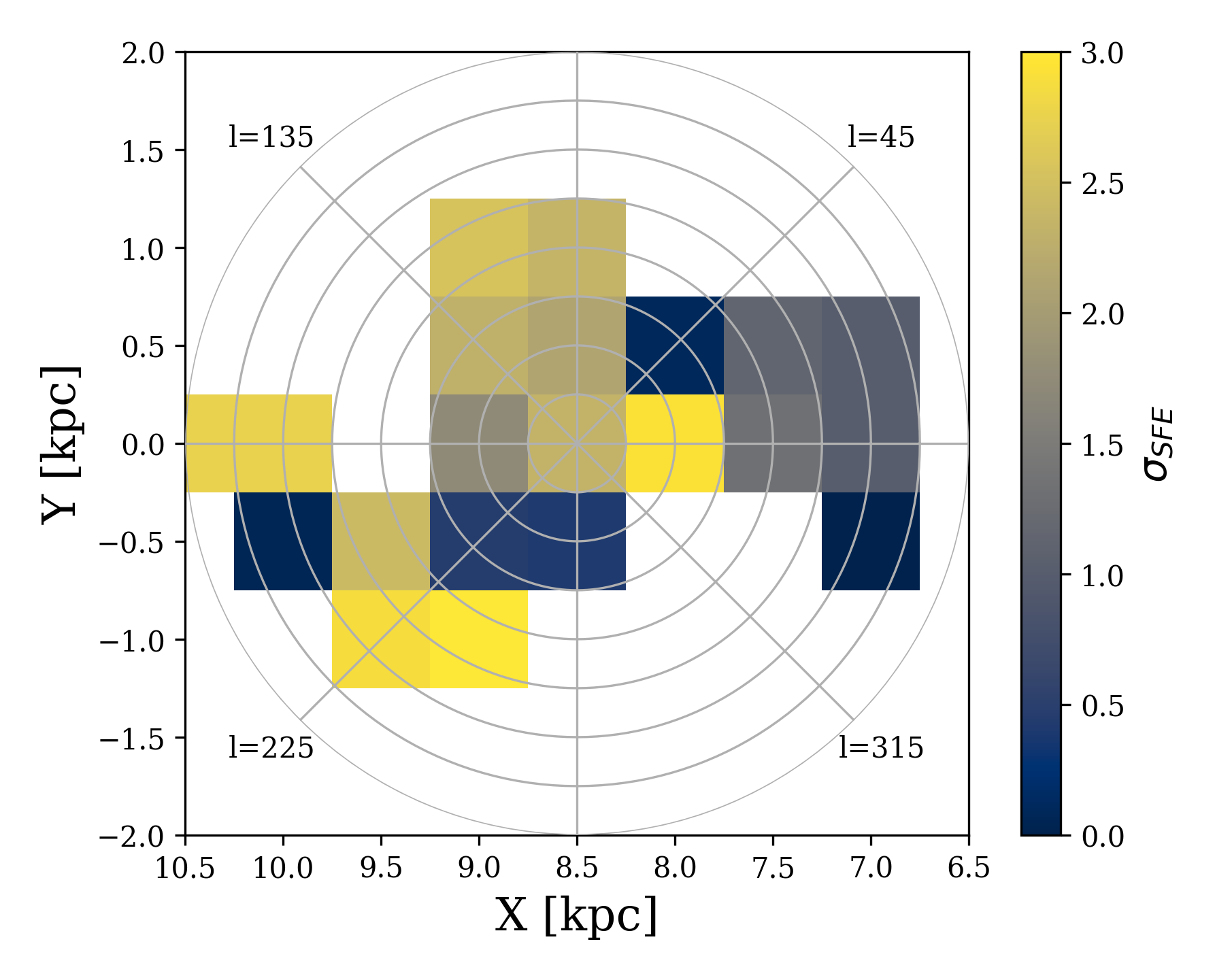}
        \includegraphics[width=0.24\textwidth]{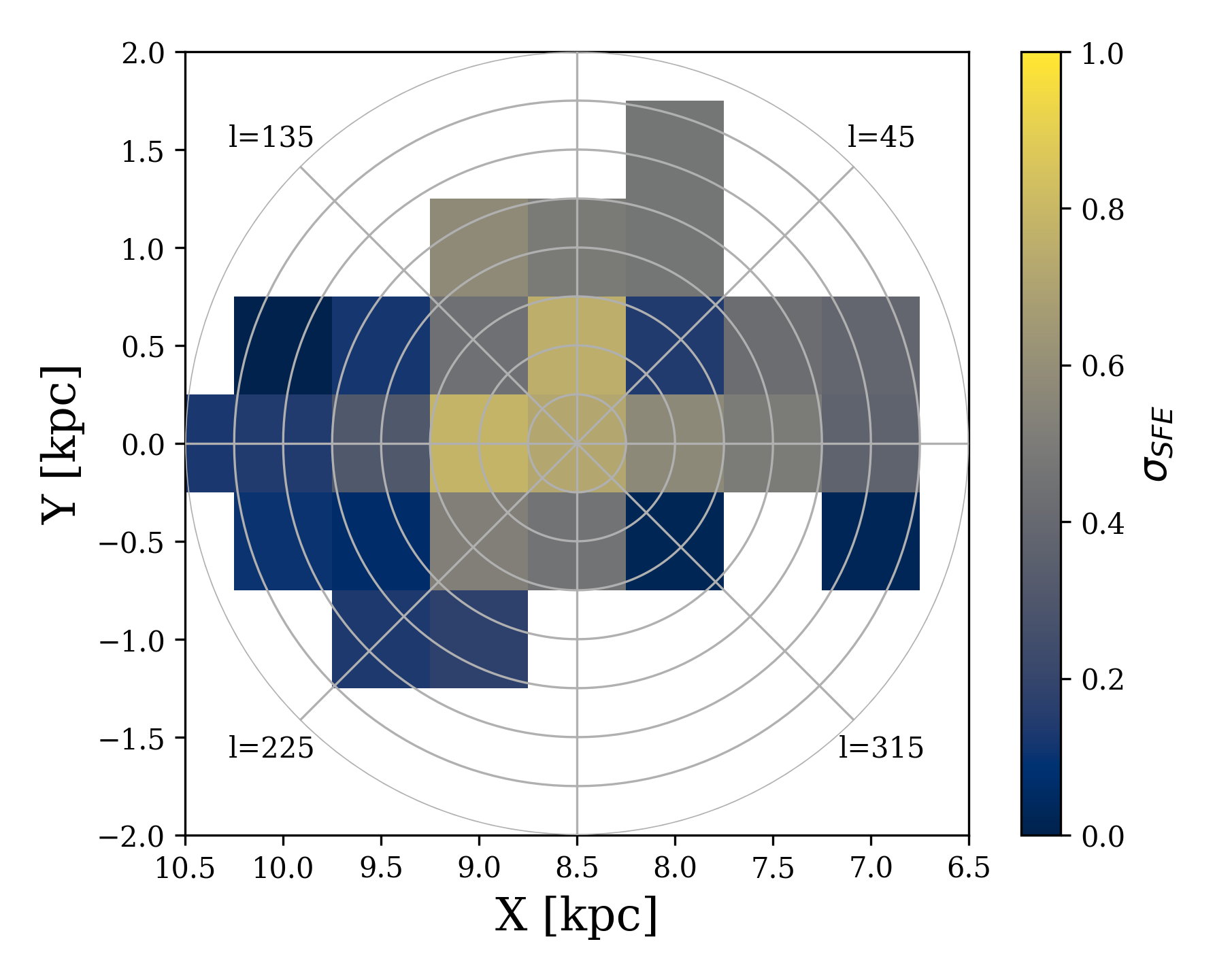}
        
        \includegraphics[width=0.24\textwidth]{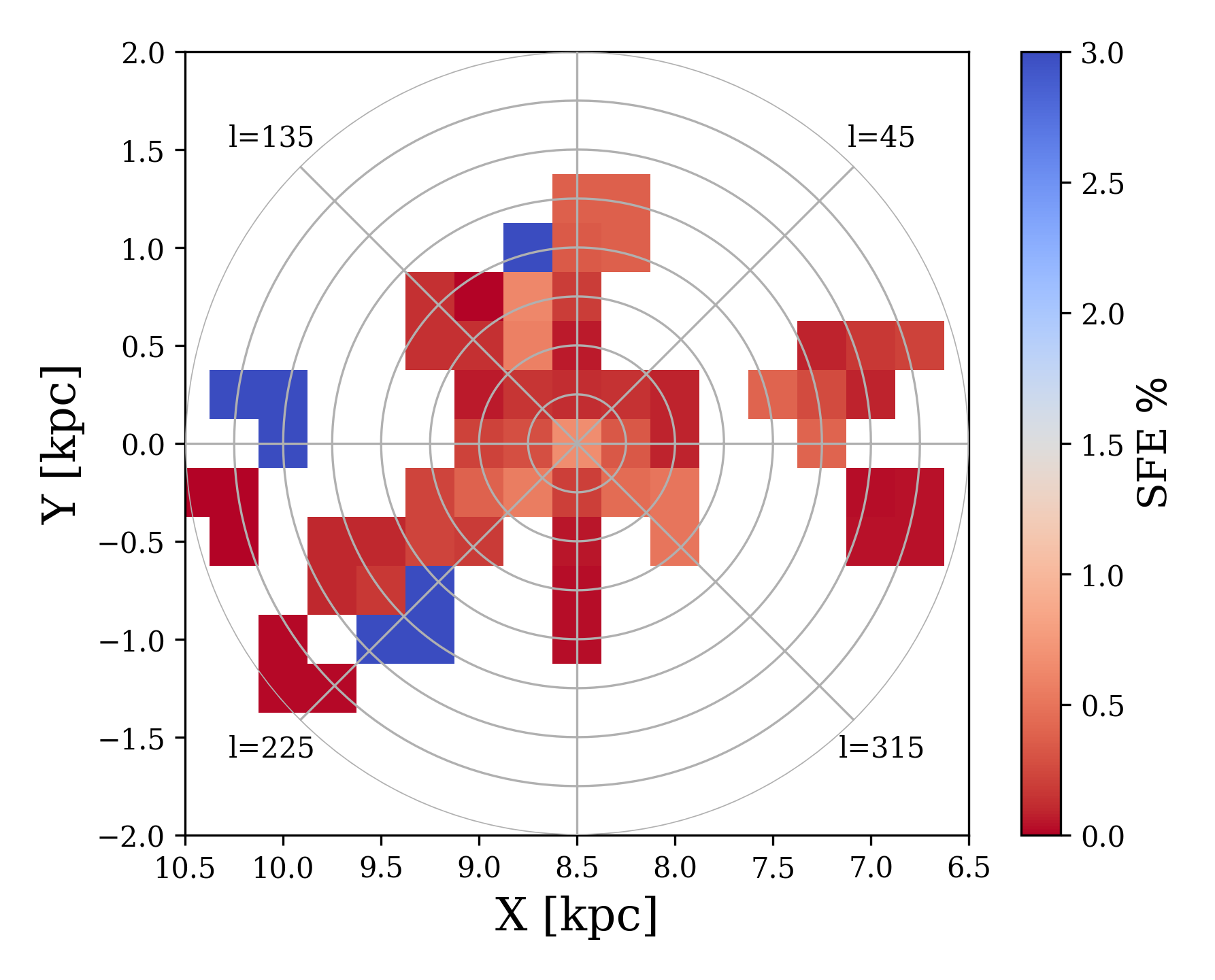}      
        \includegraphics[width=0.24\textwidth]{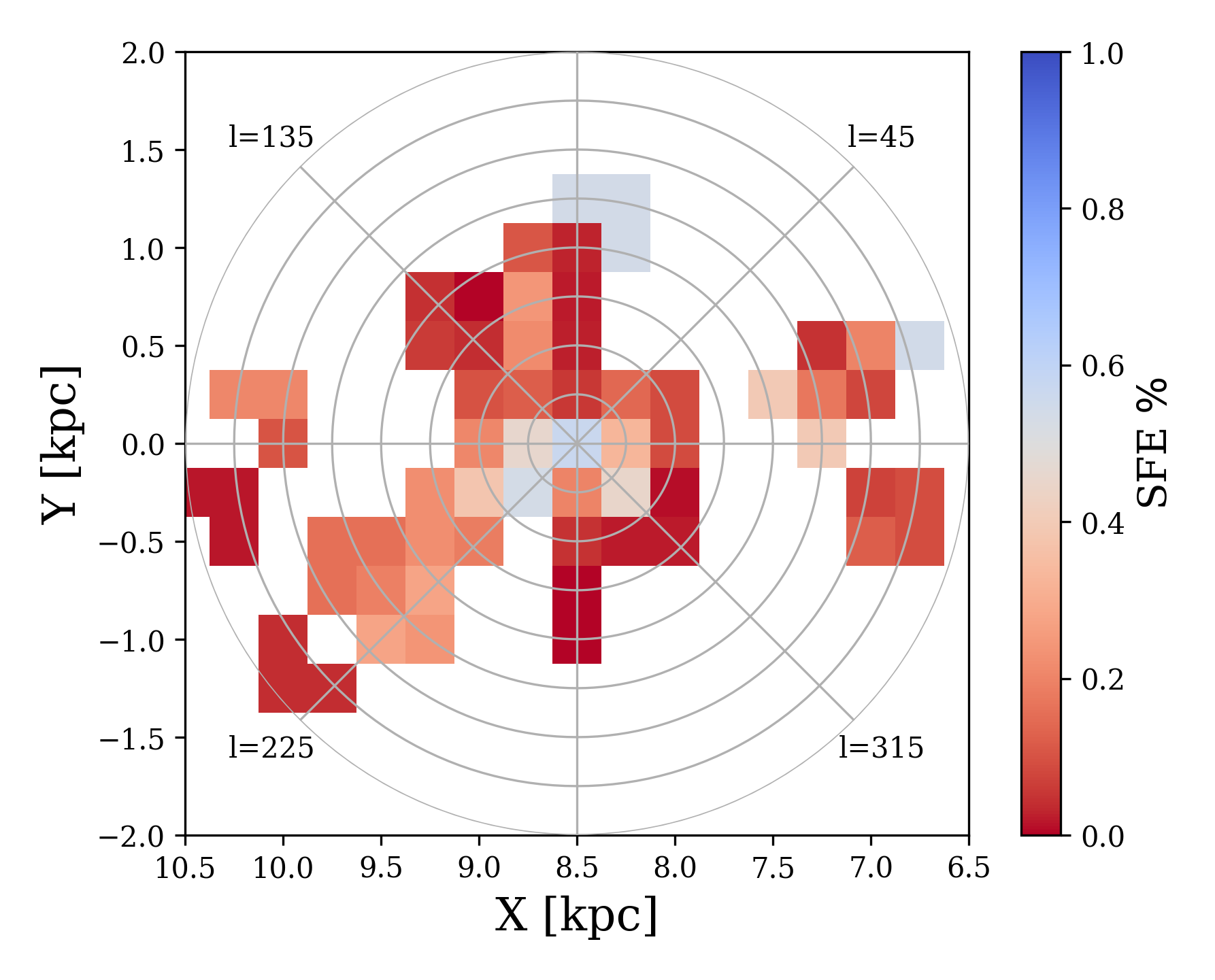}
        \includegraphics[width=0.24\textwidth]{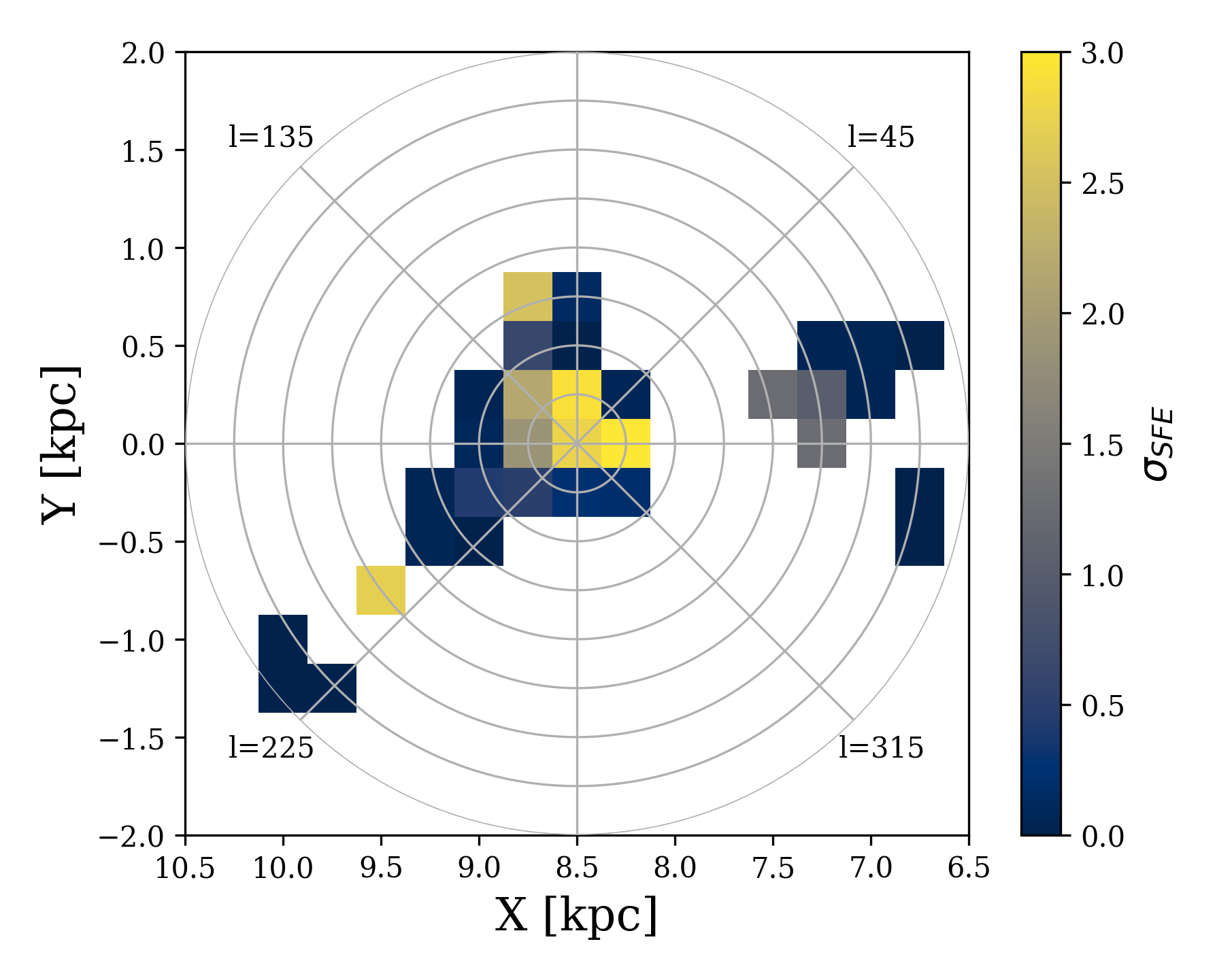}
        \includegraphics[width=0.24\textwidth]{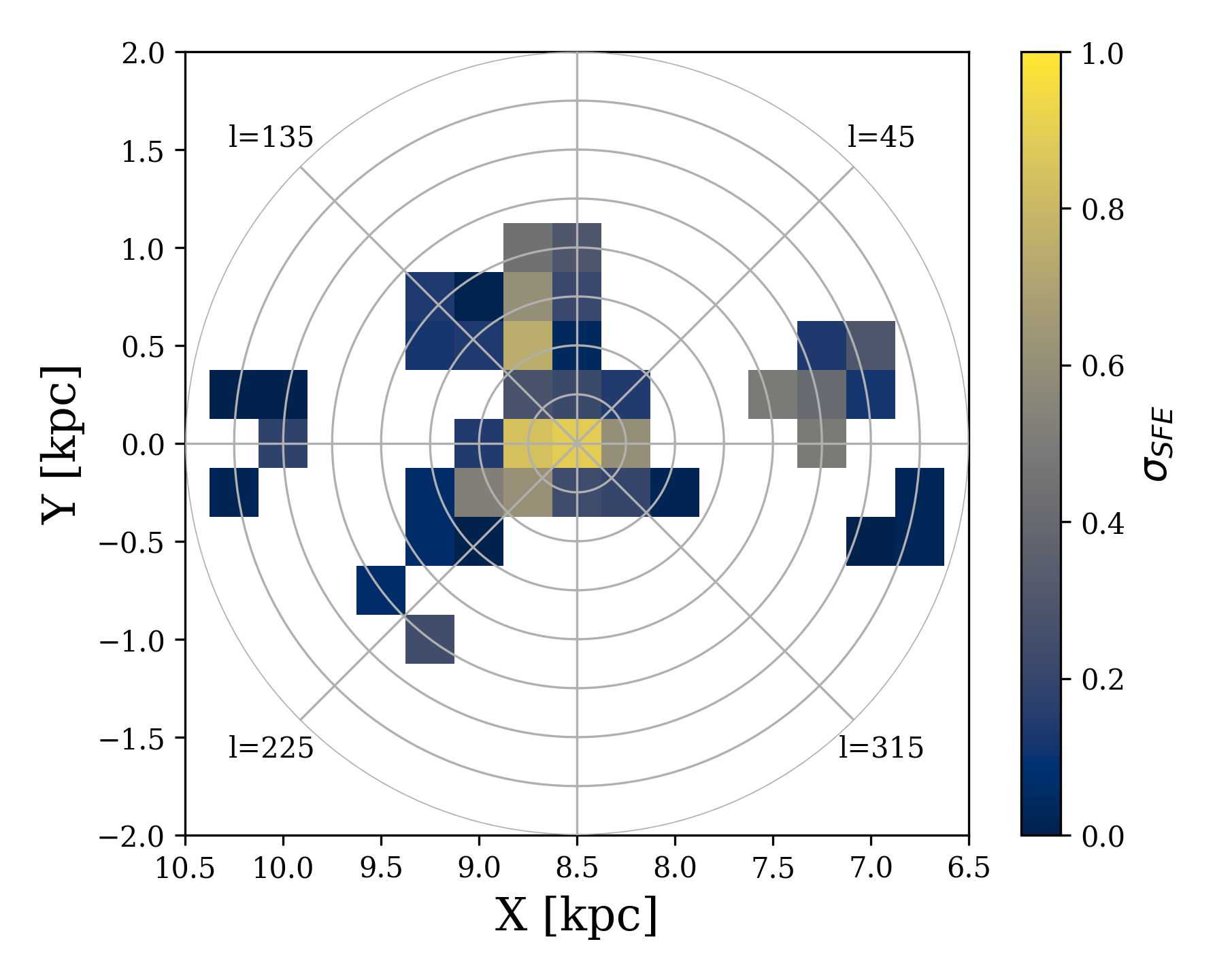}
    \caption{Maps of SFE (two left panels) and SFE variance (two right panels). Left: original. Right: Adjusted. Apertures with $R=1$ kpc (top), $R=0.5$ kpc (middle), and $R=0.25$ kpc (bottom).} 
    \label{fig:maps_ysoadjust}
\end{figure}

\begin{figure}[h!]
    \centering
    \includegraphics[width=0.4\textwidth]{newfigs/Sept/fdg_SFE_all_apmoves_fit.png}
    \includegraphics[width=0.4\textwidth]{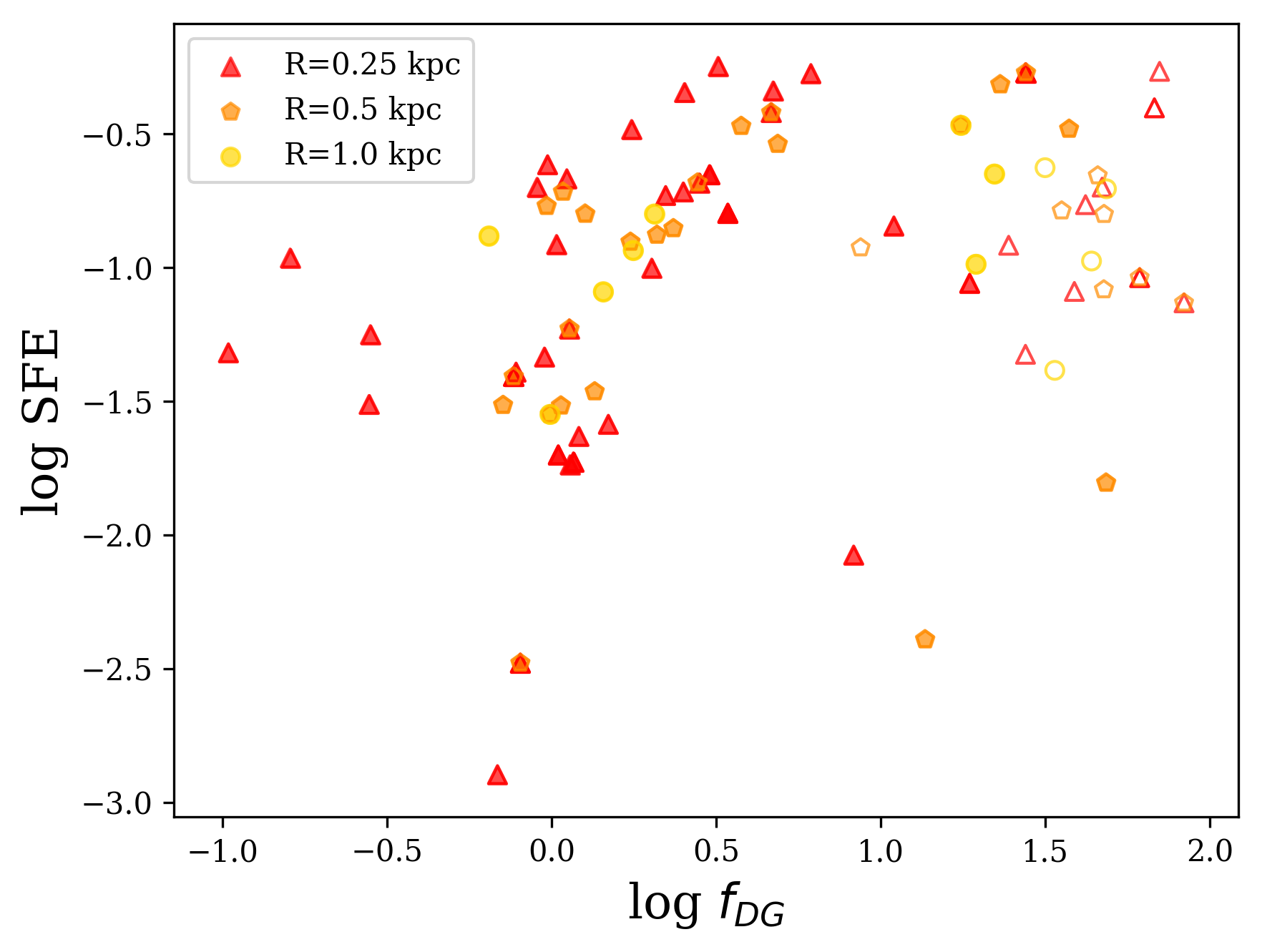}
    \includegraphics[width=0.4\textwidth]{newfigs/Sept/DelAv_SFE_all_apmoves_fit.png}
    \includegraphics[width=0.4\textwidth]{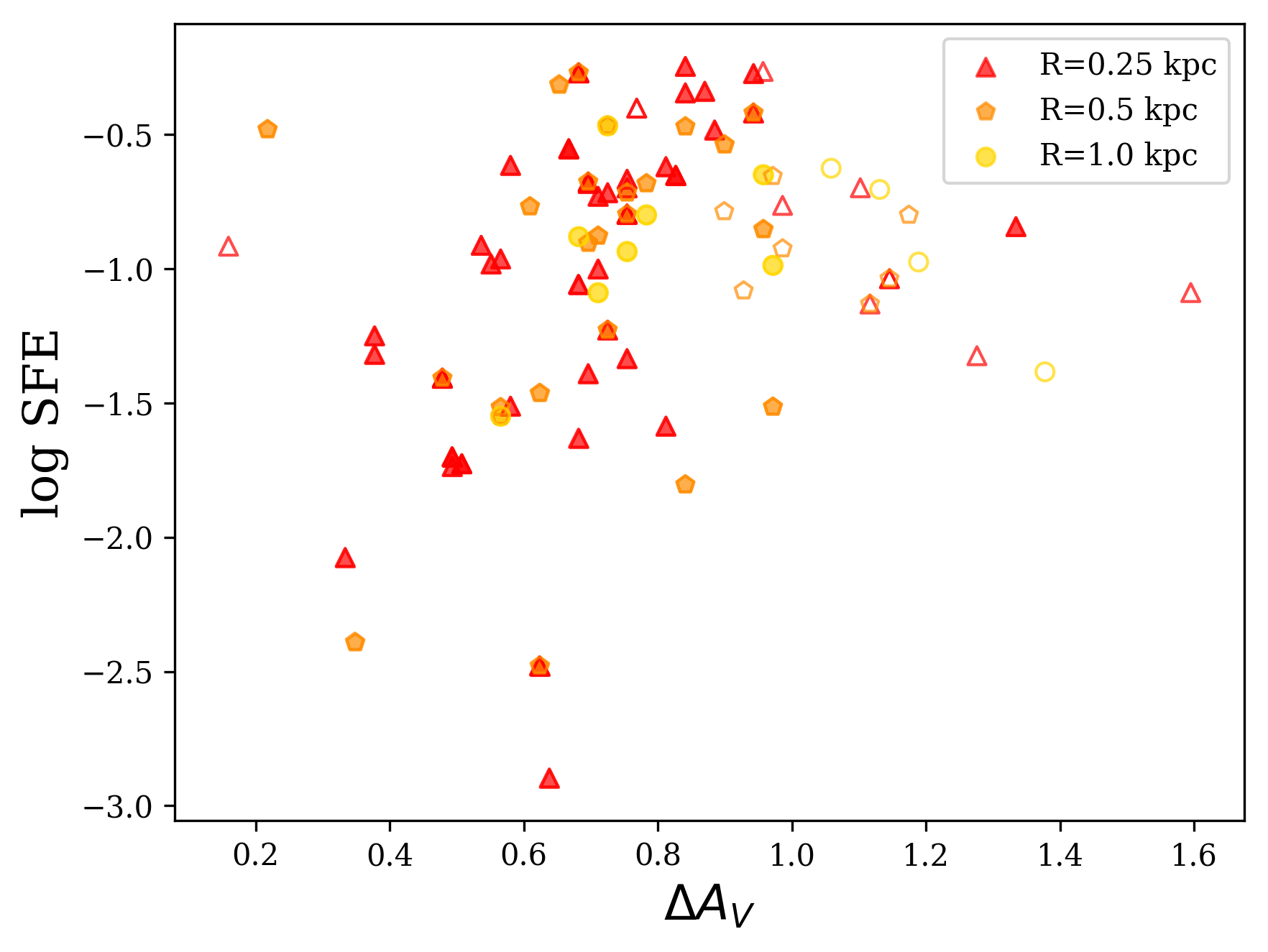}
    \caption{Relation between aperture averaged SFE and dense gas measures. Left: Original. Right: Adjusted. Empty symbols show apertures with $R_{\mathrm{Gal}}<8$ kpc. Correlations are significant for apertures of $R=0.25$ kpc for $f_{\mathrm{DG}}$ vs. SFE (p-value: 0.0004, R-value: 0.5), and for $\Delta A_\mathrm{V}$ vs. SFE (p-value: 0.01, R-value: 0.3). Thus we obtain correlations for the same apertures with and without the YSO adjustment, but with lower p-values and higher R-values for the adjusted case.
    }
    \label{fig:SFE_ysoadjust}
\end{figure}

Here we investigate the effect of adjusting for YSO completeness on our results. We increased the number of YSOs linearly for clouds between 1 -- 2 kpc, reaching a factor of three increase at 2 kpc. Figure \ref{fig:maps_ysoadjust} shows maps of SFE and SFE variance before and after adjustment. The adjusted maps show higher values for some pixels, while others seem unaffected by the adjustment. Overall, the adjustment does not significantly change our results. Table \ref{tab:ysoadjust} compares the star formation properties of the apertures with adjusted and unadjusted numbers of YSOs at far distances. It shows an increase of star formation rate in the full survey area from 19 to 27 $\cdot 10^3$ M$_\odot$ Myr$^{-1}$, and star formation efficiency from 0.2 to 0.3\%. The result that larger apertures are less efficient at forming stars (lower SFE and SFR/A) also holds true with the adjustment, however. Figure \ref{fig:SFE_ysoadjust} shows SFE versus dense gas measures for the apertures in our sample, again comparing adjusted and original versions. The correlation is significant for $f_{\mathrm{DG}}$ versus SFE for apertures of $R=0.5$ and $0.25$ kpc if the number of YSOs is adjusted, while no correlation is seen for the original apertures. 

\begin{table*}[h!]
\caption{Star formation properties of the aperture N-PDFs. Left: Original. Right: Adjusted.} % title of Table
\label{tab:ysoadjust}      % is used to refer this table in the text
%\caption{Global caption}
\begin{minipage}{.5\linewidth}
%\caption{}
\centering
%\centering 
\begin{tabular}{cccc}
%\hline
\hline\hline                 % inserts double horizontal lines
R  & SFR & $\frac{\mathrm{SFR}}{A}$\tablefootmark{*} & SFE \\   
\hline
kpc & $10^3$M$_\odot$ Myr$^{-1}$ & $10^3$M$_\odot$ Myr$^{-1}$kpc$^{-2}$ & \% \\
\hline 
 2.00\tablefootmark{**} & 10 & 0.8  & 0.13    \\
 1.00   &       6.8     $\pm$   1.9     &       2.1     $\pm$   0.6     &        0.10    $\pm$    0.02    \\ 
 0.50   &       2.2     $\pm$   0.6     &       2.8     $\pm$   0.8     &        0.08    $\pm$    0.02    \\ 
 0.25   &       2.1     $\pm$   0.1     &       2.1     $\pm$   0.7     &        0.04    $\pm$    0.01    \\ 
\hline  
\multicolumn{4}{c}{Apertures with R$_\mathrm{Gal}>$ 7.5 kpc} \\
\hline
 1.00   &       7.2     $\pm$    2.2 &  2.3     $\pm$    2.4    &        0.10    $\pm$    0.02   \\ 
 0.50   &       2.4     $\pm$    4.8 &  3.0     $\pm$    0.9    &        0.09    $\pm$    0.02   \\ 
 0.25   &       0.4     $\pm$    0.2 &  2.2     $\pm$    0.8    &        0.05    $\pm$    0.01   \\ 
\hline
\end{tabular}
\end{minipage}%
\begin{minipage}{.3\linewidth}
%\caption{}
\centering
%\quad
\begin{tabular}{ccc}
%\hline
\hline\hline                 % inserts double horizontal lines
 SFR & $\frac{\mathrm{SFR}}{A}$\tablefootmark{*} & SFE \\   
\hline
 $10^3$M$_\odot$ Myr$^{-1}$ & $10^3$M$_\odot$ Myr$^{-1}$kpc$^{-2}$  & \% \\
\hline      
 13 & 1  & 0.33 \\ 
 9.2    $\pm$   2.7     &        2.9    $\pm$   0.9     &        0.14   $\pm$    0.03    \\ 
 3.0    $\pm$   0.9     &        3.8    $\pm$   1.1     &        0.11   $\pm$    0.02  \\ 
 0.6    $\pm$   0.2     &        2.8    $\pm$   1.0     &        0.05   $\pm$    0.01     \\ 
\hline  
\multicolumn{3}{c}{$N_{\mathrm{YSOs}}$ adjusted, R$_\mathrm{Gal}>$ 7.5 kpc} \\
\hline  
9.5    $\pm$     3.2    &        3.0    $\pm$    3.4    &        0.13   $\pm$    0.03    \\ 
3.1    $\pm$     6.7    &        4.0    $\pm$    1.3    &        0.11   $\pm$    0.02    \\ 
0.6    $\pm$     0.2    &        3.0    $\pm$    1.1    &        0.05   $\pm$    0.01     \\ 
\hline  
%\hline
\end{tabular}
\end{minipage} 
\tablefoot{\\
\tablefoottext{*}{$A$ is the area of the aperture.}
\tablefoottext{**}{This is our full survey area. Therefore there is only one aperture, and no standard deviation is given.}
}
\end{table*}

\begin{figure}%[h!]
    \centering
    \includegraphics[width=0.49\textwidth]{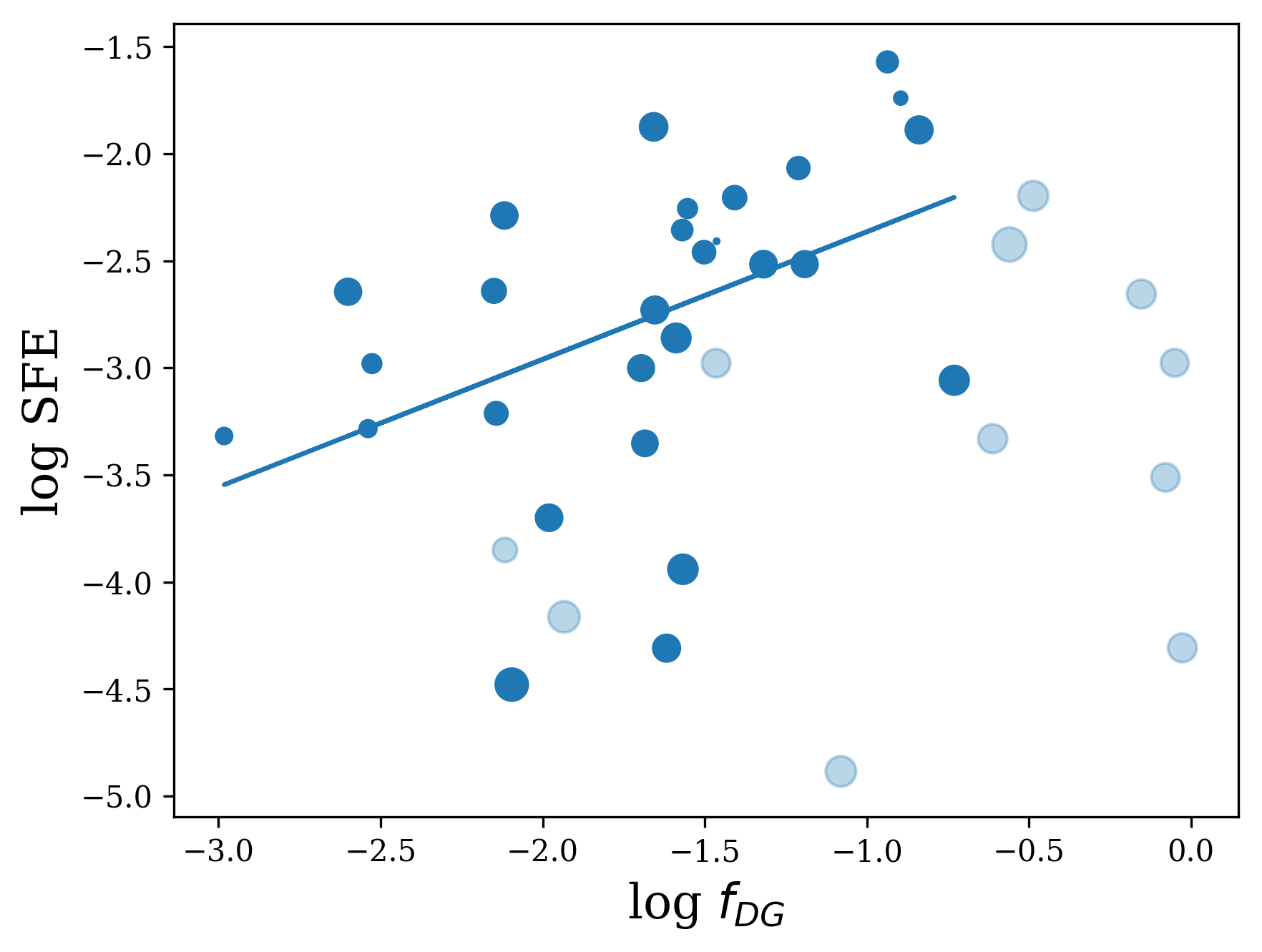}
    \includegraphics[width=0.49\textwidth]{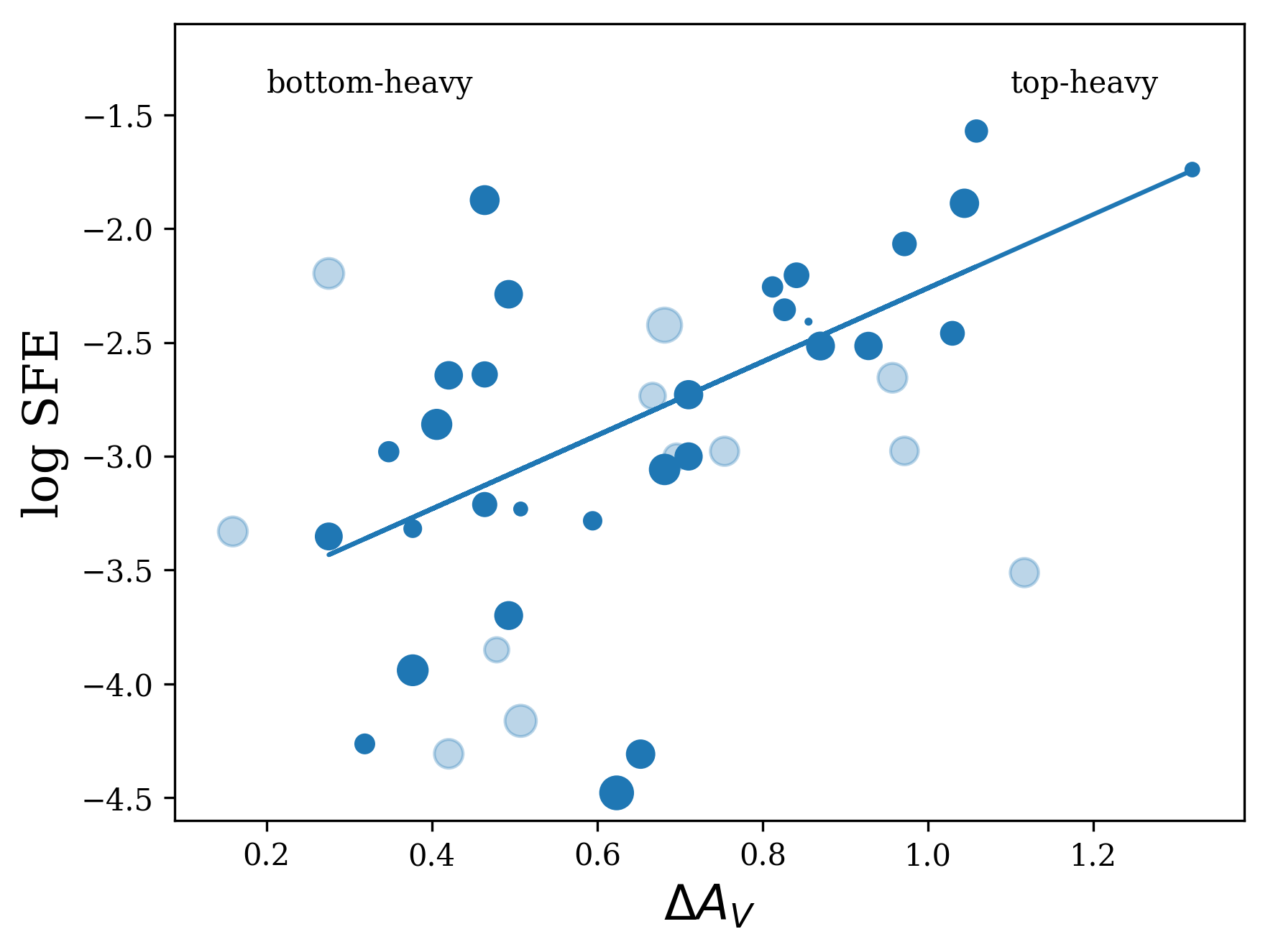}
    \caption{Relation between SFE and the dense gas measures of individual clouds. Light blue shows clouds with a distance $>1.2$ kpc. The symbol size corresponds to the cloud area $>3$mag. \emph{Left:} Dense gas mass fraction, $f_\mathrm{DG}$, versus SFE. The line shows a fit with $a = 0.6\pm0.2$, similar to the relation found previously for nearby molecular clouds \citep[e.g.][]{kainulainen2014unfolding}. 
    \emph{Right:} Relative density contrast, $\Delta A_\mathrm{V}$, versus SFE. The line shows the fit of SFE $\propto 10^{a\Delta A_\mathrm{V}}$ with $a = 1.6\pm0.4$. 
    }
    \label{fig:SF_vs_densegasfrac_12}
\end{figure}

\end{appendix}

\clearpage
\onecolumn

\setcounter{table}{0}

% THE LONG TABLES OF CLOUD PARAMETERS

%\onllongtab{
%\begin{landscape}
\begin{longtable}{llllllllll}%{@{}*{6}{>{\arraybackslash}p{0.10\linewidth}}@{}}
%table*}
\label{tab:mastertable}
\\
\caption{Clouds in our sample, with some of their properties and references.}\\
\hline\hline%\hline
Name    &       Dist    &       Mass\tablefootmark{a}   &       YSOs    & SFE &   $l$     &       $b$     &       $f_{\mathrm{DG}}$ & $\Delta$A$_{\mathrm{V}}$    &       Ext\tablefootmark{b}    \\
\hline
% & [pc] & [M$_\odot$] &  & $\log$ & [$^{\circ}$] & [$^{\circ}$] & $\log$ &  &  &  &  &  \\
 & [pc] & $10^3$[M$_\odot$] &  & $\log$ & [$^{\circ}$] & [$^{\circ}$] & $\log$ &   &  \\
\hline      

\hline
\endfirsthead
\caption{Continued.} \\
\hline
Name    &       Dist    &       Mass\tablefootmark{a}   &       YSOs    & SFE &   $l$     &       $b$     &       $f_{\mathrm{DG}}$ & $\Delta$A$_{\mathrm{V}}$    &       Ext\tablefootmark{b}    \\
\hline
 & [pc] & $10^3$[M$_\odot$] &  & $\log$ & [$^{\circ}$] & [$^{\circ}$] & $\log$ &   &  \\
\hline
\endhead
\hline
\endfoot
\hline
\endlastfoot
MBM16\footnote{\citet{larosa1999dynamical}}     &       80\footnote{\label{P13}\citet{pingel2013characterizing}}        & 0.01 &  0\footnoteref{P13}      &               &       [168,174]       &       [$-40$,$-34$]   &               &        0.26    &       N        \\ 
LDN134\footnote{\label{K9}\citet{kainulainen2009probing}}       &       100\footnote{\citet{mattila1979oh}}     &       0.11         &       5\footnoteref{K9}       &       $-$2.41 &        [3,7]  &       [34,38] & $-$1.46 &        0.86   &       N        \\ 
Taurus\footnoteref{K9}  &       130\footnote{\label{z}\citet{zucker2020compendium}}     &       3.28    &       440\footnote{\citet{rebull2011new}}     &       $-$1.57 &       [165,177]       &       [$-$21,$-$11]   &       $-$0.94 &        1.06    &       N        \\ 
Aquila-South\footnote{\label{Dame}\citet{dame2001milky}}        &       135\footnoteref{z}      &&&             &       [28.75,40.50]   &       [$-$20.25,$-$12.25]     &&       0.42    &       N        \\ 
Ophiuchus\footnoteref{Dame}     &       139\footnoteref{z}      &       5.05    &       351\footnote{\label{P20}\citet{pokhrel2020star}}        &       $-$2.07 &        [$-$12,11]      &       [8,26]  &       $-$1.21 &        0.97   &       N        \\ 
RCrA\footnoteref{Dame}  &       155\footnoteref{z}      &       0.43    &       53\footnote{\label{D15}\citet{dunham2015young}} &       $-$1.74 &        [$-$2,4]        &       [$-$24.25,$-$16.25]     &       $-$0.90 &        1.32    &       N        \\ 
Chamaeleon\footnoteref{Dame}    &       161\footnoteref{z}      &       1.56    &       111\footnoteref{D15}    &       $-$2.26 &       [290,310]       &       [$-$22,$-$6]    &       $-$1.55 &        0.81    &       N        \\ 
Pipe\footnoteref{K9}    &       180\footnoteref{z}      &       16.0    &       21\footnote{\citet{lada2010star}}       &       $-$3.35 &        [$-$5,7]        &        [2,8]  &       $-$1.68 &        0.28   &       N        \\ 
Coalsack\footnoteref{Dame}      &       187\footnoteref{z}      &       1.19    &       5\footnote{\citet{kato1999nanten}}      &       $-$2.98 &       [299,304]       &       [$-$3.5,1.0]    &       $-$2.53 &        0.35 &  N        \\ 
Lupus\footnoteref{K9}   &       197\footnoteref{z}      &       2.31    &       140\footnoteref{D15}    &       $-$2.36 &       [$-$28,$-$15]   &       [4,20]  &       $-$1.57 &        0.83    &       N        \\ 
Hercules\footnoteref{Dame}      &       223\footnoteref{z}      &       0.08    &&              &       [41,49] &        [6,11]  &       $-$2.74 &        0.49   &       N        \\ 
Camelopardalis\footnoteref{Dame}        &       235\footnoteref{z}      &       0.04    &       150\footnote{\citet{straivzys2010young}}        &       $-$2.30 &       [140.64,155.07] &       [14.62,25.88]   &&       0.33    &       N        \\ 
Aquila\footnoteref{Dame}        &       254\footnoteref{z}      &       18.0    &       403\footnoteref{P20}    &       $-$2.64 &       [12,35] &       [6,14]  &       $-$2.60 &        0.42    &       N        \\ 
Pegasus-West\footnoteref{Dame}  &       258\footnoteref{z}      &       0.02    &&      &       [89,96] &       [$-$32.62,$-$29.12]     &&       0.38    &       N        \\ 
LBN906--17\footnoteref{z}       &       260\footnoteref{z}      & 0.02 &&       &       [$-$159,$-$155] &       [$-$32.6,$-$29.2]       &&       0.42    &       N        \\ 
MBM12\footnote{\citet{hobbs1988nearest}}        &       275\footnote{\label{L01}\citet{luhman2001mbm}}  & 0.72 &  8\footnoteref{L01}      &       $-$3.28 &       [157,163]       &       [$-$37,$-$32]   & $-$2.54 &        0.59   &       N        \\ 
Perseus\footnoteref{K9} &       276\footnoteref{z}      &       7.31    &       452\footnoteref{P20}    &       $-$2.20 &       [152,162]       &       [$-$29,$-$12]   &       $-$1.41 &        0.84    &       N        \\ 
L1333\footnoteref{z}    &       283\footnoteref{z}      &   0.23 &      5\footnote{\citet{obayashi1998star}}    &       $-$3.23 &       [125.5,132.5]   &       [13.0,16.5]     &&       0.51    &       N        \\ 
Pegasus-East\footnoteref{Dame}  &       292\footnoteref{z}      &       0.12    &&              &       [100,107]       &       [$-$33.25,$-$25.25]     &&       0.33    &       N        \\ 
UrsaMajor\footnoteref{Dame}     &       330\footnoteref{z}      &       0.90    &               &       $-$4.26 &       [138.94,150.91] &       [32.88,42.25]   &       &  0.32   &       N        \\ 
Polaris\footnoteref{Dame}       &       343\footnoteref{z}      &       0.19    &&              &       [117,128]       &       [20,38] &&       0.28    &       N        \\ 
L1265\footnoteref{z}    &       344\footnoteref{z}      &   0.05 &&             &       [116.5,119.5]   &       [$-$5.5 ,$-$2.5]        &&       0.55   &       N        \\ 
CephFlare\footnoteref{K9}       &       346\footnoteref{z}      &       5.26    &       118\footnote{\citet{heiderman2010star}} &       $-$3.21 &       [97.0,116.5]    &       [7,25]  &       $-$2.14 &        0.46    &       N        \\ 
LBN968--CB28\footnoteref{z}     &       347\footnoteref{z}      &   0.20    &&           &       [$-$157.5,$-$150.5]     &       [$-$29.0,$-$22.3]       &&       0.59    &       N        \\ 
GumNeb\footnoteref{Dame}        &       348\footnote{\label{Y20}\citet{yep2020young}}   &       0.59    &       21\footnoteref{Y20}     &       $-$3.32 &       [260,273]       &       [$-$14,$-$4]    &       $-$2.98 &        0.38    &       N        \\ 
Spider\footnoteref{z}   &       369\footnoteref{z}      &    0.01   &&          &       [132.5,137.5]   &       [37.5,42.5]     &&       0.41    &       N        \\ 
LamOri\footnoteref{Dame}        &       399\footnoteref{z}      &       7.60    &       500\footnote{\citet{koenig2015spectroscopic}}   &       $-$2.64 &       [190,200]       &       [$-$18,$-$7]    &       $-$2.15 &        0.46 &  N        \\
Gemini\footnoteref{Dame}        &       400\footnote{\citet{li2015co}}  &                &&              &       [198,202]       &       [9.12,14.38]    &&       0.28    &       N        \\ 
LBN991\footnoteref{z}   &       408\footnoteref{z}      &&&             &       [$-$147,$-$145] &       [$-$30,$-$28]   &               &        0.32    &       N        \\ 
OrionB\footnoteref{K9}  &       433\footnoteref{z}      &       22.4    &       544\footnoteref{P20}    &       $-$2.52 &       [200,210]       &       [$-$17 ,$-$4]  &       $-$1.19 &        0.93   &       N        \\ 
OrionA\footnoteref{K9}  &       438\footnoteref{z}      &       40.0    &       2394\footnoteref{P20}   &       $-$1.89 &       [204,219]       &       [$-$22,$-$17]   &       $-$0.84 &        1.04    &       N        \\ 
California\footnoteref{K9}      &       466\footnoteref{z}      &       18.1    &       170\footnote{\citet{lada2017hp2}}       &       $-$3.00 &       [152,168]       &       [$-$11,$-$2]    &       $-$1.70 &        0.71    &       N        \\ 
Draco\footnoteref{z}    &       481\footnoteref{z}      &&&             &       [87.5,92.5]     &       [35.5,40.5]     &               &        0.38    &       N        \\ 
Serpens\footnoteref{K9} &       490\footnoteref{z}      &       97.7    &       222\footnoteref{D15}    &       $-$3.06 &       [25,33] &        [1,6]   &       $-$0.73 &        0.68   &       N        \\ 
Lacerta\footnoteref{Dame}       &       504\footnoteref{z}      &       0.19    &&              &       [94.25,107.50]  &       [$-$20.12,$-$9.12]      &&       0.29    &       N        \\ 
CygOB7\footnote{\citet{dobashi2019interaction}} &       561\footnoteref{z}      &       95.8    &       30\footnote{\citet{rice2012near}}       &       $-$3.94 &       [90.5,95.5]     &        [1.5,6.5]       &   $-$1.57   &  0.38   &       N        \\ 
Circinus\footnoteref{z} &       675\footnoteref{z}      &       22.9    &       47\footnote{\citet{mikami1994h}}        &       $-$3.70 &       [$-$48,$-$39]   &       [$-$8,$-$2]     &   $-$1.98    &   0.49   &       N        \\ 
CepheusOB3b\footnote{\label{Y97}\citet{yonekura1997molecular}}  &       700\footnote{\citet{allen2012spitzer}}  &       36.4    &       2188\footnoteref{P20}   &       $-$1.88 &       [106,112]       &        [0,4]   &       $-$1.66 &        0.46   &       N        \\ 
Norma\footnoteref{z}    &       721\footnoteref{z}      &       152     &&              &       [$-$22,$-$19]   &        [$-$0.3,3.5]    &   $-$0.87   &  0.33   &       N        \\ 
L1307--35\footnoteref{z}        &       741\footnoteref{z}      &   12.9    &&           &       [123.2,131.4]   &        [0.8,6.5]      &&       0.54    &       N        \\ 
MonocerosR2\footnote{\citet{carpenter2008monoceros}}    &       767\footnoteref{z}      &       31.3    &       931\footnoteref{P20}    &       $-$2.73 &       [211,224]       &       [$-$16,$-$6]    &       $-$1.65 &        0.71    &       N        \\
MonOB1\footnoteref{Dame}        &       771\footnoteref{z}      &       27.6    &       680\footnote{\citet{rapson2014spitzer}} &       $-$2.52 &       [196,207]       &        [$-$0.5,3.5]    &       $-$1.32 &        0.87   &       N        \\ 
IC5146\footnote{\citet{lada1994dust}}   &       792\footnoteref{z}      &       4.98    &       130\footnote{\citet{johnstone2017jcmt}} &       $-$2.46 &       [92,96] &       [$-$7,$-$3]     &       $-$1.50 &        1.03    &       N        \\ 
CephOB4\footnoteref{Y97}        &       850\footnoteref{Y97}    &       29.8    &       12\footnote{\citet{yang1990newly}}      &       $-$4.31 &       [116.5,122.5]   &        [1,7]   &       $-$1.62 &        0.65   &       N        \\ 
Vela\footnoteref{z}     &       866\footnoteref{z}      &       431     &       55\footnote{\citet{liseau1992star}}     &       $-$4.48 &       [$-$98,$-$85]   &        [$-$4,5]        &  $-$2.10     &         0.62   &       N        \\ 
AFGL490\footnote{\label{M12}\citet{masiunas2012structural}}     &       900\footnoteref{M12}    &       65.0    &       319\footnoteref{P20}    &       $-$2.86 &       [140,144]       &       [0,3]   &       $-$1.59 &        0.41    &       N        \\ 
L1340--55\footnoteref{z}        &       903\footnoteref{z}      &   20.3   &&            &       [128.4,135.4]   &       [7.0,12.6]      &  -3.35     &    0.62   &       N        \\ 
S140\footnote{\citet{bally2002fountains}}       &       910\footnoteref{Y97}    &   20.6 &        531\footnoteref{P20}    &       $-$2.29 &       [106,108]       &        [4.0,6.5]       &   $-$2.12   &  0.49   &       N        \\ 
\\

IC1396\footnoteref{z}   &       941\footnoteref{z}      &   13.8 &&             &       [97.5,102.5]    &        [0.5,5.5]       &&       0.36   &       N        \\ 
L1293--1306\footnoteref{z}      &       977\footnoteref{z}      &   6.42 &&              &       [119,128]       &       [$-$4.0,0.5]    &   $-$2.53   &      0.65   &       N        \\ 
Split\footnote{\label{Green}\citet{green20193d}}        &       1000\footnoteref{Green} &       1610    &&              &       [36,43] &        [$-$4,4]        &   $-$0.28   &  0.20   &       N        \\ 
Ara\footnote{\citet{arnal2003co}}       &       1064\footnoteref{z}     &       274     &&              &       [$-$25,$-$22]   &       [$-$3.5,$-$0.3] &   -2.00   &      0.32   &       N        \\ 
Cygnus\footnoteref{Dame}        &       1214\footnoteref{z}     &       2280    &       21387\footnoteref{P20}  &       $-$2.27 &       [73,87] &        [$-$4,5]        &       $-$0.56 &        0.68   &       A        \\ 
S147\footnote{\label{C17}\citet{chen2017mapping}}       &       1220\footnoteref{C17}   &   0.08 &&               &       [178,182]       &       [$-$3.5,$-$1.0] &               &        0.30    &       N        \\ 
Lagoon\footnoteref{z}   &       1220\footnoteref{z}     &       246     &       3200\footnote{\citet{strasbuger2020identification}}     &       $-$2.20 &        [6,8]   &       [$-$3.5,$-$1]   &   $-$0.49   &  0.28   &       N        \\ 
M20\footnoteref{z}      &       1234\footnoteref{z}     &       334     &       33\footnote{\citet{tapia2018star}}      &       $-$4.14 &        [6,8]   &        [$-$1,1]       &       $-$0.03 &        0.42   &       pp       \\ 
Rosette\footnote{\label{C13}\citet{cambresy2013young}}  &       1261\footnoteref{z}     &       80.3    &       530\footnoteref{C13}    &       $-$2.80 &       [204,210]       &       [$-$4.0,$-$0.5] &       $-$1.47 &        0.75    &       N        \\
CMaOB1\footnoteref{z}   &       1262\footnoteref{z}     &   18.9 &      340\footnote{\citet{sewilo2019identifying}}     &       $-$2.74 &       [$-$138,$-$133] &       [$-$3.5,1.8]    &&       0.67    &       N        \\ 
NGC2362\footnoteref{z}  &       1317\footnoteref{z}     &   2.28 &&             &       [$-$122.5,$-$117.5]     &       [$-$7.5,$-$2.5] &&       0.83    &       N        \\ 
L291\footnoteref{z}     &       1348\footnoteref{z}     &       238     &       10\footnote{\citet{molina2016stellar}}  &       $-$4.88 &       [11.0,14.1]     &       [$-$5.5,$-$1.5] &   $-$1.08   &    0.62   &       N        \\ 
GGD4\footnoteref{z}     &       1349\footnoteref{z}     &   3.14 &&             &       [181.5,185.0]   &       [$-$6.5,$-$1.5] &&       0.41    &       N        \\ 
NGC6604\footnoteref{z}  &       1352\footnoteref{z}     &       441     &&              &       [16.5,20.0]     &        [1.8,4.0]       &   3.00    &    0.30   &       N        \\ 
M17\footnoteref{z}      &       1509\footnoteref{z}     &       227     &       488\footnote{\citet{povich2010evidence}}        &       $-$2.67 &       [13.5,16.0]     &        [$-$1,0]        &       $-$0.05 &        0.97   &       pp       \\ 
S235\footnote{\label{B15}\citet{burns2015water}}        &       1560\footnoteref{B15}   &   19.1 &        230\footnote{\citet{dewangan2011infrared}}      &       $-$3.00 &       [171,175]       &        [1,4]   &&       0.70   &       N        \\
IC443\footnoteref{z}    &       1593\footnoteref{z}     &   0.98 &&             &       [186.5,191.5]   &        [2.5,7.5]       &&       0.75   &       N        \\ 
W3-W4-W5\footnoteref{z} &       1647\footnoteref{z}     &   62.6 &&             &       [132.5,137.5]   &        [$-$1.5,3.5]    &&       0.57   &       N        \\ 
NGC6334\footnote{\label{M07}\citet{munoz2007massive}}   &       1700\footnoteref{M07}   &       258     &       163\footnote{\citet{russeil2010earliest}}       &       $-$3.13 &       [$-$11,$-$8]    &        [$-$1,2]        &       $-$0.08 &        1.12   &       pp       \\ 
M16\footnote{\citet{nishimura2017new}}  &       1731\footnoteref{z}     &       234     &       1101\footnote{\citet{linsky2007chandra}}        &       $-$2.27 &       [15.0,18.5]     &        [0,2]   &       $-$0.15 &        0.96   &       pp       \\
Cartwheel\footnote{\label{L11}\citet{lopez2011massive}} &       1800\footnoteref{L11}   &       142     &       150\footnote{\citet{figueira2019multiwavelength}}       &       $-$2.92 &       [344.5,146.0]   &        [0.7,2.2]       &       $-$0.61 &        0.16   &       A        \\ 
Cyg-West\footnoteref{Green}     &       1800\footnoteref{Green} &   788 &&      &       [68,73] &        [$-$4,4]        &&       0.45   &       N        \\ 
GemOB1\footnoteref{Dame}        &       1865\footnoteref{z}     &       309     &       230\footnote{\citet{koenig2011wide}}    &       $-$3.73 &       [187,196]       &       [$-$4.0,2.1]    &       $-$1.93 &        0.51    &       N        \\ 
Maddalena\footnote{\label{M09}\citet{megeath2009detection}}     &       1888\footnoteref{z}     &       13.0    &       70\footnoteref{M09}     &       $-$3.41 &       [$-$146,$-$140] &       [$-$5,1]        &       $-$2.12 &        0.48    &       N        \\ 
\hline
%\end{tabular}
%\end{table*}
\end{longtable}
\tablefoot{\\
\tablefoottext{a}{%The masses are computed from the extinction maps, including only the pixels with an extinction higher than 3 magnitudes.
The masses are computed from the column density maps, from pixels reaching an extinction of $A_\mathrm{V} > 3$ mag.% with the conversion factor (Eq. \ref{eq:conversion}) and the physical size of the pixel (M$_{\textrm{cloud}} = \Sigma_i A_\mathrm{V}(p_i[A_\mathrm{V}>3$mag$]) \cdot \tan(\Delta l)\cdot \tan(\Delta b) \cdot $ distance$^2 \cdot $ conversion). The limit of 3 mag corresponds to where we expect that the gas is molecular \citep{heyer2015molecular}.
}\\
\tablefoottext{b}{The column density map used is noted in the last column, N for the NICEST extinction maps, A for the extinction maps derived in Appendix \ref{app:cygnus_and_cartwheel}, and pp for the PPMAP emission maps from \citet{marsh2015temperature}.
}
%\end{landscape}
}

%\onllongtab{
%\begin{landscape}
\begin{longtable}{llclc}%{@{}*{6}{>{\arraybackslash}p{0.10\linewidth}}@{}}
%table*}
\label{tab:mastertable_shapes}
\\
\caption{Clouds in our sample, with their best-fit parameters.}\\
\hline\hline%\hline
Name    & Shape\tablefootmark{a}        & $A_V$ limit\tablefootmark{b} &        Fit parameters      &       $\chi^2$    \\
\hline
% &  &  &    \\
%\hline      

\hline
\endfirsthead
\caption{Continued.} \\
\hline
Name    & Shape\tablefootmark{a}        & $A_V$ limit\tablefootmark{b}   &       Fit parameters  &       $\chi^2$    \\
\hline
% &  &  &   \\
%\hline      
\endhead
\hline
\endfoot
\hline
\endlastfoot
%Name   ands    shape   ands    Avlim   ands    sparams ands    X2      ends
MBM16   &       LN      &       0.6     &       $\sigma$= 0.38 $\pm$  0.01      &       34.53    \\ 
LDN134  &       LN+PL   &       0.5     &       $\sigma$= 0.38 $\pm$  0.01, $\alpha$=-3.06 $\pm$  0.09, x$_0=$ 1.21 &       4.89     \\ 
Taurus  &       LN+PL   &       1.5     &       $\sigma$= 0.75 $\pm$  0.03, $\alpha$=-1.87 $\pm$  0.07, x$_0=$ 3.89 &       1.57     \\ 
Aquila-South    &       --      &               &               &                \\ 
Ophiuchus       &       PL      &       1.5     &       $\alpha$=-2.33 $\pm$  0.02   &       4.10     \\ 
RCrA    &       LN+PL   &       1.0     &       $\sigma$= 0.42 $\pm$  0.03, $\alpha$=-2.10 $\pm$  0.09, x$_0=$ 1.65 &       2.26     \\ 
Chamaeleon      &       LN+PL   &       1.5     &       $\sigma$= 0.36 $\pm$  0.02, $\alpha$=-2.20 $\pm$  0.03, x$_0=$ 2.02  &       1.55     \\ 
Pipe    &       LN+PL   &       2.5     &       $\sigma$= 0.22 $\pm$  0.00, $\alpha$=-8.24 $\pm$  0.57, x$_0=$ 6.12 &       13.33    \\ 
Coalsack        &       ?       &       2.5     &               &                \\ 
Lupus   &       LN+PL   &       1.0     &       $\sigma$= 0.81 $\pm$  0.05, $\alpha$=-4.05 $\pm$  0.16, x$_0=$ 2.33 &       7.51     \\ 
Hercules        &       LN+PL   &       0.8     &       $\sigma$= 0.58 $\pm$  0.05, $\alpha$=-4.79 $\pm$  0.30, x$_0=$ 1.47  &       2.45     \\ 
Camelopardalis  &       LN      &       1.8     &       $\sigma$= 0.22 $\pm$  0.01   &       5.59     \\ 
Aquila  &       LN      &       3.0     &       $\sigma$= 0.52 $\pm$  0.04      &       6.61     \\ 
Pegasus-West    &       LN+PL   &       1.4     &       $\sigma$= 0.47 $\pm$  0.09, $\alpha$=-8.65 $\pm$  2.90, x$_0=$ 2.28  &       0.96     \\ 
LBN906-17       &       LN+PL   &       0.5     &       $\sigma$= 0.44 $\pm$  0.01, $\alpha$=-8.06 $\pm$  3.38, x$_0=$ 1.46  &       6.18     \\ 
MBM12   &       LN+PL   &       0.5     &       $\sigma$= 0.53 $\pm$  0.01, $\alpha$=-3.07 $\pm$  0.07, x$_0=$ 1.30 &       1.51     \\ 
Perseus &       PL      &       1.5     &       $\alpha$=-2.44 $\pm$  0.03      &       5.28     \\ 
L1333   &       LN+PL   &       0.5     &       $\sigma$= 0.57 $\pm$  0.02, $\alpha$=-3.33 $\pm$  0.09, x$_0=$ 1.21 &       1.10     \\ 
Pegasus-East    &       LN      &       1.4     &       $\sigma$= 0.29 $\pm$  0.01   &       1.87     \\ 
UrsaMajor       &       LN      &       2.5     &       $\sigma$= 0.25 $\pm$  0.06   &       2.64     \\ 
Polaris &       LN      &       1.5     &       $\sigma$= 0.29 $\pm$  0.01      &       2.51     \\ 
L1265   &       LN      &       0.3     &       $\sigma$= 0.59 $\pm$  0.02      &       5.68     \\ 
CephFlare       &       LN+PL   &       1.5     &       $\sigma$= 0.35 $\pm$  0.01, $\alpha$=-3.90 $\pm$  0.28, x$_0=$ 2.75  &       5.16     \\ 
LBN968-CB28     &       LN+PL   &       0.5     &       $\sigma$= 0.48 $\pm$  0.02, $\alpha$=-4.90 $\pm$  0.25, x$_0=$ 0.98  &       11.40    \\ 
GumNeb  &       LN+PL   &       1.5     &       $\sigma$= 0.28 $\pm$  0.02, $\alpha$=-5.78 $\pm$  0.09, x$_0=$ 1.92 &       2.58     \\ 
Spider  &       LN+PL   &       0.2     &       $\sigma$= 0.59 $\pm$  0.02, $\alpha$=-4.68 $\pm$  0.28, x$_0=$ 1.18 &       1.77     \\ 
LamOri  &       LN+PL   &       1.5     &       $\sigma$= 0.34 $\pm$  0.01, $\alpha$=-4.86 $\pm$  0.08, x$_0=$ 2.26 &       5.17     \\ 
Gemini  &       --      &               &               &                \\ 
LBN991  &       --      &       0.5     &               &                \\ 
OrionB  &       PL      &       1.5     &       $\alpha$=-2.33 $\pm$  0.02      &       3.30     \\ 
OrionA  &       PL      &       1.5     &       $\alpha$=-1.84 $\pm$  0.02      &       4.33     \\ 
California      &       LN+PL   &       1.5     &       $\sigma$= 0.55 $\pm$  0.02, $\alpha$=-3.11 $\pm$  0.12, x$_0=$ 3.01  &       1.02     \\ 
Draco   &       --      &               &               &                \\ 
Serpens &       LN+PL   &       1.5     &       $\sigma$= 0.55 $\pm$  0.01, $\alpha$=-2.75 $\pm$  0.17, x$_0=$ 6.21 &       2.44     \\ 
Lacerta &       LN+PL   &       1.1     &       $\sigma$= 0.35 $\pm$  0.06, $\alpha$=-7.05 $\pm$  0.32, x$_0=$ 1.46 &       5.16     \\ 
CygOB7  &       ?       &       1.5     &               &                \\ 
Circinus        &       LN+PL   &       0.6     &       $\sigma$= 0.26 $\pm$  0.01, $\alpha$=-3.62 $\pm$  0.16, x$_0=$ 2.33  &       5.77     \\ 
CepheusOB3b     &       LN+PL   &       1.5     &       $\sigma$= 0.37 $\pm$  0.02, $\alpha$=-2.93 $\pm$  0.06, x$_0=$ 2.66  &       1.80     \\ 
Norma   &       LN      &       2.0     &       $\sigma$= 0.47 $\pm$  0.03      &       22.13    \\ 
L1307-35        &       LN+PL   &       0.5     &       $\sigma$= 0.46 $\pm$  0.01, $\alpha$=-2.97 $\pm$  0.08, x$_0=$ 1.48  &       3.84     \\ 
MonocerosR2     &       LN+PL   &       1.5     &       $\sigma$= 0.42 $\pm$  0.01, $\alpha$=-2.77 $\pm$  0.10, x$_0=$ 2.63  &       1.24     \\ 
MonOB1  &       PL      &       1.5     &       $\alpha$=-2.24 $\pm$  0.02      &       1.30     \\ 
IC5146  &       PL      &       1.5     &       $\alpha$=-2.07 $\pm$  0.07      &       1.69     \\ 
CephOB4 &       LN+PL   &       1.5     &       $\sigma$= 0.62 $\pm$  0.10, $\alpha$=-3.14 $\pm$  0.07, x$_0=$ 2.30 &       1.31     \\ 
Vela    &       LN+PL   &       1.0     &       $\sigma$= 0.92 $\pm$  0.03, $\alpha$=-4.23 $\pm$  0.28, x$_0=$ 4.48 &       12.16    \\ 
AFGL490 &       LN+PL   &       1.5     &       $\sigma$= 0.97 $\pm$  0.41, $\alpha$=-3.59 $\pm$  0.11, x$_0=$ 3.47 &       1.78     \\ 
L1340-55        &       LN+PL   &       0.5     &       $\sigma$= 0.41 $\pm$  0.00, $\alpha$=-2.40 $\pm$  0.07, x$_0=$ 1.52  &       6.15     \\ 
S140    &       LN      &       0.6     &       $\sigma$= 0.44 $\pm$  0.01      &       9.27     \\ 
IC1396  &       LN      &       0.6     &       $\sigma$= 0.38 $\pm$  0.00      &       6.91     \\ 
L1293-1306      &       LN      &       0.3     &       $\sigma$= 0.66 $\pm$  0.01   &       2.44     \\ 
Split   &       ?       &       2.0     &               &                \\ 
Ara     &       LN      &       1.5     &       $\sigma$= 0.53 $\pm$  0.02      &       13.64    \\ 
Cygnus  &       ?       &       3.0     &               &                \\ 
S147    &       LN      &       0.5     &       $\sigma$= 0.33 $\pm$  0.01      &       1.69     \\ 
Lagoon  &       ?       &       2.0     &               &                \\ 
M20     &       ?       &       3.0     &               &                \\ 
Rosette &       PL      &       1.5     &       $\alpha$=-2.23 $\pm$  0.06      &       7.37     \\ 
CMaOB1  &       LN+PL   &       0.4     &       $\sigma$= 0.87 $\pm$  0.02, $\alpha$=-3.10 $\pm$  0.58, x$_0=$ 2.87 &       2.28     \\ 
NGC2362 &       LN+PL   &       0.2     &       $\sigma$= 0.59 $\pm$  0.07, $\alpha$=-1.97 $\pm$  0.10, x$_0=$ 0.81 &       1.24     \\ 
L291    &       LN      &       1.5     &       $\sigma$= 0.75 $\pm$  0.08      &       15.22    \\ 
GGD4    &       LN      &       0.4     &       $\sigma$= 0.59 $\pm$  0.02      &       13.08    \\ 
NGC6604 &       ?       &       2.0     &               &                \\ 
M17     &       ?       &       2.5     &               &                \\ 
S235    &       LN      &       0.5     &       $\sigma$= 0.65 $\pm$  0.02      &       2.31     \\ 
IC443   &       LN      &       0.3     &       $\sigma$= 0.91 $\pm$  0.05      &       3.78     \\ 
W3-W4-W5        &       LN      &       0.5     &       $\sigma$= 0.62 $\pm$  0.01   &       4.17     \\ 
NGC6334 &       ?       &       5.0     &               &                \\ 
M16     &       LN      &       5.0     &       $\sigma$= 0.85 $\pm$  0.01      &       48.70    \\ 
Cartwheel       &       ?       &       1.5     &               &                \\ 
Cyg-West        &       LN      &       1.0     &       $\sigma$= 0.53 $\pm$  0.01   &       22.13    \\ 
GemOB1  &       LN+PL   &       1.5     &       $\sigma$= 0.40 $\pm$  0.01, $\alpha$=-4.50 $\pm$  0.17, x$_0=$ 3.00 &       3.54     \\ 
Maddalena       &       LN+PL   &       1.0     &       $\sigma$= 0.34 $\pm$  0.01, $\alpha$=-3.67 $\pm$  1.30, x$_0=$ 2.56  &       1.51     \\ 

\hline
\end{longtable}
\tablefoot{\\
\tablefoottext{a}{The best-fit shape for each cloud. 'LN' denotes log-normal, 'PL' power law, and 'LN+PL' a combination. Clouds that did not reach 2 mag were not fitted, and are marked with a long dash. Clouds that were not well fitted by any model are marked with a questionmark.}\\
\tablefoottext{b}{The limit of the last closed contour. The fits were performed above this limit.}
}

\end{document}